# Sir Arthur Eddington and the Foundations of Modern Physics

Ian T. Durham
Submitted for the degree of PhD
1 December 2004
University of St. Andrews
School of Mathematics & Statistics
St. Andrews, Fife, Scotland

*Dedicated to*

*Alyson*
*for living through it all and loving me*

*Nate & Sadie*
*for being you*

*Mom & Dad*
*my heroes*

*Larry & Alice*
*for constant love and support*

*Sharon*
*for everything said and unsaid*

*Maggie*
*for making 13 a lucky number*

*Gram D.*
*for always being interested*

*Gram S.*
*for strength and good food*

*Steve & Alice*
*for making Texas worth visiting*

# *Contents*

# I
## *Preface*

Albert Einstein's theory of general relativity is perhaps the most significant development in the history of modern cosmology. It turned the entire field of cosmology into a quantitative science. In it, Einstein described gravity as being a consequence of the geometry of the universe. Though this precise point is still unsettled, it is undeniable that dimensionality plays a role in modern physics and in gravity itself. Following quickly on the heels of Einstein's discovery, physicists attempted to link gravity to the only other fundamental force of nature known at that time: electromagnetism. Both Hermann Weyl (1885-1955) in 1918 and Arthur Stanley Eddington (1888-1944) in 1921 developed field theories that in essence were early attempts at unification employing the new concept of the geometrisation of physics. Also in 1921 the German theoretical physicist Theodor Kaluza (1885–1954) attempted this by extending Einstein's field equations to five dimensions (Kaluza 1921). Essentially he postulated a five dimensional Riemannian space by adding to the four known dimensions a fifth one where particles always followed closed paths. Both electromagnetism and relativity were contained within this grand scheme but it did not contain any of the relatively young quantum theory leaving most physicists to realize it bore no resemblance to reality. The Swedish theoretical physicist Oskar Klein (1894–1977) added the quantum aspect to Kaluza's theory in 1926 (Klein 1926) and similar subsequent theories have been loosely grouped into the category of Kaluza-Klein Theories. In Klein's theory the fifth dimension was unobservable whereas Kaluza's was macroscopic in size. This unobservable dimension's physical reality was akin to a quantity that was conjugate to the electrical charge. In this way Klein also sought to explain Planck's quantum of action. The lack of sufficient mechanisms for testing such an idea and finding a practical application for the theory kept Kaluza-Klein theories largely out of the mainstream until their revival in the 1970s. This did not stop many scientists from studying unification, however. Einstein essentially devoted the final thirty years of his life to it while Eddington devoted the last fifteen.

Unification today is widely regarded as the Holy Grail of physics. Physicists have successfully unified the strong, weak, and electromagnetic forces with special relativity under the guise of quantum field theory, but any definitive link to gravity or *general* relativity remains elusive. String theory is currently the mainstream theory of choice for this but remains unproven. Unifying gravity and quantum theory then must be at the heart of this quest, and theories of quantum gravity have been at the forefront of research in physics for nearly forty years. But attempts at such a unification actually date to at least 1928, when Paul Adrien Maurice Dirac (1902-1984) derived his relativistic equation for the electron (Dirac 1928a and 1928b). Eddington, disappointed that Dirac's equation did not appear in tensor form,[1] sought to reformulate Dirac's work in 1929-1930 to put quantum theory into the language of relativity, i.e. tensor calculus (Eddington 1929). This led to the development of several theories of cosmology in the 1930s developed primarily by Eddington, Dirac, and Edward Arthur Milne (1896-1950). Several unification theories that did not directly address cosmological questions were also developed at this time.

Eddington's work rested on the premise that quantum mechanics and relativity could be united under a common framework both in the formalism and the philosophy. He began by analyzing uncertainty and became convinced that its introduction into physics heralded such a monumental change that every physicist needed to consider its philosophical implications in their work. He clearly opposed the Einstein-Podolsky-Rosen (EPR) interpretation saying that any scientist who accepted the idea of hidden variables as an explanation of indeterminacy "wants shaking up and waking" (Eddington 1935b, p. 84). He saw this fundamental indeterminacy as the foundation on which to build a unified theory of physics.

Though his work on uncertainty is clearly debatable in its validity it actually foreshadowed some later developments in physics, including the need for a quantum-mechanical standard of length. This led him to the next major component of his work: an analysis of The Pauli Exclusion Principle. He develops exclusion into a richer framework that serves, in combination with uncertainty, as the basis of later versions of his complete theory, which held that physical events depend solely on dimensionless ratios. Later, this idea was taken up by Dirac in proposing his Large Numbers Hypothesis (Dirac 1937).

---

[1] Charles Galton Darwin (1887-1962) was the first to note that Dirac's equation was not in tensor form; see C.G. Darwin, "The Wave Equation of the Electron," *Proceedings of the Royal Society* [A] **118** (1928), 654-680.

Eddington's work hinted at some of the underlying principles of modern theories including some aspects of grand unified theories (GUTs) and string theory. In fact, as this monograph discusses, *Fundamental Theory*, as it was posthumously titled, is a very early attempt at quantum field theory that quite remarkably predicts future advances in that field. It's greatest relevance to modern science is in its unique interpretation of the foundational aspects of modern physics and its philosophical implications for the underlying structure of the physical world. In fact Eddington's work has seen somewhat of a renaissance in recent years and has been studied in greater detail by a growing list of scholars.

I first discovered this aspect of Eddington's work when reading a brief account of his cosmology in Helge Kragh's *Quantum Generations*. I had known of Eddington from my work in astronomy for his many mainstream accomplishments, but this brief encounter with his unorthodox worldview turned my research from work on general problems in cosmology to addressing truly foundational problems in modern physics.

The results of my initial foray into his work on uncertainty, that reveal a deep distrust of standard measurement techniques and a worldview incorporating uncertainty into the very fabric of space-time, led naturally to his extension of the Exclusion Principle. One of the many amazing insights that continued to fuel my work was the fact that Eddington modifies the interpretation of this fundamental principle and extracts results from the new interpretation that point to a deeper philosophical meaning behind exclusion. This presented me with several fundamental questions about the nature of exclusion: could it be more than a relatively straightforward quantum phenomenon; could it reside in that fundamental area inhabited by the conservation laws, the forces of nature, and the uncertainty principle, and, if it does, what does this mean for modern physics? My conclusions in this endeavour have led to several extended pieces of research in fundamental physics that, in itself, emphasizes the surprising relevance of his work despite its chequered past.

Examining these questions is not only important for a complete understanding of exclusion and Eddington's unorthodox worldview, but they are also at the heart of the relationship between science and philosophy.  When analyzed in full compliment with his work on uncertainty, the whole of his thinking begins to unravel itself.  To say that Eddington went from being one of the subjects of my dissertation to being the *only* subject of my dissertation does not do proper justice to his influence on me.  Delving into the deep questions of uncertainty and exclusion, particularly in the context of unification and the nature of the universe itself, his work has led me into many new uncharted areas and has helped to focus my general research interests onto more fundamental and foundational questions.  But aside from my personal interest in the subject, Eddington's philosophical and even some of his mathematical work is often overlooked by modern scholars.  Bohm, Fred Hoyle (1915 - 2001), and Hermann Bondi (b. 1919) are well-known despite their controversial theories making up a large portion of their body of work, while Eddington, whose diverse work included the first observational verification of general relativity and the nearly single-handed creation of the field of stellar structure, tends to be overlooked and even marginalized.[2]  It was this historical treatment that contributed to my focus solely on Eddington.

My research concentrated primarily on what comprises the first six chapters of *Fundamental Theory* and is often referred to as his statistical theory.  These six chapters focus their efforts on reinterpreting and applying Heisenberg's Uncertainty Principle and Pauli's Exclusion Principle.  They form the philosophical and interpretive basis of his entire program of research.  I analyzed each in detail both philosophically and mathematically in search of any morsel of truth or potential application to modern physics.  The formalism of the latter chapters cannot be understood without the contextual basis the early material provides.  A direct result of his statistical theory was the derivation of many of the known constants in the universe (hence the derisive label of numerology) and I analyzed these in detail as well.  Putting all of this together I examined the impact on unification, particularly from a modern quantum field theoretic sense, attempting to determine if Eddington could have been on the right track with anything. The portions that do have relevance to the foundations of modern physics I then examined in depth. I have three published papers (Durham 2004, 2003a, and 2003b) on the subject but, other than the most recent, they bear little resemblance to what follows since my work has matured and evolved over the years as my understanding of physics itself has done the same. A robust and lengthy treatment of Eddington's statistical theory from *Fundamental Theory* comprises the nine chapters of this text.

---

[2] One slightly elderly physicist who devotes much of his time to the history of physics these days, remarked to me at a conference once that when he thought of Eddington he always thought of numerology.  Helge Kragh, a noted historian of modern physics devotes nearly all of his discussion on Eddington in *Quantum Generations: A History of Physics in the 20th Century* (Kragh 1999) to Eddington's cosmology and very little to his more mainstream, and arguably more influential, works on relativity and stellar structure.

One final note I wish to make is that, despite its title, Clive Kilmister's 1994 book *Eddington's Search for a Fundamental Theory: A Key to the Universe* is devoted more fully to an analysis of Eddington's 1936 book *Relativity Theory of Protons and Electrons* that laid the groundwork for *Fundamental Theory* published a decade later. Much was changed in the theory in the final eight years of Eddington's life and the two years between then and publication. Kilmister and B.O.J. Tupper did analyze the statistical components of *Fundamental Theory* in the early 1960s (Kilmister and Tupper 1963) but more from the perspective of their own research that built upon Eddington's. It was also prior to many of the major advances in quantum field theory. As such, my work is the only comprehensive study of the statistical portions of Eddington's *Fundamental Theory* that puts it into historical perspective and the only study that compares it to quantum *field* theory rather than quantum *mechanics* and relativistic cosmology.

I wish to thank numerous people in helping me to complete this work. In addition to everyone to whom this text is dedicated, I wish to thank my doctoral advisors at St. Andrews, Prof. Edmund Robertson, FRSE, and Dr. John O'Connor. During a recent session on advising for new faculty members at Saint Anselm College where I now teach, participants were asked about their best and worst advising experiences as students. I was the only one whose best experience was with their doctorate. Edmund and John have gone well beyond what I expected of them and made this experience a truly pleasurable one. I also wish to thank Simmons College in Boston for employing me as a full-time instructor for the vast majority of my time as a doctoral student. In addition to providing my family and I with much-needed financial resources Simmons provided excellent facilities for building my research and developing my pedagogical style. As such, thanks must also go to Saint Anselm College in New Hampshire, where I now reside as a tenure-track faculty member.  Saint A's has offered me a wonderful place to ply my interdisciplinary trade as a philosopher and historian of physics while also providing me with a fantastic group of students with whom I can discuss physics at length.

Thanks are also in order for Roger Stuewer for careful editing of one of the papers that formed part of this treatise, a process in which I learned a great deal; Kate Price for organizing the most stimulating conference of my professional career, one that actually had a greater impact on my work than anything else; John Amson, one of the founders of the Alternative Natural Philosophy Association (ANPA) with Clive Kilmister and a former Saint Andrews (the *other* Saint A's!) professor, for encouraging my interest in Eddington and welcoming me into his wonderful home on the coast of Scotland to pick his brain for an afternoon; Meg Weston Smith, the daughter of E.A. Milne, who has generously provided me with her friendship and hospitality not to mention a treasure-trove of information relating to her father and his relationship with Eddington (Milne's kinematic relativity will be a future project for me); and Alan Boufford for teaching me the organizational skills I needed to finish this.

Finally, I wish to acknowledge the numerous libraries and archives I accessed along the way including the libraries (including archives and other resources) of the University of St. Andrews, Simmons College, Saint Anselm College, MIT, the Royal Society of London, the Royal Astronomical Society, and the American Philosophical Society.  I also wish to thank the Kennebunk Free Library in Kennebunk, Maine, for a quiet place to work.  I have four generations of librarians in my family and librarians are often unsung heroes.

Ian T. Durham
October 2004
Kennebunk, Maine
9

# II

## *Eddington's Life and Worldview*

No study of Eddington's work is complete without a brief description of his life and his worldview since these were vital in shaping his research. In particular I will examine aspects of his life that are directly relevant to the development of *Fundamental Theory* and the ideas behind it. A full account of his life can be found in several texts including the standard biography of him done by his former student A.V. Douglas in 1956.

**Eddington's Life in a Nutshell**

Eddington was born 28 December, 1882, in Kendal, Westmoreland, England. He was the second child and only son of Arthur Henry Eddington, who was the headmaster of the Stramongate School, the Quaker (Society of Friends) school where the chemist John Dalton (1766–1844) once taught (Douglas 1956, p. 103). Eddington was a very intelligent child with a curious intellect and an aptitude for numbers: he attempted to count the words in the Bible and mastered the 24 x 24 multiplication table before he could read (Douglas 1956, p. 2 and Plummer 1948). He obtained a three-inch telescope a bit later thus launching his lifelong study of the heavens (Smart 1945 and Plummer 1948). His schooling began at the Brynmelyn School in Weston-super-Mare (1893-1898), where his family had moved shortly after the untimely death of his father in 1884 from typhoid. He then attended Owens College, Manchester (1898-1902), managing to circumvent the rules that prohibited those under the age of 16 from attending. His professors at Owens included mathematician Horace Lamb (1849-1934) (see Figure 1.) and physicist Arthur Schuster (1851-1934). In 1902 he received his B.S. from Owens and then moved to Trinity College, Cambridge, on a scholarship, where he studied under E.T. Whittaker (1873-1956), the person who later compiled *Fundamental Theory*, A.N. Whitehead (1861-1947), and E.W. Barnes (1874-1953), all mathematicians. In 1905 in addition to receiving his M.A., he spent a term working in the Cavendish Laboratory, then under the direction of J.J. Thomson (1856-1940), where he very nearly made a career for himself in physics. In 1907, he was awarded the Smith's Prize and elected as a Fellow of Trinity College (see Figure 2.). In January of 1906, he was appointed as Chief Assistant at the Royal Observatory, Greenwich, to succeed Sir Frank Dyson (1868-1939) who moved to Edinburgh to take up the post of Astronomer Royal for Scotland. Once there he began working on a research project that had actually started in 1900, studying photographic plates of the minor planet Eros.

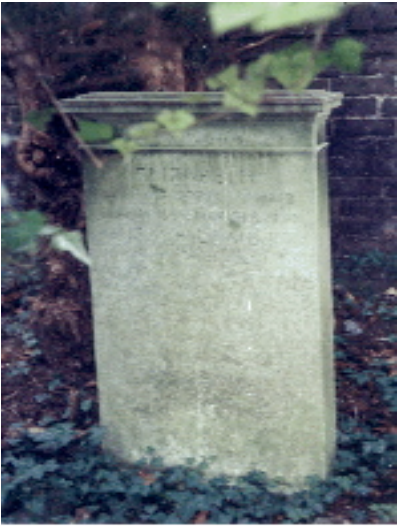

**Figure 1.** Horace Lamb's headstone in Ascension burial ground (formerly St. Giles' cemetery) in Cambridge, just a few hundred feet from Eddington's (photo by author).

Eddington completed the project determining an accurate value for solar parallax. Much of his early work at the Observatory has been overlooked but Smith gives a nice detailed account of much of this in his recent paper (Smith 2004). Once again, in 1909, he was tempted by physics but turned down Schuster's offer of a position at Manchester because, as he explained in a letter to Schuster, he preferred the observational work associated with astronomy (Eddington to Schuster 1909). He returned to physics later in his career but in the next chapter and beyond I contend that his philosophical outlook was primarily

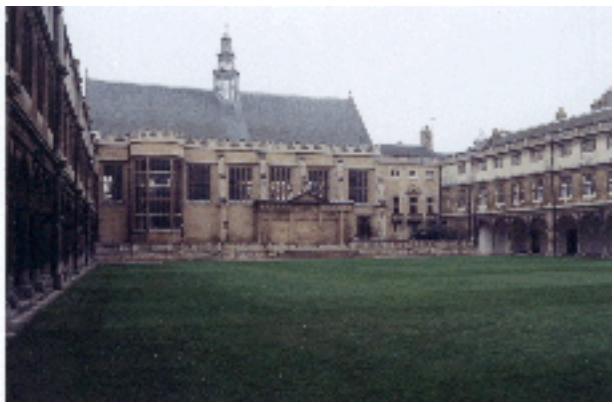

**Figure 2.** Neville's Court, Trinity College, Cambridge, 2004 (photo by author).

shaped, albeit indirectly and perhaps subconsciously, by his astronomical work.

In 1913 Eddington succeeded Sir George Darwin (1845-1912), son of evolutionist Charles Darwin (1809-1882), as the Plumian Professor of Astronomy and Experimental Philosophy at Cambridge. The following year he also succeeded Sir Robert Ball (1840-1913) (see Figure 3.) as Director of the University Observatory and was elected a Fellow of the Royal Society. As Director of the Observatory he was given the right to live in the east wing of the Observatory's main building. He moved in on Lady Day, 1914, with his sister Winifred (1878-1954) and his mother Sarah Ann (d.1924). He had previously been elected to the Royal Astronomical Society in 1906 (he was nominated by Whittaker) and served as its Secretary from 1912 to 1917, as its President from 1921 to 1923,

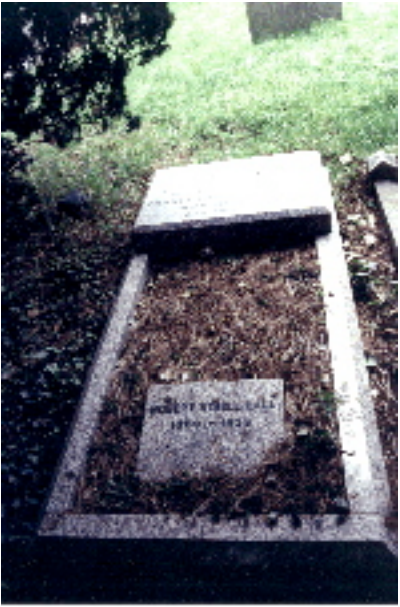

**Figure 3.** Sir Robert Ball's headstone is at the top (his son's is the lower stone). This is also in Ascension burial ground less than twenty feet from Eddington's (photo by author)

and as its Foreign Secretary from 1933 until his death. He received the Society's Gold Medal, its highest honour for achievement in astronomy and geophysics (2 are awarded each year) in 1924. In the same year the Astronomical Society of the Pacific awarded him its Bruce Medal. The Royal Society (of London) awarded him its Royal Medal in 1928, a medal originally chartered by King George IV. Shortly thereafter, in 1930, he was knighted. His knighthood status was that of Knight Bachelor as he never married. 1938 was once again a pivotal year that saw him made President of the International Astronomical Union and the recipient of yet another of Britain's highest honours, the Order of Merit.

**The Meticulous Man**

By 1917 he was fully absorbed in operations at the Observatory, though the ongoing war on the continent continued to knock on his door (see below). During this time he completed the Cambridge Zone Catalogue of Observations using telescopes housed in the Observatory. He was well-known within the astronomy community but he was about to achieve a fame and notoriety that, while not universal like Einstein's, makes him one of the pivotal figures in twentieth century science. In 1919, in part due to an attempt at deferment from active military duty, he organized and participated in an observation of a total solar eclipse (in fact, two observations) in an effort to determine if Einstein's relatively newly minted theory of general relativity was accurate. The resulting report (Eddington 1919) is legendary – it gave scientists as well as the general public physical evidence that overturned long-held concepts of space and time. Eddington followed this up by publishing in 1923 *Mathematical Theory of Relativity* described by Einstein as the "finest presentation of the subject in any language" (as quoted in Douglas 1956). His major contribution to astrophysics, *The Internal Constitution of the Stars*, that summarized his pioneering work on stellar structure, appeared in 1926. This was followed two years later by his first book aimed at a non-technical audience: *The Nature of the Physical World* based on the Gifford Lectures he had given at the University of Edinburgh. He subsequently published four more non-technical books, the impact of which has been recently discussed by Whitworth (Whitworth 2004). His attention to detail is noteworthy. His lifelong interest in cycling was well-known (see Figure 4.)

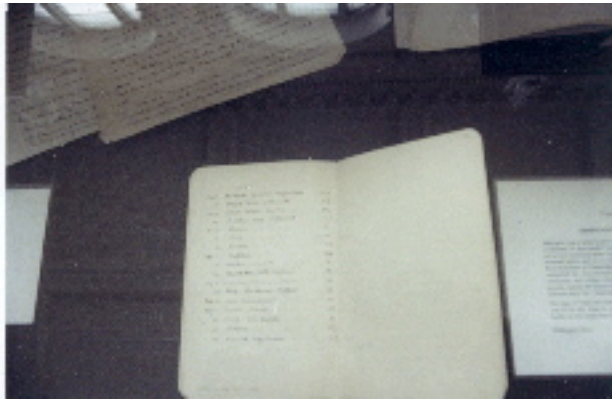

**Figure 4.** Eddington's Cycling Log at the Wren Library, Trinity College, Cambridge (photo by author).

and gives an excellent example of this in the form of his meticulous cycling logs. His meticulous nature was not confined to the taking of data: he was a master at language, in English as well as the classics.[3] He was well versed in literature and wrote poems for his own amusement that were considered quite good by those lucky enough to steal a glimpse of them. He was a fan of P.G. Wodehouse (1881-1975) and mathematician Charles Dodgson (1832-1898) who was better known by his penname Lewis Carroll. Eddington often made up grammatically correct sentences in Carroll's style that made no sense. Subramanyan Chandrasekhar (1910-1995) gives an example directly told to him by Eddington: "To stand by the hedge and sound like a turnip" (Chandrasekhar 1983). The fact that this sentence is both ridiculous and grammatically correct might serve for many as an excellent metaphor for *Fundamental Theory*, which is largely internally consistent but appears ridiculous to some upon examination (I will show it is not nearly as ridiculous as it may first appear).

Dodgon's influence on Eddington extends beyond the literary and into the realm of mathematics. He had a penchant for writing mathematical puzzles and his most famous, the Zoo Puzzle, which still circulates among puzzle enthusiasts, was written in the style and used the characters of Lewis Carroll (see Appendix A for a detailed description and solution). Some of Eddington's own more light-hearted literary fare was actually published, for example:

> There once was a brainy baboon,
> Who always breathed down a bassoon,
>     For he said, "It appears
>     That in billions of years
> I shall certainly hit on a tune (Eddington 1935b).

Harold Spencer Jones (1890-1960) and Edmund T. Whittaker (1873-1956) recalled his "retentive memory for the apposite quotation" and his fondness for Shakespeare, having been a member of The Elizabethans, a small private society devoted to The Bard at

---

[3] Though he was a consummate mathematician, he did say "If I sometimes employ pure mathematics, it is only as a drudge; my devotion is fixed on the physical thought which lies behind the mathematics" (Eddington 1939, p. 74).

Greenwich (Spencer Jones and Whittaker 1945). He was addicted to solving crossword puzzles in *The Times* and the *New Statesman and Nation* and rarely took more than five minutes per puzzle (Chandrasekhar 1983). His linguistic skill is well represented in the five popular books on science and philosophy that he had written.

But in addition to being a gifted writer, Eddington was a careful and concise mathematician, as I will explore in greater detail in the coming chapters. In truth, though Eddington's work was primarily in astronomy and physics, he was an applied mathematician. Evidence of his application of complex mathematics to even seemingly mundane astronomical problems appears fairly early in his work. For example, in an article that appeared in *Monthly Notices of the Royal Astronomical Society* in 1910, he attempted to fit three-dimensional paraboloids to fuzzy two-dimensional photographs (Eddington 1910). The attention to detail required of such a task is immense and is only one example of his ability to focus intently on certain subjects, sometimes to the exclusion of others. His meticulous nature is even apparent in his handwriting, which is extremely neat and compact, and in his arguments, which attempt to be wholly logical, though passion and zeal do creep in (see Larmor discussion below).

**Meticulous yet Much Maligned**

Despite his faith in mathematics, Eddington always searched for physical justifications for his work, despite its deductive nature. Moreover, his philosophy of science was far from unusual at the time, although in his later years his contemporaries often viewed his views unfavourably. His work was - and still is - regarded as heterodoxical, both philosophically and scientifically (Kragh 1999). His harshest critic in Britain was the astrophysicist and philosopher Herbert Dingle, who referred to the theories of Eddington, Milne, Dirac, and others as the "pseudoscience of invertebrate cosmythology" (as quoted in Kragh 1999). Eddington's mathematics, however, were fairly standard and always rigorous. Thus, the charge of heterodoxy can only be applied to his physics. But his was hardly the only physical theory before or since that was regarded as heterodoxical. As I have mentioned several times, Dirac's Large Numbers Hypothesis (LNH) was just such a theory and even Dirac's theory of electron-holes, which is still used to teach elementary quantum mechanics, is fairly heterodoxical to a quantum field theorist.

In fact to Eddington's credit he employed a number of methods that are in use today, including Clifford algebras. A little-known fact is that he introduced chirality to particle physics. He also is sometimes credited with an independent discovery of Majorana spinors (Chandrasekhar 1983), which are often described (somewhat incorrectly) as the square roots of vectors since the vector representation appears in the tensor product of two copies of the spinor representation. A Majorana spinor is actually the real representation of a non-complexified Clifford algebra (while a Dirac spinor is the fundamental representation of a *complexified* Clifford algebra). Ettore Majorana (1906-1938) is credited with their invention (Majorana 1937). Despite appearing in 1937, the substance of this paper was rumoured to have been written in 1932/33 but confined to a desk drawer until 1937 (Recami 1999). Its appearance came about a year before Majorana disappeared under suspicious circumstances on a boat ride from Naples to Palermo. Two letters left behind hint that suicide may have been contemplated, though kidnapping was also speculated as a possibility as was an intentional disappearance since he was working for the fascist regime in Italy to develop an atomic weapon (Wikipedia: The Free Encyclopaedia).[4] In any regard, his premature presumed death could have led to his receiving primary credit for the discovery as a recognition of his achievement of sorts, particularly considering that Eddington's professional reputation was beginning to suffer in the late 1930s as his work on cosmology consumed the majority of his life. Eddington's contribution could also have easily been overlooked thanks to his liberal use of novel notation, derivations, and nomenclature (something that can make reading Eddington's technical works difficult).

In other ways Eddington was ahead of his time – and is seldom recognized for it. He recognized the need for a quantum-mechanical standard of length nearly fifty years before one was adopted, and he advocated the inseparability of an object and its environment (not to be confused with the inseparability of the object and its *observer*), a philosophical stance that appears in some versions of quantum field theory. His idea of linking quantum mechanics and relativity through coordinates compares well with modern links through topology. Quite apart from his work on cosmology, however, he almost single-handedly founded the field of stellar structure (c.f. Chandrasekhar 1983). About those critics who laughed at Eddington, Einstein[5] once said: "Why should they laugh? They have never done what he has done!" (Douglas 1956, p. 146).

---

[4] Enrico Fermi said of Majorana: "*There are many categories of scientists, people of second and third rank, who do their best, but do not go very far. There are also people of first class, who make great discoveries, fundamental for the development of science. But then there are the geniuses, like Galilei and Newton. Well, Ettore Majorana was one of them...*" (Wikipedia: The Free Encyclopaedia).

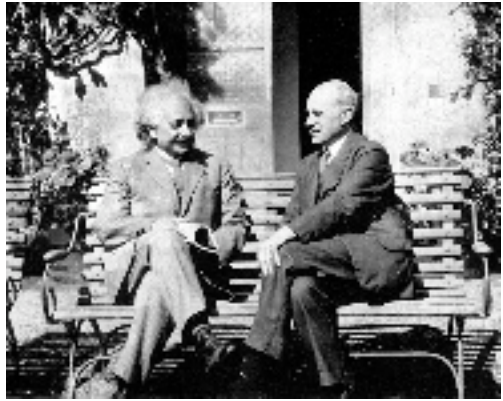

**Figure 5.** Einstein and Eddington in the Cambridge Observatory garden in 1930 (photo by Winifred Eddington, courtesy Institute of Astronomy, Cambridge).

**Formation of the Eddington Worldview**

Not even the best science occurs in a vacuum. After all, scientists are humans and subject to real human emotions, thoughts, influences, and the like. Therefore it would be remiss of me not to spend a bit of time discussing a few key influences in Eddington's career, specifically those that helped shape *Fundamental Theory*. Since his spirituality was such an integral part of his life it is important to consider the influences in this area as well since his spirituality and science are intertwined. Matthew Stanley recently completed a PhD dissertation at Harvard University entitled *Practical Mystic: Religion and Science in the Life of A.S. Eddington* (Stanley 2004b). His work discusses many of these points in much greater detail and I refer readers to that work if they are particularly interested in Eddington's spiritual side.

---

[5] Einstein visited Eddington in Cambridge in 1930, the year Eddington was knighted. He is rumoured to have played violin for Eddington, his sister, and his mother at some point during this visit (see Figure 5.).

Before venturing into the realm of religion and mysticism, more worldly influences will be discussed. Two of the more prominent influences on Eddington were his association with Sir Ralph Fowler (who eventually became Lord Rutherford's son-in-law) and his upbringing in the late Victorian age. These influences are, in fact, shared by Milne and Dirac (Durham 2003a). Fowler influenced Eddington both directly and indirectly though he was seven years Eddington's junior. In particular the two collaborated on research in super dense matter primarily as it relates to white dwarf stars, though they never published a joint paper.[6] As I show in the final few chapters of this text this work on super-dense matter was one of the key points in the development of *Fundamental Theory*, particularly the statistical aspects. Fowler's scientific contribution here was a general solution to Emden's equation of state, something Eddington had found a specific solution to. It was Fowler who introduced Dirac to quantum mechanics (and to Werner Heisenberg) and it was Dirac's paper describing the relativistic theory of the electron that helped launch Eddington's work on *Fundamental Theory* (see Ch. 4). Fowler was also a Fellow of Trinity College at the same time as Eddington and the two had frequent contact, though Eddington's work was primarily at the Cambridge Observatory while Fowler's was at the Cavendish Laboratory. Fowler's general influence on Cambridge physics is remarkable – during the period 1922-1939 he supervised sixty-four students for an average of eleven at any one time. Fifteen of these became Fellows of the Royal Society and another three became Nobel Laureates. Far from being impossible to locate due to his tremendous commitments, most of his students found themselves in a close relationship with him.

Like Eddington, Milne, and Dirac, Fowler was also raised in the late Victorian age. Two aspects of Victorianism had a direct influence on Eddington's work. The first was the strict social system that had religion at its forefront. The second was a mystical fascination with the unknown (Durham 2003a). Focusing on the latter, the late Victorian age was particularly relevant to an understanding of Eddington's formative years. H.G. Wells (1866 – 1946), who was one of the original science fiction literaries and who began writing in the 1890s, pointed out that the major change during the Victorian age was the shift in worldview from mystically religious to more cosmic in nature (Frayter 1997) incorporating new views of time, space, and evolution. Time and space began their new lives starting in 1854 with Bernhard Riemann's (1826 – 1866) publication of non-Euclidean geometry and development of hypersurfaces (Eddington was the type who might have noticed that Wells was born the same year Riemann died). To explain his hypersurfaces he created fictional creatures called flatlanders who could only live in two dimensions but who could experience a third dimension as a force or feeling. The concept was introduced to England in 1884 by Dr. Edwin Abbott Abbott (1838 – 1926), a religious scholar, in his novel *Flatland* in which flatlanders were one of many dimensionally limited species. The novel was hierarchical in that it displayed the caste-like social system of Victorian England.

---

[6] In fact, of Eddington's 265 published papers, 93 of which appeared before the famous eclipse paper, a mere 9 list co-authors.

The new evolutionary (as well as aspects of the cosmic) views of the Victorian age began appearing in a series of purely sectarian works on science known collectively as the *Bridgewater Treatises on the Power, Wisdom, and Goodness of God as Manifested in the Creation*. They were published after the death and at the bequest of the 8th Earl of Bridgewater in 1829. Paul Frayter describes Wells' *The Island of Dr. Moreau* as an anti-Bridgewater Treatise where the beneficent God was replaced by a vivisectionist (Frayter 1997). The year 1869 saw the publication of Frances Galton's (1822 – 1911) *Hereditary Genius* in which he first formulated the idea of eugenics or the improvement of the human species through selected parenthood. Wells was also inspired by Galton's work to write the dark and alien *First Men in the Moon*. Galton was a cousin of Charles Darwin (1809 – 1882) and was directly influenced by his evolutionary work. Darwin's son Horace (1851 – 1928; who is yet another famous scientist buried in the same cemetery as Eddington – see Appendix B), founder of the Cambridge Scientific Instrument Company, worked closely with Fowler, in particular, and Darwin's grandson Charles Galton Darwin (1887 – 1962) was the first person to notice (publicly) that Dirac's relativistic theory of the electron was not in tensor format, a key point of influence in Eddington's early musings on *Fundamental Theory*. Indeed one could quite possibly write an entire treatise on the varying influences of the entire Darwin/Galton clan from 1850 to 1950, but what is important here is the dramatic display of fanciful thinking, much of which was directly or indirectly inspired by the religious fervour of the time as well as the desire to reconcile the new discoveries in science and mathematics with the concept of a deity. In essence there was a sense that religious and scientific "seeking" were two parts of the same whole. This is a concept that was nearly dogmatised in turn-of-the-century Quaker reformism, something that impacted Eddington's life directly.

His family being devout Quakers, Eddington was exposed to Quakerism and its embodiment of a philosophy of peace (specifically pacifism) and inner harmony, not unlike some Eastern philosophies, early in life. Stanley (Stanley 2004a) says that in keeping the lines of communication with Germany open in difficult times "for him, Einstein and relativity were his contribution as a Quaker to world peace." Reportedly, several of Einstein's papers on relativity were smuggled to Eddington via the Dutch astronomer Willem de Sitter (1872 – 1934). As Eddington lived during two world wars it is not unexpected that it became a source of tension for him at times. For instance, during the first World War of 1914-1918, conscientious objectors were placed in camps. Due to their pacifistic creed Quakers usually became conscientious objectors, and Eddington had friends who were sent to pick potatoes in agricultural camps (Chandrasekhar 1983). Most (though not all) authors attribute his success in receiving a deferment to Dyson. The actual circumstances, though involving Dyson to some extent, also involved mathematical physicist Joseph Larmor (1857–1942) who attempted to utilize contacts at the British Home Office to have Eddington deferred on the grounds that it was not in the national interest to have a distinguished scientist in the Army.[7] Eddington was perfectly willing to go to jail to avoid serving (Stanley 2004a). Whether he was in jail, in a camp picking potatoes, or in the field fighting, the scientific establishment in Britain did not want one of their pre-eminent young members doing anything but serious science. Eddington's pacifist convictions proved too great for him to ignore and he was compelled to add a postscript to Larmor's letter to the Home Office, saying that if he was denied an exemption on the grounds of his usefulness to British science, he would claim conscientious-objector status. "Larmor and others were very much piqued" (Chandrasekhar 1983). Eddington's action thus led to a short, heated exchange of letters with Larmor, who insinuated that conscientious objectors held pro-

---

[7] Physicist Henry G. J. Moseley (1887–1915) had been killed at Gallipoli, as Larmor reminded the Home Office. Moseley had worked with C.G. Darwin and Ernest Rutherford in Manchester and had used X-ray spectra to study atomic structure, which laid the groundwork for ordering the elements in the periodic table.

German views (c.f. Eddington 1916a). Eddington never begrudged Larmor a thing, however, and the two became friends in later years (c.f. Eddington 1932).

A similar pattern was repeated with Milne, one of his strongest professional antagonists, yet one of his closest personal friends. For example, when Eddington reviewed one of Milne's books, he wrote to Milne, saying: "I realise that the review can scarcely be pleasing to you; but I hope you will recognise that it might have been worse if (holding the opinion I do) I had let myself go without regard to our friendship" (Eddington 1935a). In this way Eddington was able to separate the person from the opinion, something most of us find supremely difficult. This ability to divide people from their opinions is seemingly at odds with Quakerism's unifying principles. However, it is not really a division but simply the ability to look past the opinion and see the person beyond. As A. Ruth Fry describes in *A Quaker Adventure* there is in Quakerism a "belief in the potentiality of the divine in all men – the Inner Light, as we call it, which is in every man, no matter how hidden or darkened it may be" (Fry 1926).

It is the unifying aspect in Quakerism that promoted values that provided Eddington with a unifying approach to the world around him and the people in it. In this way it was impossible for Eddington to completely separate science and religion. Quaker reformers at the turn of the century promoted the idea that the "scientific spirit could, and should, be combined with a religious outlook." (Stanley 2004a). Similarly J. Rendal Harris, a Cambridge palaeographer and Quaker reformer said in 1896, "This theory of the detachment of science and religion from one another never has been a working theory of the universe; the two areas must overlap and blend, or we are lost" (*Report* 1896). In fact, Eddington found at least one overt way in which science and religion agreed: "The scientist and the religious teacher may well be content to agree that the value of any hypothesis extends just so far as it is verified by actual experience" (Eddington 1925a, 222). It was on this piece of overt agreement that his scientific worldview took shape. Despite the purely deductive nature of his later works his reliance on matching theory to experiment (experience) remained a central theme in his search for a fundamental theory.

Experience was one of the primary values promoted by modern Quakerism. The so-called "third way" of the Quaker renaissance maintained that religion should sprout from the individual experience of what was called the "Inward Light" and not from dogma (Stanley 2004a). He maintained that, in fact, that science is unable to quantitatively measure every aspect of human experience, specifically consciousness, yet it is just as real as the portions of human experience science *can* measure (similar ideas are explored by Bohm – see Bohm 1980 – and these similarities are discussed in Ch. 3). As Stanley puts it, "The proof of science was in empirical measurement; the proof of religion was in empirical experience" (Stanley 2004a, p. 49). Thus, scientifically, not everything is certain since science can't *measure* everything. Once again, the Quaker outlook provides guidance: "one should not try to find complete certainty because this leads to stagnation and a refusal to accept new ideas" (Stanley 2004a, p. 50). As a scientific principle and working philosophy, Eddington found this view embodied in Heisenberg's Uncertainty Principle that served as one of the foundations for his attempt to apply his unifying worldview to modern physics. This, along with the Pauli Exclusion Principle, formed the backbone of his theory. Essentially he reinterpreted the uncertainty and exclusion principles on relativistic grounds and utilized the combination as the foundation on which to build a self-consistent theory of quantum mechanics that was fully compatible with relativity – basically a theory of quantum gravity though, as I will show, it turned out to be more like quantum field theory and his motivation had its roots in his work on stellar structure and stellar motion. In order to better understand what he did it is necessary first to exam his work philosophically as well as examine the history of modern physics leading up to the development of *Fundamental Theory*, including the contributions he personally made. But first, a philosophical examination of his work is in order.

# III

## *A Philosophical Analysis of Eddington's Work*

Eddington's philosophy actually manifests itself both explicitly and implicitly, and differently in each respect.  Explicitly he quickly adopted the concept of uncertainty, as embodied in Heisenberg's Uncertainty Principle, as a cornerstone of his work.  Implicitly he worked from the notion that a truly comprehensive theory of the universe would need to be derivable purely from first principles.  In essence he felt deductive reasoning was the only path that would lead to a truly objective description of the world.  His philosophy is most fully developed in the *Philosophy of Physical Science*, an expanded version of the Tarner lectures he delivered at Trinity College, Cambridge in the spring of 1938, and in *The Nature of the Physical World*, an expanded version of the Gifford lectures he delivered at the University of Edinburgh in 1927 as well as portions of *New Pathways in Science*, an expanded version of the Messenger lectures delivered at Cornell University in 1934.  However the substance of these books implicitly announces its presence in all his major works since it is the substance that comprises his working philosophy.  I shall look specifically at a few key points from these treatises before analyzing Eddington's work from my own philosophical viewpoint.  Along the way I will occasionally contrast these.  I also continue to discuss Eddington's philosophy throughout this monograph as I discuss specific details of *Fundamental Theory*.

**The New Quantum Theory**

In the *Nature of the Physical World* Eddington was riding the crest of the new quantum theory mostly developed in the 1920s with so many important breakthroughs being made in 1926, the year he delivered the Gifford lectures that would later become *Nature*. Indeed he writes that he was preparing these lectures a mere twelve months following Heisenberg's groundbreaking paper on uncertainty.

The seeds sown early in his career by statistics are evident throughout.  He says we "must not think about space and time in connection with an individual quantum; and the extension of a quantum in space has no real meaning" (Eddington 1927 [1958], p. 201). This is because it is rather an ensemble of quanta gathered to form a "quorum" that produce statistical properties that ultimately give rise to observable quantities. Note that though he is talking about photons in this case (he is discussing one of his favourite subjects: the star Sirius) he uses the term "quanta" which *we* would just as likely apply to *any* particle, unfortunately sometimes including particles of extended size. This perhaps unintentional slip should be checked and quanta should be applied only to elementary particles that theoretically have no size (e.g. electrons). In this sense it *is* meaningless to speak of the extension of a quantum in space though one should note that Eddington was speaking about a specific example (Sirius) that was not well understood at the time.

Nonetheless he is clearly speaking in statistical terms. In such terms considerably less certainty is given to specific claims or observations – even whole models: "One must not expect too much from a model…" (Eddington 1927 [1958], p. 205). Similarly he expresses scepticism of the predictions made by the relatively new Bose-Einstein statistics in relation to wave theory saying "at least that seems to be the physical interpretation of the highly abstract mathematics of their theory" (Eddington 1927 [1958], p. 203). Though it seems odd that abstract mathematics would turn him off in light of his own mathematical ramblings in *Fundamental Theory*, his scepticism is in line with his general expressions of uncertainty.

He also finds in these new developments limitations on the measurements to be made in science. Ironically he suggests that there might be a limit to the application of numerical results to portions of scientific theory, specifically citing Dirac. Of course Eddington later perceived the numerical predictions of his theory to be one of its great triumphs. But in 1927 he was searching for a deeper interpretation of the new quantum theory regardless of its source. In Dirac's work of 1926 and early 1927 he found what Eddington termed a "non-arithmetical" calculus for the governing laws of the universe providing a purely symbolic beginning to the new quantum theory where the symbolism was unrelated to specific arithmetical operations.

To be clear there is a difference here between the numerical results Eddington is speaking of and the numerical results he later predicted in his theory. The latter were derived directly from the theory while the former were directly measured. Eddington found the latter expressed in Dirac's work. In summary,

> The fascinating point is that as the development proceeds actual numbers are *exuded* from the symbols. Thus although $p$ and $q$ individually have no arithmetical interpretation, the combination $qp - pq$ has the arithmetical interpretation expressed by the formula … [$qp - pq = ih/2\pi$]. By furnishing numbers, thought itself non-numerical, such a theory can well be the basis for the measure-numbers studied in exact science. The measure-numbers, which are all that we glean from a physical survey of the world, cannot be the whole world; they may not even be so much of it as to constitute a self-governing unit (Eddington 1927 [1958], p. 210).

The symbolic interpretation was no less physical. In fact it is here that we see Eddington's continuing use of an aether-like quantity of which I will discuss more in coming chapters. At this point Eddington refers to it as a sub-aether and describes quanta as oscillations in this sub-aether. But he asks the very pointed question, *what* exactly is oscillating? He makes it clear that it is not the sub-aether itself, rather it is something *in* the sub-aether denoted $\psi$. He gives it the classical interpretation of a probability that provides a distribution function describing the probable location of the corresponding quanta. The physical interpretation is then couched firmly in statistical considerations.

But Eddington's interest should not be construed as support. In general he disapproved of wave theory, calling Schrödinger's work "a dodge – and a very good dodge too.

> The fact is that the almost universal applicability of this wave-mechanics spoils all chance of our taking it seriously as a physical theory. A delightful illustration of this occurs

> incidentally in the work of Dirac. In one of the problems … it is found that solutions only exit for a series of special values of the frequency… In Dirac's problem the series turns out to be the series of integers… [w]e are not likely to be persuaded that the true explanation of why we count in integers is afforded by a system of waves (Eddington 1927 [1958], pp. 219-220).

Subsequent to these developments and following his delivery of the Gifford lectures, Heisenberg's Uncertainty Principle arrived on the scene, in time for its inclusion in *The Nature of the Physical World*. Immediately Eddington placed this principle on par with the principle of relativity. In fact there is a clear link here between the two principles. Since relativity demonstrates that distances must be defined by certain *operations* of measurement rather than in reference to the space between points, there appeared to be no limit to the accuracy of measurements. Similar operations can be performed for momentum. In particular a derivative of position can provide an instantaneous velocity. Mathematically there is no limit to the accuracy of numerical results but it is not clear that this is true physically. As physicists delved deeper into atoms the accuracy of measurements began to come into question. The uncertainty principle provided the limit to this accuracy. In essence indeterminacy, as Eddington called it, puts a lower limit on relativity.

More fundamentally indeterminacy is epistemological reminding us that the universe is observed from within and we cannot profess to know it in a truly objective light. As such it is difficult to say what a complete scientific description of the world would look like. It cannot include items that are not causally connected with our experience yet should not be limited to our immediate observations that are clearly incomplete. As Eddington describes,

> The description should include nothing that is unobservable but a great deal that is actually unobserved. Virtually we postulate an infinite army of watchers and measurers.

> From moment to moment they survey everything that can be measured by methods which we ourselves might conceivably employ. Everything they measure goes down as part of the complete description of the scientific world (Eddington 1927 [1958], p. 226).

Such was the state of Eddington's thoughts as the new quantum theory began to reveal itself as the 'roaring twenties' neared their end. We find his belief rooted in the idea that the quantum action is the only way we can interact with the outside world and "knowledge of it can reach our minds.

> A quantum action may be the means of revealing to us some fact about Nature, but simultaneously a fresh unknown is implanted in the womb of Time. An addition to knowledge is won at the expense of an addition to ignorance. It is hard to empty the well of Truth with a leaky bucket (Eddington 1927 [1958], p. 229).

**A New Epistemology**

We find Eddington little more than a decade later questioning the very veracity of knowledge itself: "I have said that I do not regard the term 'knowledge' as implying assurance of truth" (Eddington 1938 [1958], p. 2). Still recognizing the impossibility of making truly objective observations but faced with the undeniable interference of the observer, Eddington advocates for a 'selective subjectivism.' This is rooted in the introduction of epistemological analysis to modern physics. In fact the epistemologist, as it were, is responsible for observing the observers. This is not far from the truth of today's physics where philosophers of physics provide the cheque and balance for the physicists. As quantum information, Bose-Einstein condensates, and other formerly fringe aspects of physics find contemplation of practical applicability this cheque is becoming increasingly important and we find the lines between physicist and philosopher blurring.

        Eddington refers to selective subjectivism as providing *a priori* knowledge of the universe while empirical science provides *a posteriori* knowledge of the universe (see the discussion of *a priori* probability below).  His justification is that *a priori* knowledge in some cases has later been reconciled with *a posteriori* knowledge.  This philosophy embodies the entire purpose behind his pursuit of a fundamental theory: the ability to *derive* data from first principles that otherwise has only been known through experimentation.  Experimentation clearly plays an important role and should not be abandoned but selective subjectivism provides the *a priori* cheque on the *a posteriori* data.  The latter requires careful consideration: it often does not directly provide the desired observable data but rather provides ancillary or even undesired data that leads, by process of elimination, to the desired result.  In Eddington's terms the desired result is then referred to as an 'unobservable.'  But an unobservable is not proven as such purely by observation (or lack thereof).  There must additionally be some logical flaw discovered through the scrutiny of the very definition of the unobservable.  This would constitute *a priori* knowledge.

        Selective subjectivism finds its application in the epistemological transition from classical to quantum systems.  When classical methods are applied to microscopic transitions equations are derived that link positions, momenta, etc. at one instant to positions, momenta, etc. at another instant.  When quantum methods are applied in the same situation the equations link *knowledge* of positions, momenta, etc. at one instant to *knowledge* of positions, momenta, etc. at another instant.  The knowledge in the latter case is necessarily inexact.  Ironically the latter, though seemingly less exact, provides a more correct (I hesitate to use the word 'accurate') description of the phenomena.  This is due in part to the mathematical symbolism and the associated methodology used in the description.  Probabilistic methods actually provide an exact specification of just how inexact our knowledge is.  As such, though we may be limited in how much knowledge we may have about a given transition, we nonetheless can specify *exactly* how limited we are.  This is a powerful example of selective subjectivism: *a priori* knowledge supplies specific limitations on *a posteriori* knowledge.  Wave mechanics witnesses the

application of these ideas to unobservables: the best observation, though unable to provide a precise quantity, will narrow the range of possible values.

Epistemology then tells us that in the quantum world our knowledge is limited to probabilities. In the relativistic world our knowledge is limited to relations. The combination of these two form the foundation of *Fundamental Theory* as the bulk of this treatise will show.

**Probability and Uncertainty: Classical versus Quantum**

Traditionally statistics takes different forms in classical and quantum situations. If we imagine the roll of a pair of dice – label them *A* and *B* – classical statistics tells us that there are two possible ways of rolling a three and a two on a single roll: *A* could produce a three while *B* produces a two or *A* could produce the two while *B* produces the three. Quantum statistics, however, tells us that there is only a *single* way a three and a two can come up on a single roll because it does not recognize the difference between the two dice. Probability and uncertainty then have different interpretations in classical and quantum situations. Eddington, as I will show, tends to blur this distinction a bit particularly in his application of uncertainty to measurement.

One problem with probability is that it is often viewed as the "antithesis of fact" as Eddington puts it when, in fact, it can be a far more accurate predictor of events than other methods. Eddington expands the interpretation of quantum theory from simply the observation of probabilities to the "synthesis of knowledge which constitutes theoretical physics is connected with observation by an *irreversible* relation of the formal type familiar to us in the concept of probability" (Eddington 1938 [1958], p. 92). What does Eddington mean by irreversibility in this case? He gives the following example: imagine two bags, *A* and *B*. *A* contains two white balls and one red ball while *B* contains two red balls and one white. If we randomly draw a white ball from a bag the chances are 2 to 1 we will have drawn the ball from bag *A*. Now let's say we're actually handed this bag which we're told has a 2 to 1 chance of being *A*. What will be the result of drawing a ball? If the entire process was reversible we would *have* to draw a white ball since if we drew a red ball the chances would be 2 to 1 that the bag was *B* even though we were just told it was more likely to be *A*. Now obviously in real life we could easily pick a red ball by random chance, ruling out reversibility. In fact, in reality, given the knowledge that the bag is likely *A*, the chances are 5 to 4 the ball we draw will be white. Probability is thus an irreversible process and, of course, lies at the heart of thermodynamics that is often, more appropriately, called statistical mechanics. To clarify, in modern physics rather than determining whether or not an observation is true, we determine whether its degree of probability matches what we expect from it. That nature of this process is irreversible at its core. However, experimentally, we often test an apparatus by submitting observations with probabilities so high they are almost certain. In such cases macroscopic events can *appear* reversible but the meaning is simply that the sheer numbers involved in the process wash out the scatter. For instance, let's consider the act of removing a book from a bookshelf. Clearly this act is reversible in its most basic sense (you can put the book back). But is *everything* about the book and the shelf unchanged when you return it? If you perform this act literally thousands of times, eventually the book and perhaps the shelf will begin to exhibit some wear and tear. This is because the underlying microscopic effects are *irreversible* and not absolutely everything remains the same when you put the book back. This relates directly to the concept of multiplicity that is discussed in greater depth in chapter six. Of immediate importance is the fact that the multiplicity is directly related to probability: probabilities are ratios of multiplicities.

    Now, in general probability in physics is used in a strictly statistical manner, that is to say it is defined to be a frequency in a certain class of events (thus we see directly that multiplicity must relate to the number of events in some way). In order to be used in an actual scientific statement it must be used in this manner. However, in its more colloquial form it still may be used to *qualify* the statement as a whole. As such it really has two meanings in physics. Somewhat similarly uncertainty can have two meanings. In modern physics the meaning is determined by the uncertainty principle while classically it is a statistical quantity related to the inaccuracy of a measurement. This is an important distinction that is blurred by Eddington in several places in *Fundamental Theory* as I will show. It is a statistical quantity in both instances, but is applied in different ways. Classically we usually think of the uncertainty in measurement as being a result of the inability to make repeated measurements to 100% accuracy (ultimately this

goes back to the reversibility concept I just described). In many cases this is inherently understood since measurements may be idealized depending on the *degree* of accuracy that is desired. So, for instance, when a doctor measures a person's weight during an annual physical he or she does not usually care about changes of a pound (or even a few pounds for most full-grown adults) since body weights fluctuate on a daily cycle depending on such factors as water retention, food consumed, etc. As such these measurements are often idealized in the sense that the person is usually fully clothed each time but rarely wears exactly the same outfit year after year. In this instance there is a natural uncertainty in the measurement of plus or minus a few pounds. This is a statistical and classical interpretation of uncertainty that has no overt relation to the uncertainty principle.

  Quantum mechanically uncertainty also takes on a statistical meaning and in both cases uncertainty is a limiting factor. However, in quantum mechanics uncertainty is a result of the fact that it is *principally* impossible to make simultaneously exact measurements of certain conjugate variables. In essence there is a Gaussian spread to the measurement of each variable. Without opening a Pandora's Box, I should say there are really two ways of understanding this. In one sense if one is not measuring something that requires a specific unit a single measurement could, theoretically, be 'exact.' For example in a standard two slit (or even one slit) experiment a single particle will land on a *definite* spot on the screen (though asking which slit it passed through is another question entirely). However, *repeated* measurements will show the locations of the aggregate of particles on the screen is a *distribution* (i.e. they don't always land in the same spot). This distribution is usually Gaussian and represents the fact that repeating the measurements won't always guarantee the same results. But what about the location of any single particle among the aggregate? If the screen is graph paper, how accurately can we assign coordinate locations to a single particle? This is another problem entirely. We're no longer speaking about the statistical distribution of an aggregate of events but rather the accuracy of the measurement of a *single* event. The measurement depends on several points including the accuracy of the measuring device, the possible interference of the observer in the entire process (i.e. did the act of making the measurement change the outcome from what it would have been if the measurement had not been made), and the potential spread in the measured object itself (i.e. is it extended in space and, if so, by just how much?).

Eddington considers all of these applications of uncertainty including the classical method and often blurs the distinction between them. To be clear, including both classical and quantum conceptions, there are really two broad interpretations of uncertainty that have subtle variations within themselves: the uncertainty related to the aggregate behaviour of a number of particles or events, and the uncertainty related to the measurement of a single particle or event. It is vital to keep these concepts clear in one's mind while reading the description of Eddington's work in order to understand just how he blurs the distinctions.

**Explicit Probability**

Probability was so ingrained in Eddington's worldview it even appears in his more fanciful writing including the light-hearted limerick quoted on page 14: the bassoon-playing baboon is banking on the laws of probability to assure him that he will eventually "hit on a tune." Eddington's exposure to probability and statistics as a means for carrying out the scientific method came early when he began work at the Royal Observatory in Greenwich. One of Eddington's first assignments at the Royal Observatory was to investigate possible sources of error in a Cookson Floating Zenith-Telescope on loan from the Cambridge Observatory. As Smith says "Such investigations were fully in line with one of Greenwich's central goals over its long history: the improvement of the accuracy of observations…" (Smith 2004, p. 23).

One of Eddington's major influences in these earlier years was the work of the Dutch astronomer J.C. (Jacob Cornelius) Kapteyn (1851 – 1922). One of Kapteyn's chief goals was to chart the distribution and motions of the stars in our galaxy (other galaxies were, of course, unknown until nearly a decade after Kapteyn's death). Kapteyn argued that the motions of the stars are not actually random as many astronomers believed. In 1906, not long after he arrived at Greenwich, Eddington took to testing Kapteyn's theory. His method involved no new observations. In fact it was a purely statistical endeavour as was Kapteyn's[8]. In Kapteyn's own words from a letter to George Ellery Hale (1868 – 1938): "My studies have made me more and more of a statistician and for statistics we must have great masses of data, of course." (Kapteyn to Hale 1915). But, Kapteyn was clearly convinced that this work had nothing to do with logical or proper reasoning. In other letters he exchanged with Hale he "warned against putting deduction ahead of induction as a means of making progress in stellar studies." (Smith 2004).

Eddington's view of the role played by data and statistics was completely different. As Stanley has stated in reference to Eddington's Quakerism "one should not try to find complete certainty because this leads to stagnation and a refusal to accept new ideas." (Stanley 2004a, p. 50). Eddington's outlook embodied this view. For instance, the germination of his ideas on how to construct a unified theory are stated in his first book, *Stellar Movements and the Structure of the Universe*: "There can be no harm in building hypotheses, and weaving explanations which seem best fitted to our present partial knowledge. These are not idle speculations if they help us, even temporarily, to grasp the relations of scattered facts, and to organise our knowledge." (Eddington 1914, p. v). In the context of this quote, *Fundamental Theory* could have been seen by Eddington as nothing more than a continual work-in-progress that provided new insights every few years in the form of some published paper and it begs the question: if he had lived longer, would he had *ever* attempted to publish the entire manuscript himself – or if he did, would he have published a completely altered version years later? Many striking changes can be seen in the various manuscripts leading up to the final version (see Slater 1957 for a detailed analysis of these manuscripts).

The statistical work he was performing with the large data sets also was at least one reason he began thinking about the structure of the universe as a whole. As he stated in his obituary of Karl Schwarzschild (1873 – 1916), the "task of determining accurate data for a large number of stars inevitably leads the mind to consider the great problems of the structure of the stellar universe" (Eddington 1916b). He was speaking of Schwarzschild (and in the next sentence draws a comparison to Kapteyn), but the idea is clearly his own. His mention of 'structure' here presages his 'subjective structuralism' which is related to the 'selective subjectivism' I discussed above and was what French has referred to as "a metaphysics for the quantum age." (French 2004, p.117). French also finds a link in Eddington's ideas between uncertainty as manifested in the 'unevenness' of gravity (matter) and the structuralism that was the building block of *Fundamental Theory*: "By matter as the putative cause of irregularities in the field, … this construction is seen as eliminating substance from our ontology in favour of relational structures." (French 2003, p. 228). Here I disagree with French. As it will become apparent as I dissemble and reconstruct *Fundamental Theory* the unevenness was present in a more fundamental combination of both the gravitational *and* electromagnetic fields and thus is more than just matter.

---

[8] Kapteyn was at the poorly funded University of Gröningen and actually did not even have a telescope!

But the extension of probabilistic methods to gravity did not come until Eddington led the famous eclipse expeditions in 1919 in an effort to prove Einstein's general theory of relativity correct. Here, Eddington's observations as well as those made by the companion team, required statistical analysis and an understanding that the results could only be accurate to a certain point (as long as that point was enough to essentially prove – or at the very least, show how probable it was – that Einstein was right). This is a classic inductive argument where the conclusions to be supported are probable or probably true (Copi 1986, p. 404 and see next section). The expedition report was, in fact, one of the last articles he published dealing directly with a specific observation or statistical data set. However, he often referenced actual data even in *Fundamental Theory*, for as Batten says, "he did not turn his back on observation, but he did maintain that understanding came only when there was a theory to explain the observations." (Batten 2004, p. 169). Essentially, he went from believing observation should prove theory to be correct, to believing that theory should prove observation to be correct.

Inductive reasoning was therefore an integral part of his early work simply because it is the method by which all experimental science is carried out and Eddington's observations and statistical research were nothing more than experimental science. Thus it seems that inductive reasoning and statistical methods of analysis go hand-in-hand. Probability is a direct outgrowth of statistics and is a fundamental part of experimental science. However, in the non-aggregate form of probability discussed above there is a history of deductivism including Heisenberg's Uncertainty Principle which arose from critical thinking and not experiment. Though Eddington's work appears to turn more toward deductivism in his later career I will, in fact, show that experimental results played a large role in shaping his subsequent ideas largely *due to* the uncertainty principle which is probabilistic in nature.

But what exactly is it that we mean when we say *probability*? I have discussed probability in depth in relation to statistics but have assumed some familiarity on the part of the reader. It is often useful to return to the most basic definitions and even the history of a concept when attempting to fully understand its application in a given situation. Probability is usually thought to have begun with a series of letters exchanged between Blaise Pascal (1623 – 1662) and Pierre de Fermat (1608 – 1665) in which they argued about the proper way to divide the stakes in an interrupted game of chance. Still another possibility is that it began with Pascal's advice to seventeenth-century gambler Chevalier de Mere on how to wager in a game of dice. There also exists a study published in 1662 by Captain John Graunt (1620 – 1674) where he discusses the mortality records maintained in London since 1592 (Copi 1986). Copi suggests that the result of this "mixed ancestry" gives probability two different interpretations. The first, often referred to as the *a priori* view regards probability in its classical sense as measuring the degree of rational belief (recall my discussion of *a priori* and *a posteriori* knowledge above). In practice this simply means being able to predict the likelihood of various outcomes of an event given that these outcomes are limited in number. The alternative view to this regards probability as simply a measure of relative frequency of outcomes. The latter is often associated with statistical investigations and is thus the sense in which Eddington was first acquainted with it professionally. But Eddington recognized that one of the many goals of science is not simply to describe the world but also to be able to make reliable predictions of outcomes of events. Thus, when combined with a belief in deductive reasoning, which is essentially just "rational belief" as Copi puts it, he switches to the *a priori* interpretation. The importance of this point cannot be understated as it allowed him to maintain probability as the basis of his work while moving to a deductivist approach which is sometimes seen as incompatible with the concept of probability. In fact it may *not* be incompatible with the *a priori* interpretation since in the *a priori* interpretation one can deduce exact likelihoods based on specific limitations in the antecedent conditions which themselves could theoretically be derived from first principles. This is precisely what Eddington did.

This switch by Eddington begs the question of whether or not it was a conscious decision on his part. He certainly was aware of the history of probability theory. In *New Pathways in Science* he refers numerous times to Pierre-Simon Laplace (1749-1827), who was clearly a deductivist since his work in mathematics, including probability theory, was based on logical reasoning. But, once again, probability theory is often associated with inductive pursuits which are often quite apart from logical reasoning (though not necessarily logic itself). Laplace, though a Frenchman, is actually tied up in the history of probability theory in Britain. He was at the centre of a heated debate in Britain in the 1820s when an attempt was made to introduce the teaching of probability theory into the curriculum of the Universities of Cambridge and Oxford. At the time both were still sectarian (in contrast to the secular University of London). In particular, several religiously fervent dons at Cambridge, including William Whewell (1794–1866), argued against the introduction of probability theory to the curriculum on the grounds that it sought to answer questions better left to the Divine. Specifically the opposition singled out the teachings of Continental deductivists such as Laplace, Jean D'Alambert (1717-1783), Alexis Claude Clairault (1713-1765), Joseph-Louis Lagrange (1736-1813), and Leonhard Euler (1707-1783). Whewell favoured inductive science, and being experimental in nature was thus supposedly more supportive of religious beliefs; deductive science was apparently too mechanistic. Joan Richards has noted that "a religion that rested on evidence attested to by personal experience and conviction had no standing in probabilistic discourse" (Richards 1997). But isn't the act of taking a measurement or making an observation simply another form of personal experience? So in one sense, religion's standing in probabilistic discourse is a matter of interpretation but, as I've shown, so is the very definition of probability. Whewell's support of inductive science would suggest a recognition on his part of the inherently statistical nature of measurement, but, on the other hand, the truly difficult nature of measurement theory was not well understood in Whewell's day. Before the Uncertainty Principle and the true nature of quantum mechanics were discovered, perfectly exact measurements were thought to be possible (there is still debate regarding hidden variables – despite

Bell's Theorem, spatiotemporal non-locality has been observed in experiments). So Whewell's understanding of inductive science was built around the notion that it was in fact an *exact* science in every sense of the word. In that sense he could have potentially been persuaded to accept the *a priori* interpretation of probability on the grounds that the antecedent conditions could be known exactly (perhaps even from Divine guidance) and the outcomes were specifically constrained.

Eddington was much the same way in his seemingly inconsistent application of probability within the bounds of *deductive* science. In fact it could be said that they held the same view – probability was *really* the realm of *deductive* science and *not* inductive science and this view could be interpreted as arising from the fact that they recognized, primarily, the *a priori* definition of probability.

Eddington, however, clearly had knowledge of the "relative frequency" definition since he utilized these techniques in his work as an observational astronomer. Since he was a consummate mathematician, early on he was driven by a desire to formulate exact mathematical descriptions of astronomical observations not recognizing the inherent *inexactitude* in the observations themselves. For example he attempted to fit three-dimensional paraboloids to the envelopes of Comet Morehouse from fuzzy, two-dimensional photographic plates, finding considerable room for error (Eddington 1910). Similarly, in the eclipse expedition it turned out that only one of the numerous observations was good enough to support Einstein's theory, and only within a statistically valid error range (Eddington 1919). So, while applying the "relative frequency" form of probability through inductive science he was left with too wide a margin of error, so to speak. Note that the Comet Morehouse example is a relative frequency definition only if one considers the aggregate behaviour of the photons on the photographic plate. Otherwise it is a measurement problem analogous to the accurate assignment of specific coordinates to a single photon on a detection screen, i.e. the statistical distribution of all of the photons on the photographic plates is not overall Gaussian since, if it were, the photograph wouldn't look like a comet. In this sense, Eddington was more interested in the coordinate positions of all of the photons in order that he might *find* a mathematical distribution or form to match the photograph.

In any case, the introduction of the Uncertainty Principle gave Eddington the philosophical foundation for maintaining probability as a mathematical basis for understanding the universe, reinforcing the impossibility of exactitude that he had been observing in his analyses. But, by moving to deductive reasoning and the *a priori* interpretation of probability, he was able to put deducible limitations on any antecedent conditions thus leading to well-defined probable outcomes and limiting the randomness a bit. This is again better understood in the context of Batten's quote (see page 34). The change in his scientific approach was perhaps not as dramatic as some historians (this one included) may have previously thought. It was simply Eddington following his own rule of not being too dogmatic (see Eddington's quote on page 33) and adjusting existing methods accordingly.

By constraining the antecedent conditions he was free to extend Heisenberg's uncertainty principle, which he just recognized as a natural limit placed on mathematical probability, from microscopic (quantum) to macroscopic phenomena (see chapter three for a description of the causal nature of this principle). This is an important point: Eddington's view of the uncertainty principle was not necessarily one of a purely quantum phenomenon but rather simply as a limiting case of our usual classical inability to take repeatedly accurate measurements. This is where Eddington blurs the distinctions between all of the various interpretations of uncertainty and probability. The quantum case was, to him, simply a limit of the classical case and the individual measurement problem (e.g. the location of a single photon on a screen) is really a result of aggregate behaviour since devising a coordinate system in the first place requires at least one other object for comparison.

In any case, as Batten has said, he did not forget about observation. Being the primary way in which we interact with the universe any complete theory needs to incorporate it. The root of our interaction with the universe is essentially through the process of measurement and so he developed *deductive* methods that worked within the parameters of *a priori* probability to define standards of measurement like the meter.

Eddington thus seemingly made the dramatic switch from inductive to deductive science, yet a careful examination of the approach he used clearly shows that he really was simply changing the way in which he dealt with probability. Probability and statistics had been the mathematical foundation of his analyses from the very beginning. In an interesting and quirky little twist on all this, Whewell's inductivism incorporated what he called "fundamental ideas" (could Whittaker have gotten the title of *Fundamental Theory* from Whewell?) that were supported by observation but were found by "thinking properly." Perhaps Whewell was closer to an acceptance of probability than he himself realized.

**Implicit Reasoning**

Eddington was very explicit about his use of probability and the concept of uncertainty as I have shown. He was less explicit about his movement to deductive reasoning later in his career that I have just shown corresponds to a change in his use of probability. The deductivist nature of Eddington's later works is no secret and plenty of writers have discussed or analyzed this aspect of his method (for example see Kragh 1999). But the details of his incorporation of deductivism is much more subtle and really more *implicit* in its application, particularly in *Fundamental Theory*. I have already discussed inductive reasoning from a probabilistic standpoint, but a precise definition will be of greater use in understanding deductive reasoning. The definitions of inductive and deductive reasoning as given in Warriner's English Grammar and Composition are straightforward enough definitions to give an initial flavour of their similarities and differences (Warriner 1986).

Inductive reasoning starts from a set of observations and draws a generalization from them. Deductive reasoning starts with a generalization and draws conclusions from it. As I have shown probability can be interpreted as being either, though in early Victorian science it was thought to be more deductive in nature since its conclusions were derived from mathematical first principles and did not rely on experimental evidence (this is really *a priori* probability and not "relative frequency" probability). Experimental science, by contrast, has usually been described as a form of inductive reasoning (see for example Richards 1997 and Copi 1986). Eddington found that inductivism did not produce the most accurate answers (within the bounds of probability theory) as he felt were possible and so he turned to deductivism. But deductivism entails a definite problem, particularly in relation to science. As Bertrand Russell (1872-1970) wrote in his *Introduction to Mathematical Philosophy*:

> Since all terms that are defined are defined by means of other terms, it is clear that human knowledge must always be content to accept some terms as intelligible without definition, in order to have a starting-point for definition. It is not clear that there must be terms which are incapable of definition: it is possible that, however far back we go in defining, we always *might* go further still (Russell 1919, pp. 3-4).

If everything is deducible from first principles, where did the first principles come from? Could they ultimately be derived themselves? But, if so, from what? Eddington approached the problem from a different tack: what caused uncertainty? His answer: uncertainty arose because there was never a suitably defined starting point in measurement. So uncertainty's very existence was due to the inability to suitably produce *purely objective* first principles in any deductive argument. So it really solved two problems for him: it *required* him to use probability theory in the context of deductive reasoning and explained the problem of finding truly objective first principles in any such argument.

      Here Eddington's deductivism also finds a parallel in another passage in Russell where he notes that Gottlob Frege (1848-1925) "first succeeded in 'logicising' mathematics, *i.e.* in reducing to logic the arithmetical notions which his predecessors had shown to be sufficient for mathematics" (Russell 1919, p. 7). Eddington attempted to do the same for physics by reducing it to a logical set of arithmetical notions that had been shown to be sufficient for *physics* (or by reducing physics to mathematics in those cases where mathematics can be applied in physics). Frege was bothered by statements of identity, such as '$a = b$' which led him to develop a new theory of semantics (Copi 1986). Statements such as '$a = a$' hold *a priori* and are analytic in nature (Copi 1986). But statements such as '$a = b$' could be true or false depending on knowledge not contained

within the statement itself and thus *cannot* hold *a priori*. Eddington was similarly bothered by *physical* problems of identity, in particular the identities of indistinguishable particles (a topic still at the forefront of the fundamentals of modern physics and one I discuss in depth in several future chapters). Eddington went a step further, though, holding that in some physical senses even '$a = a$' could not hold *a priori* but required prior knowledge. I will discuss the details of this in depth when discussing his treatment of the exclusion principle, but the origin of his thinking in that regard clearly came out of his attempt to 'logicise' physics.

Any attempt to logicise physics then must deal with definitions and Eddington dealt not only with definitions of objects but also with definitions of measurements. Deductivism, however, did emerge in other ways early on in his work. In 1916, for example, he began attempting to answer the age-old question: what exactly *are* stars and what makes them shine? As Stanley has pointed out his approach was deductive: "start from completely valid premises and the conclusion is certain to be true, but without total certainty in one's premises, one has nothing" (Stanley 2004a, p. 50). He was still focused on inductive science at that time and combined this deductive approach with the inductive method of gathering data on stellar masses and brightnesses and combining it with mathematical analyses to examine the various patterns that emerged. Stanley, however, gets at the heart of why the inductive method ultimately proved unsatisfactory in Eddington's mind: it runs the risk of being disproved by new evidence or observations (Stanley 2004a, p. 51). Eddington's deductivism was *supposed* to be as objective a process as humanly possible and ultimately impervious to disproval. The irony in all this is that Eddington viewed Subramanyan Chandrasekhar's (1910 – 1995) work on white dwarf stars as being 'mindlessly deductive' (Stanley 2004a, p. 53).

But, Eddington's method, though deductive, was ultimately designed to be practical while simultaneously above speculation. In this his observational background clearly played a role. Chandrasekhar was a pure theorist. Eddington was both a theorist and an observationalist. Relativity was a key player in all of this since he was both the first person to really disseminate Einstein's theory in the English language as well as the first person to produce experimental evidence supporting it. But relativity plays a very important role in any attempt at a purely objective analysis of the universe. This is because relativity implies that there is no universally preferred reference frame. A strong interpretation of this concept would imply the impossibility of objectivity since objectivity would either imply a *single*, preferred reference frame or the same view from *every* reference frame.

**Structuralism**

Just how does Eddington, then, walk this fine line between truth and progress? Defining a purely objective point of view forces one to consider exactly how the universe is structured. Thomas A. Moore in his outstanding six-part textbook *Six Ideas That Shaped Physics* essentially reduces physics to two things: particles and their interactions (Moore 2003a). In the Standard Model, however, the four fundamental interactions (gravity, electromagnetism, strong nuclear, weak nuclear) are really represented by more particles, the carrier particles of each interaction – i.e. the particles that mediate the interaction by delivering information from one particle to another. So, ultimately, the universe is nothing but particles. But what are particles? Are they simply packets of energy? What gives them motion? Their *direction* ultimately depends on how (or if) they interact with their neighbours, but what produces them to begin with and ultimately serves as the seed of motion? Eddington has referred to them as simply being conceptual carriers of a set of variables (he used the term 'variates' – Eddington 1946, p. 30). Everything basically reduces to mathematics, then. Structure in Eddington's mind, in relativistic language, consisted of events with intervals linking the events. Point events would be relata while the intervals would be the relations or, in Moore's interpretation, the relata would be particles and the relations would be the interactions, which are simply carrier particles. As French points out, the only problem with this is a classic 'chicken and egg' issue: which came first? This is particularly troubling if the relations are simply more relata and, as I will show in chapter six, indistinguishable from the original relata according to Eddington. But as French says,

> … we can think of both the relata and the relations as equally derivative in the sense that both can be

> conceptually 'pulled out' of the package as a whole (French 2003, p. 232).

Thus Eddington took the universe and its constituents to be an indivisible whole. In his own words, "The relations unite the relata; the relata are the meeting point of the relations. The one is unthinkable apart from the other." (Eddington 1928, pp. 230-231).

The variates Eddington was talking about include such things as mass, charge, and spin. Einstein showed that in a relativistic sense mass is really just energy. Eddington, as I will show in subsequent chapters, attempted to apply the same relativistic reasoning to show that charge and spin were also frame-dependant quantities and show that *all* particles were indistinguishable (e.g. a proton is indistinguishable from an electron) since they were just packets of energy. The wholeness of the universe would then just be energy. There are subtle arguments about the nature of indistinguishability that will be worked out in greater depth as we study their applications.

Another view might be to interpret everything more fundamentally as momentum (or rather four-momentum which is a relativistic unification of energy and momentum). Moore has said that mass is simply a measure of how a particle reacts to changes in momentum (Moore 2003a). In some sense charge and spin could be related in a similar manner since, ultimately, mass' response to momentum is only through certain fundamental interactions. Any electromagnetic interaction (which technically includes all macroscopic contact interactions) ultimately is mass-independent (see Coulomb's and the Lorenz Force Laws) but *is* dependent on charge. So ultimately, momentum conservation at this level in such an interaction depends on *charge* and *not* mass, even though macroscopically it appears to depend on mass (even in the four-momentum case). A similar argument can be made for spin, particularly in the context of *angular* momentum, though this is a more complicated situation. In one sense this is true since the quantum field theoretic interpretation of charge is that it is a measure of how the *W* and *Z* bosons propagate and so a relation to momentum is natural, though that is a bit of a simplistic interpretation. As for spin, any relation to momentum may well get at the root of the exclusion principle as I will discuss in the closing chapters.

In a modern sense, much of this could be attributed to the field concept and, indeed, the field concept has been highly successful in predicting the results of particle physics experiments. But fields, or in the twenty-first century perhaps strings, are not as dogmatic a construct as is sometimes viewed. As French says "we have good reason to expect our current favoured ontology – quantum fields, strings, whatever – to … come to be replaced" (French 2004, p. 120). Despite those who search for a theory of everything, French's statement is not wholly unbelievable. The simple act of relative motion actually creates macroscopic phenomena such as magnetism that do not exist apart from the relative motion (or, perhaps more correctly, they reduce to a more familiar form without relative motion). The elegance of reducing everything via fairly simple relativistic arguments to momentum, which is essentially just motion, then is perhaps, by virtue of Ockham's Razor, a better solution than the complexities of modern theories.[9][10] Even Henri Poincaré (1854 – 1912) in 1905 in *Science and Hypothesis* indicates that knowledge of motion is ultimately at the root of many 'non-mechanical' problems:

> only the something, which we then called *motion*, we now call *electric current*. But these are merely the names of the images we substituted for the real objects which Nature will hide for ever from our eyes. The true relations between these real objects are the only reality we can attain. (Poincaré 1905, p. 162).

Eddington was thus sceptical of man-made units since for the most part they were so arbitrary. In fact, Eddington viewed the fragmentation of the universe as being manifested in the multitude of units present in science. I will discuss the details of measurement in several later chapters, but for now I want to focus on how this leads to a philosophical bypass, so-to-speak.

---

[9] Since field theories often make use of Lagrangians, action integral, and propagators one could argue that field theories already *have* reduced everything to relative motion.

[10] To debunk conspiracy theories it is often useful to make use of a variation on this known as Hanlon's Razor: "Never attribute to malice that which can be adequately explained by stupidity." (see http://catb.org/~esr/jargon/html/H/Hanlons-Razor.html).

Eddington saw a way around this by using a system of natural units and investigating unitless ratios which is essentially investigating the large-scale structure of space-time by comparing different aspects of the whole. This idea actually has some merit, though philosophically speaking its merit may rest on an interpretation of the nature of mathematics and numbers. But, ignoring *that* quagmire for a moment, it is well-known that in Euclidean geometry, which recent experiments have proven is the overall geometry of the universe as a whole (de Bernardis, et. al. 2000), the ratio of the circumference of a circle to its diameter is *always* the number $\pi$. This apparently observed 'truth' is the source of the natural measure of an angle: the radian. Technically, in fact, a radian is, of course, not a real unit at all. We just treat it as one in order to better keep track of what we're doing. The idea is that whenever a circle is wholly traversed circumferentially the change in the angle is *always* a set multiple of $\pi$ – or, in fact, a *unitless ratio* of the arc length to the radius. So, again, it is just a comparison of different aspects of the whole circle. Eddington's technique, then, is a direct application of relativity by making all objectively describable entities in the universe simply relations between different aspects of the whole. As I quoted earlier, French says of Eddington's position, "By matter as the putative cause of irregularities in the field, … this construction is seen as eliminating substance from our ontology in favour of relational structures." (French 2003, p. 228) – basically structure was formed from a relativistic worldview. The technique of employing relational structures is extended in the creation of his concept of comparison particles that will be discussed at length in the latter half of this text. The philosophical quandary Eddington gets into, though, is how to compare or relate seemingly indistinguishable particles, especially if relativity itself seems to indicate *all* particles are indistinguishable from each other (see chapter six).

Again, this brings up the ugly possibility of having to analyze the nature of mathematics and numbers. Primarily it begs the question of whether or not mathematics and number exist apart from our mental construct of it (if a tree falls in a forest and no one is around to hear it, does it make a sound?). Eddington's implied view is this: mathematics is inherent and not constructed. The symbols we use for representation are our translation of mathematics, which is simply the language of nature. But the underlying truths mathematics provides are existent apart from the language we use to interpret them. How, then, is mathematics applied to the *description* or observation of a particle, especially if the observation is direct and not deduced from some experimental apparatus? Eddington, as I stated previously, described particles as merely conceptual carriers of a set of variates. French puts this in more general terms by saying that group theory "enters as a way of expressing the relationships between relations and the important point is that whatever the nature of the entities, the use of group theory allows us to abstract away the 'pattern' or structure of relations between them" (French 2003, p. 239). In some sense, based on Cassirer's conception (see above), French says that this reduces an object to "nothing more than a node in the structure" (French 2004, p. 123). Cassirer has argued that quantum mechanics really had nothing to do with indeterminacy (and the related limitations on measurement) but really with how we conceived of objects. As I mentioned previously Eddington considered particles to simply be carriers of sets of variates. This is what he termed 'Principle of the Blank Sheet:'

> The Principle of the Blank Sheet requires that at the start we should recognise no intrinsic distribution between the particles which we contemplate, in order that we may trace to their very source the origin of those distinctions which we recognise in practical observation. The fundamental dynamics is the dynamics of indistinguishable particles; the dynamics of distinguishable particles is a practical application to be used when we do not wish to analyze the phenomena so deeply (Eddington 1936, 287).

As I will demonstrate in my technical analysis the method he employs to accomplish this is to treat particles as merely fluctuations in a uniform background. Through purely relativistic reasoning utilizing the framework of space-time, he shows that all particles (regardless of type) are indistinguishable from all other particles, which is the physical 'blank sheet' he started building everything else with. The advantage of this is that the manifestation of physical properties like mass, charge, and spin and thus the observed differences in particles could be introduced as a consequence of frame of reference or derived from the blank sheet, rather than assumed as initial conditions. He essentially treats distinguishability as a mere practical convenience for certain problems but *not* as a requirement. Mathematically, this translates as the "part [being a structural concept] is a symbol having no properties except as a constituent of the group-structure of a set of parts" (Eddington 1939, p. 145). The details go well beyond the initial analysis presented here, but, of relevance is the employment of group theory.

In a very modern sense, the group theoretic approach has led to many advances in quantum field theory. As A. Zee describes in his recent textbook *Quantum Field Theory: In A Nutshell*, the physics contained in a simple field theory can be interpreted as a source in some region of space-time "that sends out a 'disturbance in the field,' which is later absorbed by a sink" in some other region of space-time where the disturbance is referred to by experimentalists as a particle of some given mass, $m$ (Zee 2003, p. 24). Eddington's 'Principle of the Blank Sheet' and his idea that particles are simply fluctuations in the background, look very modern, then, when set against Zee's description of standard quantum field theory: all of existence, as represented by particles (including force carrier particles), can be simply reduced to fluctuations in a uniform background or field. To be more specific, Eddington's particles, which are nothing more than carriers of sets of variables, are mathematical entities set in matrix or tensor form. Transformations including rotations, translations (spatial and temporal), and permutations are achieved in many cases in *Fundamental Theory* through the application of group theory. The same is true in quantum field theory where the introduction of group theory actually helped lead to profound discoveries. In Eddington's case group theory was used most liberally in *Relativity Theory of Electrons and Protons* but, particularly in the first six chapters of *Fundamental Theory*, not as much. Perhaps some of this stems from the debate he had with Braithwaite regarding his use of group theory in the former as well as in the *Philosophy of Physical Science* (see Eddington 1936 and 1940). In any case, group theory is not a major component of the statistical portion of *Fundamental Theory*. It saves its grand entrance here until his discussion of $E$-number theory near the end.

**Deducing the Structure**

There are two problems here in both perceiving the structure of the universe as well as deducing it from first principles. The two problems are intertwined and it is difficult to say which is the causal result of the other. The *logical* problem with Eddington's reasoning here is that in assuming that uncertainty permeates absolutely everything, then if all of structure – and not only its mathematical description – is deducible from first principles, how can one deduce a definite conclusion from an indefinite starting point? Related to the *logical* problem then is the *epistemological* problem of our ability to even *gain* objective knowledge about the universe. If we cannot separate the observer (ourselves) from the observed, is it possible to deduce *any* truth?

Russell has said: "In one sense it must be admitted that we can never *prove* the existence of things other than ourselves and our experiences" (Russell 1952 as quoted in Feinberg 1986, p.193). Russell's idea is that the difficulty arises if we perceive everything as sense-data – data that is relayed to our brain through our five senses – which, in effect, is all data. But is all data merely a set of sense-data? Do things exist apart from our senses – i.e. do they exist while not being observed? Eddington side-stepped this by flat-out rejecting the atomistic view of sensation by saying that a single sensation really doesn't provide any useful information about the world at all and by requiring a group-theoretic structure that applied to both the universe and the mind. This addresses the Mind-Body Problem and was a point Russell actually made in 1919 in his *Introduction to Mathematical Philosophy* and again in following up his statement above by saying that although the inability to prove the existence of anything other than ourselves and our experiences "is not logically impossible, there is no reason whatsoever to suppose that it is true" (Russell 1952). Technically in one sense, particularly in his earlier work, he was referring to cardinality. This poses a problem since it reduces nearly everything to number and set theory. As the mathematician M.H.A. Newman (1897 – 1984) pointed out, this structuralist 'ontology,' as it were, admits that "*nothing* can be known that is not logically deducible from the mere fact of existence, except ('theoretically') the number of constituting objects" (Newman 1928, p. 144). This meant that knowing only the structure of the universe was useless information. Notably Eddington did derive the number of particles in the universe, i.e. the "number of constituting objects" in the structure and *Fundamental Theory* is built on purely mathematical analyses of an aggregate of particles. So in some sense 'number' is fundamental here. But the problem is that Eddington derived this number *from* the structure and not the other way around which is: a.) why his application is wholly different from Russell's since for Russell the number *was* the structure, and b.) why Newman's criticism still fits and leaves open the question of the truth of any observation (except, perhaps in Newman's view, the number of constituent objects in some structure unless the nature of pure number is called into question – yet another quagmire).

Eddington recognized that the separability of observer and observed is the major problem with *inductivism* since inductive science relies on observed data. But he does not seem to realize that this is a problem for *deductivism* as well, missing the point that the mental constructions of a theory are built on pre-existing sense-data already stored in the brain: with absolutely no knowledge of any aspect of our universe whatsoever (ignoring, for a moment, any multiple-universe theories), its structure – even its mere existence – could not be derived. This is the essence of Newman's argument. Eddington appeared on the verge of this realization as early as 1920: "the distinction of substance and emptiness is the mind's own contribution, depending on the kind of pattern it is interested in recognising" (Eddington 1920, p. 420). Everything Eddington does employ given factual knowledge not just about the universe but also about mathematics, physics, astronomy, etc. all of which are *parts of* the universe and known through sense-data (see the discussion on stabilization in chapter six).

This all falls under the larger question of certainty. Can we really be certain of anything, including end points and boundary conditions? The uncertainty principle itself seems to imply we can't and Eddington even extended this principle to include uncertainty in the actual reference frame of observation (see chapter five). And yet Eddington reaches definite conclusions in his work. How can he reach definite conclusions beginning with an indefinite starting point? Even his introduction of the 'Blank Sheet' is not necessarily logically consistent with this idea since, even though physical differences (and thus initial conditions) of particles are derivable from the 'Sheet,' the sheet itself, according to Eddington, contains some uncertainty. It would be more logically consistent to say that, rather than assuming that particles take on very well-defined properties that clearly separate them into various groups, they have the probability of taking on *any* property whatsoever (e.g. there exists the small but finite probability that a fundamental particle with a mass of 1 kg could come into being). This appears to be inconsistent with observation, however it is worth keeping this in mind when we encounter Eddington's use of multiplicity since multiplicity is simply a ratio of probabilities.

In the context of group theory and the underlying mathematics, French points out (interpreting R.B. Braithwaite) how this manifests itself:

> It is only by specifying the group relation, or mode of combination, that we actually have a group to begin with. But such a specification introduces a non-structural element into our structuralism, because we have to have some ground – which clearly cannot be structural itself – for selecting one combination over another (French 2004, p. 129).

In extending this to relations between groups (for instance, let us say all electrons are one group and all protons are a second group), the groups become the relata and the relevant transformations (rotations, permutations, etc.) become the relations that, by Eddington's own admission, cannot be considered apart from one another. The relations between the groups (for instance the ratio of the masses of the proton and the electron), then, are essentially non-structural content since they are supplied by the mathematician or physicist, according to Braithwaite (French 2004 and 2003).

Eddington may have recognized part of this limitation since he responded directly to Braithwaite's criticism by pointing out that group theory is simply used in physics as a tool for expressing the relationships between various relations (which, in the Standard Model, might be viewed simply as more relata) thereby bringing out the underlying structure or pattern – perhaps a bit like a litmus test or hidden ink (Eddington 1940). In fact Eddington referred to the group structure as simply the "pattern of interweaving" or the "interrelatedness of relations" and insisted that pure structure is only attainable by considering these (Eddington 1939). French points out that the error in Braithwaite's argument is the separation of the relations from the group elements that are the original relata. Eddington's point was that one could not have one without the other. This is in fact true in the modern sense if the relations are simply other particles, which immediately puts them into the same category as the group elements themselves that are particles (or particle properties). Once again Braithwaite is trapped by fragmentation by assuming that content produces the structure while Eddington takes the reverse tack and assumes structure actually produces content. Any supposed non-structural components are really just caused by the mind being led astray. Such non-structural components could include initial conditions such as values for mass, charge, and spin. But this argument does not properly provide a mathematically consistent way to deduce these values within the context of uncertainty and Eddington himself could not find a proper way to do this.

As such he introduced the concept of stabilization by assuming or taking for granted certain quantities (like the mass and charge of the electron), but, again, where do these values come from? How can one assume these quantities have any particular values? The values are merely sense-data again or averages of sense-data and he is back to inductivism that he was attempting to shun. In fact the circular nature of all of this can be demonstrated when considering the masses of sub-atomic particles, for example, which are often found through experiments that include the uncertainty principle in their analysis. Eddington's half-hearted attempt at getting around this is noted in *New Pathways in Science*: "we have been concerned to show that probability is always relative to knowledge (actual or presumed) and that there is no a priori probability of things in a metaphysical sense, i.e. a probability relative to complete ignorance" (Eddington 1935b, p. 133). He basically attempted to put a limit on probability by making it relative but was unable to do so in a logically consistent way despite the fact that uncertainty is consistent with special relativity (see chapter three). This, then, is the biggest flaw in Eddington's philosophical reasoning: his assumption that a proper theory of quantum gravity could be deduced from logical reasoning alone if the theory itself implied that there could be no definite or certain starting point for logical reasoning.

# IV

## *The Roaring Twenties: Dawn of the New Quantum Theory*

As if deducing the structure wasn't difficult enough, Eddington firmly latched onto the uncertainty principle, a foundation of the 'new' quantum theory, as a defining philosophy making deduction even more difficult as I have just shown. On the surface this appears to be a contradiction – how can certain results be derived from uncertain first principles? A detailed look at uncertainty and Eddington's mathematical treatment of it (see chapter five) is necessary to fully understand his reasoning and to ultimately determine if that reasoning is valid. In analyzing uncertainty and Eddington's interpretation of it, we must do what Eddington did and shed any preconceived notions we might have regarding the implications of uncertainty or even the principle itself. In particular, as I will show, it will be most helpful to shed the preconception that uncertainty is what ultimately leads to the particulate nature of matter. But other preconceived notions surrounding uncertainty are also a hindrance in some instances particularly when we begin to deal with it in relation to structuralism. I will discuss these as they arise, though some of them can be dispensed with simply by looking at the historical development of the principle itself. In addition it is necessary to then extend this to include a historical analysis of the exclusion principle, which is intimately intertwined with uncertainty. Related, again, to both is the history of the development of wave and matrix mechanics, of which I will have something to say.

**The Origins of Uncertainty**

There are countless histories of the development of Werner Heisenberg's (1901 – 1976) Uncertainty Principle from single articles to entire texts. For a basic historical treatment Helge Kragh's *Quantum Generations: A History of Physics in the Twentieth Century* shows its relation to the rest of modern physics. Max Jammer's comprehensive text *The Conceptual Development of Quantum Mechanics* contains a more technical account, particularly in relation to the Copenhagen Interpretation. However, a radical reanalysis of the history of the Copenhagen Interpretation that profoundly alters the view of the context within which many of the original points were made, was recently put forth by Mara Beller and bears close scrutiny (Beller 1999). Another classic critical analysis of Bohr's work is *The Philosophy of Niels Bohr: The Framework of Complementarity* by H. Folse. In addition Walter Moore's *Schrödinger: Life and Thought* provides a new perspective on the impact of Schrödinger's wave mechanics on the development of Heisenberg's Uncertainty Principle.

The roots of the mathematical aspect of uncertainty – the uncertainty relations themselves – date to the mid 1920s and ultimately Max Born's (1882 – 1970) probabilistic interpretation of the wave function. The germination of the general concept of uncertainty in physics has its roots in philosophy. Jammer has argued that Heisenberg was influenced early on by philosophy having first come into contact with atomic theory through Plato and was later influenced by the writings of Immanuel Kant (1724 – 1804) and Ludwig Wittgenstein (1889 – 1951) (see Figure 6.). But, perhaps even more philosophically influential, was Einstein's concept of an observable relativistic time that brought with it the very important rejection of the absolute simultaneity of events as indefinable (Jammer 1966). This first put the emphasis on the concept of observable quantities in relation to quantum mechanics. Heisenberg even confided this latter point to Einstein in 1926 saying "the idea of observable quantities was actually taken from his [Einstein's] relativity" (*Archive* 1963 as quoted in Jammer 1966, p. 198). Jammer's points can be contrasted with the views of Beller, who maintains the germination of Heisenberg's ideas grew out of dialogues he had with certain scientists including Bohr, Schrödinger, Pauli, Dirac, and Jordan (Beller 1999, pp. 65-102), and with Moore, who points to debates about causality and determinism in the early 1920s as appearing to "foreshadow the much more cogent attack on causality derived from quantum mechanics by Werner Heisenberg in his *indeterminacy principle* of 1927" (Moore 1989, p. 154). These debates included the likes of Schrödinger, Franz Exner, and Georg Hamel.

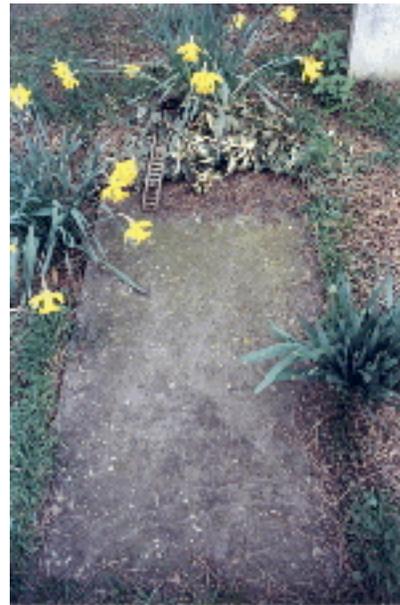

**Figure 6.** Wittgenstein's headstone – yet another person appearing in this narrative buried less than fifty feet from Eddington. Note the reference to 'Wittgenstein's Ladder'. More curious were the hundreds of pennies (difficult to see in the photo) near the ladder end of the headstone. No one in Cambridge (nor any visiting philosophers I ran into) had any explanation for this (photo by author).

Heisenberg also employed Niels Bohr's (1885 – 1962) Correspondence Principle in a unique way. This principle's versatility alone implied a non-rigid formulation of quantum theory. But Heisenberg used it to formulate a new mathematical approach to be used in a new theory of mechanics. The idea was to make correspondence a root foundation of the new theory so that the mathematics could be built upon that. In 1925 Heisenberg set about developing a solution for an anharmonic oscillator that was built around correspondence and utilized the notion of observable quantities. The resulting paper was sent to Wolfgang Pauli (1900 – 1958) whose encouraging comments prompted Heisenberg to give the paper to Born in July of 1925. The unique solution included the use of matrices and complex numbers, though the matrix multiplication he used was not immediately obvious to Born. Matrices had been rarely used by physicists prior to this point (what physicist could imagine living without them now?) with minor exceptions including Born's own work on the lattice theory of crystals. In 1924, just prior to Heisenberg's paper, Richard Courant (1888 – 1972), armed with the lecture notes of David Hilbert (1862 – 1943), published the remarkable volume *Methods of Mathematical Physics* which just happened to contain exactly those parts of algebra and analysis that would later serve as the basis for quantum mechanics. As Jammer notes: "it seems almost uncanny how mathematics … prepared itself for its future service to quantum mechanics" (Jammer 1966, p. 207).

Born took on the task, then, of putting Heisenberg's new matrix mechanics in a more logically consistent framework. Originally his assistant in this endeavour was to be Pauli, but ironically, Pauli, whom shall cross our path again later in this chapter, turned down Born's offer. In a particularly providential scenario, Born was on a train to Hanover discussing his work with a colleague from Göttingen. He mentioned the problems he was having with the matrix calculations and was overheard by the twenty-three year-old Pascual Jordan (1902 – 1980) who just happened to be sitting in the same compartment. Jordan and Born did not know one another, but upon arrival at Hanover Jordan introduced himself to Born being certain to point out his experience with matrix algebra and offered his assistance. By September of 1925 Born and Jordan submitted their paper (Born and Jordan 1925) for publication. It was the first rigorous development and formulation of the new matrix mechanics.

**The Commutation Relation and Born's Matrix Mechanics**

This paper utilized classical Hamiltonian methods combined with matrix mechanics in the manner now standard in both classical as well as quantum mechanics (see Goldstein 1980) to show that finding the extrema of the trace of the Lagrangian
$$\mathbf{L} = \mathbf{p\dot{q}} - \mathbf{H(pq)}$$
leads to the canonical equations:
$$\mathbf{\dot{q}} = \partial \mathbf{H} / \partial \mathbf{p} \qquad \mathbf{\dot{p}} = \partial \mathbf{H} / \partial \mathbf{q}.$$
Rather than using only $q$ as Heisenberg did, Born and Jordan used both $q$ and $p$ as independent matrices to write for the first time what is now known as the *commutation relation* in quantum mechanics:

$$\mathbf{pq} - \mathbf{qp} = \left(\frac{h}{2\pi i}\right)\mathbf{1} \qquad (4.1)$$

where **1** is the unit matrix. Their derivation was based on the correspondence principle thus planting the seed of uncertainty through the non-rigid, versatile nature of the principle (see above).

In November a sequel to this paper appeared with Heisenberg as a co-author with Born and Jordan and it was in this famous paper that a logically consistent method for solving quantum mechanical problems was developed (Born, Heisenberg, and Jordan 1925). The commutation relation itself led to several important results including Dirac's discovery that it led directly to Bohr's relation that connected the frequencies of atomic vibrations with the differences in atomic energy levels. Thus it paved the way for a more complete understanding of atomic radiation. In an interesting twist that is only a tiny morsel of the vast amount of evidence supporting the idea that Dirac was a genius, he developed the same theory without using or even knowing about matrices – and he was only twenty-three (Kragh 1999).

**De Broglie, Schrödinger, and the Dawn of Wave Mechanics**

Simultaneous to this was the development of the wave model of light and matter. The wave-particle debate for light dates to the seventeenth century and the works of Isaac Newton (1642 – 1727), who was a proponent of the particle nature of light, and Christiaan Huygens (1629 – 1695) who was a proponent of the wave nature of light. The breakthrough in this debate was Louis de Broglie's (1892 – 1987) discovery of the wave nature of matter in his doctoral thesis research in 1923 (his thesis was presented in 1924 (de Broglie 1924) but several papers had already been published on the subject). The key result of de Broglie's work was what is known as the 'de Broglie relationship' linking the effective wavelength of the wavefunction of a beam of quantum particles with their corresponding classical momentum: $\lambda = h/p$ where $p$ is the momentum. This led several physicists to develop a new theory of wave mechanics where the kinematics of particles could be described by treating them as waves rather than particles. De Broglie's work laid the foundation for Erwin Schrödinger (1887 – 1961) to derive what is now popularly known as the Schrödinger time-independent wave equation, nearly at precisely the same time as Born, Heisenberg, and Jordan were first developing matrix mechanics. The irony, of course, is that Born was really wrapped up in the development of both theories since his probabilistic interpretation of the wave equation was a foundation piece of Heisenberg's uncertainty principle (see Beller 1999). In any case, Schrödinger began with the standard Hamiltonian equation for energy:

$$H\left(q, \frac{\partial S}{\partial q}\right) = E$$

and replaced the separable function $S$ by $K \log \psi$:

$$H\left(q, \frac{K}{\psi} \frac{\partial \psi}{\partial q}\right) = E .$$

In simple cases involving only $\psi$ and its first derivatives this can be easily written as a quadratic equation. As Jammer explains:

> Schrödinger then replaced the quantum conditions by the following postulate: $\psi$ has to be a real, single-valued, twice continuously differentiable function for which the integral of the just-mentioned quadratic form over the whole configuration space ($q$ space) is an extremum. The Euler-Lagrange equation, corresponding to this variational integral, is the wave equation (Jammer 1966, p. 259).

Schrödinger applied this to a hydrogen atom with the potential energy $-e^2/r$ to produce the following familiar form of the wave equation:

$$\Delta\psi + \frac{2m}{K^2}\left(E + \frac{e^2}{r^2}\right)\psi = 0. \tag{4.2}$$

Schrödinger then applied his new theory to a linear harmonic oscillator, a rigid rotator with both fixed and free axes, and a vibrational rotator that was equivalent to a diatomic molecule (see his general result below which, when combined with the de Broglie relation, matches the Born-Heisenberg-Jordan result precisely). Schrödinger's results were thus in complete agreement with Heisenberg's matrix mechanical results (Jammer 1966).

      The importance of the this entire process cannot be understated both in its relevance to Eddington's formulation of his theory and its standing in the history of modern physics. To underscore this fact I will now present a somewhat simpler derivation of the wave equation. This will also serve to demonstrate the pivotal nature of the de Broglie relation in the development of modern quantum mechanics.

**A Simple Derivation of the Time-Independent Schrödinger Equation**

The primary application of the Schrödinger equations is to find the energy eigenfunctions in the Hamiltonian equation for energy. Prior to the 1920s Newtonian mechanics was the only method used to study non-relativistic particles. Since Newtonian mechanics is still applicable in many situations it must be recoverable from any subsequently broader theory such as relativity (in fact it is not entirely recoverable from quantum mechanics, a point that still haunts physicists and philosophers). But let's at least attempt to compare a Newtonian particle with a quantum particle and see what happens.

Assume we have some classical particle with some energy $E$ moving in a straight line in some direction such that as it moves it slows down. Thus the potential energy of whatever interaction is causing it to slow down is increasing while the kinetic energy of the particle itself is decreasing. Basically the particle transfers its kinetic energy to the potential energy of the interaction that, by the Standard Model, is itself a particle – essential one could interpret this as the carrier particle gaining some kinetic energy that manifests itself to an observer as potential energy. The original particle's kinetic energy is given as $K = E - V(x)$. If the particle is nonrelativistic, its kinetic energy is:

$$K = \frac{1}{2}mv^2 = \frac{(mv)^2}{2m} = \frac{p^2}{2m}$$

where $p$ is the particle's momentum (Moore 2003b, p. 176). The particle's momentum is then $p = \sqrt{2mK}$. Obviously this implies that as the kinetic energy decreases, so does the momentum. This is where the previously mentioned de Broglie relationship becomes useful since it includes the classical momentum: $\lambda = h/p$. The implication here is that if the particle's classical momentum is decreasing its wavelength should be getting longer.

The problem here, of course, is that the wavelength is now not a constant – it is changing in time. But initially, we're only interested in handling the time-*independent* wave equation so we would like to 'freeze' the particle at a specific point (ignoring, for a moment, the problem of uncertainty here). This allows us to define something Moore refers to as the 'local wavelength.' Through simple calculus applied to a simple sine-wave with a changing wavelength, Moore shows that this is:

$$[\lambda(x)]^2 = \frac{-4\pi^2 f(x)}{d^2 f / dx^2} \quad \text{or} \quad \frac{1}{[\lambda(x)]^2} = \frac{-d^2 f / dx^2}{4\pi^2 f(x)} \tag{4.3}$$

where $d^2f/dx^2$ locate the extrema of $f(x)$ (Moore 2003b, p. 179). Thus we now have a means of calculating the wavelength of some function at some point $x$ even if the wavelength is *also* dependent on $x$ since the function and its second derivative can be evaluated at any $x$-position (simply find the second derivative and plug an $x$-value into this and the original function!).

Once again ignoring problems with uncertainty for a moment, we now have a way of finding the wavelength of some decelerating particle at a given position and by the de Broglie relation we know that this is related to the particle's classical momentum. Mathematically this involves combining equation (4.3) with the de Broglie relation beginning with:

$$\frac{h}{p} = \lambda \quad \text{or} \quad \frac{h}{\lambda} = p = \sqrt{2mK}.$$

We can isolate the kinetic energy on the right side by squaring both sides and dividing by $2m$:

$$\frac{h^2}{2m}\frac{1}{\lambda^2} = K = E - V(x).$$

The inverse of the square of the local wavelength we determined was:

$$\frac{1}{\lambda^2} = \frac{-d^2\psi_E / dx^2}{4\pi^2 \psi_E(x)}.$$

Substituting this into the previous equation and recognizing that $\hbar \equiv h/2\pi$ we find the time-independent Schrödinger equation:

$$\frac{-\hbar^2}{2m}\frac{d^2\psi_E}{dx^2} - [E - V(x)]\psi_E(x) = 0$$

which is simply a slightly rearranged version of Schrödinger's original equation:

$$\Delta\psi + \frac{2m}{K^2}\left(E + \frac{e^2}{r^2}\right)\psi = 0.$$

Note that in Schrödinger's form he has simply given the potential energy the value $-e^2/r$. In fact Schrödinger himself deduced a more general form using the de Broglie relations:

$$\Delta\psi + \frac{8\pi^2 m}{h^2}(E - U)\psi = 0. \tag{4.4}$$

The beauty in this derivation is that it demonstrates the link between classically measurable quantities such as momentum and energy, and the more typically quantum quantity of wavelength. It is quite simply found by combining ordinary calculus, classical energy conservation, and the de Broglie relation. It is clear from this just how pivotal a discovery the de Broglie relation was.

**The Time-Dependent Schrödinger Equation**

Schrödinger, of course, recognized that the energy eigenvalue $E$ varies from position to position as the particle decelerates (or accelerates). In a time-dependent wave equation, then, it would be most useful to find some way to eliminate $E$. The wavefunction in four dimensions is defined as $\Psi \equiv \psi(x,y,z,t)$. Schrödinger was able to eliminate $E$ by setting $\Psi = \psi(q)\exp[2\pi i(E/h)t]$ which is simply the full equation combining the real and imaginary parts of a sinusoidal wave travelling in time (one-dimension relativistically speaking) with $kq = 0$ and $\omega = 2\pi i \frac{E}{h}$. Substituting this into his more general form of the time-*independent* wave equation he produced the Schrödinger time-*dependent* wave equation:

$$-\frac{h^2}{8\pi^2 m}\Delta\Psi + U\Psi = \frac{h}{2\pi i}\frac{\partial\Psi}{\partial t}. \qquad (4.5)$$

I will leave it in the form presented here since some of the combinations of constants present in this form became integral parts of Eddington's work.

Once again, the simplicity of the derivation is beautiful. Simple pre-calculus level mathematics describing sinusoidal curves combined with the already fairly simple time-independent wave equation produced one of the bedrock equations of modern physics. Often the complexity of the broader subject overshadows the simplicity of the basic assumptions and derivations.

**A New Interpretation of the Wave Function**

Let me return for a moment to 1926 and Schrödinger's attempt to interpret his own work. He ascribed to $\psi$ electromagnetic physical attributes and defined $\psi\psi^*$ as something he called the 'weight function' of this charge distribution. This is a term borrowed much later by Eddington and adapted to somewhat different purposes (see equation 8.7). But it is worth remembering this particular interpretation of this term. Schrödinger used it to draw a relationship between microscopic phenomena and macroscopic phenomena showing that the macroscopic behaviour of a linear harmonic oscillator can be described using microscopic wave packets. Unfortunately this led to several erroneous results on his part. His first flaw was assuming that the wave packets did not spread out in space. In addition he incorrectly proposed that $\psi$ was a cause of some radiation phenomena. Ultimately, everything hinged on the question of how a particle could remain stable in wave dispersion phenomena if it was simply a bunch of waves. A new interpretation of the wave function was needed to bridge the gap between matrix and wave mechanics.

It was Born who picked up the yoke on this task as I mentioned briefly above. He was studying quantum mechanical scattering processes, specifically those between free particles and atoms. He was actually employing Schrödinger's wave mechanics in this work, rather than his own matrix mechanics. Given an electron with an energy of $E = h^2/2m\lambda^2$ that approaches an atom whose unperturbed eigenfunctions are $\psi_n^0(q)$, he assigned it the following eigenfunction: $\psi_{nE}^0(q,z) = \psi_n^0(q)\sin(2\pi z/\lambda)$ (assuming it was approaching from some positive direction, $z$). He then assigned a potential energy to the interaction between the two particles and applied perturbation theory to obtain the following expression for a scattered wave a great distance from the centre of scattering:

$$\psi_{nE}^{(1)}(x,y,z,q) = \sum_m \iint d\omega \psi_{nm}^{(E)}(\alpha,\beta,\gamma)\sin k_{nm}^{(E)}(\alpha x + \beta y + \gamma z + \delta)\psi_m^0(q)$$ where $d\omega$ is an

element of solid angle in the direction of the unit vector with the components $\alpha$, $\beta$, and $\gamma$. This expression also contains the wave function that determines the differential or scattering cross-section for this direction. Born said that if this equation allowed for the particle interpretation of matter then there is only one possibility here. Originally Born said this possibility was $\psi_{nm}^{(E)}$. After reading galley proofs of his paper he changed this to $\left|\psi_{nm}^{(E)}\right|^2$ which measures the probability that the electron is scattered in the direction defined in the scattering expression (Jammer 1966). To emphasize the significance of this result in the history of quantum mechanics here are Born's own words: "The motion of particles conforms to the laws of probability, but the probability itself is propagated in accordance with the law of causality" (Born 1926, p. 804). Probability had found its home deep in the heart of quantum theory, though still imbued with the spirit of causality (thereby minimizing the heart palpitations of strict causal proponents such as Einstein). Born's result restricted classical, macroscopic phenomena to a limiting case of the new interpretation giving fundamental forces new meaning while still allowing them to predict results, as they had done successfully for centuries, consistent with the classical framework.

This, of course, was something Eddington wholeheartedly adopted as the core of his philosophy. In fact, the root of his ability to be both a strident relativist and a proponent of uncertainty is found right there in Born's statement. The preservation of causality satisfied his relativistic (classical) sensibilities, but the use of probability brought in a mathematical tool he could employ to deal with the problems that arose in measurement theory (see chapter five).

There is still a bit of related history that could be told here including the use of matrix mechanics to develop the transformation theory. The importance of the development of transformation theory by Jordan and, from a different point of view by Dirac and Fritz London (1900 – 1954), is that it generalized Born's interpretation and brought all the previous formalisms of quantum mechanics together in a single formulation (Jammer 1966). Both Jordan and Dirac's formalism utilized the idea of probability amplitudes and suggested that the correspondence principle might be irrelevant as a foundational principle in quantum mechanics to be replaced instead by statistical considerations. The full story of this stage in the development of quantum mechanics also includes, then, the story of the work of the mathematician David Hilbert (1862 – 1943) who, again in 1926 and continuing into 1927, together with his assistants Lothar Wolfgang Nordheim (1899 – 1985) and John (Johann, János, or Jancsi) von Neumann (1903 – 1957), elaborated in a purely mathematical way on the transformation work of Jordan and introduced the mathematical concept of Hilbert spaces to quantum mechanics.[11]

Subsequent to his work with Hilbert and Nordheim on transformation theory, von Neumann tackled the as-yet unsolved problem of fully merging matrix mechanics and wave mechanics. The major epistemological problem was that matrix theory consisted of a space of discrete variables while wave mechanics consisted of a space of continuous variables. Von Neumann contended that unifying the two theories required transforming a differential operator into an integral operator using 'improper' functions such as Dirac's $\delta$-function that in 1927 was considered mathematically illegitimate (it still is if it is considered an ordinary function). As such, von Neumann rejected the idea of unifying these two spaces and developed a completely new mathematical framework for quantum mechanics between 1927 and 1929, based on Hilbert's work on linear integral equations. Von Neumann's formalism is particularly suitable to nonrelativistic mechanics and its offshoots including relativistic quantum mechanics and quantum field theory. In developing it he contended that the set of all sequences in the discrete space of matrix theory, called the sequence space where the sum of all the probability amplitudes is equal to one, is essentially identical to the set of all summable and square-integrable complex-valued functions in the continuous space of wave mechanics. Essentially he took advantage of the fact that integrals are simply summations. The discrete space is Hilbert space. He first presented these ideas in May of 1927. His formalism, though new, simply recognized the meeting point of matrix mechanics and wave mechanics. He did not propose any new physical or epistemological assumptions (beyond those proposed by Dirac and Jordan in their transformation theories). This is largely since the uncertainty relations had only just been proposed in March and hadn't played any role in his early work. But they would have a profound impact on the later expansion of his theory and on his work in measurement, the latter a point he shared with Eddington.

**The Birth of the Uncertainty Relations**

---

[11] For a brief description of Hilbert spaces see Liboff 1998, pp. 101-106. For a more thorough description see van Fraassen 1991, pp. 144-146 and *Introduction to Hilbert Space and the Theory of Spectral Multiplicity* by Paul R. Halmos (1957).

The introduction of probabilistic methods to quantum mechanics brought in the now familiar idea of expectation values – rather than knowing quantities with certainty, the theory merely predicted what a quantity was *expected* to be. Mathematically the expectation value can be represented in modern notation as:

$$\langle C \rangle = \int CP(C)dC$$

where *P(C)dC* is the probability of finding the observable *C* in the interval *C, C +dC*. The integration is over all values of *C*. Born used expectation values in his probabilistic interpretation of the wave function, but no direct links between this idea and actual measurements had yet been developed, even though Bohr's frequency relation was actually part of the formalism and there was agreement with spectroscopic observations. There was still much left to interpret, in particular the meat of the formalism itself. As Jammer points out:

> This unusual state of affairs was due, of course, to the peculiar development of the formalism which at first was dominated by the correspondence principle but later detached from it (Jammer 1966, p. 323).

From an interpretational standpoint there were also problems with the very quantities and their linguistic representation in the theory. These problems were exacerbated by the many languages within which the theory was developed. From a purely mathematical standpoint, however, there was one crystal clear implication: the basic commutation relation *pq – qp = h/2πi* required that these quantities could not be solely defined by their classical meanings.

These problems were hardly an afterthought. They reared their ugly head right from the beginning reaching an early crescendo during Schrödinger's 1926 visit to Bohr's institute in Copenhagen where the clash between Schrödinger's continuous interpretation and Bohr's discrete interpretation stimulated extensive discussions in Copenhagen long after Schrödinger returned to Germany (Jammer 1966). It was nearly a miniature preview of the great Einstein-Bohr debate that was still a decade away. Ironically perhaps, the person who was able to see through the dogmatic adherence of both sides to their opinion (for that was all it was at the time) and realize that the root of the conflict was the lack of a definite interpretation of the actual formalism, was Heisenberg. Once again Heisenberg looked to relativity for guidance. He was inspired by the thought of what would have happened if the Lorentz transformations in relativity had been combined with the descriptions of space and time prior to Einstein's developments. Einstein had to completely reinterpret space and time in order to make his theory work and did so by turning the tables a bit: rather than describe nature by mathematics which is a top-down scheme of taking known mathematics and forcing it to apply to the real-world, assume, rather, that mathematics is the language of nature and we merely need to learn to interpret it.

      With that in mind, Heisenberg sought the ways in which nature has its voice in quantum mechanics and found that it was through the observables $p$ and $q$ (basically nature speaks to us through the things we can observe via our senses – see chapter three for a full discussion – and ultimately the interpretation of those observations involves mathematics at the most fundamental level). So, he simply assumed there was only the possibility of having $\Delta p \bullet \Delta q \geq h/2\pi$ as the relationship between these two observables. If that was true, could it make a consistent statement? Could experiment prove it to be true? These questions were retrospectively asked by Heisenberg himself (Jammer 1966). He reasoned that the formalism of quantum mechanics does not allow for classical descriptions of space and time nor does it completely preserve causality. Rather than develop a new conceptual framework, Heisenberg maintained the classical ideas but restricted their applicability to limiting cases.

      In late October of 1926 Heisenberg first stated the principle in ordinary language in a letter to Pauli. In it he states that the basic commutation relation clearly shows that it is "meaningless to speak of the place of a particle with a definite velocity" (Heisenberg in Fierz and Weisskopf 1960, p. 42). In February of 1927 Heisenberg, after much thought, sent another letter to Pauli who encouraged him to elaborate. Pauli then passed on the elaboration to Bohr who recommended a few changes. At the end of March the paper was submitted to *Zeitschrift für Physik*. It was not the first realization that $p$ and $q$ cannot be simultaneously known – both Dirac and Jordan made statements to that affect

(Jammer 1966). The difference was that Heisenberg performed a mathematical investigation into the relationship between the distributions of *p* and *q*. His line of reasoning began with the transformation theory and assumed that the dependence of *q* was described by a Gaussian error curve[12]. Heisenberg's conclusion was that the product of the uncertainties $\delta q$ and $\delta p$ was $h/2\pi$. This is a "direct intuitive interpretation" of the basic commutation relation (quoted in Jammer 1966, p. 328). As early as 1924 his mind had been preoccupied by a thought experiment with a $\gamma$-ray microscope. Notably, in 1923, Wilhelm Wien (1864 – 1928) asked Heisenberg about the resolving power of such a microscope at Heisenberg's oral PhD defence; Heisenberg was unable to answer the question. He used some of the knowledge he gained from this experience while thinking about uncertainty, though when actually discussing the experiment he did not account for the angular aperture of the objective lens (something Bohr soon corrected).

The thought experiment consisted of illuminating an electron with radiation of a reasonably short wavelength and then observing it under a microscope. However, the Compton effect immediately means that the impact of the light (quanta) used to observe the electron changes its (the electron's) momentum dramatically (this is also one of the first descriptions of the idea that an observer has no choice but to interfere with the observable in order to make the observation). The electron's momentum cannot be completely determined, however, since the angle of the scattered photon cannot be determined by the microscope (within the range provided by the wave packets entering the microscope). But the electron's position, however, is now known definitely. Quantitatively, if an identical observation is made on an electron multiple times to locate its position, even if the experimental setup is identical each time, the results will not always be the same.

Heisenberg also applied this reasoning to a Stern-Gerlach experiment (which will be discussed in the next section) and found that the faster an atom crossed the deviating field, the less precise was the measurement of its energy: $\delta E \cdot \delta t \sim h$. As such he concluded that *any* two canonically conjugate quantities observed simultaneously carry with them some level of uncertainty. Introducing one of the most fundamental interpretations in all of quantum mechanics, he held that a classical notion such as a path or trajectory could be retained since "the path comes into existence only when we observe it" (Heisenberg 1927, p. 185). This also presents the idea that the wave and particle natures of particles are not simultaneously observable using the exact same apparatus for when a quanta (photon, electron, etc.) exists in a well-defined spatial location it acts like a particle whereas if it is not confined to a well-defined spatial location its momentum can be defined more precisely and it behaves like a wave (Liboff 1998). A simple thought experiment that I use when first presenting this idea to my students, though not entirely accurate, at least gives the flavour of the problem: imagine you have a videotape of a ball falling to the ground next to a ruler. In order to measure the momentum one needs velocity which requires several measurements (several frames of the film) since velocity is defined as a *change* in position during a given time interval (instantaneous velocity is simply a limit here) so there is no single position. Conversely, if you want to see the ball at a specific point along the ruler you must pause the tape at the point and the ball has no velocity – it's frozen in place for the purposes of your measuring its position.

---
[12] Eddington relied heavily on Gaussian distributions as will become apparent in subsequent chapters.

Returning for a moment to the idea that this relationship holds for any two canonically conjugate quantities a generalized version of the uncertainty relations can be written. Suppose that two observables $\hat{A}$ and $\hat{B}$ anticommute:

$$\left[\hat{A}, \hat{B}\right] = \hat{C} \neq 0 \quad (4.6)$$

then if the measurement of A is uncertain in some state the measurement of B will also be uncertain such that:

$$\Delta A \cdot \Delta B \geq \frac{1}{2} |\langle C \rangle| \quad (4.7)$$

It will also be helpful to note the relationship between this and probability since the two are inextricably linked. As Liboff explains:

> If $\hat{A}$ and $\hat{B}$ do not commute, then the eigenstate $\varphi_a$ of $\hat{A}$ which the system goes into on measurement of A is not necessarily an eigenstate of $\hat{B}$. Subsequent measurements of *B* will give any spectrum of eigenvalues of $\hat{B}$ with a corresponding probability distribution *P*(*b*) (Liboff 1998, p. 146).

The probability amplitude (i.e. the probability of some measurement of *B* occurring) is $P(b) = |\langle \varphi_b | \varphi_a \rangle|^2$. This generalization of the uncertainty relations, appearing here in more modern notation, is the result of work by Schrödinger who improved upon the results of H.P. Robertson (1903 – 1961) and Edward Condon (1902 – 1974) (Jammer 1966).

A particle such as an electron, then, is represented by a wave packet in configuration space that is composed of the eigenfunctions of a collection of states. Its size is determined by the precision of the position measurement. In Heisenberg's thought experiment with the microscope this precision, and thus its size, is determined by the wavelength of the light illuminating the electron. The packet can describe an orbit if the electron is bound to an atom just like a classical particle except that the packet spreads out with time. Each consecutive observation of the packet essentially shows a smaller and smaller packet. This is basically a temporal sequence of the locations of the wave packet (or, rather, where it was observed) and it produces the orbit. Heisenberg utilized these results in reaching several conclusions that have tremendous philosophical implications. The first is that quantum theory, despite employing statistics, is *not* limited to only statistical conclusions. He pointed to the experiments of Hans Geiger (1882 – 1945) and Walther Bothe (1891 – 1957) as examples of this. The second is that the problem with causality in quantum theory is *not* the inability to predict the future from knowledge of present events, but rather the inability to actually *have a full and complete knowledge of the present* (this is the origin of the completeness debate in quantum mechanics). In a rather titillating hint of things to come he says:

> it may be suggested that behind the statistical universe of perception there lies hidden a 'real' world … ruled by causality. Such speculations seem to us – and this we stress

> with emphasis – useless and meaningless. For physics has
> to confine itself to the formal description of the relations
> among perceptions (Heisenberg 1927, p. 197).

In this statement he clearly eschews interpretive work such as complementarity as outside the scope of physics. Conversely, as Folse points out, Bohr viewed complementarity and the uncertainty principle as consequences of the quantum postulate where his was "the consequence for the conceptual framework …, while Heisenberg's discovery was its formal, mathematical consequence" (Folse 1985, p. 128). The influence of Bohr and Heisenberg on Eddington is less direct since Eddington reinterpreted the new quantum theory in his own unique way.

**Interpretations of Uncertainty and Bohr's Complementarity**

The philosophical implications of uncertainty were not of immediate importance in 1927. Physicists were focusing on the definability and measurability of the various aspects of quantum mechanics. Approaches to the quandaries of the theory primarily involved practical laboratory results that could be obtained from their proper consideration. Heisenberg's paper had plenty to say on these subjects apart from its statements on the philosophical nature of quantum mechanics. Of particular relevance to Eddington's treatment of the theory is Heisenberg's view that something is defined if it is measurable. So, to Heisenberg, objects that could not be measured were useless since they couldn't be defined. In addition Heisenberg's use of light as both a measuring tool and an object to be measured refocused the attention of physics to the importance of measurement, something Eddington treats in depth.

Philosophically it took some time for his conclusions about causality to be fully appreciated. Even in the philosophy community, upon which Heisenberg's conclusions had huge effects, didn't realize initially what had happened and, when they did, they were essentially taken by surprise (Jammer 1966). E.H. (Earle Hesse) Kennard (b. 1885) who happened to be visiting Copenhagen at the time of Heisenberg's discovery was, apart from Bohr and Pauli who proof-read Heisenberg's manuscripts, the only person who really understood the implications of Heisenberg's work calling the uncertainty relations "the core of the new theory" (Kennard 1927 quoted in Jammer 1966, p. 333). Hermann Weyl's (1885 – 1955) *Theory of Groups and Quantum Mechanics* that first appeared in 1928 was the first text on the subject that gave the uncertainty relations a central role in the theory.[13] The relations were criticized for a time (for instance the philosopher Karl Popper (1902 – 1994) remarked that Heisenberg has tried "to give a causal explanation why causal explanations are impossible" (Popper 1935, p. 184)), but that did not stop continued work on them by the likes of Kennard and Arthur E. Ruark (b. 1899), who focused on the experimental aspects, and Robertson and Condon whose mathematical work was discussed above.

As we have already seen, related to the uncertainty relations was Bohr's idea of complementarity – the fact that two mutually exclusive descriptions of the world, one discrete (particle) and the other continuous (wave), are not only allowed but required by the laws of quantum mechanics. Though Bohr's work was partially aimed at Heisenberg's principle, Folse maintains that Bohr never intended it to be limited to *only* the uncertainty principle (Folse 1985). Since he viewed both his and Heisenberg's work as consequences of the quantum postulate it is natural to assume he felt his work stood well on its own not to mention the fact that it was Bohr himself who proposed the quantum postulate in the first place (to explain the stability of atoms). In this light complementarity appears to logically precede uncertainty and, in fact, Heisenberg began his work on uncertainty on the somewhat isolated island of Helgoland after an intense visit with Bohr (Beller 1999).

Jammer maintains that complementarity in Bohr's mind originated in his final acceptance of the wave-particle duality of quanta. Experimentally there seemed to be no explanation for the apparently paradoxical behaviour of quanta under differing conditions – why would an electron behave like a particle in one experiment and yet behave like a wave in another, even if it was the same exact electron in both experiments? Bohr was originally opposed to Einstein's idea that light also came in packets or quanta (photons). But the Bothe-Geiger experiments provided overwhelming evidence in support of

---

[13] Another connection in the complex web of relations among physicists is that Weyl's text was first translated into English by Robertson in 1931.

Einstein. Jammer maintains that the first evidence that truly supports Bohr's move to acceptance of wave-particle duality appeared in the postscript of a 1925 paper of his where he discusses the reconciliation of the Bothe-Geiger experiments with work performed by himself, Hendrick Anton (Hans) Kramers (1894 – 1952), and John C. Slater (1900 – 1976). In it he says that "one should not be surprised if the required extension of classical electrodynamics leads to a far-reaching revolution of the conceptions on which the description of nature has been based so far" (Bohr 1925 as quoted in Jammer 1966, p. 346). Beyond this the true germination of his ideas is difficult to track. But it is clear that during the time between this comment and his first presentation of complementarity in 1927 he struggled primarily with the logic that was underneath the formalism itself. For instance, both the equation for the energy of a photon, $E = h\nu$, and the equation relating the momentum of a photon to wave number, $p = hk$, contain particle and wave elements. Simply glancing at these two equations should reveal that the answer must relate in some way to Planck's constant.

Bohr found no answer to this quandary in existing logical conjectures and thus determined that a new logical framework was required to make sense of these results. He called his logical framework 'complementarity' in reference to the fact that two different descriptions of phenomena that are mutually exclusive are nonetheless required for a full description of the situation. The uncertainty relations simply provided him with a concrete measure of what had to be sacrificed in violating the normally rigorous exclusion of conceptual ideas. However, they also provided him with a mathematical assurance that complementarity wouldn't lead to a logical contradiction since if one quantity is measured to a great degree of accuracy the other complementary one would be nearly immeasurable – no physical situation can simultaneously *and* rigorously display both complementary quantities.

I have always personally had trouble with the solidity of the logical foundation upon which Bohr's idea rests – it almost seemed like a bit of a cop out to me, though I have never gone so far as to support a theory involving hidden variables. Bell's Theorem and subsequent supporting experiments such as those performed by Alain Aspect (based on an idea originally put forth by Bohm) at Orsay in the early 1980s seem to offer unquestionable support for the notion that uncertainty is inherent in the fabric of the universe and that complementarity correctly interprets the wave-particle duality and so I am cautious in my criticism. But at the time Bohr argued that the reason it works is due to our inability to truly define what we mean by 'observation' because the observation itself interferes with the object under observation (see Heisenberg's thought experiment with the γ-ray microscope discussed previously). In essence, this presents the problem of truly objective observation as I discussed in the preceding chapter: our subject view may be behind wave-particle duality. Eddington may, in fact, have been dissatisfied with Bohr's idea as well and was subsequently inspired to search for a completely objective way to observe the universe.

**Exclusion and Spin**

Parallel to these developments in uncertainty and the new mechanics there was quite a bit of work being done in spectroscopic and particle beam analyses. In the early 1920s one of the major experimental problems that was at the forefront of physics research was finding a full physical explanation for the fine-structure of spectral lines. In late 1921 Otto Stern (1888 – 1969) and Walther Gerlach (1889 – 1979) used evaporation in a heated oven to produce a beam of silver atoms that was directed through a nearly perfect (high) vacuum through collimating slits that were situated along the gradient of the magnetic field at the sharp edge of the pole piece of what is known as a DuBois electromagnet. Through a series of similar experiments they were able to show that the atomic beam split into two 'beamlets' in the presence of a magnetic field. Classically the resulting distribution of atoms on the collecting glass should have been Gaussian since classical mechanics implies a continuum of possible results including atoms that would have not been affected by the magnetic field at all.[14] However, Stern and Gerlach's result displayed *no* atoms unaffected by the magnetic field – all the atoms appeared in one of two locations on the glass, a clear indication of the quantization of space. Additionally they were able to show that the atoms all had a magnetic moment aligned with the field direction. In and of itself these results were actually not unexpected. Arnold

---

[14] This experiment is an excellent demonstration of the different meanings given to probability in classical and quantum situations (see chapter three).

Sommerfeld (1868 – 1951) and Alfred Landé (1888 – 1976) developed what is often referred to as the 'magnetic-core hypothesis' between 1921 and 1923 in which the nucleus and inner (nonoptical) electrons (often referred to collectively as the atomic core) possessed an angular momentum that was some multiple of $h/2\pi$ along with a magnetic moment (the connection between the two was tenuous at that time – see below). Thus the angular momentum vector of the optical electron can only assume discrete inclinations with respect to the axis of the core. The assignment of angular momentum to the electron, which was considered by many to be a point particle, was not universally accepted. Pauli, as I will show, did not actually use this exact concept when initially developing the exclusion principle. Regardless of the truth of the Sommerfeld – Landé theory, the Stern-Gerlach experiment did raise some important questions. What mechanism produced the two beamlets found by Stern and Gerlach and was it related to magnetic moments or, in the 'magnetic-core hypothesis,' the angular momenta?

The 'magnetic-core hypothesis' or Sommerfeld – Landé theory, developed in part to explain the anomalous Zeeman effect in multiplet spectral lines, had earlier that year provided a theoretical framework that apparently fit the Stern – Gerlach results perfectly, another example of theory running ahead of experiment. But there was one nagging problem that was not answered by either the theory or the experiment and was first pointed out by Einstein and Paul Ehrenfest (1888 – 1933) – just *how* do the atoms orient themselves in the field? Since the field was in a high vacuum collisions could be ruled out as a possible explanation. The same could be said for radiative energy exchanges since they require a considerably longer timespan than was observed in the alignment. That left only two solutions: either the atoms never assume states in which they are not completely quantized (i.e. they're quantized to begin with) or states could occur during rapid changes that violate normal quantum rules. Einstein and Ehrenfest showed that *either* assumption leads to conceptual problems.

In a somewhat providential circumstance, Pauli completed his PhD in 1921 under Sommerfeld and was thus directly exposed to the early development of the Sommerfeld – Landé theory. In the winter of 1921/22 he was in Göttingen as an assistant to Born. It was there that he met Bohr who was giving a series of guest lectures on his research on the periodic 'system' (table) of elements. In the fall of that year Bohr invited Pauli to Copenhagen to assist him in the German edition of his works. It was then that Pauli first undertook to explain the anomalous Zeeman effect.[15] In two papers, the first one written in Copenhagen and the second in Hamburg where he accepted a position in 1923, he attempted to generalize some of the results produced by Sommerfeld regarding the Zeeman effect in alkali and alkaline-earth spectra. Without any theoretical justification at all he directly associated magnetic moments with angular momenta and fully traced the splitting. This was the first instance of this association and helped explain some of the results of the Sommerfeld – Landé theory. But Pauli felt his work was unfinished since he had been unable to explain why electron shells in atoms became filled or closed – why couldn't any number of electrons exist in a given shell? In the fall of 1924 Pauli, in fact, showed that the 'magnetic-core' hypothesis as proposed by Sommerfeld and Landé was actually inconsistent with experimental results (this is perhaps why he initially resisted the idea of assigning rotational quantities directly to an electron).

Based on these results, both theoretical and experimental, Pauli assumed that closed shells must have an overall angular momentum of zero and an overall magnetic moment of zero – i.e. in Sommerfeld – Landé theory if any of the electrons in them had angular momenta and magnetic moments, they all must cancel in order to preserve the balance of the shell. As such, Pauli concluded that, at least for alkali atoms, the outermost shell or valence electron must be the only part of the atom that contributes to its angular momentum and its energy changes in external magnetic fields. But, Pauli did *not* recognize the fact that the angular momentum here really *belonged* to the electron. Since it was assumed to be a point particle in this instance, angular momentum was a meaningless concept. It definitely affected the core's angular momentum, but exactly how was not yet known. Here yet *again* the theory of relativity played a role in the furthering of quantum theory: Pauli was an expert on the subject since he had written an article on it for an encyclopaedia at the age of twenty and it was a theory he ardently supported and believed in. This led him to actually reject the Sommerfeld – Landé 'magnetic core' hypothesis and opened the door to spin (Jammer 1966).

---

[15] In an oft quoted comment, Pauli said in 1945 while addressing the Institute for Advanced Study in Princeton, "A colleague who met me strolling rather aimlessly in the beautiful streets of Copenhagen said to me in a friendly manner, 'You look very unhappy'; whereupon I answered fiercely, 'How can one look happy when he is thinking about the anomalous Zeeman effect?" (Pauli 1946).

By this time (the fall of 1924) an acceptable description of the periodic system of elements had been devised largely by Bohr building on the ideas of Walther Kossel (1888 – 1956). He had proposed a basic scheme by 1921 and subsequently developed more detailed schemes over the next few years. This work deviated slightly from experimental progress made around the same time by de Broglie and Alexandre Dauvillier (1892 – 1979). But, in the fall of 1924, Edmund Stoner (1899 – 1968) succeeded in merging the experimental results with the theory, mostly provided by Bohr but including some of Landé's 1922 suggestions such as the idea of assigning three quantum numbers to each sublevel (Stoner 1924).

Stoner's article proved to be tremendously influential on Pauli in his further deliberations on the subject. Stoner's scheme has been shown to be the correct scheme quantum mechanically, but well before that it was still considered an improvement over the work of Bohr (and Landé whose theory, though a solid foundation for these subsequent studies, did not enjoy the wide-spread popularity of Bohr's). The primary difference between Stoner's scheme and Bohr's is that in Stoner's there is a greater concentration of electrons in the outer sublevels. Six months earlier chemist J.D. Main Smith (dates unknown) found results consistent with Stoner's from purely chemical considerations.[16] Stoner's paper, however, connected the distribution of the electrons with the problem of multiplet structure. So, now the stage was set for Pauli to put everything together – angular momentum, magnetic moment, multiplet structure, and electron shell restrictions.

Pauli added one component to Stoner's work (and thus the work of Main Smith, Bohr and Landé) – a fourth quantum number to represent the component of angular momentum of the *atom* (and not yet the electron – really this represented how much the electron supposedly contributed to the atom's overall angular momentum) that is in the direction of an external magnetic field applied to the atom. Pauli then realized that the shell structure of atoms is perfectly explained if each possible orbit or state is assigned a

---

[16] Main Smith wrote a letter in response to Stoner's publication claiming priority. Subsequent papers in both the physics and chemistry community often refer to both Stoner and Main Smith as the correct modifiers of Bohr's scheme.

set of four quantum numbers and only *one* electron is allowed to occupy each of these states.[17]

There was still one unsolved problem. Bohr had asked why don't *all* the electrons occupy the innermost shell when the atom is in the ground state? Pauli suggested that his solution of one electron per state (or orbit) could be interpreted as a principle. In essence it explained the shells and the resulting periodic system of elements because it *excluded* more than one electron from occupying a state or orbit. Pauli found a simple test for the principle was the case of triplet *s* terms for two essentially equivalent electrons (same *n* and *l* values) in an alkaline-earth atom. In this case both electrons would have to share the same exact set of quantum numbers. Pauli's principle supposedly excluded this from happening so the fact that cases of these electrons do not appear in nature seemingly vindicates the principle. Since the exact meaning of the third quantum number was not established at the time Pauli could only show that the principle worked in the presence of strong external magnetic fields, but he was able to employ thermodynamics to show that at least the *number* of states didn't change when transitioning to a weaker field. As such the principle was validated generally, if not specifically.

In that spectacular year of 1926, while the formalism of quantum mechanics and the philosophy of uncertainty were just coming together, Heisenberg , Dirac, and Pauli, among their many other accomplishments, were able to extend the essence of the principle to include the requirement that all state functions (including spin) of a system of similar particles, assuming they are fermions (meaning they obey Fermi-Dirac statistics which was in the process of being developed that very year in relation to the transformation theories discussed above), must be antisymmetric with respect to the particle exchange – i.e. the particles must be dissimilar in some respect. As such exclusion became a root principle by which other physical laws were generated and its applications are wide. In 1921 Compton had concluded "that the electron itself, spinning like a tiny gyroscope, is probably the ultimate magnetic particle" (Compton 1921, p. 155) but wasn't certain if it was a highly important or completely useless conjecture. Spin was proposed in 1925 by Samuel Goudsmit (1902 – 1978) and George Uhlenbeck (1900 – 1988) based on spectroscopic evidence as well as Pauli's development of the exclusion principle. Primarily, however, they had pondered Pauli's work and found they could only understand his additional quantum number (for spin angular momentum) if the electron were not a point particle but rather a small, rotating sphere. Unfortunately, in this scenario they found that the rotational velocity at the surface of the electron exceeded the speed of light several times over. But, if the electron held the rotational properties of a sphere yet had no size, being a point particle, one needn't worry about this issue since there would be no surface for which to calculate this rotational velocity. Their paper, published in 1925 only on the insistence of Ehrenfest, suggested that the angular momentum of the atom was really the angular momentum of the valence electron that imposed a stress on the atomic core enabling it to assume two orientations, like the electron itself, rather than one (Kragh 1999 and Jammer 1966). The speed of light problem was mentioned briefly in a footnote.

---

[17] The four quantum numbers are, in modern notation: $n$, for the energy level, $l$, for the number of angular cycles of the wavefunction around the atom (also the number of bumps in the squared wavefunction), $m$, for the number of possible orientations of the wavefunction in space, and $m_S$, for the spin orientation (see Moore 2003b, p. 163). The physical meaning of $m$ was not well-defined when Pauli was first developing the exclusion principle and the spin was only associated with the electron later in 1926 by Goudsmit and Uhlenbeck.

      Classical electrodynamics shows that magnetic fields arise from changing electric fields, thus the rotation of the electron is what produces the magnetic moment that thus forces the electrons to align with an external field. In retrospect, then, the principle also explained the Stern – Gerlach results – the atomic beam split because half the atoms had valence electrons that aligned with the field direction while the other half antialigned with the field direction. For instance, think of a compass needle. It always aligns itself with a magnetic field line with the south pole of the magnet being attracted to the north *magnetic* pole of the earth and the north pole of the magnet being attracted to the south *magnetic* pole of the earth.[18] In the Stern-Gerlach experiment there was no telling which way the valence electron was spinning and thus which direction (by the old right-hand rule) the magnetic moment was pointing (i.e. which end of the electron was north or south as it entered the magnetic field). As it turns out, statistical calculations of the multiplicities show that it's about half and half.

---

[18] This is a common misconception about the Earth – the north geographic pole is actually the *south* magnetic pole so when a compass points north that end of the compass really is the north pole of the compass needle.

So, from about 1921 to 1926, it was established that valence electrons account for the quantum properties of the atoms they are in and that they spin on an axis. Since they represent charge and thus have an electric field, this rotation produced a magnetic field whose direction was given by simply applying the right-hand rule. When subjected to an external magnetic field, then, they would line up their magnetic fields with the external field – the north pole of the electron would be attracted to the south pole of the field and vice-versa. Pauli was finally convinced in late 1926 after L.H. (Llewellen Hilleth) Thomas (1903 – 1992) corrected an error of a factor of two in the calculation of the doublet separations, a result subsequently verified by Yakov (Iakov or Jakov) Frenkel (1894 – 1952). Pauli's opposition stemmed from the view that rotation (spin) was a wholly classical phenomena and so could not be represented by a quantum number. His eventual acceptance of spin, however, would lead him in 1927 to develop, in nonrelativistic quantum mechanics, a consistent theory of spin for the electron utilizing his now famous spin matrices. In doing so he reinterpreted spin to be a wholly quantum mechanical property. In modern notation, the projection of the spin angular momentum on the $z$-axis for any quanton (atom, electron, etc.) is:

$$S_z = s\hbar, (s-1)\hbar, ..., -s\hbar \qquad \text{where } s = 0, \frac{1}{2}, 1, \frac{3}{2}, ...$$

and, quantities containing $\hbar$ are quantized, so spin indeed is a quantum mechanical property.

**The Merger of Uncertainty and Exclusion**

I have just traced, then, the early history of the uncertainty principle and the exclusion principle (along with the related development of the new mechanics) up to about 1928. Both were separate threads but often explored by the same people. As such it was natural for someone to attempt to find a link between the two quantum phenomena. This happens to be where Eddington fully enters the picture.

In 1928 Dirac had published a generalization of the Schrödinger equation that consisted of a set of four simultaneous first order partial differential equations (Dirac 1928a) that is now known as the Dirac equation. In 1926-27 Oskar Klein (1894 – 1977) and Walter Gordon (1893 – 1939) independently attempted to construct a relativistic wave equation (as had Schrödinger himself privately). The Klein-Gordon equation, as it is now known, had two problems, however. It did not correctly predict the fine structure of hydrogen (it is still useful for spin-0 particles). In, what appeared to be a separate problem at the time, it also could not be combined with Pauli's 1927 theory of spin. Of course we now know that these two problems are, in fact, two faces of the same single problem since fine structure is related to spin. Dirac's follow-up to Klein-Gordon in 1928, despite not directly introducing the idea of a spinning electron, actually contained the correct spin in the result (Kragh 1999). His paper contained a generalization of the Schrödinger equation that consisted of a set of four simultaneous first order partial differential equations (Dirac 1928a) that is now known collectively as the Dirac equation.[19]

Spin turned out to be a monumental discovery on par with the de Broglie relation, particularly in relation to Eddington's work since it led directly to the exclusion principle that was the second pillar of his theory (uncertainty being the first). I have already shown that spin was considered a fundamental part of quantum mechanics fairly soon after its discovery. As such any further work in the field needed to account for it. As I previously stated the Klein-Gordon equation, which was the first relativistic attempt at a wave equation, did not account for spin, but Dirac's equation did. In fact, Pauli's spin matrix development of 1927 turned out to be a nonrelativistic limit of Dirac's equation. Since Dirac's equation was a form of the wave equation, spin, which is so vital to the core of the exclusion principle, was merged with the core of wave mechanics that held as a basic tenet the uncertainty principle. The uncertainty principle was derived from the basic commutation relation that corresponded to the time-independent wave equation. Dirac's relativistic wave equation was described by Bohm as

> a first-order relativistic wave equation [containing] four complex wave functions. The extra wave functions correspond to additional variables, which can be related to the spin and charge of the electron. In this way, he is able to obtain conserved probabilities, as well as an accurate description of many of the relativistic properties of the electron, not treated correctly by any other theory (Bohm 1951 [1979], p. 90).

---

[19] Dirac's paper was communicated to the *Proceedings of the Royal Society of London* by none other than Ralph Fowler (see chapter two).

Schrödinger's wave equation was thus a nonrelativistic limit of Dirac's equation. This link between uncertainty and exclusion is tenuous in this form, however. It would be better to find a direct link between the two principles and, indeed, one does exist – a rather simple one, in fact.

Pauli's spin matrices

$$\sigma_x = \begin{pmatrix} 0 & 1 \\ 1 & 0 \end{pmatrix} \qquad \sigma_y = \begin{pmatrix} 0 & -i \\ i & 0 \end{pmatrix} \qquad \sigma_z = \begin{pmatrix} 1 & 0 \\ 0 & -1 \end{pmatrix}$$

are simply a way of representing the coefficients in front of $\hbar$ in matrix form in the general equation for the components (axis projections) of spin angular momentum, generalized here for any direction: $S_n = s\hbar, (s-1)\hbar, ..., -s\hbar$. $S$ here is actually an operator and the components satisfy the following commutation relation:

$$\left[S_x, S_y\right] = i\hbar S_z.$$

The Pauli matrices also satisfy a commutation relation:

$$\left[\sigma_x, \sigma_y\right] = 2i\sigma_z$$

(see Gasiorowicz 1996, p. 242). Both $S$ and $\sigma$ satisfy equations (4.6) and (4.7) and thus obey the uncertainty principle. Experimentally this can be confirmed by a standard experiment utilizing a Stern-Gerlach device (see Moore 2003b for a simple demonstration of such a device) and is something quantum physicists nowadays take for granted.

    The year following these developments Heisenberg and Jordon incorporated spin into quantum mechanics to correctly derive hydrogen's fine structure as well as explain the anomalous Zeeman effect.

**Enter Eddington, Structuralism in Tow**

Eddington was actually present in these developments from the very beginning. Like Heisenberg he was influenced by relativity, but to a much greater extent. In fact in general the history of the development of quantum mechanics rests partially on the shoulders of relativity. Some of this development actually arose from early attempts at a unification of fundamental forces, something Eddington was active in from the start.

One of the earliest attempts to unite gravity and electromagnetism using a field theoretic approach was proposed by Hermann Weyl (1885 – 1955) in 1918. Some three years later Weyl's trail was picked up by Eddington who published a generalization of Weyl's theory (Eddington 1921). As Eddington explains in the beginning of his article, Weyl

> has shown that, on removing a rather artificial restriction in Riemann's geometry, the expression for the metric includes terms that are identified with the four potentials of the electromagnetic field (Eddington 1921, p. 104).

Eddington felt, however, that Weyl's approach was still too restrictive and worked to elaborate on it. In doing so he began to contemplate the structure of the universe in a more concrete way. This paper could be interpreted as Eddington's first mathematical attempt at describing a universal structure. In Eddington's own words:

> The natural geometry of the world … is the geometry of Riemann and Einstein, not Weyl's generalised geometry or mine. What *we* have sought is not the geometry of actual space and time, but the geometry of world-structure, which is the common basis of space and time and things (Eddington 1921, p. 121).

This foreshadows his concept of a universal background (uranoid) upon which everything is built and gave him the freedom to make uncertainty inherent in the fabric of space-time itself (see chapter five). It establishes his view of the unity and continuous nature of the universe through geometry and structure and develops a link between the very large (gravity) and the very small (electromagnetism) that was to be a cornerstone of his *Fundamental Theory*. In a letter to Bohr in 1923, Einstein remarked "I have finally understood the connection between electricity and gravitation. Eddington has come closer to the truth than Weyl" (Einstein 1923, reprinted in French 1979).

It is to be remembered that relativity here is essentially a classical theory and Einstein a classical thinker who put his faith in strict causality over probability. As he wrote to Born in the following year "I find the idea quite intolerable that an electron exposed to radiation should choose *of its own free will*, not only its moment to jump off [the atom], but also its direction" (Einstein 1924, quoted in French 1979). Einstein's endorsement of Eddington's attempt at unification appears to put Eddington in the mould of a classical thinker. As I will show, one of Eddington's most unique qualities was his ability to transcend classical and quantum labels. Whether he was successful or not is immaterial; it was simply his ability to not be dogmatically trapped by a single ideology that is admirable.

This early attempt at unification followed on the heels of a few investigations he had performed into the astronomical applications of electricity (Eddington 1917 and 1918a). However in generalizing Weyl's theory he crept dangerously close to considering, for the first time, microscopic phenomena. Up to this point in Eddington's career he had been entirely concerned with macroscopic phenomena, especially on astronomical scales. Thus, with these ideas firmly planted in his mind, already teetering on the edge of the microscopic world, he began investigating electrons themselves in astronomical situations (see for example Eddington 1923a and 1925b). Simultaneous to this he began studying rotational motion in primarily stellar situations (see for example Eddington 1923b and 1925c). In 1926 he considered a solely microscopic phenomena for the first time by merging his two lines of thinking: electrons and rotational motion. The result was a letter to *Nature* simply titled "Spinning Electrons" (Eddington 1926). This followed on the heels of Goudsmit and Uhlenbeck's initial description of spin in 1925 and actually *preceded* Pauli's 1927 theory of spin (perhaps he was inspired by Compton's conjecture of 1921 or by Goudsmit and Uhlenbeck – see above).  The fact that Eddington's work is not remembered speaks for itself, but it really is the first paper he wrote on a quantum mechanical topic.

The following year Eddington published his second paper on quantum mechanics (this one memorable enough to deserve mention in some histories on the subject – see Jammer 1966).  It was an integral part of the scientific debate on the problem of the completeness of the eigenfunctions in Schrödinger's wave equations.  It was a problem Schrödinger himself had recognized.  Schrödinger employed the use of the Laplace transformation to find the solution of the radial equation in the standard Kepler problem of radial orbits.  This solution, however, says nothing about the completeness of the eigenfunctions.

Eddington was able to show that the radial equation actually can be solved in a much more elementary way and that the solution was actually implicitly contained in several other standard treatises on the subject (Eddington 1927 and Jammer 1966).  Eddington specifically mentions the work of Edmund Whittaker (1873 – 1956) who is precisely the person who assembled *Fundamental Theory* from Eddington's notes after his death.  Whittaker and George Watson (1886 – 1965) had published the book *Modern Analysis* whose fourth edition appeared in 1927 and included what is now known as the 'Whittaker integral' (Eddington referred to it as the 'Whittaker function').  It was a generalization of various special functions (i.e. the special functions were special cases of this integral) and resulted from work he performed on partial differential equations that also produced a new solution for the wave equation and a general solution of the Laplace equation that included Eddington's solution to the radial equation in the Kepler problem.

**Eddington Out of the Twenties**

In the beginning, then, Eddington latched onto Schrödinger's wave mechanics rather than Born's matrix mechanics. This was a pivotal moment for Eddington.  Considering the heavy influence relativity had on him and his strong mathematical background he was naturally disturbed by the fact that Dirac's relativistic wave equation was not presented in tensor calculus form.[20] But he was equally influenced by wave mechanics. He immediately produced a sort of rebuttal to Dirac and began the process of creating a 'wave-tensor' calculus that took the merger of matrix and wave mechanics to another level by using tensors rather than vectors that are the basic components of matrices (Eddington 1928). This is often considered the first paper in the development of his 'Fundamental' theory or 'fundamentalism' as Kragh puts it (Kragh 1999) and will be elaborated upon in chapter five.

So by 1928 wave mechanics, matrix mechanics, the uncertainty relation, and the exclusion principle (via spin) had all been related and could all be found either directly in, or by taking some sort of limit on, Dirac's relativistic wave equation. As I have already hinted, this was a watershed moment for Eddington: the paper in which Dirac formally proposed his equation was perhaps the single most influential moment in Eddington's career as it put him on the path of his ardent 'fundamentalism,' something that would change his life and his stature among physicists.

---

[20] In fact, Charles Galton Darwin (1887 – 1962), grandson of the evolutionist and collaborator with Ralph Fowler (see chapter two), was the first to officially note that Dirac's work was not in tensor form (Darwin 1928).

# V

## *Probability Leads to Uncertainty*

Dirac's paper took on the challenge of obtaining a "wave equation … which shall be invariant under a Lorentz transformation and shall be equivalent to [the Klein-Gordon equation] in the limit of large quantum numbers" (Dirac 1928a). The most general form in which Dirac presented his equation was

$$\left[ p_0 + \frac{e}{c} A_0 + \rho_1 \left( \sigma, \mathbf{p} + \frac{e}{c} \mathbf{A} \right) + \rho_3 mc \right] \psi = 0 . \tag{5.1}$$

Eddington was particularly bothered by the fact that this equation was not in tensor calculus form, though as I have noted before, Darwin was the first to notice this (Eddington credits Darwin with this point at the beginning of his paper). Darwin, however, did not take steps to rectify the apparent problem:

> here we have a system invariant in fact but not in form. Should it not be possible to give it formal invariance as well … ? It is so possible, but it is not hard to show that it requires no less than 16 quantities to do it … and even so each will have a real and imaginary part, so that we may say that 32 quantities are required! … it is rather disconcerting to find … that physical quantities exist which would be … very artificial and inconvenient to express as tensors (Darwin 1928, p. 657).

He had assumed that these physical quantities simply could not be expressed in tensor form. Eddington, on the other hand, was determined to find a way around this difficulty. In *his* follow-up to Dirac's work Eddington presents the following modification.

He begins with a tensor containing five components that are regarded as the coordinates in a 5-space. By doing this he is essentially employing the methods of Kaluza and Klein less than a decade after their introduction of a fifth dimension in order to more easily incorporate the complicated terms. He calls the 5-space, *t*-space since he is working with a hypothetical tensor *T*. The original 4-space we are all familiar with is referred to as $\psi$-space. The additional dimension accounts for a rotation of the 4-space itself. This latter point is an important one to keep in mind when considering just how he incorporated uncertainty into his theory: he has essentially treated the standard, intuitive, and purely observable (and classical) 4-space as a singular unit and imbued it with its own property (the ability to rotate). This requires the fifth dimension, though Eddington's interpretation is purely mathematical. This foreshadows his use of the standard 4-space reference frame as a singular unit in *Fundamental Theory* that he imbues with uncertainty.

He the uses *T* to transform $\psi$-space as $T_\mu^\nu \psi^\nu = \psi_\mu$ which is valid for any linear transformation of the $\psi$'s. By utilizing the second-order equation and assuming that the required form of the first order equation is $T\psi = 0$, he arrives at the following *tensor calculus* form of the relativistic wave equation, (5.1):

$$T\psi = \left\{(ih\mathbf{grad} + (e/c)V)(E_1, E_2, E_3, -iE_4) + mcE_5\right\} = 0 \qquad (5.2)$$

(Eddington 1928). The *E*'s are associated with various combinations of coordinates or matrix values that satisfy the invariant condition and are, thus, easily assignable to the 16 quantities described above by Darwin. The only requirement is that they are mutually perpendicular – no actual physical phenomena depend on the choice of what quantity to assign to what *E* value. The advantage is that the *E*'s present a symmetrical method while Dirac's original formalism does not. The disadvantage is a common one with much of Eddington's later work – the formalism was unusual, complex, and, perhaps, a bit inaccessible even to clever physicists. However, he did prove that his solution was equivalent to Dirac's.

The important historical aspect of this is that he employed an essentially geometric approach utilizing coordinate dimensions to present the Dirac equation in a tensor form. This also served as the seed paper for his development of *E*-numbers within the context of *Fundamental Theory* and his treatment of standard 4-space as an independent object with its own properties. But the coordinate dimensional form served as the basis for uniting relativity and quantum mechanics under a single framework.

This makes sense when considering the history I have developed so far: Heisenberg and Pauli's admitted reference to relativity. Coordinates are simply a mathematical way of defining a point-of-view. Relativity was – and still is – the ultimate *declaration* on points-of-view. As such it makes sense to utilize coordinates in a fundamental way since they are a natural aspect of relativity and since they had *already* served as a foundation for two of the most fundamental principles governing quantum processes.

The seeds of Eddington's interpretation of the exclusion principle are also contained in this formulation. As I will show, Eddington's form of exclusion ultimately made no distinction between *any* type of particle (regardless of spin, charge, or mass) since *all* particle properties were *frame-dependent* in Eddington's interpretation. I will discuss this in greater depth in seven through nine. For now I will use this discussion of coordinates as an introduction to the final version of *Fundamental Theory* as it posthumously appeared in 1946.

**Introducing *Fundamental Theory* Through Coordinates**

The historical development of quantum mechanics up to Eddington's 1928 paper responding to Dirac is of great importance to understanding the larger framework in which he worked. Many of these highlights leave an indelible impression on *Fundamental Theory* in its form, order, and philosophy. Eddington's own development of the theory, however, is not necessarily to be looked at purely chronologically. Important early drafts will be discussed as they find relevance to the discussion at hand, but this is not an historical analysis. Rather it is a technical analysis of an historical document, particularly in relation to modern quantum field theory since the similarities are striking at time. With that said, coordinates served as the first stone on which *Fundamental Theory* was built.

In relativity coordinates serve as a way of locating objects in reference frames. In quantum mechanics, their role is similar, though the constraint of the uncertainty principle is an added limitation. In fact, one view is that quantum mechanics doesn't (or shouldn't) change relativity at all – it simply clarifies it by putting limitations on what sets of coordinates can be effectively used in concert.  It also makes the point that coordinates are simply one observable quantity among many.  So, in that sense, quantum mechanics should be a generalization of relativity.  Eddington actually took the opposite view – all observables are reducible to coordinates via relativity and so, with the exception of the inherent nature of uncertainty, relativity should be a generalization of quantum mechanics.  Either way the marriage of relativity and quantum mechanics through coordinates seems natural.

To be rigorous with the definition of coordinates, due to the Principle of Equivalence, which ultimately implies that there is no global inertial reference frame, the assumption is made that a reference frame is equal to a coordinate system or a set of observables since a coordinate is simply an observable.  Relativity's condition for observability begins with a denial of absolute motion such that all observables are necessarily measured relative to another observable (technically they should at least come in pairs, then).  In quantum mechanics the condition for observability is the uncertainty principle meaning an object cannot be precisely located in a geometrical frame or as a world-line in 4-space (Eddington 1946).  Regarding the view of a set of observables as a coordinate system, in general relativity it also can be equivalent to a manifold that is an m-dimensional 'hyperplane' in n-dimensional Euclidean space (m $\leq$ n).  Basically, a manifold is any set that can be continuously parameterised where the number of independent parameters is the number of dimensions and the parameters themselves are coordinates.  Metrics are often introduced onto manifolds in order to carry information.  In current language we often relate a manifold to a quantum field, which is a collection of position dependent operators (often modelled as simple harmonic oscillators), where the field matches the manifold point-for-point.

Eddington thus sought a more general form of the uncertainty principle that would incorporate both observability conditions – the relative point-of-view that requires a minimum of two observables (ignoring just for the moment the relation between observer and observed) and the quantum requirement that observations of certain pairs of observables have a certain limit in accuracy.  Eddington's combined principle states that "*a coordinate $\xi$ is observable only if it is a relative coordinate of two entities both of which have uncertainty of position and momentum in the geometrical frame*" (Eddington 1946, p. 1, Eddington's emphasis).  So, for instance, let's say we wish to observe the position of an object.  The obvious question is: its position relative to what?  If you assign an object specific coordinate values *x* and *y*, what you're also doing is specifying the location of the origin since each coordinate is really a displacement vector (the concept of a vector is key to this illustration).  Thus it is *impossible* to specify any location without specifying at least one other location (usually the origin).  So the displacement is in fact the observable quantity here, *not* the actual coordinates of the two objects.  However, assuming that these two objects are measured by observables that obey a commutation relation in the form of equation (4.6), they will also have some uncertainty in the form of equation (4.7).  What I will show is that Eddington, by assuming that every observable is frame-dependent, demonstrates that *like* variables (such as two position variables) obey a commutation relation (though he is not explicit in his derivation of this point).

From the point of view of statistics, an observable is actually a *statistic* based on a double probability distribution (one for each of the 'end-points,' if you will, of the observable).  This means that an observable coordinate is not measured in reference to some abstract, observer imposed mathematical origin, but rather from something that is physically intrinsic to the measurement (of the coordinate) itself i.e. the origin must have a physical nature.  Since the reference here is not some mathematical origin but an actual physically real secondary object (so to speak), it must obey the uncertainty principle as well.  As such, Eddington lays down the rule that there must be *two* origins – one physical, as just described, and one geometrical as imposed by the observer.  The latter should be eliminated from "observationally verifiable results; being therefore aloof from the rough-and-tumble of observational inquisition, it has a sharpness of definition which contrasts with the blurring of all physical landmarks by probability scatter" (Eddington 1946, p. 2).  Essentially, given a geometrical origin applied by the observer, the physical origin that it corresponds to fluctuates.  Eddington then applies this principle to wave mechanics by stating that the coordinates in wave-mechanical equations must be measured from a physical origin since both they and their conjugate momenta are observables and *not* mathematical constructs.  This presents a problem: what *is* this supposed origin and what sort of distribution function does it have?

This idea can be potentially rectified with standard quantum measurement theory. In such a situation the observer interacts with the object under observation and thus the Hamiltonian operator consist of two pieces: one for the object and one for the observer (Bohm instead refers to the observing 'apparatus', see Bohm 1951). Once the observation begins, there is an interaction between the observer and the object and, since interactions are nothing more than a different type of object (see chapter three), the interaction must also contribute to the Hamiltonian of the entire process (Bohm 1951). It is possible to associate the observer (apparatus) in Bohm's example with the physical origin in Eddington's example. Then, as Bohm explains, during the instant of observation itself (which is interaction corresponding to the act of making the observation), the Hamiltonians of the observer and observable produce such negligible changes in the wavefunction that they can be ignored. As such the only contributing portion of the Hamiltonian for this situation is from the interaction (Bohm 1951). In Eddington's formulation, then, the observable coordinate, which is a relative coordinate, corresponds to the interaction in Bohm's example and represents the act of making an observation or taking a measurement.

Bohm makes the point, however, that in order to make a measurement (called an 'impulsive measurement') short enough to only include the interaction's Hamiltonian, a large interaction energy is needed (Bohm 1951). Bohm shows that it is possible to account (or correct) for changes in the observed quantity introduced by the process of interaction itself. He also shows that by including the apparatus coordinates (the physical origin in Eddington's language) it is possible to turn this into a completely objective measurement entirely free of human interaction (see Bohm 1951, pp. 606-607). This is precisely the point of Eddington's formulation. Once again, it appears that Bohm's presentation borrows a bit from Eddington, though, in the case presented here, Bohm's ideas are a standard treatment of the subject.[21]

Eddington's description of the physical origin corresponds to *either* an actual particle included simply for the sake of measurement, *or* the centroid of a set of particles.

---

[21] Ironically, in this 1951 treatise (*Quantum Theory*), Bohm presents an argument *against* hidden variables (Bohm 1951, p. 622). Just a few years after the publication of this book Bohm's theories had begun crossing the line into the realm of hidden variables. Now, Bohm's theories, though clearly based on hidden variables, are often viewed as a third alternative to traditional Copenhagen versus hidden variable theories since they allow for unknowable quantities.

The centroid here, though not a physical object, is nonetheless, frame-independent in that it's location relative to the constituent particles is set regardless of any geometrically imposed coordinate system (you can't shift a centroid around without also shifting the mass distribution). But, since it isn't actually a real particle it can be simply treated as a geometrical origin as well. As such the use of the centroid provides us with a way of transforming between physical and geometrical coordinates. In this way Eddington has answered half his question: the physical origin is a centroid of a system of particles. But what is its spatio-temporal distribution function if it's not a real particle? Statistically, the centroid of a large number of particles is *always* Gaussian regardless of the distributions of the individual particles (this is an important step in moving from microscopic to macroscopic objects and phenomena). Given this *a priori* knowledge of the geometrical distribution of coordinates (and neglecting one supposedly disposable constant in the Gaussian – Eddington gives no valid reason for this), the physical origin then has, in three dimensions, the distribution function:

$$f(x_0, y_0, z_0) = (2\pi\sigma^2)^{-\frac{3}{2}} e^{-(x_0^2 + y_0^2 + z_0^2)/2\sigma^2} \qquad (5.3)$$

The standard deviation, $\sigma$, is referred to by Eddington as the 'uncertainty constant' and here is applied to the physical reference frame.

The origin of Eddington's thinking on this matter clearly has its roots in his statistical astronomy work from early in his career. The particles in question will be considered to be the particles in the universe (or could be the galaxies – or stars if considering something in relation to a single galaxy) and are assumed to be spherically symmetric. The centroid of this distribution is then essentially the centroid of the universe (this obviously clashes with the standard cosmological presumption that there *is* no centre to the universe which really says that the universe has no preferential axis, though there have been recent attempts to calculate the birefringence of the universe and thus produce a preferential axis) and this corresponds to the physical origin for all measurements. The particles will later become the standard background environment Eddington refers to as the 'uranoid' of any small system to be studied. The uranoid itself will be explored in coming chapters.

Eddington does admit that the rest of his theory which is derived from this physical frame could very possibly be wrong since quantum theory leaves its ultimate frame-of-reference (master frame, perhaps) undefined and thus it is impossible to know if his physical frame is *the* physical frame for quantum mechanics.  He offers this qualification:

> The reader interested in logical rigour should bear in mind that the development of the theory turns partly on strict deduction and partly on ultimate saving of labour.  The former part requires proof, the latter part success (Eddington 1946, p. 4).

Certainly, the validity of his theory, then, would rely partly on its predictions of experimental results (though that in itself is not a proof of its actuality – see Ockham's Razor for clarification).  He does offer an argument *reductio ad absurdum* showing that quantum theory must be based on a physical origin associated with such a centroid in order to validate his attempt, but the argument is simply an expansion of the ideas already presented, specifically as compared to Bohm's description of measurement, and is somewhat circular.  His physical argument in support of the transformation from geometrical to physical origins hinges on the standard deviation, $\sigma$, being used to 'put the scale into' the physical frame (and thus everything constructed in it).

**Physical Interpretation of the Standard Deviation**

To find a physical interpretation of the standard deviation, then, begin with a very large number of particles $N$ all having the same coordinate probability distribution and assume a large number of them exist in a volume $V_0$ that is fixed in some geometrical (i.e. mathematical and non-physical) frame.  Each particle has some probability $p$ of being in $V_0$ and the expectation value (mean number in $V_0$ in Eddington's terminology) is $n_0 = pN$.  The *actual* number in $V_0$ is $n = n_0 + y$ where $y$ is some fluctuation in one coordinate axis.  Note that $y$ (and $x$ and $z$ as well) is *not* a coordinate length in this instance since the argument of the exponent in equation 5.3 must be unitless. Since the square of the standard deviation would be $\sigma^2 = n_0(1 - n_0/N)$ by Eddington's definition and is unitless, $y$ must also be unitless. In essence, since number (in particular, particle counts) is the root of probability and statistics, and is in some senses the simplest of measurements to make (aren't all measurements really just ways of counting something?) it makes sense for Eddington to reduce everything to this basic idea (see my discussion of Russell in chapter three).

In any case substituting Eddington's definitions into equation (5.3) with $x$ and $z$ set to zero yields:

$$f_N(y) = \{2\pi n_0(1 - n_0/N)\}^{-\frac{1}{2}} e^{-y^2/2n_0(1-n_0/N)} \qquad (5.4)$$

Assuming that $N/n_0 \to \infty$, the fluctuation $y$ (i.e. the difference between the number of particles actually *in* the volume at a given time and the *expected* or *mean* value) has the distribution law:

$$f_\infty(y) = (2\pi n_0)^{-\frac{1}{2}} e^{-y^2/2n_0} . \qquad (5.5)$$

Both equations (5.4) and (5.5) are Gaussian distributions and thus equation (5.4) is actually equation (5.5) compounded with:

$$f_e(y) = (2\pi n_0^2/N)^{-\frac{1}{2}} e^{-Ny^2/2n_0^2} . \qquad (5.6)$$

Now define a new coordinate $\zeta = y/n_0$ such that $n = n_0(1 + \zeta)$. The distribution of $\zeta$ is then:

$$g_e(\zeta) = (2\pi/N)^{-\frac{1}{2}} e^{-\frac{1}{2}N\zeta^2} . \qquad (5.7)$$

This effectively resolves equation (5.4) into two completely independent distributions or, as Eddington calls them, fluctuations. He calls (5.6) the 'ordinary' fluctuation that is a result of the finiteness of $n_0$ and (5.7) the 'extraordinary' fluctuation that is a result of the finiteness of $N$ (Eddington 1946). The latter is a negative fluctuation meaning it is subtracted from the ordinary fluctuation. The ordinary fluctuation is the standard distribution of an expectation value (or mean) and is thus given no further clarification by Eddington. However, the extraordinary fluctuation must be considered in spherical space for the following reason: according to relativity theory the only distribution of matter that can be in self-equilibrium is a uniform one that fills some spherical space. This is known as an Einstein universe and requires a cosmological constant in its description. This space has a finite volume, which means that $N/n_0$ is finite. In a Euclidean space $N/n_0 \to \infty$ and the extraordinary fluctuation vanishes from equation (5.4) producing equation (5.5).[22] As such the introduction of curvature gives rise to the extraordinary fluctuation. Thus, in utilizing the Einstein universe as a basis for reasoning, Eddington seems to have attached himself to a *static* interpretation of the universe that appears odd more than fifteen years after Edwin Hubble's (1889 – 1953) discovery that the universe is expanding. But, as I will show shortly, this is far from the truth. It is simply yet another case of Eddington straddling the line – he applied the basic tenets of a *static* universe to an expanding one just as he took a middle-of-the-road stance on the basic interpretation of quantum theory. Note that it was recently proven that the large-scale geometry of the universe *is* Euclidean (de Bernardis, et. al. 2000) while, at the same time, the cosmological constant has returned as the preferred explanation of the source of the mysterious dark energy that is accelerating the universe's expansion. If Eddington's *only* reason for utilizing an Einstein universe was the ease with which it explained curvature, he may have been delighted to discover it might not have been necessary. In fact, it turns out that he utilizes Minkowski (flat) space-time (derived from a transformation) throughout *Fundamental Theory* and derives a cosmological constant, making his work at least hypothetically consistent with factual data. As such this entire line of reasoning might be amenable to simplification.

Returning to a discussion of the extraordinary fluctuation, the above arguments imply that there is a certain level of uncertainty in the number of particles contained in the volume. The particle density $s = n/V_0$ can be equivalently defined as $n_0/V$ where $V$ is now a fluctuating or uncertain *volume* and the number of particles is held constant and is defined as: $V = V_0/(1+\varepsilon)^3$ and the uncertainty is now contained in the linear scale factor $1+\varepsilon$.

At this point it is instructive to recall that (5.6) and (5.7) are distribution functions with discrete values since they are based on a Bernoulli distribution (Eddington 1946).

---

[22] This also implies that $p$ is zero indicating that, in Euclidean space, objects would have a zero probability of occupying a specific volume.

The distribution in the volume scale factor $\varepsilon$ is continuous so some transformation between continuous and discrete is required. Eddington derives this as:

$$(1+\zeta)^2 = (1+\varepsilon)^4 \tag{5.8}$$

for equation (5.7) where the values of $\zeta$ that have what he calls 'sensible' probabilities are on the order of $N^{-1/2}$ making them extremely small (he uses the value $10^{39}$ for $N$ giving $N^{-1/2}$ a value on the order of $10^{-20}$ – $10^{39}$ is half of the number of particles, $10^{78}$, that he derives as inhabiting the universe as a whole). Equation (5.8) can then be approximated ("amply" in Eddington's description) as $\zeta = 2\varepsilon$. Equation (5.7) gives the standard deviation of $\zeta$ as $N^{1/2}$ meaning the standard deviation of $\varepsilon$ is:

$$\sigma_\varepsilon = \frac{1}{2}\sqrt{N}. \tag{5.9}$$

The ultimate point of the above derivation is that the extraordinary fluctuation of the particle density is represented as a *scale* fluctuation with a standard deviation given by (5.9). Basically, since $N$ is now taken to be an exact number, the uncertainty is transferred to the volume, meaning it manifests itself in an uncertainty of the scale of measurement. If the volume is spherical, the only uncertainty is in the scale of measurement of $r$.

In summary, if considering a point some distance $r$ from the origin, the difference between its physical and geometrical coordinates consists of a fluctuation with a standard deviation of $\sigma$ in all directions that is due to the uncertainty of the position of the physical origin, and a fluctuation with a standard deviation of $\sigma_\varepsilon r$ in the radial direction due to the uncertainty in the scale of measurement of $r$. The extraordinary fluctuation it represented by the latter while the ordinary fluctuation is represented by the former. As such the total standard deviation consists of both radial and transverse components:

$$\sigma_{radial} = \sqrt{\sigma^2 - \sigma_\varepsilon^2 r^2} \tag{5.10}$$

$$\sigma_{transverse} = \sigma. \tag{5.11}$$

Equations (5.10) and (5.11) are referred to as the 'local uncertainty' of the physical reference frame. It is essentially the uncertainty of a local physical origin relative to a local geometrical origin. If the difference between the physical and geometrical origins, $r$, is small (and it stands to reason it would be almost *immeasurably* small since it would be ridiculous to have geometrical coordinates assigned to an object that was nowhere near it!), then both (5.10) and (5.11) are simply $\sigma$ which is the original value of the standard deviation for the physical coordinate frame.  Eddington concludes, "Independently of coordinate systems, *the local uncertainty in a given direction defines an extension which might be adopted as the unit for measuring lengths in that direction in that locality*" (Eddington 1946, p. 6 his emphasis).  This is then referred to as the $\sigma$-system of defining lengths, or the '$\sigma$-metric.'

Finally, Eddington derives the length of a line-element in the $\sigma$-metric as:

$$ds^2 = \frac{dr^2}{1-\left(\sigma_\varepsilon^2/\sigma^2\right)r^2} + r^2 d\theta^2 + r^2 \sin^2\theta d\varphi^2 \qquad (5.12)$$

which is the standard line-element in a spherical space with a radius of $R_0 = \sigma/\sigma_\varepsilon$ (note that in this definition $R_0$ is a *dimensionless quantity*).  Here he employs equations (5.10) and (5.11) to find the proportionality constants for (5.12).  The radius of spherical space can then be combined with (5.9) to find the following simplified equation for the standard deviation:

$$\sigma = R_0 / 2\sqrt{N}. \qquad (5.13)$$

This gives a direct relation between the uncertainty in an observable coordinate measurement and the two cosmological numbers $R_0$ and $N$.

Pausing for a moment to consider the significance of the preceding, rather lengthy, derivation, Eddington has connected the microscopic measurement problem with the macroscopic cosmological problem.  What he is assuming is that the cosmological parameters are *measurable* quantities that, due to their enormous scale, minimize uncertainty *in their own measurement* (i.e. Eddington interpreted this as the fact that the uncertainty principle rarely plays a role in any large-scale measurement, blurring the distinction between classical and quantum uncertainty – see below).  But, what this provides is a way of finding the microscopic uncertainty (the standard deviation) from macroscopic measurements – large-scale phenomena provide information about small-scale phenomena.  In addition it makes the uncertainty a measurable quantity (at least for a given or measured value of *N*).

The standard deviation, then, plays multiple roles in the formulation of the complete theory. First, it links the macroscopic to the microscopic. Second, it provides a way of measuring the uncertainty (perhaps even eliminating, for a specific *N*) since it is known to be negligible for macroscopic quantities. Third, it provides a potential unit of length measure in a given direction. Fourth, it has the added effect of transferring uncertainty to geometrical coordinates (i.e. if $r$ is small the physical and geometrical origins essentially coincide meaning that, even if the geometrical coordinates are supposed to be set without error by the observer, if they are equivalent to the physical coordinates then they equivalently have some uncertainty). And, finally, it associates the extraordinary fluctuation with the curvature of space and, since the extraordinary fluctuation can be reduced or eliminated in many cases, so can the curvature of space (at least locally – note that this is how he approximates Minkowski space).

Notice here Eddington's blurring of the concept of uncertainty (see chapter three). He uses the term 'uncertainty' in several places here but in these cases they clearly imply a classical notion since they are simply various statistical quantities. But it is *not* clear that Eddington really recognizes a difference between classical and quantum notions of uncertainty. It appears from his continuous application of the underlying philosophical *principle* developed by Heisenberg that quantum uncertainty was simply a limiting case of classical uncertainty. He draws no clear distinction in his application of Heisenberg's principle between microscopic and macroscopic situations. In fact he uses them to indisputably (in his mind) *link* microscopic (quantum) and macroscopic (classical) phenomena. In fact Eddington would rather refer to macroscopic and microscopic phenomena than to classical and quantum. In this way there is no clear distinction in *Fundamental Theory* between classical and quantum situations since macroscopic (classical) blends into microscopic (quantum) as multiplicities decrease (see chapter six). Quantum phenomena then become a limiting case of classical phenomena. The trouble with this is that it does not account for the fact that complementarity (or some comparable explanation) is required in quantum cases while it is not in classical cases, i.e. there is no clear and purely statistical explanation for complementarity.

Regardless, Eddington proceeds with this blurring of classical and quantum regimes and argues for a radical change in measurement in general. This leads to his requirement of a standardised unit of length that was his best attempt at eliminating units entirely. Once again the derivation originates in the standard deviation.

**A Standard for Length**

Focusing on the third role the standard deviation plays as a potential unit for measuring length brings to the surface one of Eddington's deepest troubles with mid-twentieth century science. With the increasingly microscopic measurements that were being obtained in physics, the limitations of the standard units of measurement became increasingly apparent. Eddington had a particular interest in finding a proper definition for length, calling it "the most urgent requirement of all; for when we come to examine what is actually measured in any kind of experiment, it is nearly always a length or a spatial measure – the length of a thread of mercury in a thermometer, the shift of a bright spot on a galvanometer scale, the displacement of a dark line in a spectrogram, etc." (Eddington 1939, p. 71). Prior to 1960 the standard measure of length was the Paris Metre that was, quite literally, a bar of metal in Paris that was defined to be a meter in length. The history of the Paris Metre dates to 1790 when Talleyrand (Charles Maurice de Talleyrand Perigord, 1754 – 1838) recommended reform in France and the National Assembly ordered the Académie to create a commission on the subject of standardizing measures. Two probabilists already mentioned in chapter three were on the commission: Laplace and the Marquis de Condorcet. Despite what appears to be the arbitrariness of *any* measurement unit, in 1791 the commission gave three possibilities for a scientific basis of any new measuring system: "the length of a pendulum, a quadrant of the circle of the equator, finally a quadrant of the earth's meridian" (quoted in Linklater 2002, p. 122). An excellent and sometimes humorous look at measurement in history is Andro Linklater's *Measuring America* (2002) where he describes how the commission chose from the three options. They "frivolously" chose the last option in what Linklater describes as an almost "capricious fashion" (Linklater 2002, p. 123)[23]. Thus, the meter was defined to be one ten-millionth of a quadrant of the earth's meridian.

---

[23] Linklater's assertion has been strongly disputed by E.F. Robertson (Robertson 2004).

The arbitrariness with which the commission chose the unit of length is, perhaps, not all that surprising considering two probabilists were on it. After all, how could someone who helped father the mathematics of *in*exactitude be expected to commit to a scientifically reasonable description for an *exact measure*? In essence, what difference did it really make at the time since all measure would be inherently inexact in some cases? On the other hand, they did their best to maximize the certainty by which they measured the meridian so as to obtain as accurate a result as possible. Nonetheless this still left the problem of duplication. Certainly, in order to accurately duplicate the meter, one could simply retake the measurements performed by the French, but the arduous task was not simple from a logistical point of view, nor was it easy from the point of measurement technique. The only other option was to attempt to copy the metal bar directly, which is what was most often done, but the possibilities for error are obvious.

This was the state of affairs during Eddington's entire lifetime and it is not difficult to see how problematic this becomes when attempting to carry out microscopic measurements. For instance, if the uncertainty between the original meter and some duplicate somewhere was a centimetre, it would be wholly unrealistic to measure anything at the millimetre level without introducing a great deal of uncertainty. Since physicists were working many, many orders of magnitude smaller than that by the 1930s, there obviously was a problem.[24] In the absence of anything better, Eddington defined his standard of length by the standard deviation described above. Alternatively he described it in terms of the periods of light waves and the amplitudes of vibrations of crystal lattices, finding, as a consequence, that the speed of light was a constant (this was universally accepted from the mid-twentieth century until only recently when varying speed of light (VSL) theories legitimately entered the mainstream accompanied by supporting astronomical evidence – see Magueijo 2004).

---

[24] In 1960 an atomic standard was adopted based on the wavelength of a particular red-orange spectral line emitted by krypton-86 in a gas discharge tube. Even here, however, the reproducibility of krypton-86 limited the accuracy. This standard was replaced in 1983 by the current one, which defines one meter as the length of the path travelled by light in a vacuum during 1/299,792,458 second. This implies that the speed of light is known exactly and that the second is well-defined. This is a quantum-mechanical definition of just the type that Eddington wanted.

There is a branch of physics known as metrology that is devoted to the study and improvement of measuring techniques that actually predates relativity (which brought many of these problems to the fore). In any definition of length the most crucial part is the existence (or creation) of a standard that is readily available for comparison anywhere at any time. Regarding the Paris Metre itself as a standard, he says "Metrologists do not look upon a particular bar of metal, such as the Paris metre, as an ultimate standard; the mere fact that they feel anxiety as to its permanence shows that they have in mind a more ideal standard with which it might be compared" (Eddington 1939, p. 74). Eddington says that some physical structure that is not necessarily permanent but *is* unique and reproducible is what is required to truly define some length. He gives as an example a calcite crystal that has a certain number of lattice intervals – nature requires this particular type of crystal to be the same everywhere (otherwise it's actually something else entirely). Ultimately this suggests requiring some type of measure for length that is actually unitless – the number of lattice intervals is simply that – a pure number. Thus any structure that is easily reproducible *from* some quantum specification can properly serve as a standard. Finally, all "such standards are equivalent, being definite numerical ratios to the unit of length $h/mc$ which appears in the fundamental equations of quantum theory" (Eddington 1939, p. 75). The standard of time (or time-extension in Eddington's terms) is similarly defined as a time-periodicity of the same structure that provides the standard for length.

The standard of length, however, prior to Eddington's writing, comes from relativity. But relativity appeals to theories external to itself in order to define length. Eddington, here, enumerates another crucial requirement for *his* view of a unifying theory – it must be completely self-contained such that the theory *itself* produces its own standard, i.e. the standard is derivable from the theory itself. In this regard I will show that Eddington was unsuccessful. However, he recognized that since certain processes were only representable in a relativistic quantum theory, relativity and quantum mechanics cannot be considered separately and thus must have the same standard of length (Eddington 1939). Furthermore, through an examination of a *changing* velocity of light in a vacuum in relation to a hydrogen atom, he concludes that "either the ratio of the period of the emitted light to the time-period intrinsic in the emitting atom varies with time, or the ratio of the length of the emitted waves to the spatial scale of structure of the emitting atom varies in time" (Eddington 1939, p. 78). This makes little sense. In order for the former ratio to vary, the period – and thus the frequency – of the emitted light must be changing, or the intrinsic time-period in the atom must be changing. But these quantities are linked – a different intrinsic time-period should produce a correspondingly different frequency. In the case of the latter ratio a similar principle holds true: if the spatial size of the emitting object were to change the emitted wavelength should have to change. Modern VSL theories are certainly rigorous and may even be partial truths, but the state of affairs in 1946 was not nearly as advanced as it is now (however, hindsight is always 20/20): since atomic transitions were seen as the *only* way to emit a photon, if the value of the constants changed it would imply that there was no link between the photon and its source which is absurd.

The natural extension here then is to consider a quantum-specified standard, which is what we essentially have in place today (Eddington clearly had excellent foresight). However, he recognized two limitations in such a standard. The first is that the standard is not fully reproducible in very strong electric and magnetic fields. What this means is that if the standard is based on the properties of a photon, say in a field-free situation, that photon most likely will have different properties while in a field since the field can affect the photon. In view of the fact that a photon is the carrier particle for the electromagnetic interaction *itself*, the information carried by the photon will vary depending on the nature of the interaction. In addition, subsequent to Eddington's work, electromagnetism was united with the weak nuclear interaction in electroweak theory and would need to be reconciled with the W and Z bosons that are the carrier particles for the weak interaction. As such, a standard of this nature would also be artificial when measuring in a gravitational or strong interaction – gravitons and quarks would then need to be brought into the fold. Obviously, a fully unified theory might be able to accomplish this, but, at least from Eddington's point-of-view, this is a serious limitation (quantum field theorists and string theorists can ponder this *ad infinitum* should they wish). Eddington even points out that "at least a dozen different 'unified theories' of the gravitational and electromagnetic fields have been put forward each implying a slightly different definition of length" (Eddington 1939, p. 80).

The second limitation on the standard is that it must be *short*. Eddington gives the following example to demonstrate why:

> Suppose we try to measure the diameter of the earth with a long crystal standard stuck through it like a knitting needle through an orange. It is well known that the earth is strained out of shape by the tide-raising forces of the sun and moon; the long crystal will likewise be strained … We cannot always remove the bodies that are causing the strain … Thus, in general, we have to be content with short standards which are proportionately less affected by strain (Eddington 1939, p. 82).

I briefly mentioned another example earlier in this section: imagine one wishes to measure something on a millimetre scale but the standard by which one is measuring is only accurate to the centimetre scale. Measurements in millimetres could be small enough that they are *less than* the actual uncertainty in the standard – it's like attempting to measure something on the millimetre (or smaller) scale with a wooden meter-stick (or yard-stick) like one might find in a fabric store (a micrometer would be far more useful in this situation).

Eddington points out that measuring large distances or lengths is simply accomplished by integrating the short distances assuming that they are unambiguous. As such, it could be argued that the meter is an inadequate length measure in physics, particularly since such a standard could undergo length contraction in even fairly simple situations. As Eddington points out, "the failure to define long distances observationally, or in mathematical language the non-integrability of displacement, is the *foundation* of Einstein's theory of gravitation" (Eddington 1939, p. 83 my emphasis). Eddington's conclusion, then, is that since the length and time intervals "are the basis of nearly all other physical definitions [to] avoid circular definitions it is essential that the standards of length and time interval should be the extensions of structures completely specified by pure numbers" (Eddington 1939, p. 84). The standard deviation resulting from uncertainty fits this profile even though equation (5.12) contains $R_0$ – from equation (5.13) we know that it is dimensionless and, thus, so is $\sigma$ (see 5.13). Eddington continues by specifying that the general specification of physical structures by pure numbers (e.g. numbers of elementary particles in states defined by quantum numbers) is developed in quantum mechanics meaning that the standard of length must be quantum-mechanical in nature. Again, $\sigma$ fits the bill, though the latter point needs to be proven. It *can* be proven if any single quantum-specified extension can be shown to have a fixed ratio to $\sigma$ (the local uncertainty). This is the primary point of *Fundamental Theory* – investigating ways in which extensions of various structures can be related to $\sigma$. In essence, the entire treatise is a justification of the use of $\sigma$ as a standard for measure.

**Problems in Measurement Theory**

The standard deviation also plays the role of providing a value for the uncertainty in any measure (this is one of the standard definitions of it). The mere fact that there *exists* uncertainty in measurements, particularly in quantum mechanics, raises a host of issues to be resolved. In quantum mechanics a measurement is usually referred to as an observation. As early as 1932 Von Neumann felt the problem of observation and measurement was not being properly considered (Jammer 1966). His explanation rested on the notion that every quantum mechanical measurement *process* involves some aspect that can't be analyzed. This aspect involves the interference of the observer in the measuring process. He suggested that in addition to the continuous causal propagation of the wave function by the usual wave equation, the function also experiences a *dis*continuous, *non*causal, and instantaneous change due to the intervention of the observer. A similar situation was proposed by Heisenberg in relation to the $\gamma$-ray microscope thought experiment he used in his derivation of the uncertainty principle (see chapter four). But in Heisenberg's case the interference of the observer was entirely causal (as I mentioned earlier the uncertainty principle is completely causal in nature). Von Neumann's idea was based on irreversibility (the second law of thermodynamics reminding once again us that there *is* no fountain of youth) in that given some statistical operator $U$, a measurement of an observable $R$ containing a set of orthonormal eigenfunctions $\varphi_1$, $\varphi_2$, etc. alters $U$ as follows:

$$U' = \sum (\varphi_n, U\varphi_n) P_{[\varphi_n]} \tag{5.14}$$

where $P$ represents the probability and, thus, irreversibility of the process (Jammer 1966). There is a corresponding causal, continuous, and reversible change as well. Given a state $\psi = \sum a_n \psi_n$ that is expanded in the eigenfunctions of $R^{25}$, measuring $R$ reduces $\psi$ to a specific eigenstate $\psi_k$ whose eigenvalue is the result of the measurement. So the act of measuring an eigenvalue reduces the general state to a specific state not allowing the observer to have full knowledge of the general state $\psi$. This is known alternately as the 'reduction of state,' 'reduction of the wave packet,' or, in more modern terms, 'wavefunction collapse.'

The last description provides a simpler physical explanation for what von Neumann was postulating: given a probability distribution of possible values for some measurement, the act of taking the measurement collapses the wavefunction to correspond to the single measured value. Von Neumann's analysis attacks the doubly difficult problem of proving whether statistical quantum mechanics is logically self-consistent or whether it could be completely and deterministically described using hidden variables. He concluded that statistical quantum mechanics would have to be false if any other explanation were introduced. Another way of looking at this is via the Schrödinger Cat analogy. In 1935 Schrödinger proposed the following thought experiment: imagine a cat locked inside a box containing a Geiger counter which itself contains a small amount of radioactive material that, in one hour, has an equal probability of decaying and not decaying (Schrödinger 1935). If it decays, the counter tube discharges and, through some form of relay system, triggers a hammer that smashes a bottle of hydrochloric acid, thus poisoning the cat. An observer can have no knowledge of *anything* inside the box until it is opened. Thus, before opening the box the cat could be said to be alive or dead. In fact the wavefunction for the entire system, since it's an exponential, can easily be broken into superposable pieces; it's a superposition of two states: one in which the cat is alive, and one in which it is dead. The reality is that prior to observation, the cat is described as

---

[25] The $\psi$'s are the eigenstates *of* the eigenfunctions (the $\varphi$'s).

*both* alive and dead, simultaneously. On a macroscopic level this appears contradictory, though it has been repeatedly verified on a quantum level. Schrödinger's idea is similar (and most likely based on) Von Neumann's idea that homogenous (ensemble) states cannot be eliminated from consideration simply by resolving them into a mixture of various substates each of which would have a set of hidden variables, where the homogenous state is an average of the substates, since this is contrary to the definition of homogenous states (Jammer 1966). Essentially he proved that it is impossible to construct a deterministic description of physical processes with the existing formalism of quantum mechanics.

Von Neumann's conclusion has served as the foundation for many additional arguments both in favour of and opposed to hidden variables. Experiments in quantum non-locality originally based on the theoretical work of Bohm and John Bell (1928 – 1990) have sparked renewed interest in hidden variable theories. Thus, for a slightly alternative (though still acceptable) interpretation of the measurement problem, I turn back to Bohm (recalling that Bohm's 1951 work actually precedes his research in hidden variables). Earlier in this chapter I described Bohm's splitting of the Hamiltonian in any observation into three pieces – one for the observer (or observing apparatus), one for the object under observation, and one for the interaction between the object and the observer. In this formulation Bohm was able to show that the interaction always multiplies each part of the wave function that corresponds to a measured value, by a random phase factor $e^{i\alpha_a}$ (Bohm 1951). All wavepackets are really a superposition of waves that interfere both constructively and destructively. Introduction of this phase factor destroys the interference and is analogous to the wave collapsing.

**Observational Uncertainty**

Once again the physical problem here is the fact that an observation actually interferes with the object under consideration. This is often interpreted as introducing uncertainty into the measurement. Related to this is the fact that objects have wave-particle duality, behaving as one or the other based on the type of observation that is made. In addition, prior to measurement an object or a system (such as Schrödinger's Cat) can exhibit wave-like properties that are reduced to particle like properties by the measurement process (see above). So the interference of the observer with the object collapses the wavefunction and produces an uncertainty that is related to the possible states the wavefunction *could* have assumed before the measurement. This relationship is demonstrated in the general equation for uncertainty (4.7). By defining a standard of length as a dimensionless ratio based on uncertainty itself Eddington sought to circumvent this problem. Conversely, Bohr proposed the idea of complementarity saying that two mutually exclusive interpretations were required. This is a fragmented view of the universe in the sense that it automatically requires a division of the universe into a minimum of two things – waves and particles. This, then, ties in directly to Eddington's view that the fragmentation of the universe was manifested in the various units of measure (meters, seconds, Coulombs, etc.). Going further still, this can be reconciled with Heisenberg's original causal interpretation of the uncertainty principle: the uncertainty is in our knowledge of the *present* (are there hidden variables or not?) rather than the future and this, perhaps, is a result of our *fragmentation* of the present as represented in a multitude of units of measure. Eddington's approach was rooted in his Quakerism which is very Eastern-like in its world-view (see chapter two) and he found division not in religion but in measure: "the division of the external world into a material world and a spiritual world is superficial … the deep line of cleavage is [actually] between the metrical and non-metrical aspects of the world" (Eddington 1925a, p. 204). He thus advocates a purely deductive theory based on pure numbers or dimensionless ratios of fundamental constants. He believed that there "was an entire realm of human experience that could not be measured or quantified – the realm of consciousness – and this was *just as real*" (Stanley 2004a, p. 48).

      Eddington expresses his theory well in his popular books in addition to his technical books. As Whitworth says "he anticipates the opinions of people who trust in the evidence of the senses, who trust in the reliability of measuring devices, and who trust in standard measures" (Whitworth 2004, p. 70). In Eddington's own words he professes the view that reality is basically subjective: "because this real world is undetectable we do not as a rule attempt to describe it. Not merely in everyday life, but in scientific measurements also, we describe the world of appearance" (Eddington 1918b, p. 16). In

perhaps his clearest statement of the basis for his entire line of reasoning, he says later in the same treatise that "the theory of relativity offers not an 'explanation of gravitation,' but 'an explanation of the real nature of our measures'" (Eddington 1918b, p. 36 and Whitworth 2004, p. 74).

In one final note on this subject, Eddington played his own devil's advocate regarding objective measurement as demonstrated in the following syllogism he constructed (Eddington, et. al. 1937, p. 1000 and Batten 2004, p. 166):

*a.* It is impossible to have *a priori* knowledge of an objective universe.
*b.* The mass-ratio [of the proton to the electron] has been found by an *a priori* method.
*c.* Knowledge of the mass-ratio is not knowledge of an objective universe.

He obviously, as suggested by this syllogism, struggled with the dichotomy of a need for objective knowledge and the impossibility of gaining that knowledge *a priori* (see chapter three on probability).

**Applying $\sigma$ in Physical Coordinates**

Now that the foundations of Eddington's system of physical coordinates have been established with a length standard given by the standard deviation of the probability distribution of the given coordinates, a simple application can be developed. Consider two particles very close together such as two protons (or neutrons) in a nucleus or two protons involved in a scattering experiment. Assume they have physical coordinates $\xi_r$ and $\xi_s$ respectively. Since the only physically relevant position is a coordinate difference, their positions can be described by $\xi_{rs} = \xi_s - \xi_r$ in reference to some geometrical origin. So, in this case the coordinate difference is measured relative to a geometrical origin that is separate from the two objects under consideration. One can also measure the coordinate difference directly without reference to any origin (which basically corresponds to placing the origin at one of the object) and this measure is given by $\xi'_{rs}$.

Both coordinate differences are observables that obviously have the same mean value (since they measure the same relative position, just in different coordinate systems). However, they have different probability distributions. Since $\xi_{rs}$ is in reference to a geometrical origin (which is essentially a third point, versus the two points used in a direct measure) the probability spread is larger. Basically, using the formalism developed earlier in relation to the separation of the geometrical and physical origins which, for the sake of keeping the notation straight, will be called *d* for the time being (it was *r* in the previous formulation but that is confusing for obvious reasons since I am sticking with Eddington's notation for the two particles, *r* and *s*), the result is that if *d* is large, we use $\xi_{rs}$; if *d* is small, we use $\xi'_{rs}$.

In first considering measurements made in reference to the geometrical origin (but assuming, based on previous arguments, that this origin must have an uncertainty (probability distribution) to it since it needs to be assigned a physical meaning), a measurement of $\xi_r$ provides a distance from some point in the probability distribution of particle *r* to some point in the probability distribution of the origin. Similarly, if $\xi_s$ is measured, it provides a distance from some point in the probability distribution of particle *s* to some point in the probability distribution of the origin that is *not necessarily the same point as that measured for r*. As such, $\xi_{rs}$ includes *two* random points in the probability distribution of the origin. Any two random points can be given a Gaussian probability distribution with the standard deviation of $\sigma\sqrt{2}$. This is the scatter introduced when using a geometrical origin that is apart from either particle (corresponding to a large *d*) and is an error that should be reduced or eliminated. By taking the *direct* measurement $\xi'_{rs}$ this error can be eliminated. Mathematically this is written as:

$$\xi_{rs} = \xi'_{rs} \pm \sigma\sqrt{2}. \tag{5.15}$$

To be completely rigorous a distance between two particles in three dimensions must explicitly state whether it is in reference to the physical origin and is thus a direct measure from one particle to another, or whether it is in reference to the geometrical origin and would then include the extra error in (5.15). Specifically, either

$$r_{12} = \left(\xi_{12}^2 + \eta_{12}^2 + \zeta_{12}^2\right)^2 \quad \text{or} \quad r'_{12} = \left(\xi'^2_{12} + \eta'^2_{12} + \zeta'^2_{12}\right)^2.$$

**The Range Constant of Nuclear Forces**

So in considering two particles very close together in a nucleus or a scattering experiment, nuclear forces are of particular interest and the range of their action is required for detailed calculations (recalling that in the early 1940s the strong and weak nuclear forces were not fully developed and the electromagnetic force was still considered the primary force acting on a nuclear level)[26]. To be rigorous in the analysis, the investigator must specify $r_{12}$ or $r'_{12}$ as the range. Since the ideas presented here are unique to Eddington's analysis, quantum theorists obviously were using $r_{12}$. As such something such as the electrical potential would be proportional to what Eddington refers to as the Coulomb energy, $q^2/r_{12}$. He refers to the *non*-Coulombian energy as a singular energy corresponding to $r'_{12} = 0$ which occurs when the particles actually collide, entangle, or occupy the same position (see chapter six for a full discussion of the exclusion principle). The full electrical energy is then given as $e^2/r_{12} + B\delta(r'_{12})$ where $\delta$ is the Dirac delta function and $e = q$. In this case, then, the second term only exists when the two particles coincide (collide, entangle, etc.). Since $r'_{12} = 0$, the individual coordinates will be zero, $\xi'_{12}, \eta'_{12}, \zeta'_{12} = 0$, which by (5.15) implies that $\xi_{12}, \eta_{12}, \zeta_{12} = \pm\sigma\sqrt{2}$. Thus, even though $r'_{12} = 0$, it has a Gaussian probability distribution with a standard deviation of $\sigma\sqrt{2}$. Essentially, this is, once again, the transference of the physical uncertainty to the geometrical coordinate system as described above. When the two particles coincide, even though there is technically no separation between them, there is a probability distribution for their separation due to uncertainty (i.e. they may *look* like they're in the same place, but they may be slightly separated).

    The non-Coulombian energy term, $B\delta(r'_{12})$, that I analyze in greater depth in chapter nine, then must be transformable into some form that will produce a Gaussian probability distribution. The development of quantum mechanics outlined in chapter four, provides the solution. Born's formulation of a scattered wave far from the centre of

---

[26] Hideki Yukawa's (1907 – 1981) meson theory was just being considered in Western circles in late 1937 and it was not until the early 1960s that real progress was made in understanding nuclear forces despite the ability to harness them for energy. Yukawa and his theory are discussed below.

scattering (4.6) indicates that a Gaussian probability distribution can be obtained from the exponential for the time-dependent wave equation (4.5). It is convenient to then assign to $B\delta(r'_{12})$ the form $Ae^{-r'^2_{12}/k^2}$ where $k = \sqrt{2} \cdot (\sigma\sqrt{2}) = 2\sigma$. By (5.13) this gives $k = 2\sigma = R_0/\sqrt{N}$ and is referred to as the 'range constant' of nuclear forces and is interpreted by Eddington simply as being the effect of the uncertainty of the reference frame that turns what normally would be a singularity, $r'_{12} = 0$, into a Gaussian distribution of $r_{12}$. But $k$ is an actual experimentally available value independent of Eddington's interpretation (or derivation) of it.

Historically, the first value for $k$, or what can be considered equivalent to $k$, was given in 1930 by George Gamow (1904-1968) in what is perhaps the earliest suggestion of the liquid drop model of the nucleus (Gamow 1930).[27] In a bit of notational confusion, he refers to Eddington's $k$ as $R_0$, though it is clearly the same quantity as $k$ and is completely unrelated to Eddington's $R_0$. In order to maintain consistency and minimize confusion, I will continue to refer to it as $k$.

Gamow gave the following definition for $k$:

$$k = \left(\frac{Am}{h^2}\right)^{1/(n-3)} \quad (5.16)$$

where $A$ is a constant. Gamow, however, explicitly states that $k$ is not calculable since the nature of nuclear forces was not well-known (essentially $A$ is unknown). He thus assumes a value gained from several previous investigations (Gamow 1930).[28] The value he uses is $k = 2 \times 10^{-13} cm$. Eddington uses the similar value $k = 1.9 \times 10^{-13} cm$ with the same justification. This begs the question of whether Gamow and Eddington were actually talking about the same thing. Gamow's definition (5.16) is derived from what he refers to as the Debye formula for surface tension which is then given as being equivalent to the inside pressure (of the 'liquid drop' nucleus). The resulting equation for the radius of the nucleus includes the constant given by (5.16) as a multiplier for the number of $\alpha$-particles in the nucleus. Thus $k$ needs to have units of length since $N$ is simply a number.

---

[27] For a full discussion of this model, see Roger H. Stuewer, "The Origin of the Liquid-Drop Model and the Interpretation of Nuclear Fission, *Perspectives on Science* **2** (1994), 76-129, especially 78-87.
[28] The investigations he references include work by Bieler in 1924, Hardmeier in 1927, and Gamow himself in 1928. See Gamow 1930, p. 635 for details.

In addition, since it defines the size of the nucleus, the constant governs the range of whatever forces hold the nucleus together. Eddington's definition, $k = 2\sigma = R_0 / \sqrt{N}$, also has units of length and governs the range of the forces holding the nucleus together or governing a strong collision interaction. It stands to reason that Gamow and Eddington were both referring to the same thing (i.e. the range of nuclear forces).

Eddington, however, as is his penchant, takes this a few steps further and again attempts to make a link between microscopic and macroscopic phenomena by showing that the microscopic value, $k$, can be defined in terms of the cosmological (macroscopic) values $R_0$ and $N$. This is a result of his application of $\sigma$ to spherical space in (5.12). Since $\sigma$ is a general quantity that is independent of scale, it can be equally well applied to the very large or the very small. As such it is really $\sigma$ that serves as the link between the microscopic and macroscopic through the following relation:

$$\sigma = \frac{k}{2} = R_0 / 2\sqrt{N} . \tag{5.17}$$

Thus, Eddington has shown that the standard deviation, if employed as a standardized quantity (unit of length), is a powerful tool for linking quantum mechanics and relativity.

**Empirical Evidence**

Finally it was left to Eddington to show that when actual values were introduced into (5.17) it yielded consistent results. For example, anyone could come up with a strange enough theory such that they found the masses of the electron and proton to be equal, $m_e = m_p$ but every student of basic science knows from empirical evidence that this is not true. Since the value for $k$ has already been stated as being given by experiment, Eddington gives the equation for the mass $M$ of an Einstein universe:

$$\kappa M / c^2 = \frac{1}{2}\pi R_0$$

where $\kappa$ is the constant of gravitation (we know this as $G$). Since the universe is almost entirely composed of hydrogen, say $M = \frac{1}{2} N m_h$, where $m_h$ is the mass of a hydrogen atom and $N$ is the total number of all protons and electrons in the universe (hence the factor of 1/2). Combining this with Einstein's equation gives:

$$R_0 / N = \kappa m_h / \pi c^2 = 3.95 \times 10^{-53} cm. \tag{5.18}$$

Since the value for $k = 1.9 \times 10^{-13} cm = R_0 / \sqrt{N}$, this result can be combined with (5.18) in a system of two linear equations with two unknowns to give $R_0 = 9.14 \times 10^{26} cm$ (which is 296 MPc) and $N = 2.31 \times 10^{79}$. In contrast, currently accepted values give $R_0 = 1.3 \times 10^{28} cm$ and (though this is a *very* rough estimate) $N \sim 10^{79}$. Obviously Eddington's value for the radius of the universe is incorrect, but his value for the number of electrons and protons in the universe is tantalizingly close. In one final comment that shows he was at least aware of Yukawa's work (see footnote 28) and the existing debate over its validity (see Kragh 1999, pp. 201-202), he says that since observational data implies that $r'_{12} = 0$, the non-Coulombian energy is definitely associated with a singularity (again, see chapter nine).

> Thus we need not hesitate to reject the 'meson-field' hypothesis altogether. It is in any case quite unnecessary in genuinely relativistic quantum theory. It is not an alternative way of taking into account the uncertainty of the origin, because it gives an energy distribution $Ae^{-\lambda r_{12}}$ instead of $Ae^{-r_{12}^2/k^2}$ (Eddington 1946, p. 10).

In a footnote Eddington noted that it was his hope that the shape of the non-Coulombian potential well would be found experimentally thus validating either his or Yukawa's theory. Yukawa's theory was one of the first to suggest that forces are mediated by carrier (exchange) particles (see Kragh 1999, p. 201). So while Eddington interpreted $k$ as arising from the spreading of the combined wave packet of the interacting nucleons (protons and neutrons), i.e. it roughly defined the physical limits of the probability distribution, Yukawa, interpreted $k$ as the furthest distance a carrier particle could travel. Discovery of such a particle would serve as fairly irrefutable proof of Yukawa's theory. Discovery of the pion ($\pi$-meson), as it is now known, came in 1937, though Yukawa's theory was proposed several years earlier (Kragh 1999). As it turns out there are several mesons and other heavy particles that play similar roles. Understanding how this fully relates to Eddington's theory requires an explanation of the basic mechanism governing Yukawa's particles.

**The Range of Fundamental Interactions**

The range limit imposed on carrier particles is due to their need to transfer their information and return 'home' before they are 'missed,' so to speak, i.e. the *carrier* particles are, in fact, a fundamental part of the *interacting* particles and in order not to fundamentally change the character or nature of the *interacting* particles, the carrier particles have to have a round-trip time small enough that the parent particle isn't altered in any way. Essentially they have to have an independent lifetime that is short enough that no one really notices they exist. What would a logical estimate be for such a lifetime? The uncertainty principle in combination with Einstein's famous equation for the rest energy of a particle gives us the answer.

Recall that the time-energy version of the uncertainty principle is given as:

$$\Delta E \bullet \Delta t \geq \frac{\hbar}{2}. \quad (5.19)$$

Einstein's equation for rest energy is, of course, $E = mc^2$. Quantum mechanically a particle can only raise its energy in discrete amounts and so $\Delta E$ is limited to $mc^2$. Thus:

$$\Delta E \bullet \Delta t \approx mc^2 \Delta t > \frac{\hbar}{2}. \quad (5.20)$$

Ultimately we're interested in the range rather than the particle lifetime since that is what Eddington calculated. It makes sense then to isolate $c\Delta t$ which has units of length, where the roundtrip time (lifetime) for the carrier particle is $\Delta t$, and call it the range of the interaction. Note that these particles must sneak in and out based on the uncertainty principle (i.e. there is some non-zero probability that they actually *didn't* sneak out which keeps the parent particle happy or their lifetimes are so short they may *or may not* have existed) and thus they are referred to as virtual particles since their entire existence falls under the restrictions imposed by uncertainty. Rearranging (5.20) and making use of Eddington's notation of *k* for the range, we can write:

$$k \approx c\Delta t \approx \frac{\hbar}{2mc}. \tag{5.21}$$

It is thus a simple matter to calculate the range of the four fundamental interactions. Since the photon is considered massless (see any quantum field theory text for the caveat to this) *k* is obviously infinite for the electromagnetic force, which makes sense since the interaction equation (Coulomb's law) follows an inverse-square law. The graviton has never been observed, but since Newton's Law of Gravitation is also an inverse-square law, *k* must be infinite here as well implying that the graviton is also massless. Now, since the masses of the *W* and *Z* bosons are known from empirical evidence, the range can be directly calculated for the weak interaction and it turns out to be on the order of $10^{-15}$ *cm*. That leaves us to calculate the strong interaction, which is the interaction of primary interest in the present discussion (recalling, however, that the true nature of the strong interaction was unknown at that time – it was popularly thought that the electromagnetic interaction governed nuclear processes and it was even thought that electrons existed inside the nucleus – see Kragh 1999).

Strong interactions at the most fundamental level occur between quarks, with the carrier particle being the gluon (here this interaction is often called the 'colour' force). However, since quarks and gluons are bound tightly inside nucleons (protons and neutrons) it turns out that the minimum emission from a nucleon is a quark-anti-quark pair. The positively charged pion, which is the meson Yukawa originally predicted, is composed of a quark-antiquark pair. It is not a stretch, then, to use the pion as the carrier particle for the strong force *between nucleons*. Note that this is only a minimum requirement. Thus more massive particles actually exist that carry information about the strong interaction between nucleons. Since the pion is the smallest possible emission from a nucleon, equation (5.21) implies *it* would have the longest range of any strong carrier particle between nucleons. Yukawa estimated a theoretical value of $2 \times 10^{-13} cm$ which is in the range of one fermi. The actual value for the pion is $0.73 \times 10^{-13} cm$ so Yukawa's prediction was fairly accurate and netted him the Nobel Prize in 1949.

Clearly Yukawa's theory was correct (though incomplete), but the root of his reasoning was the same as Eddington's – the uncertainty principle. The difference simply was the way (and the form) in which it was applied. While Eddington used the position-momentum interpretation of the uncertainty principle, Yukawa's theory employed the time-energy interpretation. In light of this realization it is conceivable that Eddington's application is *also* correct, though incomplete in its lack of a heavy carrier particle. But in Eddington's formalism, a heavy carrier particle would go unnoticed *anyway* since it would be completely hidden in the overlapping probability distributions of the interacting particles. As such, at least in extremely short range situations where $r'_{12} = 0$, the two descriptions are complementary (though both are incomplete). The trouble comes when $r'_{12}$ is closer to its maximum range of $10^{-13}$ *cm*. In this case a third probability distribution would need to be introduced in Eddington's theory to represent the carrier particle and the calculations become a bit laborious. But they might be worth pursuing, even for purely historical reasons, since the basic reasoning makes sense.

This comparison to Yukawa's work serves to demonstrate every aspect of Eddington's basic thoughts on uncertainty and its implications for measurement theory. It is a clear application of one of the very first statements he made in *Fundamental Theory* on the relativity of measure (see quote p. 82). It is also a demonstration of his commitment to the application of basic probabilistic methods to both quantum mechanics and relativity, a commitment derived from the philosophical stance that a good theory of measurement must strive to be as objective as possible by specifying new units of measure and holding fast to the notion that an observer fundamentally alters the result of an observation simply by making it. With these basic ideas in mind it will now serve best to investigate some of the ancillary aspects of applying uncertainty in measurement and how these extensions lead to a fundamentally new interpretation of the exclusion principle. Beyond that I will investigate how Eddington was able to make some remarkably accurate predictions (and other not-so accurate ones) and how his theory compares to modern quantum field theory. In any case these first few pieces of the puzzle are now in place.

# VI

## *Filling in the Gaps*

In the last chapter the exact point of intersection between the microscopic and macroscopic came when Eddington found the equation for a line element in terms of the standard deviation, (5.12). His assumption was that if the line element for the microscopic σ-metric was the same as the line element for the macroscopic universe, assuming both are globally flat, then one can substitute $R_0$ for $\sigma_\varepsilon / \sigma$ in the metric. But obviously there is local curvature to the universe and that must be accounted for particularly since observational analyses (based on our relative view of the universe) can, and often are, carried out in spherical space. On a microscopic level, the nucleus was considered at the time to be roughly spherical as well, so an application of the coordinate system transformation (physical to geometrical) needs to be applied to a spherical space.

**Projecting Uncertainty into Spherical Space**

Eddington's method of accomplishing this is to project the points of a spherical space *orthogonally* onto a flat space tangent at the origin. It is vital to remember here that Eddington was working in four-dimensional space and not our usual three-dimensions that can be visualized easily. Eddington's reason for doing this was a result of the fact that he found curvature and the extraordinary fluctuation to be the same. So, in essence, to him curvature was simply a result of natural background fluctuations and, were there no fluctuations, the universe would be flat (it's the same thing as saying that if there was no matter in the universe it would be entirely flat).

Geometrically, then, a theoretical (mathematical) sphere (call it a hypersphere) is used to represent some phenomena that is more easily analyzed in a spherical situation. In the usual flat space there is a geometrical origin, $P$. Consider, then, a particle at a point $T$ that has the coordinates $x_r$, $y_r$, and $z_r$. Representing the extraordinary fluctuation by curvature does not alter these three coordinates, but it does add the coordinate $u_r$ that represents the displacement of the particle from $T$ to $S$, which is on the hypersphere. The origin, $P$, then must also be displaced to the geometrical origin of the sphere, $O$. Obviously this displacement is equal to the radius of the sphere, $R_0$ since the flat space is tangential to the hypersphere. The hypersphere then has the equation:

$$x^2 + y^2 + z^2 = R_0^2.$$

However, the particle that has been displaced from $T$ to $S$ is actually represented by a probability distribution. So it's mean values in all coordinate directions are

$$\overline{x_r^2} = \overline{y_r^2} = \overline{z_r^2} = \overline{u_r^2} = \frac{1}{4} R_0^2.$$

The standard deviation, then, of any one coordinate of the particle from its mean value on the hypersphere is $\frac{1}{2} R_0$.

Let's now assume there are $N$ particles on the hypersphere. Denote the centroid of these particles as $O'$. The standard deviation of the centroid then is

$$\frac{1}{2} R_0 \cdot \frac{1}{\sqrt{N}} = R_0 / 2\sqrt{N} = \sigma.$$ $O'$ is simply a projection of the physical origin $P'$ just as $O$ is a projection of the geometrical origin $P$. From the discussion in chapter five regarding the range of nuclear forces, if both the geometrical and physical origins are represented by particles, they cannot be more than $10^{-13} cm$ apart. Thus $P'$ is close enough to $P$ that it can be approximated as being tangent to the hypersphere just as $P$ is (so it is a point both on the sphere and on the flat space tangential to the sphere). Since $P'$ is on the tangent flat space, there is no need to represent it with a $u$ coordinate (this coordinate displacement is zero, essentially, since the point lies in both spaces). As such the definition of the physical origin as the centroid of $N$ particles can be extended to spherical space by ignoring the $u$ coordinate. The coordinate differences between $O$ and $O'$ can then be represented by $x_0$, $y_0$, and $z_0$ that can also be used to describe the coordinate difference (and thus position) of the physical origin $P'$. Any $u_0$ that might exist is said to be the scale fluctuation. As such, a direct ratio of the line between the physical origins and the line between the geometrical origins is given by:

$$\frac{O'P'}{OP} = \frac{(R_0 - u_0)}{R_0}. \tag{6.1}$$

The scale fluctuation, $u_0$, is simply the standard deviation, $\sigma$. As such, (6.1) can be rewritten as

$$\frac{O'P'}{OP} = \frac{R_0 - \sigma}{R_0} = \frac{\sigma/\sigma_\varepsilon - \sigma}{\sigma/\sigma_\varepsilon} = 1 - \sigma_\varepsilon. \tag{6.2}$$

This essentially serves as a proof of the substitution he makes in (5.12) and implies his reasoning at least is mathematically correct. Physically some extended reasoning is required.

**Physical Justification for Linking Microscopic and Macroscopic**

In *The Nature of the Physical World* Eddington points out that it has usually been taken for granted that the usual theory of knowledge applied to macroscopic or large-scale surveys of the universe can equally be applied to microscopic situations and therein lies the problem. But there still must be a link between the two if the description of the universe is to be self-consistent. Again he appeals to general relativity where local irregularities that produce local curvature are simply superposed on the hypersphere that represents the universe. But can this appeal have any true physical meaning when using it to connect macroscopic to microscopic phenomena? Eddington's solution is to find a situation where general relativity and quantum mechanics actually agree (which is a rare find indeed). Such a situation is provided by a steady distribution or an Einstein universe (Eddington 1946). Eddington's interpretation of quantisation is that it is a "complication which arises from uniformity and symmetry" (Eddington 1946, p.12). Slight *non-uniformities* in quantum theory were often treated in Eddington's day (and still today in many situations) as perturbations. There is a subtle interplay here in the mathematics since quantum theory consists of dynamical integrals that produce the quantisation (most often describing angular momenta). When the non-uniformities are introduced they do not *alter* the integrals, though they do reduce the time that they persist. So a continual increase in non-uniformities in quantum theory reduces the time of persistence of the dynamic integrals thus reducing the quantisation. Since particles are non-uniformities in an empty universe (or new particles are non-uniformities in a steady distribution) the more particles that are present in an aggregate the less quantisation has any relevant effect. This is Eddington's explanation for the lack of quantum behaviour at macroscopic sizes – objects such as buildings, people, planets, etc. are huge aggregates of particles and thus the dynamical integrals of the individual particles act over such short periods that quantisation is small enough to go unnoticed.

Of course the stability of matter was not proven until 1967 by Freeman Dyson (b. 1923) of the Institute for Advanced Study in Princeton and Andrew Lenard (b. unknown) of Indiana University (as legend has it, driven by the offer of a bottle of wine to anyone who could prove matter's inherent stability). Dyson and Lenard's proof was later simplified by Elliott Lieb (b. unknown) of Princeton University and Walter Thirring (b. unknown) of the University of Vienna. The basis of the proofs lay in the exclusion principle that was first applied to the problem by Ehrenfest in 1931. Fully developed quantum statistics is more subtle – and beautifully simple in its almost Bohmian physical features (the simplicity of the mathematics, I suppose, is relative to one's mathematical acumen). It is a subject upon which much of the rest of this monograph is built, though from a distinctly Eddingtonian view. Ultimately it holds that all elementary particles and thus all atoms are absolutely identical to and therefore indistinguishable from one another. Quantum field theory explains this quite elegantly as the fact that particles are simply fluctuations in universal fields. For instance, every electron in the universe is simply a fluctuation in the same electron field, $\psi$, often called a psi-field (Zee 2003). (There is a subtle interplay in actual quantum field theory between the statistics of the field and the spin that gives rise to the various fundamental interactions.) What is striking is that Eddington's development, as I will show, presages many of these points, as well as others, in quantum field theory.

For example, consider Eddington's treatment of fundamental interactions. Since Eddington did not hold to the Yukawa interpretation of exchange or carrier particles acting as mediators for the transmission of forces, he was forced to find another way to explain how force information could be transmitted. He imagined an atom in a gravitational field where the non-uniformity was the gravitational field itself. In standard quantum mechanics the non-uniformity would treated as a perturbation that would not change the eigenstates of the atom but, rather, would induce transitions between them. The eigenstates are contained in a wave equation which, when represented in tensor form (see chapter five), contain $g_{\mu\nu}$. The eigenstates, however, are the same regardless of the presence of the non-uniformity. That means that the wave equation and thus the coefficients $g_{\mu\nu}$ are also unaffected by the perturbation. The inherent problem here is that general relativity modifies $g_{\mu\nu}$ in order to represent the gravitational field. Eddington needed a different interpretation, then, in order to properly represent the gravitational field in the wave equation. As such he viewed any attempt at extending Dirac's wave equation to general relativity as being misguided since the principle of equivalence does not apply here. Building on his interpretation of quantisation at macroscopic levels, he assumes that the $g_{\mu\nu}$ used in extensive structures is considerably different from the $g_{\mu\nu}$ used in the wave equation. The wide deviation in $g_{\mu\nu}$ then implies frequent transitions between the eigenstates which makes the wave analysis useless just like reducing the time of persistence of the dynamic integrals can reduce the effect of quantisation. Thus any problem involving $g_{\mu\nu}$ likely involves structures extensive enough to ignore wave-like or quantized effects (Eddington 1946). But, since Eddington introduced curvature via the extraordinary fluctuation rather than the presence of matter, he suggests a third form of relativity be conceived known as 'intermediate' relativity. Whereas special relativity deals with flat space-time and general relativity deals with non-uniform curvature, intermediate relativity would deal with *uniform* curvature.

**The Uranoid**

In a letter to Schrödinger in November of 1937 Eddington says

> I have sometimes thought of using the term <u>uranoid</u> instead of "universe" – corresponding to the <u>geoid</u> in … geodesy. The actual universe is irrelevant, because the experimental measurement is understood to be carried out in conditions represented by the <u>uranoid</u>; and the experiments will apply any corrections necessitated by the actual irregularities or disturbing conditions present (Eddington 1937).

Once again Eddington returns to the problem of observer and environment that is so much a part of quantum theory. As I will show this has a remarkable effect on the so-called 'intermediate' relativity he speaks of.

He takes a decidedly non-Bohmian view in dividing the universe into two parts called the 'object-system' and the 'environment.' The terms object-system, object-particle, object-field, etc. will refer to the object under immediate study while the environment is everything else (including things that might not just surround the object-system, but could also permeate it). Eddington suggests that it could be referred to as the 'background' though that term does not have the force of connotation that Eddington's uranoid does. Any object-system under consideration cannot be considered apart from its environment. Eddington's reasoning here is clearly based on relativity where "we do not recognise the concept of an atom as a thing complete in itself" (Eddington 1946, p. 13). Simple object-systems include elementary particles, simple atoms, etc. while simple environments include specific idealizations such as uniform, electrically neutral, etc. Eddington gives the name 'uranoid' to these simple environments. So far, the uranoid I have discussed in the greatest detail is a uniform probability distribution of particles or an Einstein universe occupying a hyperspherical space. A further clarification can be made if it is referred to as a 'zero-temperature' uranoid such that the particles are all nearly at exact rest. By limiting the environment to an effective zero-temperature radiation is eliminated from the consideration and only matter is present (this would obviously be unsuitable as a uranoid for considering the early universe). This uranoid is also electrically neutral so that any introduction of an electromagnetic field must accompany the object-system itself. This standard uranoid is the entire universe as a whole and is included in every problem. Eddington rationalizes this since dividing the universe into pieces would require calculating boundary conditions on the edges of the pieces.

Previously the $\sigma$-metric served as the connection between macroscopic and microscopic where a large-scale system determined the uncertainty in the physical reference frame that then determined the scale of the various small-scale structures in that frame (including microscopic structures). In the present discussion the connection appears to be one of mechanics where the physical interactions of the particles in the assemblage need to be considered. The former is a metrical effect since it involves measurement while the latter is a mechanical effect. But Einstein already united metrical and mechanical aspects of the universe by uniting geometry and mechanics in relativity theory. Here Eddington takes a decidedly geometric interpretation of the structure of the universe that implies it is, indeed, whole and continuous. Both the metrical and mechanical aspects are contained in the field of $g_{\mu\nu}$ and it can thus either influence the measured characteristics of the object-system through a perturbation or disturbance, or determine the standard of measurement used for reference in a measurement.

In a final nod to Newton, the field of $g_{\mu\nu}$ is usually just called the gravitational field but sometimes is known (not so much anymore, but certainly moreso sixty years ago) as the inertial-gravitational field. In this context the standard uranoid provides the inertial part of the field whereas the gravitational part arises from some disturbance or perturbation that is described as a deviation from the standard uranoid. The gravitational part obviously does not affect individual particles that much, particularly when they are simply interacting with each other. However, inertial considerations can be important in this instance. Einstein, of course, had assumed that inertial and gravitational affects were identical (principle of equivalence) and, though tests of this fact continue to this day, no experiment has proven otherwise. But, in attempting to adjust one's thinking to coincide with Eddington, it seems that he is making, here, a distinction between special and general relativity where special relativity plays the inertial role in these considerations and general relativity plays the gravitational role. As such, he is completely correct in assuming that inertial (special relativistic) considerations are important on the microscopic level but, as I will discuss in the coming chapters, he appears to abandon the equivalence principle as a result of this interpretation.

As for Eddington's precise views about the existence of an aether chapters eight and nine discuss this in greater depth, but suffice it to say he certainly assumed that space had some definite structure apart from the pure vacuum. Perhaps the earliest indication of this appears in a 1932 letter to Sir Joseph Larmor (1857 – 1942) where he asked Larmor to examine some calculations he had made on an enclosed sheet of paper describing the radiation emitted by a rotating ring of *n* electrons. He asked Larmor what would happen if one electron were removed from the ring. The ring would then become discontinuous as would the emitted radiation since the "propagation of a discontinuity is a discontinuous process" (Eddington 1932). He introduced a vector he called the "aether displacement" to describe the discontinuity that linked the electron to the aether so that any measurement on the electron also required consideration of the aether. One can clearly see the basis of the uranoid concept in this. He also extended this idea to instantaneous states such as the present instant, "now": the "world-wide instant 'now' is created by ourselves and has no existence apart from our geocentric outlook.…" (Eddington 1922, p. 17). In a tie once again to uncertainty, a four-dimensional world view removes such instantaneous states (see the discussion of spherical space above). Thus, uncertainty is inherent in space-time because the reference frame cannot be separated from the object under observation as I've previously shown. He did make the point that the aether was *not* a field. In an early draft of *Fundamental Theory* he asserted that to use the term "field" in place of "aether" was "ill-advised" (Slater 1957, p. 72). The strictly mechanical properties described both by second-rank tensors and wave mechanics allowed matter to look more like a field but was simply a way of describing its behaviour. As for the aether itself, a further discussion begins in chapter nine.

**Beating Around the Bush**

As the puzzle continues to be filled in, the idea of the uranoid releases again the nasty problem of the interconnectedness of observer, environment, and object all manifested in the painful act of measurement. Thus we return once again to the question of objectivity. Eddington is forced now to finally consider what to do about the plethora of units floating around in the fragmented universe. Ideally there should be one measurable unit that corresponds to the standard deviation, $\sigma$ (see chapter five) but which unit? Is it not somewhat of an arbitrary choice particularly since the standard deviation as used by Eddington is *unitless*? If it truly is arbitrary then it loses its objectivity. Eddington addresses this issue by adopting a system of natural units.

Eddington adopts the standard practice of setting $c = 1$. Now rather than adopting the standard $\hbar = 1$ and stopping (thereby relating all the needed units), or perhaps adopting another semi-standard, $G = 1$, he puts the two together in an unusual way by defining $8\pi G\hbar^2 = 1$. Eddington's rationale for doing this can be demonstrated by writing the fundamental equations for the energy tensor and momentum vector:

$$-8\pi GT_{\mu\nu} = G_{\mu\nu} - \tfrac{1}{2}g_{\mu\nu}G \quad \text{and} \quad p_\mu = -i\hbar \partial/\partial x_\mu.$$

$8\pi GT_{\mu\nu}$ is a spherical curvature that can be represented by the inverse square of a length (see (4.5) for the origin of $8\pi$). As such it has the same dimensions as $(p_\mu/\hbar)^2$. With Eddington's system of natural units $T_{\mu\nu}$ has the same units as $p_\mu^2$ or $p_\mu p_\nu$. Therefore, "An energy tensor is, both dimensionally and tensorially, the product of two momentum vectors" (Eddington 1946, p. 15). This, of course, is a key point in the derivation of four-momentum which is the relativistic union of energy and Newtonian momentum. In addition, Eddington's system is designed to define a particle density (number of particles in a unit volume or the corresponding probability of a particle in such a volume) as a momentum vector, once again linking relativity and quantum mechanics.

As with the usual system of natural units it leaves a single measurement (with a single unit) to be taken, be it a length, mass, density, etc. Each physical quantity has, then, what Eddington refers to as a "dimension-index" that shows how the physical quantity varies with the unit of the single measurement (Eddington 1946). So, for example, if the single unit, called the extraneous standard, is a length, the dimension-index shows how various physical quantities vary with length.

Once again in statements that presage Bohm's idea of fragmentation, Eddington holds that the *internal* structure of a system can still be entirely described by ratios of whole numbers, but in order to take a measurement at least one standard is necessary. This is perhaps a nod to the impossibility of a truly objective universe.[29] He recognizes that the an ideal system would be the whole universe that would make an outside standard unnecessary. But, as Eddington says, "the analytical method of physics divides the universe into simple systems of various types which are studied one by one" (Eddington 1946, p. 15).

---

[29] Bohm's holographic universe is an attempt at a truly objective viewpoint and has gained support in recent decades. It remains to be seen which view is correct or even if we can *make* a choice – lots of philosophical pitfalls here.

So, working with a standard of length ($k$, $R_0$, etc. measured in *cm*) as was developed in chapter five, any standard will have an associated uncertainty since the extraordinary fluctuation can be represented as a fluctuation in the scale with an accompanying standard deviation, $\sigma_\varepsilon$. Since every observable is referred to in this reference frame, they must be measured with an extraneous standard that has a standard deviation, $\sigma_\varepsilon$. For instance, rather than assigning an extraneous standard the value of 1 *cm*, we would assign it $1 \pm \sigma_\varepsilon cm$. Using the system of natural units as defined (and assuming we're still working with our example where the extraneous standard is a length), since any physical quantity would have dimensions (length)$^y$ it would be assigned a 'scale uncertainty' (transformation to length units) of $1 \pm y\sigma_\varepsilon$. This indicates that some form of scale-free physics would be most useful.

The standard uranoid consists of two linear characteristics $\sigma$ and $R_0$ that are assumed, for the purposes here, to be independent. Scale-free physics would consist of a system that is completely unconcerned with these two characteristics. In such a system structures can be adjusted to any scale meaning there really is no difference between macroscopic and microscopic unless they are directly compared. Ultimately, many practical applications can be simplified to scale-free physics by letting $\sigma$ approach zero or $R_0$ approach infinity. Ultimately this presents three distinct branches (fragments) in physics: scale-free involving neither $\sigma$ nor $R_0$, cosmological ("cosmical") physics involving only $R_0$, and quantum ("quantal") physics involving only $\sigma$. Obviously, there are a few areas that involve both as I have already outlined, but much (though clearly not all) of Eddington's emphasis in *Fundamental Theory* is on developing a relativistic version of quantum mechanics through scale-free methods. As such he refers to quantum ("quantal") physics as being scale-fixed in order to better compare it to the scale-free theory. Really quantal physics only includes those parts of quantum physics that involve quantisation and discrete eigenstates. Eddington's formal statement on scale-free physics is as follows:

> If we specify the characteristics of a system in terms of an extraneous standard, and consider the series of systems formed by varying the standard but keeping the specification the same, then (for a scale-free system), if one system of the series in physically possible, all are possible (Eddington 1946, p. 17).

**Wavefunctions and Observables in Eddington's Theory**

In standard quantum mechanics corresponding here to scale-fixed theory, wavefunctions are discrete and self-normalising in the sense that they are built on what Moore calls the state vector rule: "The state of a quanton [electron, photon, etc.] at a given time is described by using a *normalized state vector* $|\psi\rangle$ having a certain number of complex components" (Moore 2003b, p. 101). Every possible numerical value an observable might have will be associated with a normalized state vector that is called that value's eigenvector. A function $\bar{\psi}(q)$ can be built out of a generalized state vector $|\psi\rangle$ and an observable, $q$, by setting $\bar{\psi}(q_i) \equiv \psi_i = \langle q_i | \psi \rangle$. This function is interpreted by performing a measurement to determine the value of $q$. If such a measurement is carried out the probability that the measurement of the quanton in state $|\psi\rangle$ will give the value $q_i$ is:

$$\Pr(q_i) = |\langle q_i | \psi \rangle|^2 = |\bar{\psi}(q_i)|^2. \qquad (6.3)$$

One major problem with (6.3) is that as $dq \rightarrow 0$ (where $dq$ is the step size) the probability that a measurement $q$ will be made also decreases to zero. One way around this problem is to define a rescaled function as:

$$\psi(q) \equiv \frac{\bar{\psi}(q)}{\sqrt{dq}}. \qquad (6.4)$$

If this rescaled function describes a quanton's state it is referred to as the wavefunction. If it describes an eigenvector it is called an eigenfunction. The probability that a measurement of the quanton will yield a value $q_n$ is then:

$$\Pr(q_n) = |\bar{\psi}(q_n)|^2 = |\psi(q_n)\sqrt{dq}|^2 = |\psi(q_n)|^2 dq. \qquad (6.5)$$

The probability of being within a range of values is given by:

$$\Pr(q_{n-1} \leq q \leq q_{n+1}) = \int_{q_{n-1}}^{q_{n+1}} |\psi(q)|^2 \, dq. \tag{6.6}$$

Normalization means that as $dq \to 0$

$$1 = \langle \psi | \psi \rangle = \int_{-\infty}^{+\infty} |\psi(q)|^2 \, dq \tag{6.7}$$

which simply means that the probability of finding a value somewhere between -∞ and +∞ is 1 (i.e. a measurement is guaranteed to yield *a* value – maybe not the expected value, but some value of some sort). Now any measurement to determine $q$ will collapse its wavefunction to an eigenfunction before a result is given (this is wavefunction collapse as described in the section on measurement). This is a result of the interference of the observer (or apparatus) in the measurement. So, to some extent, the value is predetermined (this point is debatable among physicists) by the nature of the interference. In addition, the eigenfunctions of real observables involve $h$ making them discrete.

Another way of describing the above discrete wavefunction is as a particle density that "rapidly decreases outwards so that the integral over space converges" (Eddington 1946, p. 17). Being normalized it is then said to correspond to "unit occupation." Each eigenfunction is assigned an occupation factor $j$ that has an associated density. The occupation factor gives either the number of particles in the given state or the probability that there *is* a particle in that state. Sometimes the occupation factor is written as an operator, $J$, that reduces to an eigenvalue when there is definitely an integral number of particles in the given state.

Eddington draws a distinction between this relatively standard description of wavefunctions and what he refers to as pseudo-discrete wavefunctions. He associates the latter with the infinite plane waves commonly associated with elementary wave mechanics. This is *not* the same thing as a psi-field in quantum field theory since Eddington describes these infinite waves as not truly infinite – they are large when compared with $\sigma$ but small when compared with $R_0$ allowing both quantities to be ignored making the wavefunctions scale-free. A particle said to occupy a pseudo-discrete state represented by a pseudo-discrete wavefunction is said to be an "*unidentified* member of a large assemblage" (Eddington 1946, p. 17).

Part of the problem with wavefunctions is that they contain a phase factor that contributes to the interference properties of two interacting waves. This means that rather than replace a series of distribution functions for discrete values of some parameter with a single continuous distribution function, we replace it with a single wave function with the domain of the parameter being divided into small ranges each with an associated wave function. The latter are the pseudo-discrete wavefunctions just described. One way to interpret this is by saying that if the entire occupation is concentrated in one of these small ranges the system is nearly exact (e.g. a system, perhaps, nearly at rest).

The observables of wavefunctions are the eigenvalues. The eigenvalues correspond to various characteristics that can either be scale-fixed or scale-free. For instance, the proper mass of an elementary particle is obviously a fixed characteristic (theoretically there shouldn't be any variation in these characteristics but that is a debatable topic). Proper density, however, can be varied by varying the volume over which the density's distribution function extends (see the previous discussion on the physical interpretation of the standard deviation for an example of this). In general terms the energy tensor (and any associated particles) is scale-free and the momentum vector (and any associated particles) is scale-fixed. Once again Eddington draws a link between the macroscopic and the microscopic by representing the energy tensor as a wavefunction made up of pseudo-discrete wavefunctions as described above where the pseudo-discrete wavefunctions represent the particles of the larger assemblage (that is represented by the energy tensor). Each pseudo-discrete wavefunction can be thought to represent a small portion of the overall energy tensor labelled $\Delta T_{\mu\nu}$. This is a bit analogous to the superposition of wavefunctions in the usual sense.

In summary each particle has a wavefunction that is really a probability distribution. All of these wavefunctions are part of a single collective wavefunction. The particle wave functions are pseudo-discrete. We can only know the probability of each particle being in various states which means the occupation factors of the pseudo-discrete states are interpreted as probabilities of an individual particle or, alternatively, frequencies in the assemblage. So, for instance, even though masses are not scale-free quantities, a ratio of masses is equivalent to a ratio of densities where density, as described above, is not a fixed characteristic (and thus is scale-free). This will become useful in an analysis of the ratio of the masses of the proton to the electron.

It may help to pause here to digest this rather complicated discussion. Eddington himself offers some words that may help smooth out the rough edges a bit. In an early fragment described in Slater (1957), Eddington notes that in wave mechanics an observable is described by a product of two functions, while in relativity an observable is a relationship between two or more bodies (an observation point and observed object, or a reference point for a measurement between two observed objects). The problem is that in either case the observable properties really belong to the *relationship*. There must be some way to transfer the observable properties to the individual bodies themselves. In quantum mechanics, the self-properties of two observables together with their conjugates, $\varphi^*\varphi$ and $\psi^*\psi$, are observationally equivalent to $\varphi^*\psi$. Thus, if the observation is made and $\varphi^*\psi$ is measured, $\psi$ is given directly while $\varphi$ is found by noting that its complex conjugate is $\varphi^*$. If a wave function represents the definite momentum of a particle, its position is entirely uncertain. In relativity these self-properties are represented by the stress-energy tensor $T_{\mu\nu}$. The goal is to derive the eigenvalues for the observable that also gives a value for $\Delta T_{\mu\nu}$. In essence one can build the universe up from the smallest parts by building $T_{\mu\nu}$. In analyzing the problem from the quantum mechanical point-of-view, one sees that in essence the wavefunctions of the interacting particles are essentially correlated through coordinates. For example a distance between the two can be given by coordinate (vector) differences. A more physical example is given by Eddington: consider a box containing a proton and an electron. Each has a distribution function such that, if they are unobserved, they are equally likely to be anywhere in the box. At some point they will combine to form a hydrogen atom, thus emitting a photon. They are still equally likely to be anywhere in the box since neither has been directly observed. But their probability distributions and thus their coordinates are now correlated in a single function. As such an atomic wavefunction (e.g. for the hydrogen atom) is a correlation wavefunction in coordinates as opposed to a distribution wave function that describes a single particle's probability distribution. This description parallels that first employed by Schrödinger in his first paper on wave mechanics (Schrödinger 1926).

In one final comment on scale-free versus scale-fixed theory, quantisation appears only in the scale-fixed theory and is considered by Eddington to be a wholly electrical phenomenon since $\hbar$ is simply the "electrical unit" $e^2/c$ multiplied by a factor of 137 (see chapter nine for a full discussion). Eddington thus describes scale-free theory as "mechanical" (Eddington 1946, p. 19).

**A Problem with Observables**

There still is a problem with observables. From a theoretical standpoint an observer is rarely confronted with an object or system that is wholly foreign and presents absolutely no prior knowledge. For instance the mass and charge of the electron are now considered known quantities thanks to immense amounts of experimental data. These values are considered to be essentially free information by Eddington. However, mass is frequently experimentally derived from experiments employing conservation of momentum – indeed, Moore (2003) has described mass as a way of gauging a particle's reaction to momentum shifts since velocity is both relative and can be held approximately constant in elastic situations. Momentum in this instance, though, is governed by the uncertainty principle and thus its exact value – and, by inference, the exact value of mass – can be unknowable in such situations. Holding the velocity approximately constant in these situations can reduce the uncertainty but not eliminate it entirely.

A truly objective observer really can't obtain free information, then, since if they were presented with an unknown particle they would have to perform experiments relying on the uncertainty principle to infer what the particle's characteristics are (and thus infer what type of particle it is). Once again Eddington is blurring the distinction between classical and quantum uncertainty and is thus forced to beat around the bush a bit since there doesn't seem to be any way around this paradox. Thus any free or tabular information such as the mass or charge of an electron is called a 'stabilised characteristic.' Stabilised characteristics are really not observables since they have exact tabular values with no uncertainty or probability distribution (some tables will give standard errors, though, which could be considered the same thing). The act of stabilising the characteristics of a particle or system has the effect of reducing its degrees of freedom since any characteristic that is stabilised is no longer a variable. A note of caution, however: stabilising the components of a vector or tensor forces the abandonment of tensor transformation properties. However, stabilisation can be applied to invariant conditions that must be satisfied by the tensor as a whole, e.g. impose the condition on a second rank tensor that it is antisymmetrical, that it is the outer product of two vectors, or that it is the outer square of two vectors. Such conditions are invariant for tensor transformations. In addition they reduce the number of independent variables needed to specify the tensor, thus reducing the number of dimensions in the probability distribution.

**Introducing Fields**

In taking the above a step further Eddington very nearly reproposes Yukawa's meson-field theory, which he had previously dismissed, when discussing a particle's interaction with the environment. Particles (object-particles in Eddington-speak) disturb the distribution of surrounding particles through fundamental interactions (as I have previously mentioned, Eddington studies the only two interactions known with certainty at that time: electromagnetism and gravity). As such the surrounding environment cannot have the simple specification of the standard uranoid. Eddington treats the disturbed environment as the sum of a uniform environment such as the standard uranoid and a disturbance. As such the universe, for Eddington, consisted of particles, disturbances, and the standard uranoid.[30] In relativity theory the disturbance is considered part of the environment (it's an actual curvature of space-time) and is called, then, a field theory. In wave mechanics the disturbance is part of the particle (object-particle) itself as represented in its wave-nature. As such, Eddington refers to the disturbance in this instance as the "object-field" and prefers to work entirely with this representation of disturbances, leaving the standard uranoid as the background.

---

[30] I say 'disturbances' while, in actuality, Eddington uses the singular, 'disturbance' – as if he intended there to be only a single disturbance in the universe. I have mulled this over in my mind and I have concluded that he, in fact, meant the plural. However, if he had meant the singular it would imply that all localized disturbances – interactions – are really different manifestations of some overall 'disturbance field' so-to-speak. Since he does not elaborate on this latter point I suspect he is using it much like one might use 'trouble' in places of 'troubles.'

It is worth contrasting these ideas with those considered current. In the words of Zee:

> We thus interpret physics contained in [a] … simple field theory as follows: In region 1 in space-time there exists a source that sends out a "disturbance in the field," which is later absorbed by a sink in region 2 in space-time (Zee 2003, p. 24).

In current jargon experimentalists refer to the disturbance in the field as a particle. This goes back to Yukawa's original idea that the meson mediated the force between two nucleons that was analogous to the photon in electromagnetic fields (non-quantum fields, of course, being a nineteenth century construct). There are two universal fields, then: gravity and electromagnetism. There are also limited range fields associated with the strong and weak interactions. In any case, Zee describes the field as a sort of mattress (as on a bed). Lumps in the mattress are particles and standing on the mattress induces a disturbance that causes the lumps to move toward one another (it's worth trying this at home if for no other reason than an excuse to jump on the bed). The disturbance, then, produces an attractive interaction between the particles (represented here by lumps)[31] (Zee 2003). The lumps here are the source and sink described by Zee and the attractive interaction is the exchange or carrier particle.

Before 1959 quantum field theory generally considered space-time to be continuous on the microscopic level. The idea of 'quantizing the field' had not yet been born. But at a conference in Kiev that year, Soviet physicist Lev Landau (1908 – 1968) proposed replacing the conventional theory with one based on observables and "equally elementary compound particles" (Kragh 1999, p. 337). Landau's suggestion was driven by the inability to extend the field-like properties of quantum electrodynamics to the strong and weak interactions. Present quantum field theory includes a mix of pre- and post-1959 (Landau) interpretations – continuous or continuous-like fields that are quantized (there is still some philosophical debate about whether the fields are truly continuous – advocates for both sides can be vociferous). Eddington clearly held the pre-Landau interpretation of a continuous space-time.

---

[31] The astute reader will note that like charges *repel*. In calculating the photon propagation terms one finds that it is necessary to perform an integration by parts that flips one of the signs in the equation for the path integral (really in the equation for the action). As with nearly every sign-flip in physics, this indicates a change in direction – the photons cease to propagate *toward* another charge and instead propagate *away* from it. The mechanism that drives this ultimately turns out to be spin. See Zee 2003, pp. 31-35 for a fuller discussion.

Another problem that presented itself early in the development of quantum field theory is the fact that particles interact with their own fields. Alternatively one could say they interact with the psi-field (of which they are a part), for instance, immediately surrounding themselves regardless of the presence of other particles. Quantum field theories, then, must account for such self-interaction. The problem goes back to the first attempts to formulate a quantum theory of the electromagnetic field. Both Dirac and Jordan independently developed the first theories of quantum electrodynamics (QED) in 1927. Two year later Pauli and Heisenberg developed a version of QED that was relativistically invariant and quantized radiation as well as matter waves (Kragh 1999). The trouble was that the self-energy of the electron, which is the energy associated with the electron due to its own electromagnetic field, was infinite. As QED developed in the late 1930s more infinities crept out of the woodwork. The meson-field theories born largely out of Yukawa's work turned out to have their own diverging results. The problems were not resolved until well after Eddington's death.

But Eddington's formulation of object-fields partially addresses these issues, though obviously not necessarily in an accurate way. For instance, he defines two types of object-fields: extraneous and complementary. An extraneous field is a field external to the object under study, perhaps introduced intentionally for the sake of studying certain behaviour. The complementary field accounts for the readjustment of the environment due to the object's presence – essentially it's a self-energy-type field, though it is handled in a slightly different manner. As an example, if the object has some charge the complementary electromagnetic field represents the induced charge in the particle's surroundings (i.e. the presence of charge of one type must induce the presence of an opposite charge in the surroundings if one does not already exist since charge is

conserved[32]). Conversely, it could also represent any substitutions made in the equations in order to neglect induction effects in the environment. The latter representation is simply a mathematical difference between the disturbed and undisturbed environments measured either through energy tensors or momentum vectors. In this sense it follows on Eddington's idea of ideally measuring locations via coordinate differences directly between objects (i.e. with the origin essentially at one of the particles)[33]. Since the difference is measured through tensors or vectors it follows that the field has the same variables (energy, momentum, etc.) as a distribution of particles. Eddington acknowledges the description of a field in terms of potentials (something any second-semester physics students should be familiar with) but only employs this version when grouping the disturbance with the environment as in relativity (Eddington 1946).[34]

Particles, then, would carry two bits of information. For instance, they would have both a particle energy tensor that is associated with their fundamental properties as well as a complementary energy tensor that is associated with the energy they have as a result of their disturbance of the surrounding environment (i.e. their interaction with their own disturbance in the field). Eddington's use of the term 'environment' here indicates the strange way in which he divided up the universe. It acts as a middle layer, in a way, between the object-fields (particles) and the standard uranoid (zero-temperature Einstein universe). True fields in Eddington's sense, particularly quantum ones, can be associated with any of his three parts of the universe since fields are merely associated with the average characteristics of an ensemble of particles rather than with the characteristics of the individual particles themselves. Complementary fields (or self-energy fields in modern terms) are, of course, object-fields. Eddington felt that they had been ignored by quantum physics (Eddington 1946, p. 23). Though QED was dealing with a plague of self-energy issues in the late 1930s these issues were directly linked to unwanted infinities which was not something Eddington had encountered in his self-energy fields.

---

[32] There really is only *one* fundamental conservation law in this sense: the conservation of four-momentum. The conservation of momentum and energy are explicitly contained within this single law while the conservation of charge is implicitly contained within the conservation of energy. Since mass is the magnitude of the four-momentum vector in some sense conservation of mass is the most fundamental.
[33] This was developed at the beginning of chapter five. It very loosely resembles the Green's function portion of a path integral in the sense that, for example, a Green's function $G(x_1,x_2)$ does not depend on $x_1$ and $x_2$ separately but rather on the *difference* between them, $x_1 - x_2$.
[34] In an historical 'missed-it-by-*that*-much' event, in 1948, just four years after Eddington's death, Hendrik Casimir (1909 – 2000) proposed disturbing the vacuum such that there would be a shift in its energy density. While the energy density itself is not observable, its shift should be since the method of disturbing the vacuum is controllable. This shift leads to a small force (since experimentally discovered) known as the Casimir force. In this case the disturbance was grouped with the environment – in fact it was *in* the environment. Had Eddington lived a bit longer he likely would have made extensive use of the Casimir effect.

**Working with Fields**

In beginning his work with fields Eddington considers the treatment of gravitational fields in quantum theory where the movement (or even complete removal) of particles doesn't alter $g_{\mu\nu}$. The field in this case is then referred to as a rigid field. As such eigenstates can then be specified while the occupation can be assigned later. Since the occupation is flexible it allows physical change to be represented as the transitions between eigenstates with the field remaining unchanged. The method is only an approximation, however, with the condition that "the field must be stationary for small changes of the occupation factors of the eigenstates" (Eddington 1946, p. 24). Of interest is the fact that Eddington makes a distinction between quantum particles and 'relativity' particles. The former are the usual particles treated in standard physics while the latter are singularities. Singularities are not discussed much in *Fundamental Theory*, particularly in the early chapters, for a variety of reasons including the prevailing notion at that time that they were simply ugly infinities that needed to be eliminated. But perhaps the clearest reason is simply that it didn't fit the theme of Eddington's theory. Quite simply, rigid fields serve as the foundation for further development of his theory and singularities arise from a non-rigid field (since any movement or removal obviously changes $g_{\mu\nu}$).

      As described above, the field acquires its own characteristics including its own energy separate from the particle energy. In modern interpretations there are several ways of looking at this (without getting overly philosophical) – either the particle is simply a fluctuation in the field and so is a subset of the whole field energy, or the particle's energy is produced separate from any one individual field and is produced by interacting fields. The trouble with all of this comes when considering the graviton. All fields and particles "live" in space-time but the graviton *is the quantization* of space-time. The graviton had not been formally postulated by the time Eddington had died, though

quantizing space-time is an idea that had been bantered about (though very informally) since the beginning of quantum theory.

Eddington separates out the particle energies by discrete eigenstates where each state is assigned an occupation factor. Changes in the system are manifested by changes to the occupation states that are represented as generalized coordinates or momenta. The total energy of the system is then a function of the occupation factors, $H^0(j_1, j_2, j_3,...)$. $H^0$ is not necessarily a linear function. The energy of a particle in a given state is given as:

$$E_r = \partial H^0 / \partial j_r .\qquad(6.8)$$

Eddington's rationale for this is that any change in the energy of the system must be wholly accounted for by the particles if the field is to be rigid – i.e. the field's energy does not change. Presumably this makes the calculations simpler since the field assumes a role analogous to the uranoid by remaining stable through various system transitions. The collective energy of *all* the particles is then given by:

$$E^0 = \sum j_r E_r = \sum j_r \partial H^0 / \partial j_r .\qquad(6.9)$$

The field energy is then the difference $W^0 = H^0 - E^0$. If the total energy of the system is a homogenous function of the *n*th degree then:

$$E^0 = nH^0, \quad W^0 = (1-n)H^0 .\qquad(6.10)$$

In this analysis Eddington has outlined the treatment of energy. A similar process can be performed for other characteristics that are additive such as energy-density, momentum, pressure, angular momentum, and even, presumably mass, though the interpretation of the latter could be tricky. Such a process, however, assumes the system is an unidentified member of a large assemblage and that the majority of other systems in the assemblage are in some initial state.

When this process is applied to a scale-free system, however, a complication arises in that the eigenstates are no longer discrete meaning they no longer are specified by quantum numbers. As such Eddington develops a generalized characteristic $X_\alpha$ ($\alpha =$ 1,2,...,n) where any such characteristic will always have the same physical dimensions with an associated extraneous standard. For instance $X_\alpha$ could be a set of coordinates in some *n*-dimensional 'representation space' (Eddington 1946). Points in this space don't necessarily correspond to possible states (so it's not simply the location of particles in some assemblage as described above). What actually is specified is some relation *between* the characteristics $X_\alpha$. Again, this simply builds on the idea developed earlier that it is really *relational* coordinates (or properties) that have physical meaning – i.e. you always have to have some reference *object*. The actual number of possible states, $k$, will be less that than the number of dimensions in representation space, $n$; in essence the possible states for a $k$-dimensional locus in the representation space that is referred to as the phase space of the system. The number of dimensions $k$ is referred to here as the 'multiplicity factor' (Eddington 1946). I will have more to say regarding multiplicity factors in the next section. For now it will serve as a useful tool in adopting the previous formalism to scale-free systems.

Phase space is, of course, a term culled from classical mechanics where each dimension of the space corresponds to an independent state variable of a system (position, velocity, momentum, etc.). It is a purely mathematical concept that simply provides a tool for better analyzing a system. In the present discussion it serves as a tool for making scale transformations such that $X_\alpha \to \lambda X_\alpha$ transforms some $k$-dimensional element of a volume $d\tau$ into a new element of volume $\lambda^k d\tau$. The discrete occupation factors are now replaced by a continuous occupation factor that is a function of the coordinates. As such equations (6.8), (6.9), and (6.10) can be rewritten by replacing summations with integrations and ordinary differentiation with respect to $j_r$ with Hamiltonian differentiation with respect to $j(X)$ – i.e. differentiation with respect to a function rather than a variable. Equation (6.10) becomes:

$$W^0 = -\left(\frac{l+k}{l}\right)E^0, \quad H^0 = -\frac{k}{l}E^0 \qquad (6.11)$$

where $l$ is the dimension-index of $H^0$. As described above, the scale-free condition makes $H^0$ a homogenous function, in this case of degree $-l/k$, of the pseudo-discrete occupation factors. If $l = 1$ then (6.11) becomes:

$$W^0 = -(k+1)E^0, \quad H^0 = -kE^0. \qquad (6.12)$$

As an example, the energy tensor is a scale-free characteristic and thus the simplest kind of scale-free particle will only have the energy tensor as a characteristic.

The components of the energy tensor are then the $X_\alpha$. For such particles, the total energy $H^0$ is the energy tensor:

$$W_{\mu\nu} = -(k+1)E_{\mu\nu}, \quad T_{\mu\nu} = -kE_{\mu\nu}. \tag{6.13}$$

where $k$ (the multiplicity factor) is the number of independent components of $T_{\mu\nu}$. The value of $k$ can be reduced by introducing stabilized characteristics thus altering the numerical division of (6.13). Through a redefinition of the energy tensor he is able, once again, to make a direct link between the quantum (discrete) world and the relativistic (continuous) world: a relativity particle is formally a quantum particle with a multiplicity of –1. In order to understand this result a bit better a deeper look at multiplicity is required.

**Multiplicity**

Multiplicity is most often defined in relation to Einstein solids and monotomic ideal gases though it can be applied to virtually anything where the multiplicity of a macrostate is the number of possible microstates for that macrostate. Eddington's use is fairly standard in that $k$ in both cases is the number of possible states (microstates) in some larger system or space (macrostate). Multiplicity can be described as being a function of both the total energy of the state (system) and the number of particles (units) in the state (system).[35] An Einstein solid, for example, can be modelled as a system of $3N$ independent oscillators where $N$ is the number of atoms[36]. Since the factor 3 is due to the dimensionality of the space (degrees of freedom) this could be altered in Eddington's theory since he does make alterations to the total number of dimensions in space-time. But, for the time being I will continue to assume a three-dimensional space-time.

    The total energy in an Einstein solid, $U$, is an integer multiple of some basic energy unit $\varepsilon$. The total number of energy units in the solid are defined as $q \equiv U/\varepsilon$. In Eddington's theory, rather than dealing with a solid, one would simply deal with a large assemblage of particles. In any case, the multiplicity of an Einstein solid's macrostate, in

---

[35] For an indepth discussion of multiplicity and how it can be fully derived in numerous situations see Schroeder (1999).
[36] This is similar to Zee's description of quantum fields (Zee 2003).

modern notation (dropping Eddington's use of *k* for reasons that will shortly become obvious), is given by:

$$\Omega(N,U) = \frac{(q+3N-1)!}{q!(3N-1)!} \tag{6.14}$$

(Moore 2003c, p. 66). In Eddington's theory the factor of 3 would obviously be different as would the exact value of *q*. But the basic idea holds: multiplicity is the number of possible states in a system.

There is an additional macroscopic step that can be taken with Einstein solids. If two such solids are in contact there is a total energy to the system made up of the energies of the two solids. The total energy of the system can be partitioned by multiples of the basic energy unit in a number of ways that are not necessarily confined by the solids themselves. The total multiplicity of the system, then, is the product of the multiplicities of the two solids. In adapting this to Eddington's theory, one could macropartition some larger system like the universe into smaller macroscopic systems in order to find the multiplicity of some larger system. Since we're dealing with exponentials at the fundamental level here this makes sense. As long as the total energy *U* remains fixed the combined system of two solids (or subsystems) can (and will) randomly shift between different microstates. This is actually known as the 'fundamental assumption' of statistical mechanics and is formally stated as: "All of a system's accessible microstates are equally likely in the long run" (Moore 2003c, p. 68). The term 'accessible' basically means that the microstates must shift such that *U* remains constant. This is the same idea as Eddington's description of the electron and proton in a box: at some point they will join to form hydrogen thus correlating their wavefunctions but both still have a non-zero probability of being anywhere in the box.

Some implications of (6.14) that can easily be transferred to Eddington's theory without knowing *q* or the multiple of *N* include the fact that larger systems will have huge multiplicities – in fact *ridiculously* huge multiplicities (see Moore 2003c, p. 78) – as well as the fact that the shape of the macropartition distribution becomes less and less Gaussian as the system gets larger to the point where the probability distribution is essentially an infinitesimally narrow spike. Moore concludes for Einstein solids that 1.) if the system is not in the most probable macropartition to begin with it *will* fairly rapidly move toward that macropartition and 2.) it will *stay* at that macropartition regardless of random exchanges in microstates between the solids. This is simply a statistical way of pointing out that systems tend to move toward equilibrium, which is just the Second Law of Thermodynamics. As such there's no reason this can't be applied to a situation such as Eddington's – in fact it's fundamentally *required*. In standard statistical mechanics this is simply interpreted as the fact that random and/or quantum processes even out on a macroscopic level (hence the reason Newtonian mechanics works fine as a predictive tool for macroscopic objects).

One final definition that theoretically could be extended to Eddington's theory (though more research would be needed to determine the appropriateness of the use of Boltzmann's constant) is the following relationship between entropy and multiplicity for a macrostate, where $k_B$ is Boltzmann's constant (not to be confused with Eddington's notation for multiplicity, $k$):

$$S \equiv k_B \ln \Omega. \tag{6.15}$$

Again, since the fact that the entropy of an isolated system never decreases is merely a restatement of the second law, it follows logically that this can be applied to Eddington's situation. In a frequently misleading interpretation, entropy is often deemed a measure of disorder in a system. In reality, the basic meaning of entropy is *multiplicity* and *not* disorder as is commonly held (Moore 2003c). In Eddington's theory multiplicity is interpreted in three different ways: the number of degrees of freedom of the system, the number of phase-space dimensions, and the number of components in the energy tensor. All three interpretations are equivalent.

**Probability in Statistical Mechanics**

Obviously the mere name 'statistical mechanics' implies that probability has a formal role in any such development. But probability actually has a *direct* measurable relationship with multiplicity in a way consistent with Eddington's philosophy of whole-number ratios. It is a direct application of equation (6.15) and leads to what is known as the Boltzmann factor.

Consider, then, a small system with two quantum states having energies $E_0$ and $E_1$ where $E_1 > E_0$. If, for instance, the lower energy corresponded to the vacuum energy and the higher energy corresponded to the ground state of some particle, in a highly simplified (and incomplete) way this situation could be interpreted to be the probability that a particle exists at all in some region. Fundamental statistical mechanics states that the ratio of the probabilities is equal to the ratio of the system's multiplicities in each case:

$$\frac{\Pr(E_1)}{\Pr(E_0)} = \frac{\Omega_1}{\Omega_0}. \tag{6.16}$$

From (6.15) we get:

$$\frac{\Pr(E_1)}{\Pr(E_0)} = \frac{e^{S_1/k_B}}{e^{S_0/k_B}} = e^{(S_1 - S_0)/k_B} = e^{\Delta S/k_B}. \tag{6.17}$$

In reality the change in entropy here is for a reservoir that supplies the small system with a theoretically inexhaustible supply of energy. But the combined system of the reservoir and the small system together must conserve energy if it is isolated. Thus whatever energy is lost by the reservoir must be gained by the small system, $\Delta U = (E_1 - E_0)$. Since $1/T = \partial S/\partial U$, for the reservoir $\Delta S \approx \Delta U/T$ as long as $T$ remains approximately constant. This is the well-known definition for change in entropy that is given in most standard introductory physics texts. The assumption that $T$ will remain approximately constant is fitting for Eddington's theory since he assumes in most cases an unchanging environment (and/or uranoid). Equation (6.17) can then be written as:

$$\frac{\Pr(E_1)}{\Pr(E_0)} = e^{\Delta U/k_B T} = e^{-(E_1 - E_0)/k_B T} = \frac{e^{-E_1/k_B T}}{e^{-E_0/k_B T}} \tag{6.18}$$

where the minus-sign is a result of determining the values for the reservoir rather than the small system. This relationship must hold true for all pairs of small-system quantum states meaning that the probability that a small system in contact with a reservoir will be in a quantum state with energy $E$ is:

$$\Pr(E) = \frac{1}{Z} e^{-E/k_B T}. \tag{6.19}$$

The proportionality constant $1/Z$ must be the same for all the quantum states of the small system. The value of this constant follows from the knowledge that the *total* probability of the system being in *some* quantum state must be 1. Thus:

$$1 = \sum_{all} \Pr(E_i) = \sum_{all} \frac{1}{Z} e^{-E_i/k_B T} = \frac{1}{Z} \sum_{all} e^{-E_i/k_B T} . \tag{6.20}$$

Leading to the following value for $Z$:

$$Z = \sum_{all} e^{-E_i/k_B T} . \tag{6.21}$$

Equation (6.19) then becomes:

$$\Pr(E) = \frac{e^{-E/k_B T}}{\sum_{all} e^{-E_i/k_B T}} \tag{6.22}$$

where $e^{-E/k_B T}$ is known as the Boltzmann factor (Moore 2003c).

Interpreting this in terms of Eddington's theory, the argument is very similar to the reasoning that follows from equations (6.9) and (6.10), though from a probabilistic standpoint. The denominator (the assemblage) in (6.22) is an averaged sum over the field that acts as the reservoir in this case and the numerator acts as the particle (system). This is a bit like Zee's source and sink description of fields. The equation obeys Eddington's two rules: 1.) the system under consideration is an unidentified member of a large assemblage and 2.) most of the other systems in the assemblage remain in their initial state. The multiplicity tells us the number of *possible* microstates (vacuum fluctuations, particle masses, etc.) in a given macrostate (environment, uranoid, etc.) and (6.22) tells us the probability that a system will be in one of these microstates (a quantum state) with a given energy relative the macrostate's energy. In terms of observables like momentum, these probabilities must account for the differences in physical and geometrical coordinates. In fact, the probability of a *physical* momentum is actually the combined probability of the *geometrical* momentum of the particle and an opposite (recoil) momentum of the physical origin. As such "the distribution of geometrical momenta is turned into a distribution of physical momenta by weighing the ranges *dp*" with a statistical weight function (Eddington 1946, p. 76). We've basically pulled the field and particle energies apart probabilistically, though this is a simplification. True QFT gets much more sophisticated than this (as does true quantum mechanics), but the gist here is to understand that Eddington's thinking, despite the unusual notation and nomenclature, is remarkably modern in many ways.

One final note on the use of multiplicities is the consideration of quantum particles that have (reverting to Eddington's notation for multiplicity) multiplicities of $k_1$ and $k_2$. The relation between their masses is:

$$\frac{m_1}{m_2} = \frac{k_2}{k_1}. \tag{6.23}$$

The application of this to protons and electrons will be presented in chapter eight but it is worth mentioning here since a relation could be drawn between (6.23) and (6.14) (and extended to include (6.22)) such that multiplicities, and thus probabilities, can be used to determine mass ratios.

**Coordinates in Rigid Fields**

The above arguments make it clear that the field energy is fairly large from an ordinary point of view, but Eddington points out that, for instance, gravitational fields can be obtained by a transformation of coordinates and that this is how such large field strengths are reached when working with rigid fields. A set of 'rigid coordinates' is then introduced that must satisfy the condition that the field remains stationary for small changes of the occupation factors. The relation between Galilean coordinates (unprimed) and rigid coordinates (primed) are:

$$x' = x, \quad y' = y, \quad z' = z, \quad t' = -kt. \tag{6.24}$$

Since $g'_{\mu\nu}$ have Galilean values,

$$g_{44} = k^2, \quad \sqrt{-g} = -k. \tag{6.25}$$

The spatial coordinates here remain unchanged so the number of particles per unit coordinate remain unchanged. As such the only energy-momentum coordinate that changes is the temporal one,

$$p_4 = -kp'_4.\text{[37]}$$

---

[37] In four-momentum this is the energy component so Eddington is talking about a system change *without* a change in Newtonian momentum.

Eddington gives this as the reason the momentum vector in wave mechanics is a covariant expression $p_\mu = -i\hbar \partial/\partial x_\mu$ [sic] (rigid representation) while in relativity it is a contravariant expression $p^\mu = m dx_\mu/ds$ [sic] (non-rigid representation). He also gives this as the reason that the velocity of a particle in wave mechanics is really the group velocity of the waves rather than the wave velocity. Creating a gravitational field through a coordinate transformation, then, satisfies (6.13) (Eddington 1946).

The minus sign here is less problematic than it appears since it is accounted for in the *i* as in normal quantum mechanics. But an unrecognized problem takes its place – the constants measured by experimentalists are not, in fact, the 'true' values but are distorted by our non-objective view. The 'true' values can only be obtained by a coordinate transformation such as the one just described.

**Energy Issues**

Since *k* introduces a minus sign the particle energy always has the opposite sign of the total energy. Eddington is clearly advocating in favour of Dirac's hole theory that suggests there is a 'sea' of negative energy states that are almost entirely filled (see fuller discussion in chapters seven through nine). Eddington's application of this to gravitation means that by inserting a particle with a mass *m* at some point *P* the energy in the region containing *P* is increased by the amount *m* (employing natural units, of course). Gravity, though, is treated as a potential well and particles in its presence have a small amount of negative energy. So the introduction of the particle introduces a small gravitational potential that supplies a negative amount of energy from all the particles in this environment. It is conceivable that the addition of the energy *m* from the addition of the particle is actually more than counteracted by the negative energy arising from all the surrounding particles due to the new gravitational potential – i.e. adding a particle seems to *decrease* the net energy. Thus to increase the net energy one would have to add a particle with a negative mass. In a classical sense this poses a serious problem but from the point of view of wave mechanics, the positive and negative energies are merely superposed uniform distributions (Eddington 1946). There may also be a link here to the somewhat artificial separation of energy and momentum in non-relativistic situations where in four-momentum the sign convention is opposite for energy and momentum (if one is positive, the other is negative).

How is a particle of negative mass (energy) introduced? Given a particle with mass *m* it must have an associated field energy of –2*m* so that the total rest energy is –*m* thus satisfying the need for a particle with a negative energy (mass). Now put this particle in motion such that it gains some kinetic energy. Since the field is rigid the kinetic energy is only added to the particle's energy. What happens if we simply flip the signs here? The kinetic energy is now $-\frac{1}{2}mv^2$ and the total rest energy is now *m*. The whole energy, including both rest and kinetic portions, is $m - \frac{1}{2}mv^2$. To avoid what Eddington refers to as a dynamical paradox (how can something have a *negative* amount of kinetic energy?) this needs to be rewritten as $m + \frac{1}{2}m(iv)^2$. So the particle in a rigid field has a velocity *i* times the classical velocity and thus a momentum *i* times the classical momentum. This is Eddington's explanation for the presence of *i* in quantum theory, basically through an argument involving particle self-energies and the separation of field and particle energies. A physical explanation can be found in the Ahronov-Bohm effect where a quantum wave-like particle is split into two partial waves that then pass on either side of a region of a magnetic field. The recombination must involve a phase shift and introduces the need for complex algebra. But this effect was not discovered until nearly a decade after Eddington's death.

This all presents one significant problem that should be evident by now considering everything I've introduced about Eddington's theory so far – how is any of this measurable, particularly since the measured quantities are really relative values between a particle and some reference particle (see discussion of measurement theory above)? Really any measurement made determines characteristics that belong both to the particle under consideration *and* some reference particle *jointly*. So consider an object-particle (the one under observation) and a reference particle that will be called the 'comparison particle' (Eddington 1946). Their momentum vectors are $p_\mu$ and $p'_\mu$ respectively. The so-called 'mutual' energy tensor that describes the two as a single system is of the form:

$$M_{\mu\nu} = \tfrac{1}{2}C\left(p_\mu p'_\nu + p'_\mu p_\nu\right). \tag{6.26}$$

The self-energy tensors for each take the form:

$$T_{\mu\nu} = Ap_\mu p_\nu, \quad T'_{\mu\nu} = A'p'_\mu p'_\nu. \tag{6.27}$$

Here $C$, $A$, and $A'$ are dimensionless constants. For particles nearly at rest the three energy tensors reduce to densities where the densities of the two particles sum to the mutual density:

$$\rho_m = Cmm', \quad \rho = Am^2, \quad \rho' = A'm'^2$$
$$\Rightarrow \quad Am^2 + A'm'^2 = Cmm'. \tag{6.28}$$

Values for the constants will be explored in chapter nine while comparison particles will be discussed in chapter seven. The important point here is to recognize the relationship between the measured quantities (representing mutual quantities) and actual desired quantities (representing the object particle related to a comparison particle). Basically Eddington is making the point that one of the reasons measured values are never exact and always contain some amount of uncertainty is because *they have not been properly corrected to account for the field*.

Adding spin to the mix can make the problems more difficult since spin is not only a measurable quantity but it is also a purely quantum quantity in the sense that it has no real classical analogue (it's not really just angular motion in the classical sense). The Riemann tensor in (3+1)-dimensional space-time has 256 independent components from 16 indices (the "energy tensor is both dimensionally and tensorially the product of two momentum vectors" – see Eddington 1946, p.15), though this number can be reduced to 20. The complete energy tensor has 136 independent components when including spin and only 10 without it. Eddington calls a particle that carries only a complete energy tensor a standard carrier. If a particle carries an additional 'permutation variate' it will have 137 dimensions in phase space. This additional variate adds energy to the system via the idea of interchange, which will be identified with the Coulomb energy (see chapter seven). As such it adds a degree of freedom to the system. There is also a non-Coulomb energy that is an adjustment of any initial energy in the system that allows it to be reduced from four particles to two.

**Scale as a Variable**

Since Eddington has equated the uncertainty of the scale with curvature through the extraordinary fluctuation, by dealing directly with the scale uncertainty curvature is automatically considered. As such all equations can be treated in a flat space so that local curvature is simply overlaid via the scale uncertainty. The scale is now treated as a separate observable with a probability distribution that is specified by the momenta and coordinates. The variables in probability distributions occur in conjugate pairs so the scale must have a conjugate. The conjugate of scale is called the 'phase' (Eddington 1946). If scale reduces to an eigenvalue it is considered a momentum while the phase would be considered a coordinate. This phase coordinate is now considered the fifth dimension normal to space-time. This makes the scale and phase invariant for rotations and transformations (including Lorentz). Incidentally the gravitational potential between two particles in $(n + 3 + 1)$-dimensional space follows a $1/r^{1+n}$ dependence assuming that $r \ll R$ where $R$ is the scale associated with any extra coordinates. This gives rise to the possibility of 'large' (at least on a particle physics scale) extra dimensions (Zee 2003). The graviton, in this situation, is actually a fluctuation throughout *all* dimensions rather than just the (3 +1) we're familiar with. This makes the graviton special since it is the only particle we know of that exists in more than four dimensions.

    The scale uncertainty, as stated before, is primarily a fluctuation in the extraneous standard that is reflected in the measured characteristics of the system. This fact compares rather well with the Casimir Effect (see footnote 37) where the shift in the energy density could be interpreted as a fluctuation in the extraneous standard. In the Casimir Effect the measured characteristics all come from the *shift* in the energy density since the energy density itself is unmeasurable. As I will show in chapter seven, the extraneous standard (among other quantities) is represented by a 'comparison particle,' a description not unlike the modern particle physics representations of nearly everything, including forces, as particles (this is rather ironic since Eddington disavowed Yukawa's meson-field theory – see chapter five). The scale and phase dimension are equivalent to this use of comparison particles and is introduced first in my presentation in order to maintain the cohesive field-theoretic argument to this point.

    Curvature as represented in the extraordinary fluctuation makes its measurable appearance through something known as the scale momentum that must be given a

Gaussian probability distribution with a standard deviation of $\sigma_\varepsilon$. In many cases this is a stabilised characteristic. If it is *de*stabilised it opens a new series of investigations to us. As an example, consider a scale momentum that is an angular momentum. The phase coordinate in this case is an angle since scale and phase are conjugate. For angular coordinates with a uniform probability distribution between 0 and $2\pi$ there is an infinite uncertainty, i.e. as the uncertainty in the angular momentum decreases, the angle "tends toward a uniform distribution over the range $2\pi$" (Eddington 1946, p. 47). Or, another way of looking at it is by assuming that an exact scale is extended to a slightly fluctuating scale "by spreading the distribution uniformly over a thickness $2\pi$ in an extra phase dimension" (Eddington 1946, p. 47). Eddington calls this the 'widening factor.'

What we have then is two equivalent ways of representing curvature. The first is the familiar way of having a stabilised scale in spherical space. The second is as a fluctuating scale in flat space. The widening factor is used in a comparison of the two approaches. For example, consider a spherical space with volume $V = 2\pi^2 R_0^3$. This can be reconsidered as a volume $V_3 = \pi R_0^3$ in three-dimensional space with a thickness of $2\pi$ in an extra phase dimension. A flat (Euclidean) sphere has a volume $V_4 = \frac{4}{3}\pi R_0^3$. Comparing these two gives $V_3 = \frac{3}{4} V_4$ but since in (Eddington's) natural units the inverse of a volume is a mass, this gives:

$$m_3 = \tfrac{4}{3} m_4. \tag{6.29}$$

This is an example of the comparison of masses and multiplicities given by equation (6.23) where $V_3$ is stabilised (exact) and $V_4$ is destabilised, thus giving it an extra degree of freedom (multiplicity). This is, in fact, just Eddington-speak for saying that adding extra dimensions adds degrees of freedom (1:1) and the multiplicity is just a measure of the degrees of freedom of a system which is perfectly consistent with our standard treatment.

*Fundamental Theory* **Redux**

Before moving on to discussions of the exclusion principle and calculations of the fundamental constants in *Fundamental Theory*, I will pause for a moment to summarize what we have learned so far in Eddington's theory. Despite its often abstruse formalism, the basic ideas are fairly simple. Eddington builds quite simply on the idea that everything – all particles – can be represented by Gaussian probability distributions. Relativity then requires that no particular point anywhere in the universe is special so the only relevant measurement is *between* particles (probability distributions). Since at least one of the particles involved in a measurement must be considered the reference body (at least in terms of placement of a coordinate origin), there are actually two locations of this point: the particle, which is subject to the uncertainty principle, has a probability distribution associated with its location, while the observer then places a geometrical coordinate origin as near as possible to the mean for this distribution. The geometrical origin, though, being limited by the ability of the observer (among other problems) has its own probability distribution. There is a small difference between the location of the physical origin and the location of the geometrical origin. There is also a standard deviation to the physical origin's probability distribution and this is used to furnish a measuring standard in the hope of developing a truly objective theory. This standard deviation becomes a standard unit for measuring length. There is also an uncertainty in the scale of measurement called the scale uncertainty that is due to the inaccuracies of standard measurement techniques. It is associated, once again, with the idea that no observation can be truly objective and, despite the use of natural units, at least one unit must be specified – called an extraneous standard – and will contain this scale uncertainty (since the extraneous standard is a scale). Specifying values of observables then becomes known as stabilisation. It is possible to define a scale-free form of physics by not referring to either the extraneous standard (standard deviation) or the radius of spherical space (i.e. working somewhere in between).

    Fields are introduced then since no particle exists purely in and of itself (at the very least it has its own self-energy producing fields) – i.e. particles cannot be considered apart from their environment (particles *are* the fields or field sources, as it were). In order to simplify calculations situations must be arranged such that the fields stay relatively unchanged for any transitions in them and, thus, the concept of a rigid field is

created.  It becomes necessary, then, to separate the field and particle energies so one can be sure the calculations are being performed on the particles alone.  It also becomes necessary to define a set of rigid coordinates to apply to the rigid field.  In performing these two operations (separating the energies and supplying the coordinates) the true depth of the interaction of the field and particle energies must be considered along with any energy contributed to the system from any reference particle.

All of this introduces the idea of scale being a measurable quantity meaning in wave mechanics it must have a conjugate which Eddington calls the phase.  In considering these issues in relation to tensor quantities one finds that there is an additional degree of freedom in electrical problems.  As such there are also both Coulomb and non-Coulomb energies associated with such systems, the latter arising from the reduction of a system to a two-particle system (including comparison particles which will be discussed in the next chapter).

With fields and particle-field interactions taken care of, the next logical step is to address that thorny issue of measurement again from the point of view of reference bodies.  This will lead directly to the exclusion principle.

# VII

## *Uniqueness*

It is somewhat ironic that one of the primary purposes of quantum field theory is to explain individual particles making QFT the particle physicist's primary mode of analysis. As such a discussion of fields frequently appears to break down into a discussion of particles once again, but with a different formalism than that we're used to from quantum mechanics. This new formalism is the result of QFT. The same thought process is evident in *Fundamental Theory* where the discussion of fields breaks down into a discussion of particles. In truth, Eddington simply seeks to eliminate surrounding fields by including them in the environment or background – to some extent they are a nuisance, but they cannot be ignored since a particle can't be completely removed from its surroundings.

Having considered primarily how particles can be measured individually in chapter five and how they then individually interact with fields in the previous chapter, it is now time to consider in greater depth how particles interact with each other and how they can best be measured. Ultimately this leads directly to the exclusion principle and Eddington's somewhat unusual (and rather modern) interpretation of it.

**Carrying Information: A How-to Guide**

In the previous chapter I introduced the concept of the 'standard carrier.' To reiterate, Eddington's view of particles was that they were nothing more than conceptual carriers of a set of variates. The standard carrier is the simplest type of carrier in scale-free theory and, again as I've stated before, carries a complete energy tensor including spin and nothing more. As such they have 136 independent components and are referred to as $V_{136}$ particles. What we're really describing here is a mathematical form – a blank sheet as I described in chapter three – on which various characteristics and properties can be superimposed. Basically one starts with a complete energy tensor whose components are empty. How the components are filled determines what type of particle it becomes. This definition does not simply apply to elementary particles (why should it?) but also to composite particles such as atoms and, perhaps, even molecules (in fact the definition could be extended to organisms, theoretically, but the components would be so phenomenally complex it would likely be impossible to write them down!). Eddington, though, recognizing the difficulty with more complex combinations, refers to anything beyond simple combinations as systems and does not extend the definition to include these.

      Eddington's conception of the standard carrier as an empty mathematical tool able to be filled in any number of ways in order to describe various phenomena, indicates his belief that there likely were more particles than were known in the early 1940s (and we know this has been proven true), but that, theoretically, there are many more particles than are even known today. As he describes:

> We shall freely invent particles to carry the sets of variates that our form of analysis groups together. The provision of a carrier is not so much a necessity of thought as a necessity of language (Eddington 1946, p. 31).

Eddington finds that the necessity of language may indicate that it is

> desirable to distinguish the 'mathematical fictions' from the 'actual particles'; but it is difficult to find any logical basis for such a distinction [since] '[d]iscovering' a particle means observing certain effects which are accepted as proof of its existence; but it seems to be a matter of fashion or convention that one sort of effect rather than another is accepted as critical for this purpose (Eddington 1946, p. 31).

He gives the discovery of the companion of the star Sirius as an example of the last point. The companion's existence had been inferred fairly early from the elliptic motion of the star itself, but it was not 'confirmed' until it was actually seen visually in 1862. But, he asks, why is the visual confirmation better than the gravitational? The visual observation is mediated by radiation or, more clearly, the electromagnetic interaction. The gravitational observation is mediated by the gravitational interaction. Both observation systems have their own intricate levels of analysis between the actual fundamental interaction itself and the processing of the result by the brain. Certainly, if the visual observation is made without the use of CCD cameras (i.e. with the naked eye), then there is *less* of this analysis, but still, what makes gravitational evidence better than electromagnetic? As Eddington explains, experimental tests, regardless of their type, are concerned not with the conceptual carrier itself (the fundamental particle of the interaction) but rather with the information it carries. Thus, trusting the photon's information over the graviton's appears completely arbitrary (although we have never observed gravitons while we observe photons every day). This foreshadows Eddington's interpretation of the exclusion principle where he doesn't distinguish between particles of any type – a particle is a particle is a particle. This is also the reasoning behind the choice of a standard carrier. By substituting different values in the complete energy tensor the carrier can manifest itself as different particles. It's as if each particle in the universe is simply a different set of clothes for a single mathematical function.

      A subset of the standard carrier is the 'vector carrier' which is a particle specified not by a complete energy tensor but rather by a complete momentum vector (including angular momentum and thus spin). As such it only has 10 independent components and is referred to as a $V_{10}$ particle. Vector carriers can be derived from standard carriers by making a standard carrier the *outer* square of a complete vector. This is a stabilising characteristic meaning it reduces the degrees of freedom (multiplicity), $k$, (from 136 to 10) and is somewhat subjective (see the discussion on stabilisation in chapter six). However, Eddington makes the point that such particles *are* in fact observable in nature (any spin-0 particle could theoretically be represented this way). He concludes that $V_{136}$ and $V_{10}$ particles are what might be called 'actual particles' while the $V_3$ and $V_4$ spinless particles don't have the possibility of including spin, which is a very real effect. Thus these particles must be simplified mathematical fictions.

**Relatively Speaking**

In chapters five and six I introduced the idea that measurements only really make sense when there is a reference body present for comparison (e.g. I could say that a particle has 15 acks – well, what's an ack and what (and where) is *the* ack?). In most experiments a measurement is taken and then compared to a reference object that in all likelihood was itself derived from another reference object, and so on, all the way back to the original definition of the particular unit of measurement under investigation (which is likely arbitrary to begin with; see again chapter five). This is a rather long chain of connections and each link in this chain will include its own probability distribution. As such, the longer the chain, the greater the standard error in the final measurement. The ideal situation, then, would be to make a *direct* comparison between the object under investigation and *the* standard of measurement for a particular unit or quantity (e.g. the Paris Metre). Since this is impractical an idealised experiment would be one in which the standards of measurement were embodied in a particle that the object under investigation could directly interact with. In fact this is, to some extent, precisely what spectroscopy has always done and is similar to Heisenberg's thought experiment with the γ-ray microscope. Granted, wavelengths, spectral line widths, frequencies, and other features are often represented by comparison to a length, inverse time, etc. that is not part of the problem, but ratio comparisons are always possible since the method of measurement is through a photon for which the same quantities can be determined. So, for instance, a more accurate length measurement might really be a dimensionless ratio of the length of the object under investigation and the wavelength of the photon detecting the object. Spectroscopy may not always perform this mathematical method, but a comparison is often made with the wavelength (frequency, etc.) of the light involved in the experiment thus partially eliminating the propagation of error.

The idea, then, of having some standard that actually interacts with particles as they do with one another is given form in the idea of 'comparison particles.' A comparison particle is simply a carrier that includes the standard of measurement and interacts directly with the object under investigation. Specifically, the comparison particle carries the extraneous standard and is thus outside the object-system. However, in order to make a truly objective measurement the comparison particle needs to be included in the object-system and thus the object-system is expanded to include it. Eddington refers to this extended object-system somewhat optimistically as the 'perfect object-system' in that it is self-contained: the standard for measuring the system is actually *within* the system itself.

Once inside the object-system, the comparison particle is set to carry only a single variate, specifically the scale variate, meaning that all other possible variates are stabilised (recall that scale is now a measurable variable on par with momentum and energy). Eddington is careful to point out that the stabilised characteristics of the comparison particle must be chosen so that they do not affect the measurement of the desired quantity at all. But, this means that the only inexact (non-stabilised) quantity in the comparison particle is then the scale used for comparison, which appears paradoxical. However, his point is that the stabilised characteristics are imposed by the observer (and are thus subjective) while the scale, though inexact, is *measurable*. Again, this is similar to the example I gave above regarding spectroscopy. Via the use of natural units extraneous standards have fixed relations to the fundamental scale uncertainty of the physical reference frame (which is embodied in the comparison particle anyway). By embodying the scale in a particle (a comparison particle representing the scale) we also embody the scale uncertainty.

There is another important property of comparison particles that will soon become useful. If the comparison particle is outside of the object-system then it is a *mean* (average) particle meaning its attributes are based on the collective distribution of all the particles in the uranoid (essentially they are average quantities). Since we know that the larger an assemblage of particles we have, the smaller the effect of the uncertainty principle or probability in general (i.e. large data sets are always better since they minimize error), then the uncertainty or fluctuation in an external comparison particle must be fairly small. In fact it is $\sigma_\varepsilon$.

However, if the comparison particle is *inside* the object-system its attributes are no longer averages of a large assemblage; it is an individual and thus has an uncertainty or scale fluctuation on the order of $\sqrt{N}$ times $\sigma_\varepsilon$ by (5.13) and the fact that $R_0 = \sigma / \sigma_\varepsilon$. This seems to indicate comparison particles would have a ridiculously large uncertainty in any of their characteristics. According to Eddington this problem is apparently overcome by replacing the continuous probability distribution of the scale (internal to the system) with distribution that is over what he calls 'eigenscales' (Eddington 1946, p. 45). A discrete eigenscale is simply a convention by which the scale for each resulting (measured) eigenstate is exact. The scale uncertainty is manifested as an uncertainty in the state of the *system* as a whole unless it is stabilised to exist in a particular eigenstate.

By that token, measuring the mass of a comparison particle will not serve as a standard for measuring other masses that are not in the object-system, but when it *is* employed as a standard *within* the object-system its measured mass is assumed to be the same as the measurement that would be obtained if it were to trade places and be treated as the object-particle itself, rather than the comparison particle – i.e. the measurements for a particle are assumed to be the same regardless of whether it is treated as an object-particle in one instance and a comparison particle in a different instance. As such the comparison particle is, like everything else, just some modification on the standard carrier meaning that it begins life as an unspecialised element of the uranoid's energy tensor.

Eddington assumes that *all* particles are essentially identical in their most basic form and relativity simply makes them *appear* different. As such in any object-system there really only needs to be a single comparison particle regardless of how many object-particles there are. This is how Eddington analyses atoms, then, by considering one of the elementary particles in the atom to be the comparison particle while the other (or others) is the object-particle(s). In fact, much of this is simply complicated jargon – Eddington-speak, if you will – and all he is really saying is that the most effective measurements, particularly in atomic situations, are ratios between like characteristics of the constituting particles. For example, rather than measuring a mass for the electron or the proton by comparing it to some standard kilogram, why not simply compare the electron and proton to each other? Could such a measurement be as instructive? Certainly; imagine comparing these masses to each other as in a hydrogen atom. Now, how would we know whether the lighter mass, for example, had any comparison to anything outside of the atom if we're not measuring it in reference to anything outside the atom? If we're simply measuring a ratio, who's to say (assuming, for a moment, we are ignorant of the periodic system) there isn't some other potential two-particle atom with the same exact ratio but different individual masses? The logical thing to do in this case is to then compare the ratios of the masses of the *two* atoms under consideration. If this ratio is unity they have the same total mass meaning the constituent particles must be the same. When working with particles that appear to be isolated (e.g. observing a lone electron in interstellar space) the comparison particle could be the photon (or other carrier or exchange particle) used to make the observation. Rather than measuring in terms of mass, natural units and Einstein's brilliance allows us to equivalently measure in terms of energies making the fact that the photon, gluon, and graviton are massless, of little concern; a ratio of energies would be taken rather than a ratio of masses with the resulting energy for the isolated particle being resolved into a rest mass. It is a rather elegant system of measurement, in my humble opinion, that will be explored further in chapter eight.

**Interchange and Fermi-Dirac Statistics**

I will present here Eddington's interpretation of interchange. There are subtleties inherent in the concept that I do not discuss until the next section, specifically the notion of permutation invariance. Since Eddington blurred the distinction between quantum and classical applications of statistics it is best to look at his theory on interchange as a whole before looking at recent work in permutation invariance that was not influential to Eddington. Its subsequent analysis it does bring Eddington's ideas into a new light. But first, Eddington's version of interchange is presented.

By measuring all positions (and thus velocities) as being relative, observable coordinates and momenta must involve two physical entities. In the standard method of experimental science, measurements actually involve *four* quantities – two involving the objects and two involving the comparison. For example, a length measurement involves the two end-points of the physical distance between two points as well as the two end-points of the comparison distance on the yard-stick (tape-measure, ruler, etc.) used to take the measurement. Eddington early on made the point that true *observables* are relative measurements such as this. So when measuring the mass of the proton, say, as a ratio to the mass of the electron the observed quantity is really the ratio here. In order to make sense of this number a comparison ratio must be provided (much like I described in the previous section regarding measuring quantities for the hydrogen atom). The true *measurable*, to be distinguished from an observable, is the combination of the two ratios. A measurable, then, is a characteristic that arises from a set of four entities each of which must, of course, have a probability distribution. The overall error then grows larger as the system gets more complex which seems to contradict the notion that errors grow smaller the larger an assemblage becomes since they average out. Eddington attempts to sidestep this problem by taking one or more of the four entities out of the overall object-system, which I have previously shown reduces their variance, and makes it Gaussian, which, if it is not negligible, can be accounted for with simple corrections.

His goal here is to reduce systems to a maximum of two probability distributions. My example with the two hydrogen atoms is another way of stating this. By measuring the ratio of the masses of the two particles in both atoms, one deals with only two entities (not four since it's a ratio), each of which has a probability distribution. So by performing a ratio operation rather than a direct comparison measurement, a maximum of two probability distributions is all that one encounters. When comparing to another hydrogen atom, the atoms are treated as single entities each having a probability distribution, or, the values for the mass ratios each follow some probability distribution and it is really these that one is initially comparing. Either way you still have only two probability distributions. The danger is the possibility of losing important correlations by averaging too soon (analogous to a rounding error caused by excessive rounding in successive calculations).

Looking at the hydrogen problem yet another way, imagine a hydrogen atom initially under consideration that includes the two entities $A_1$ and $A_2$ representing the proton and the electron respectively. A comparison hydrogen atom would have two entities $A_1'$ and $A_2'$ for its proton and electron respectively. A true measurable (*not* observable) could be written as $[A_1 A_1' A_2 A_2']$ where each particle from the first atom is compared to a similar particle in the second atom (electron to electron, proton to proton). One can perform the following transformation

$$[A_1 A_1' A_2 A_2'] \rightarrow [A_1 A_2' A_2 A_1'] \tag{7.1}$$

where the first proton is now compared to the second electron and first electron is compared to the second proton. This process is called 'interchange' and, again, builds on the assumption that all particles are ultimately the same in their basic form.

The idea of interchange arises from Fermi-Dirac statistics (that describe fermions). In *Relativity Theory of Protons and Electrons* (the prelude to *Fundamental Theory*) Eddington makes the argument that there is no real difference between the Coulomb (electromagnetic) interaction and Fermi-Dirac statistics (presumably he would have had to modify this statement to include the colour (strong) force between quarks in the same category if he had known of its existence). Both describe an interaction meaning they assign to the fermions a probability distribution for position, momentum, and spin that are different than those for non-interacting particles. In Fermi-Dirac statistics symmetrical wave functions have zero probability thus preventing two electrons from sharing the same set of quantum numbers. Conversely the Coulomb force modifies the wave functions such that they satisfy a modified wave equation that is also asymmetrical (it contains an extra term called the Coulomb energy). This these two mechanisms explain the same interaction. Eddington concluded that it

> cannot be seriously maintained that the Coulomb force, which prevents two slow moving electrons from approaching one another, is an altogether distinct phenomenon from the exclusion principle (contained in Fermi-Dirac statistics) which achieves the same result by forbidding them to occupy the same phase cell (Eddington 1936, p. 282).

He does not mention how to reconcile this with fast moving electrons which can collide, thus overcoming the Coulomb force, yet still obey Fermi-Dirac statistics. The discovery of bosonic charge clearly refutes Eddington's claim. In another example, which is a demonstration of quantum non-locality (the non-existence of any objective local theory) and was first experimentally proven in 1982 by Alain Aspect, colliding electrons exchange portions of their wavefunctions such that after the collision, regardless of their separation distance, they cannot be considered as individual particles but rather as a system. As such, for example, if one is spin up then the other must be spin down in order for them to obey the exclusion principle. This actually implies that these particles are observationally distinguishable since they are forced to have at least one quantum property that is different. Particles in large ensembles cannot be truly considered to be individuals based on Aspect's experiment (which was an idealization of Bohm's modification (found in his book *Quantum Theory*) of the famous Einstein-Podolsky-Rosen (EPR) thought experiment) since they share common properties. This fits Eddington's requirement that particles, unless directly under observation, must be *unidentified* members of a large assemblage – they're ultimately indistinguishable.

The philosophical implications and arguments resulting from Aspect's experiment run the gamut from those that believe the experiment (or rather experiments – there were multiple runs) indicate a flat-out denial of any possibility of hidden variables in physics (this seems to at least be true locally) to those that have found the peculiar behaviour of the entangled electrons at great distances (seemingly at odd with special relativity) to be disconcerting enough to possibly believe in hidden variables (string theory with its 11 dimensions may ultimately resolve this problem in favour of the indeterminate subjective view that denies the existence of hidden variables). Since these experiments and even Bell's inequalities occurred long after Eddington was gone, he was thus unaware Einstein, Podolsky, and Rosen were wrong (especially in 1936, just one year after their paper was published).

In any case there is now sufficient evidence to support the fact that the Coulomb force and Fermi-Dirac statistics were not necessarily the same thing since the Coulomb force could be overcome while Fermi-Dirac statistics could not. But in 1936 the situation was considerably different and, though he does not explicitly say as much, Eddington must have assumed that the relativistic situations that could overcome the Coulomb force must also have overcome Fermi-Dirac statistics in the process since he considered the two equivalent. What happened to the neutron which had 'won' its independence, so to speak (from the electron-proton composite view) by Eddington's death? Since charge was a relativistic phenomenon our perception of Coulomb repulsion (i.e. the inverse square law every elementary physics student is familiar with) was purely relativistic and the exclusion principle itself was the *true* origin of the force. In his own words:

> This separation of the interaction of electrons into two effects strongly resembles the separation of gravitation and inertia in Newtonian mechanics. The latter taught that a body tends to move uniformly in a straight line by its inertia, but is pulled into a different path by the gravitational field. Similarly today quantum physics teaches that electrons tend to take up the probability distribution corresponding to Fermi-Dirac statistics, but are forced into a different distribution by their electrical repulsions. There is need for the same kind of treatment that has proved so successful in the unification of gravitation and inertia (Eddington 1936, p. 282).

Since Fermi-Dirac statistics arose, in Eddington's view, from the indistinguishability of the fermions involved in a given process, then the Coulomb force, if it is indeed simply another manifestation of the same fundamental interaction, must also arise from the indistinguishability of the fermions involved in that same process. As I have mentioned before one reason fermions are indistinguishable is that they are all fluctuations in the same universal field – so they are *not* separate objects, necessarily, rather they are like different wrinkles on the same sheet. Eddington goes on to show that this indistinguishability produces the extra Coulomb energy term in the wave equation.

The fundamental idea enumerated above then indicates that, since all fermions (protons, electrons, etc.) are indistinguishable from one another, they can be interchanged and any equations applied to the system must be invariant for the interchange of indistinguishable particles. Since invariance in this sense (as Eddington applies it) refers to Lorentz transformations, it is a purely relativistic phenomenon – interchange is simply a new type of relativity transformation viewed as a rotation of the system (e.g. a transition could change a mass value). Interchange involves the transfer of probabilities from one object to another and can be carried out gradually via a 'permutation coordinate' that has a conjugate momentum known as the 'interchange energy' which is really the Coulomb energy term in the wave equation (Eddington 1936).  Here Eddington pauses in a somewhat uncharacteristic display of frustration at his colleagues.  He laments:

> Now that interchange energy is regularly used in practical problems, it is difficult to see why the author's theory of the Coulomb energy of electric charges is still looked upon as a dubious excrescence on wave mechanics.  In the equations in current use the identity of interchange energy and Coulomb energy is accepted (Eddington 1936, p. 283).

He then directly quotes from Dirac's *Quantum Mechanics*:

> The interchange energy is given as $\frac{1}{2}V_{rs}\left\{1+\left(\overline{\sigma}_r, \overline{\sigma}_s\right)\right\}$, whose eigenvalue is the Coulomb energy $V_{rs}$.  The unitary matrix factor depends on the circumstances of the problem to be treated, and does not affect the identification (Dirac, 1935, 2nd edition, p. 228 as quoted in Eddington 1936, p. 283).

**Permutation Invariance and the Indistinguishability Postulate**

All of this is very suggestive of modern quantum field theory and one of the major hallmarks of both QFT and Eddington's theory is summed up by Zee who says it "should be recognized as a triumph of quantum field theory that it is able to explain absolute identity and indistinguishability easily and naturally" (Zee 2003, p. 117).  Eddington's theory obviously has numerous holes but it at least shares this fact with QFT.

However the problem of indistinguishability is not nearly as simple as Zee implies and has been studied in great depth by a number of current philosophers of physics who have concluded that particles formerly thought to be indistinguishable might actually not be in their entirety. The subject strikes deep at the heart of quantum statistics and ultimately begins with the notion of permutation invariance (PI). The basic idea "is that dynamical laws depend only on the distribution of states, not on which 'individual' possesses which state" (Huggett 1999, p. 326). If permutations are considered invariant it means that they are undetectable by any measurement whereas transformations are usually covariant meaning they are only indistinguishable to observers that are likewise transformed. The origin of PI is in a paper by Messiah and Greenberg where they explained that "dynamical states represented by vectors which differ only by a permutation of identical particles cannot be distinguished by any observation at any time" (Messiah and Greenberg 1964, p. 248). A basic definition and lemma for PI is given by van Fraassen:

> $P$ is *symmetric* on set $S$ of events if and only if, for all members $E_1, \ldots, E_n$ ($n = 1, 2, 3, \ldots$), the probability of the sequence $[E_1, \ldots, E_n]$ is invariant under permutation of indexes; i.e., $P([E_1, \ldots, E_n]) = P([E_{t1}, \ldots, E_{tn}])$ for any permutation $t$ of $\{1, \ldots, n\}$.
>
> *Lemma*: If $P$ and $P´$ are symmetric on $S$, so is their mixture
> $cP + (1 - c)P´$ with $0 \leq c \leq 1$
>
> (van Fraassen 1991, p. 62).

This is essentially known as De Finetti's theorem. Van Fraassen gives a more concise definition as follows:

> Given a sequence of exchangeable random variables $f(i)$, we can represent their individual distributions $P([f(i) < r])$ as all produced by integration on a set of distributions

> deriving from probability measures for which these variables are independent (that is, measures $p$ such that $p([f(i) < r] \cap ([f(i) < q]) = p([f(i) < r]) \, p([f(i) < q]))$ (van Fraassen 1991, pp. 64-65).

In equation (7.1) Eddington essentially permutes the indices of the variables involved. In this case the variables are four particles – two electrons and two protons (an extra for each for comparison). Clearly Eddington intends for electrons and protons to be interchangeable with one another and this arises from his Principle of the Blank Sheet – all particles are variations on a single mathematical form known as the standard carrier. Does this imply indistinguishability? Not necessarily. All atoms exhibit PI in relation to exchanges of their electrons (i.e. if two electrons in a sodium atom swapped places no one would know the difference) but can be individuated by other properties such as momentum. So the sodium atoms of my example might be identical (in one sense) but not necessarily indistinguishable. This is an important distinction: indistinguishable particles are all particles in the *same quantum state*. Identical particles might be all free electrons moving at $2/5c$ but with some other differing property (i.e. they must have *different* quantum states even if every other attribute appears to be the same).

There is now another distinction that needs to be made. There are actually three *types* of quantum statistics: Bose-Einstein (BE), Fermi-Dirac (FD), and parastatistics (for particles that are neither BE or FD). In Bose's original work the identity of particles was ignored; each possible assignment of occupation numbers to the different energy levels was equiprobable. Fermi-Dirac statistics, on the other hand, does not assign equal probabilities to all occupation numbers with the most glaring example being spin states – if an electron with a given spin state exists in an orbital there is zero probability that another electron will exist in that orbital with the same spin state. By contrast, Maxwell-Boltzmann (MB) statistics are used in classical situations and make a full distinction between *all* particles in an ensemble (see van Fraassen 1991, p. 378 for a comparison of MB with BE).

Let us apply all three statistics, MB, BE, and FD, to a specific example, that of two dice. If the exclusion principle holds, then on a given roll the dice cannot have the same value in FD. This reduces the possible outcomes from 36 to 30 and the maximum possible value becomes 11. For MB the exclusion principle does not hold so as van Fraassen puts it, a gambler (who would perhaps unknowingly use MB) would assign a probability of 1/18 to a roll of 11. But assuming exclusion applies to both BE and FD but in different ways, FD gives the probability of 1/15 to a roll of 11 while BE gives the same roll a probability of 1/21. What's the difference between FD and BE? FD favours combinations of *distinct* numbers while BE favours *doubles*. So, for example, the probability in FD of a roll of 12 is 0 while in MB it is 1/36 and in BE it is 1/21. Thus in the case of a roll of 11, BE < MB < FD, while in the case of a roll of 12, FD < MB < BE.

It should be clear from this description and Eddington's definition of interchange that Eddington reduced FD to BE. For example, since Eddington insisted that particles come in pairs (a particle and its comparison since all measurements must be relative), his statistics favour doubles in a sense similar to BE. This is actually consistent since *pairs* of fermions actually obey BE and single isolated particles were irrelevant (or, better, uninteresting) to Eddington. There is also a way to bridge the gap between quantum and classical statistics here. As van Fraassen says, BE is the "natural probability function, for a perfectly simple and natural combination of ignorance and chance" (van Fraassen 1991, p. 417) and in the case of maximal ignorance agrees with MB. The case of maximal ignorance corresponds to a uniform statistical distribution and represents the sole point of contact between quantum and classical models.

To put a more intuitive spin on this idea we might say that bosons tend to aggregate in the same cells (we'll see this is at the heart of Eddington's version of the exclusion principle) while fermions tend to aggregate in different cells. Distinguishable (classical) particles show no tendency either way (van Fraassen 1991). However, a word of caution is in order. There *are* other possible aggregations for the various particle types and the preceding is simply meant as a mental aide in understanding the most basic difference between FD and BE.

We can clearly see from this intuitive example, however incomplete, that fermions clearly exhibit individual identity. In addition bosons clearly exhibit some sense of identity in pairs, though not as individuals per se. So is there really such a thing as indistinguishability? Again we must be certain to mark the difference between indistinguishability and identity. In the discussion of the various types of statistics particle attributes were generally limited to the usual quantum numbers. Clearly, however, spatial location should be included in any consideration of indistinguishability, i.e. it is obvious that a blue car parked in your driveway is distinguishable from a blue car driving simultaneously by your house because they are spatially separate. In regard to permutation invariance, a strong version of PI would *exclude* properties of spatial location (making it easier to say particles are indistinguishable) while a weak version would include them (French and Redhead 1988). Since relativity reminds us that time is just another dimension, what can we say about *temporally* separated particles? Your car in your driveway now versus two hours from now seems to clearly be the same car. But if time and space are of the same basic structure why is one case different from the other? The answer requires knowledge of *where* an object obtains its attributes. The statistics I introduced above don't exactly answer that question but they do acknowledge a difference for different types of particles. But even in the case of BE where particles seemingly appear to have little or no individuality it is still clear that they are different particles, i.e. how can we speak of multiple particles if they don't have some sense of individuality? We might as well be speaking of just a single particle. This is an argument that has been used in some PI discussions regarding fermionic states. To better understand this, let us return briefly to quantum statistics and consider a two-particle system. Following French and Redhead, suppose there are four possible states:

(1) Both particles are in the state $|a^r\rangle$
(2) Both particles are in the state $|a^s\rangle$
(3) Particle 1 is in state $|a^r\rangle$ and particle 2 in state $|a^s\rangle$
(4) Particle 1 is in state $|a^s\rangle$ and particle 2 in state $|a^r\rangle$

The exclusion principle corresponds to the fact that the first two arrangements are not allowed for fermions. For quantum particles, if they are individuals then it is clear that (3) and (4) are *not* identical. But these states aren't really the ones we use to discuss quantum statistics. The relevant states for quantum statistics really are:

(5) $|a^r\rangle \otimes |a^s\rangle$

(6) $|a^s\rangle \otimes |a^r\rangle$

(7) $(1/\sqrt{2})(|a^r\rangle \otimes |a^s\rangle + |a^s\rangle \otimes |a^r\rangle)$

(8) $(1/\sqrt{2})(|a^r\rangle \otimes |a^s\rangle - |a^s\rangle \otimes |a^r\rangle)$

These four states are mutually orthogonal and span the same subspace as the first four states *but* they are chosen so that (5), (6), and (7) are symmetric under an exchange of particles labels, i.e. they satisfy PI. (5) and (6) are really the same states as (1) and (2). The difference is really between (3) and (4) and (7) and (8). Of these four states all are symmetric *except* (8). So bosons are restricted to the three possible symmetric states while fermions are restricted to the antisymmetric states of which there is only one – (8)! Huggett likens this to the imposition of something akin to a boundary condition (Huggett 1999). As French and Redhead say, states without the correct symmetry are "eliminated because they are not *accessible* to the joint quantum system, not because there are no such states!" (French and Redhead 1988, p. 237). Recall here that we are speaking of a *joint* quantum system meaning two fermions or two bosons.

But what do these states represent? Well, at some point if we wish to avoid mindless mental exercises there ought to be something practical at the end and in quantum mechanics that is usually an observable. I can now introduce the Indistinguishability Postulate (IP) in quantum mechanics:

(9) $\langle P\phi|Q|P\phi\rangle = \langle \phi|Q|\phi\rangle$, $\forall Q$, $\forall \phi$

where $|\phi\rangle$ is an arbitrary N-particle state and Q is a possible observable (French and Redhead 1988). What (9) does is restrict the possible *states* for the N-particle system. In fact it limits it to only the fermion or boson possibilities. However, it *could* be interpreted as a restriction on the number of possible *observables* for the N-particle system. This has the effect of reducing the *accessibility* of the states rather than their existence. But how does that affect the individuality of the given particles?

Take (8), for instance. In this case the particles are *not* in separate states. Rather they *both* exist partially in *both* states |a$^r$⟩ and |a$^s$⟩. This is the situation entangled particles encounter. They do not have an identity unto themselves because they are in fact a mixture of two states. For example, imagine a can of white paint and a can of black paint. Let us say they interact such that in the end we still have two cans of paint, but each can has an equal amount of *both* black and white paint. This is a superposition of states. But, if Q is our measured observable standard interpretations of quantum mechanics refer to $Q_1$ and $Q_2$ as expectation values in accordance with statistical probabilities and *not* as values actually possessed by the corresponding particles. However, French and Redhead have clearly shown that PI can be violated in such a situation (French and Redhead 1988). If these particles – or, rather, the observables corresponding to the measurement of these particles – are *not* invariant to permutations meaning that we *can* tell if they are suddenly switched, then even though they consist of a mixture of states, they *can* be discerned as individuals. The simplest way to do this harkens back to Eddington as we'll see: spatio-temporal separation. If nothing else, these particles are *spatio-temporally* different.

**Fermi-Dirac Statistics and the Coulomb Force in *Fundamental Theory***

Let's return, then, to Eddington's claim that the Coulomb force and the exclusion principle, which results from Fermi-Dirac statistics, indicates that there must be some relation between spin and electrical charge. I present two arguments here. The first is related to the above discussion of permutation invariance and indistinguishability while the second is more heuristic.

Eddington essentially is claiming that in an atom the reason there is a limit to the number of electrons in an orbital (i.e. exclusion) is because Coulomb repulsion forces the additional electrons into another orbital. Obviously this implies that the repulsion between electrons with like spin is greater than the repulsion of electrons with *dis*like spin since two electrons can exist in the same orbital if they have different spin states. Clearly this also implies there ought to be an additional term in Coulomb's law (and Eddington, in fact, has added this term). Since electrons are fermions they would obey

FD and tend to congregate in their own cells making them fairly easily distinguishable on an individual basis. But, in addition, they tend to aggregate in groups of two since that is the maximum number of electrons possible in a given orbital (or suborbital). In that sense they appear to obey BE at least in pairs. Recall that in cases where the exclusion principle applies, BE < MB < FD, while in cases where the principle does *not* apply, FD < MB < BE. In the case of the atom, then, starting with an orbital, FD has the highest probability and the orbital is thus filled with two fermions, the limit provided by the exclusion principle. The orbital (or suborbital) gets filled in pairs and thus, exclusion being fulfilled, BE now gives the highest probability (this explains why pairs of fermions behave like bosons). However, none of this provides a clear mathematical interpretation of the existing version of Coulomb's law. So, at least from a statistical standpoint linking Coulomb repulsion to FD would require an additional term in the force law equation which Eddington has supplied in order to account for the relative strength in excluding and non-excluding situations. So now a more heuristic argument is made.

      We already know what spin is, but what exactly is *charge*? We return once again to the fundamental description of the universe as particles and their interactions where the interactions are actually mediated by more particles. The electromagnetic interaction is, of course, mediated by the photon. Obviously, then, there must be some exchange of photons when charges are in close proximity (in fact the electromagnetic interaction, as I've shown, is infinite in range as evidenced by the fact that we can observe quasars on the edge of the universe). The photons are exchanged in a different manner depending on whether the charges are alike or not. One of the most common ways of describing particles is through a mathematical term known as a propagator. A propagator describes how a particle moves and through some fairly complex math can also describe how particles are created an annihilated. Charge can thus be described as being a measure of how photons propagate.[38] Note that the same analogy cannot necessarily be made for the

---

[38] A major question any serious believer in probabilistic methods should ask at this point is, why are the proton and electron charges *exactly* equal? Why doesn't the uncertainty principle dictate minor differences between the two? The answer is fairly complicated but boils down to the fact that the renormalization of the charge is directly related to the photon and can be derived from something known as gauge invariance. Ultimately since the photon *has* to propagate between positive and negative charges, they must have the same value; any fluctuation would throw the photon off course. It's a bit like imagining the photon is a dart and the charges are dart boards: moving a board might cause the dart to miss it entirely! This can be explained macroscopically by the fact that atoms must be electrically neutral to an amazing degree of

relationship between the graviton and mass since the graviton propagator can be written in flat space-time in which case it contains no mass terms, whereas the renormalized photon propagator *does* contain charge (or, rather, charge-related) terms.

So we now know that charge is a measure of how the photon propagates, but how is it related to spin? Charge is not a wholly fermionic phenomenon since, for instance, a pair of electrons has a charge of –2*e* but obeys BE. What, then, is the difference between a pair of (coupled) electrons and a single electron? Quite simply, the spin. A pair of coupled electrons (e.g. a Cooper pair) must obey the exclusion principle and thus, with one spin-up and the other spin-down, they will have a total spin of zero. We can rule out a direct relation then between charge, in general, and FD since it is possible to have particles with charge that obey BE. The problem is that we also can find particles *without* charge that obey FD (e.g. *n* (the standard neutron), $\Lambda^0$, $\Sigma^0$, $\Delta^0$, and $\Xi^0$ are all neutral baryons and neutrinos are neutral leptons). It seems that Eddington's correlation between Coulomb repulsion and exclusion as formulated in FD is simply incorrect, though in deference to him, many of these particles were not known then (though the neutron *was*). In addition, Pauli proposed his famous spin-statistics theorem in 1940 that linked spin values to statistics (half-integer spin to FD and integer spin to BE; Pauli 1940).

Eddington could have been onto something here, however, since there *is* a correlation between the weak interaction and particles obeying FD. The weak interaction essentially transmutes quarks and leptons (e.g. electrons, neutrinos, muons, and taus) from one form to another (essentially this is parity violation) and these are the most basic fermionic particles. In essence the weak interaction allows particles to exchange characteristics. This raises the question of indistinguishability once again. However, if this is merely a permutation of observables between particles I have shown above that in some senses these particles are nonetheless still distinguishable. In any case, the weak interaction is the mechanism by which particles can permute various indices. In fact, neutrinos *only* interact via the weak interaction. It turns out that all half-integer spin particles interact via the weak interaction and the weak interaction only occurs between

---

accuracy if standard cosmology is to work. If not, electrostatic forces between large objects would literally tear the universe apart.

particles with half-integer spin. All half-integer spin particles must have antisymmetric wavefunctions meaning they obey FD. As such it stands to reason that there is a link between the exclusion principle (or FD) and the weak interaction. At phenomenally high energies (e.g. at levels found in the early universe) the electromagnetic and weak interactions were simply different manifestations of the same interaction and so through electroweak unification there might be some way to reconcile Eddington's proposal. One final note, however, is that indistinguishability in Eddington's theory does not result from quantum effects but rather from relativity where all particle variables are frame-dependent and interchangeable. Again this reminds us that Eddington really blurs the lines between quantum and classical effects.

**Fundamentals of Eddingtonian Interchange**

Imagine, then, a proton and an electron along with their corresponding comparison particles. When brought together into a single system this is a measurable. But due to their common origin in the standard carrier of protons and electrons (Eddington's reasoning for this is vastly different than convention as I'll show) there is only a need for a single comparison particle in the combined system since both are represented by $V_{136}$ (there is no need to duplicate the extraneous standard). As such, when a proton and an electron are combined there are now a total of *three* particles rather than four in the measurable (just one comparison particle is needed). However, some fourth object must remain in order to preserve (7.1). As such the permutation variate is introduced to carry the extra information required by (7.1). The permutation variate in Eddington's theory adds an interchange energy to the system that *is* the Coulomb energy. This transforms the standard carrier $V_{136}$ into a $V_{137}$ by adding an extra degree of freedom (the multiplicity increases by one). Basically Eddington accounts for the electromagnetic interaction here simply through the concept of interchange – protons and electrons are bound together via interchange which, ultimately, is simply some combined form of FD and BE (I will discuss this in great depth later).

In bringing two particles together into a single system their two separate scale momenta are replaced by a join scale momentum and a permutation momentum and their two phase coordinates are replaced by a joint phase coordinate and a permutation coordinate. The permutation coordinate is defined to be an angle for several reasons (basically he works up to everything here from relativity as I mentioned previously). First, it incorporates natural units without any difficulty. Second, a transformation through $2\pi$ restores the original measurable and limits the range of uncertainty to $2\pi$ (this is actually a permutation and exhibits PI). This limit to the uncertainty is the widening factor discussed in the previous chapter in relation to the treatment of scale as a variable. A change in configuration via (7.1) can either happen as a result of actual spatial motion or simply by an increase in the permutation coordinate by $\pi$. Eddington uses the term 'interchange' solely in regard to the latter which, in the 'old' quantum theory (before the advent of wave and matrix mechanics), would be considered to be a quantum jump. In the basic original form of wave mechanics the jump is represented as a continuous flow of probability from one state to another. The discretisation appears as previously described in the eigenvalues for given eigenstates. Eddington pictures this jump as a circulation in an extra spatial dimension normal to (3 + 1) space-time (this circulation is just the permutation coordinate I just described). Ultimately, as I have described before, he is working in a (4 + 1) space-time. Mathematically the extra interchange (Coulomb) term results from the inability of the probability distribution to satisfy the continuity equation if only spatial transitions are accounted for. The interchange term then is added in order to preserve continuity. Since quantisation requires that there are no eigenstates with zero angular momentum in a plane of degeneracy (which the interchange rotation is in), interchange is unavoidable (Eddington 1946). The link between Eddington's interchange and PI is the indistinguishability of particles in large ensembles. As I have mentioned before, however, recent has pointed to weaknesses in PI. If Eddington's interchange is simply an alternate formulation of PI then it is subject to the same criticisms.

Interchange in *Fundamental Theory* is ultimately relativistic in character since it relies on the indistinguishability of particles that is due to the frame dependence of all characteristics. Since relativity implies that there is no absolute frame of reference and, thus, motion is a completely relative term, any particle can be brought to rest (or similarly accelerated) simply through a transformation. Since energy is related to velocity as well as mass, it makes sense that it might be possible to change the mass of a particle through a relativistic transformation (this is *not* the same thing as changing the mass through a weak interaction). In fact the known masses of particles are really their *rest* masses since, as a particle is accelerated, its mass is converted to energy. As such, as far as masses are concerned, relativistic transformations can theoretically make protons and electrons indistinguishable. Though this is jumping ahead just a bit, Eddington assumes the same is true of spin and charge. For spin, despite the point-like nature of fundamental particles, he assumes a frame rotating with spin-up could be transformed into a frame rotating with spin-down just as a frame embedded in the Earth could be transformed into a frame with opposite spin (this is pretty simple to imagine in a classical sense – who says 'north' is 'up?' – there's really nothing special about either north or south). The only difference quantum mechanically is that it essentially takes two complete rotations to return to the original state. But this does not affect the relativistic view Eddington imposes – two rotations are the same regardless what direction you're viewing it from. So spin and mass are both frame-dependent quantities. However, one must remember that any transformation that changes spin-up into spin-down, also changes the original spin-down into spin-up, i.e. the interchange must occur in pairs. In the case of spin this is true because as soon as an axis of rotation is introduced into anything it immediately creates a duality in that there are two possible ways to rotate around that axis.

Charge is a trickier matter, however, since there is no simple intuitive way of thinking about it in frame-dependent terms. As I've shown charge is essentially a measure of how the photon propagates. But, obviously it propagates differently depending on the sign of the charges involved – if they're alike then the photons produce repulsion, while if they're different the photons produce attraction. Ultimately in the equation for the effective action of the photon (in general the massive vector meson), which contains the propagator, there exists a source (current) term, $J^\mu(x)$ that can be modified to accommodate both positive (*p*) and negative (*n*) charges as $J^\mu = J^\mu_p - J^\mu_n$. When the charges are alike the effective action is positive producing repulsion; when the charges are *not* alike the effective action is negative producing attraction. Thus, if charge is to be considered frame-dependent, then a rotation of the frame of $J^\mu(x)$ must change its sign, i.e. $J^\mu(-x) = -J^\mu(x)$ must be true. However, a simple derivation, following Lorentz' own reasoning, can show that $\partial_\mu J^\mu = \nabla \cdot \mathbf{j} + \frac{\partial \rho}{\partial t}$. Since the conservation of charge requires that $\frac{\partial \rho}{\partial t} + \nabla \cdot \mathbf{j} = 0$ then $\partial_\mu J^\mu = 0$ meaning charge can neither be created nor destroyed (Zee 2003). Thus any change in reference that flips a sign, say, from positive to negative *must* flip some *other* sign from negative to positive in order to maintain charge conservation. This is the same situation we have just seen occurs with spin (your right hand is your mirror image's left hand and your left hand is your mirror image's right hand).

Eddington confronts this with the idea that particles are simply carriers of variates. Charge, mass, and spin are all simply the contents of the carriers themselves. As such a particle $V_{10}$ as defined by Eddington could carry the information for either an electron or a proton. This is born out of the idea that interacting particles (at least to Eddington) cannot be considered as individuals. In some sense this is similar to the weak interaction's ability to transfer characteristics from one particle to another thus reinforcing the notion that particles are simply 'empty' structures that can be filled in various ways in order to *appear* different.

The one topic I have not touched much on yet that requires discussion is how particles behave in atoms – can the proton and the electron in a hydrogen atom interchange attributes and, if so, is the result still the *same* hydrogen atom? Could we tell such an interchange had occurred? Eddington would say the answer is yes and would cite PI (if he had been aware of it). French and Redhead, van Fraassen, and Huggett would say the answer is no because PI is not strong enough to completely make the two particles indistinguishable.

**External v. Internal**

There is an alternate way of looking at atoms. Rather than treating them as a collection of particles, one can redivide the system into what Eddington referred to as 'external' and 'internal' particles. This is not uniquely Eddingtonian; it is common practice among nuclear scientists and others, though it is not referred to in the same jargon. For example two-particle systems are often represented by examining the centre of mass on one hand and the internal energy interactions on the other hand (Gasiorowicz 1996). So Eddington's external particle is the hydrogen atom as a whole represented by its centre of mass, while the internal particle is a representation of the internal energy of the atom or the interactions between the electron and proton. The usual notation for the masses of these two 'particles' is

$$M = m + m', \quad \mu = mm'/(m + m') \tag{7.2}$$

where $M$ is the mass of the external particle (the hydrogen atom as a single unit) and $\mu$ is the mass of the internal particle referred to in modern texts as the 'reduced mass' (Gasiorowicz 1996). In typical Eddington form he develops new jargon for these particles: 'extracules' and 'intracules.'

By analysing the system in this manner interchange becomes quite simplified since, if one only considers the extracule, interchange is irrelevant since it is an *internal* property. Eddington verifies this by considering the coordinates of the extracule in the usual mass-of-mass manner, but then treating the coordinates of the intracule in *his* usual manner by taking a *difference* between the particles making up the intracule. So the coordinate 'position' for the intracule is really a coordinate difference of the two participating particles, which follows Eddington's reasoning that only relative coordinates have meaning. He uses similar reasoning to show that the total volume of the bi-particle (extracule/intracule system) is the same as the standard hydrogen atom, thus certifying his approach (and as I've noted this is all fairly standard practice today, though without the unusual jargon). However, the Hamiltonians in each system do not quite match. This is because when the system is transformed from a proton-electron system to an extracule-intracule system the zero-level for the energy is adjusted to account for the additional density of the distribution – essentially the datum is shifted. This transformation is referred to by Eddington as "freeing the intracule" (Eddington 1946, p. 53).

The Hamiltonian is then written in the form (for the energy)

$$E = \mu_1 + \frac{p^2}{2\mu_2} \tag{7.3}$$

where $\mu_1$ is the rest mass and $\mu_2$ is call the mass-constant of the particle.  He makes the point that the two agree in a purely inertial field.  However "the particle possesses potential energy in a gravitational or electrical field, and this constitutes the difference $\mu_1 - \mu_2$" (Eddington 1946, p. 53).  *What happened to equivalence?*  Has Eddington abandoned it entirely?  He clearly has a point that there does exist a potential in the presence of a gravitational field that is not there in an inertial situation, but the equivalence of gravitational and inertial frames is one of the great breakthroughs Einstein made with relativity!  Clearly something is amiss.  It seems unfathomable that such an ardent supporter of Einstein's theory would abandon one of its key results.  In addition, Eddington makes one more rather subtle error that is made by many modern physics text authors (and this could perhaps simply be Eddington using the colloquial terminology as most authors still do): he attributes the potential energy to the *particles* themselves.  In reality potential energy is associated *only* with interactions (thus it could be said that exchange particles contain potential energy).  Potential energy is entirely absent in situations where a particle is completely isolated.  In the case of the hydrogen atom there is, of course, potential energy associated with the interactions *inside* the atom, but this is considered to be a contribution to the *internal* energy of the atom if the atom is treated as a single object.[39]  This seems to be a fatal flaw in reasoning, but, for the sake of completeness, let us continue to follow Eddington's course.

Eddington associates the mass-constant with the given mass values in tables of physical constants.  When $\mu_1 = \mu_2$ the particle is referred to as 'free' and when $\mu_1 = 0$ it is referred to as 'bound' (Eddington 1946).  Intracules are considered to be bound particles in classical theory since they are made up of other particles bound to each other (as such nucleons could be analysed using the external/internal particle model).  In scale-fixed or quantal theory intracules are to be treated as free particles.  Eddington argues that Dirac's wave equation for the hydrogen atom is actually an equation for a free intracule.  As such there must be some way to connect situations in which the intracule is free to those in which it is bound.  Obviously in passing from the bound state to the free state the intracule gains $\mu_1(=\mu_2)$ amount of energy meaning relativistically that the datum for zero energy has been shifted.  Since in relativity energy determines the gravitational field and thus the metric this shift is important since it appears to naively create energy.  Once again Eddington appears to demonstrate a fundamental misunderstanding of the principle of equivalence.  On the other hand, philosophically, Eddington could have been pointing out a fundamental flaw in the understanding of equivalence at that time.  Equivalence is much easier to understand in this particular context when it is known that the *interactions* contain the potential energy and that the interactions themselves are particles.  In that sense the difference between gravitational and inertial situations is simply the presence of an additional particle, the graviton.  Simple relativistic transformations can shift the kinetic and potential energies around in such a situation.  Since the exchange particle description of interactions was first proposed by Yukawa, as I have discussed in depth before, and was not really even discussed in Europe until much later, the acceptance of these ideas came about long after Eddington had died.  As such, from *his* point of view there did appear to be a fundamental philosophical problem with equivalence.

---

[39] Since potential energy is associated only with interactions and it is *not* present in purely inertial situations, equivalence implies that potential energy is a relative quantity.  In fact this is indeed true since a conversion of potential to kinetic energy simply requires a Lorentz transformation – shifting a particle to its rest frame eliminates kinetic energy but since four-momentum must be conserved it must be transferred to another form, namely potential.  In the case of an isolated particle this would require the emission of a photon that would carry away the potential energy.

Continuing with his line of reasoning, however, the double wavefunction of the $V_{136}$ particle will have to be broken into separate wavefunctions for the extracule and the intracule.  The energy given by the Hamiltonian (7.3) is then resolved into

$$E_e = m_0, \qquad E_i = \mu + p^2/2\mu \tag{7.4}$$

where $m_0$ is the proper mass of the external particle (in this case the hydrogen atom).  As Eddington explains, due to the change in datum, the rest energy of the standard particle (standard carrier $V_{136}$) is now

$$m_0 + \mu = \tfrac{137}{136} m_0 = \beta m_0. \tag{7.5}$$

The factor $\beta$ accounts for the ratio of degrees of freedom that arises from making the transition to the extracule/intracule representation.  Basically what is happening here is that interchange introduces an extra energy term as I've shown previously.  The extra energy term is equivalent to the Coulomb energy.  Up until this point I have offered no explanation for this.  The reasoning becomes self-evident, however, when one considers that the extra term is *eliminated* by the shift to the extracule/intracule representation.

What's different here? The proton and electron in the hydrogen atom are now considered together as an intracule and the electromagnetic interaction occurring between them becomes internal energy for a single particle. So the interchange energy must be the Coulomb energy since it is the Coulomb energy that is eliminated (or rather transformed into internal energy) in the transition to the new representation.

Particles represented by (7.5) are then referred to as 'hydrocules' (Eddington 1946). This becomes the second system for representing energy. The first system that I will label *A* includes a standard particle uranoid with bound intracules while the second system that I will label *B* includes a hydrocule uranoid with free intracules. So in the first system the standard reference system for the universe (see discussion in chapter five) is as we have seen up to this point and is particularly useful when working with large systems of particles (or even galaxies if they are treated as particles). Since this is not always useful in microscopic situations it pays to have a second standard reference system and the hydrocule is it.

The change in standard reference for the move to the microscopic has a few consequences. The first is that in moving from a standard particle to a hydrocule $\frac{1}{137}$ of the mass is converted to energy (the interchange energy represented by the extra degree of freedom). In macroscopic situations a change in density of a large number of particles disrupts gravitational equilibrium. In relativity only the density of a steady distribution of particles at rest (the density of a pressureless Einstein universe) is compatible with the gravitational constant *G*. As such the hydrocule, if used extensively in converting back and forth between microscopic and macroscopic situations would require an adjustment. Since it is the density that requires matching to *G* one could simply redistribute the particles over a smaller volume in order to maintain the density. Conversely *G* (and, as a result of the system of natural units, $\hbar$) could be varied (this is a result both Milne and Dirac found explicitly in their own cosmologies). Eddington finds a third way by coupling with a change from *A* to *B* a change in the extraneous standard. Essentially the measured densities are multiplied by *β*. The particle density, *s*, of both the standard uranoid and the hydrocule uranoid are kept the same thus requiring dividing the mass-density by *β*. But, by changing the standard as well the mass-density is restored to its original value keeping the uranoid compatible with G. As such $\hbar$ remains unchanged. In summary, s, G, and $\hbar$ do not change in moving from A to B.

The above is logically consistent with Eddington's previous statements: the extraneous standard is furnished by the uranoid itself (see chapters five and six) and thus any change in uranoid will be accompanied by a change in standard making the entire transformation relativistic. The result for lengths and times is that in moving from A to B they are multiplied by $\beta^{-\frac{1}{6}}$. However this can *only* be applied to quantities that are directly measured. Scale-free particles can only carry a density since they must be free of unit-dependence (and density can be developed as a dimensionless ratio with the proper system of natural units). Scale-free particles are *not* carriers of mass and any masses attributed to them are not directly measurable. Technically, then, $m_0$, M, and $\mu$ are really densities that have been converted into masses when passing into scale-fixed theory. Ultimately we find that the rest masses $m_0/\beta$ and $m_0$ of the hydrocule and standard particle (carrier) in A are converted to $m_0$ and $m_0 + \mu$ respectively. As an example consider the hydrogen atom at rest both externally and internally as the standard particle. Results for system A, which is the standard system we are normally familiar with, have already been determined in chapters five and six. In system B, however, the rest mass is made up of the mass of the extracule $m_0$ and the mass of the intracule $\mu$. The corresponding energies become

$$E_e = m_0 + p'^2/2m_0, \quad E_i = \mu + p^2/2\mu. \quad (7.6)$$

Compare this to (7.4) and we find an additional term present in the extracule's energy. In (7.6) both external classical motion $p'$ and the motion of the transition from A to B, $p$, are accounted for.

**Interpreting the Energy Transfer**

What has happened here is that energy in the form of $\mu$ is added to the system in order to free the intracule. In the process the *rest* mass of the atom changes to $m_0 + \mu$ though its mass (or mass constant as Eddington calls it) $m_0$ for classical motion (by a change of space-time axes) does not change. The added energy must be potential since it does not alter the position of the atom and is not put into motion if the atom is put into motion (i.e. relative to the atom itself – it moves with the atom).

As I have already discussed extensively Eddington has interpreted this added energy as electrical (the interchange energy is really the Coulomb energy). In addition I have pointed out faults in this line of reasoning, though such faults were aimed primarily at the link between Coulomb energy and the exclusion principle. Building on that idea I have suggested that it is really the energy associated with the weak interaction that can be related to the exclusion principle. However, Eddington was unaware of the weak interaction's existence while he was working and he attributed this relationship to the electromagnetic energy instead, which makes sense for charged particles given the knowledge of the day.

I have already shown that by adding the permutation variate and thus the interchange energy a $V_{136}$ particle (standard carrier) becomes a $V_{137}$ particle thus adding a phase space dimension (multiplicity). I have also shown that the rest masses of these are related by $m_0' = m_0/\beta$. Since I have just shown that the latter is the mass of the hydrocule we can conclude that a $V_{137}$ is a hydrocule. Both particles are hydrogen atoms. The difference is related to the observation taking place. In dealing with the standard carrier an observation (that, of course, must interact with the atom itself) is made to determine (within the uncertainty limitation) the location of the electron and proton relative to each other. In dealing with the hydrocule an observation is made that seeks to determine exactly *which* particle is at which location. The latter is a deeper understanding of the given situation and thus atomic, nuclear, or quantum physicists would be inclined to use the hydrocule while physicists dealing with large-scale phenomena will be inclined to use the standard uranoid as first developed. Once again Eddington has developed a system of moving from macroscopic to microscopic. One other interpretation that he adds is to view the standard particle as the $V_{136}$ of system *A* and the hydrocule as the $V_{136}$ of system *B*. Or, in atomic physics (*B*), the hydrocule consists of an extracule and a bound intracule while the standard particle consists of an extracule and a free intracule (both equivalently consist of an electron and a proton).

One of the limitations in Eddington's interpretation comes from the fact that experimental nuclear physics was unable to really probe the contents of the atom in a manner he would have found sufficient until well after his death. Certainly the nucleus had been artificially broken apart in 1919 by Rutherford and Chadwick's discovery of the neutron came in 1932. However, when Chadwick made his momentous discovery it was still thought that the neutron was actually a proton-electron composite. The neutron was not recognized as elementary by anyone but a few until 1933 when Heisenberg proposed that exchange forces occurred between protons and neutrons in the nucleus (which was treated quantum-mechanically). Even then Heisenberg's theory still held that the neutron was an electron-proton composite. At the 1933 Solvay Congress the debate about the elementary nature of the neutron was prominent. By October 1934 most everyone was convinced of its elementary status by new data showing that the neutron was actually more massive than an electron and proton together and that it was unstable spontaneously decaying into a proton and an electron. At this point electrons ceased to be considered nuclear particles. Note that Eddington's first crack at a fundamental theory was *Relativity Theory of Protons and Electrons* that was published a mere two years later (1936) with much of the sourcework performed earlier. Clearly nuclear physics was not well understood phenomenologically then. Linear accelerators and cyclotrons did not make a large-scale appearance until the late 1930s and then mostly in the United States. Lise Meitner (1878 – 1968) and Otto Frisch (1904 – 1979) first realized that a uranium nucleus could split when capturing a slow moving neutron only in 1938 and their fission hypothesis was first reported in early 1939. The basis for a true understanding of atomic energy was laid later that year when Bohr and John Wheeler (b. 1911) completed a semiempirical study that appeared as a paper in *Physical Review* on September 1, 1939, the same day World War II began (Kragh 1999). Needless to say work on atomic energy and in physics in general focused less on general understanding than on wartime applications for many years. Since Eddington died in 1944, nearly a year before Hiroshima and Nagasaki, he never witnessed the atom and nucleus truly probed to a level that would have met his satisfaction.

  As such Eddington was forced to conclude that system *A* corresponded to observed quantities since most observations of hydrogen were as a gas consisting of countless atoms and thus were *macroscopic*. Since linear accelerators and cyclotrons were so new no one was really experimentally studying single atoms yet, at least not to the level required. Experimental physics was firmly system *A* and system *B* was wholly theoretical. In order to make any proper experimental investigation of system *B* the associated quantities would need to be transformed into system *A* for comparison to observables and other data. Or, conversely, data observed under *A* would have to be transformed to *B* and compared with predicted theoretical values. He gives specific conversion factors for $e$, $\hbar$, and $\mathfrak{F}$ (Faraday's constant). He also derives an exact procedure (anchorage as he calls it) for obtaining theoretical values from experimental measurements. Secondary methods (anchors) exist, of course, and he outlines three including the spectroscopic determination of the proton-electron mass ratio, the deflection determination of $e/m_e c$, and the direct determination of $h/e$. Eddington is blunt in basically saying that, since theory and measurement obviously have to agree, if a theory does not match measurement something must be done to *force* it to match. In his own words:

> We have to accept the convention that any quantity that has been extensively used in the systematisation of observational knowledge has acquired thereby the status of a *vested interest*. It if [sic] does not arise naturally in the theory we have to go out of our way to introduce it in order to avoid talking a different language from everyone else. Unless an internal inconsistency is detected the established procedure of reduction of the measurements – which is the definition of quantity – must be accepted without amendment (Eddington 1946, pp. 57-58).

Eddington's derivation of the Coulomb energy is far more extensive than is presented here and continues well beyond the subject of this monograph that is interested primarily in uncertainty and exclusion in his theory. He gives several early arguments for it in the various chapters leading to the exclusion discussion but does not fully investigate it until he develops a wave equation for the intracule much later. But he does make the point that the addition of the Coulomb energy changes the multiplicity to 137 that is roughly the fine structure constant (he argues, as I'll show in chapter eight, that it is *exactly* 137). More specifically he points out that the derivation of the fine-structure constant produces the $k$ in $V_{137}$ rather than $k + 1$ in $V_{136}$. Utilizing the mechanics of a large ensemble of particles in relativity theory Eddington derives the expansion energy for a wave packet as being $3\hbar^2/4\sigma^2 m$. This is twice the value for the mean kinetic energy of the particle and thus the particle's total energy consists entirely of this and mass energy since, if it is isolated or non-interacting, there is no potential energy term. In order to stop (or slow) the expansion, a negative potential must be introduced. The result produces an equation describing a spinless $V_3$ particle. The macroscopic form of this process involves an assemblage of particles attempting to disperse and a negative potential energy must be introduced to prevent (or at least inhibit) the dispersal. As such Eddington has introduced a method of argument seemingly supportive of steady-state cosmologies. Since he calculates the rate of expansion he obviously does not believe in a static universe, but his methods introduce the possibility. The expansion energy, then, is countered by a negative potential energy. But he argues throughout *Fundamental Theory*, particularly in direct response to Dirac, that negative potential energy is simply a result of shifting the datum. As such the negative potential energies, as described in chapter five, are really positive energies opposing the expansion energy. The additional energy must be 'created' somehow, seemingly in violation of energy conservation, and it is through interchange that this occurs. Thus the additional potential opposing expansion of the wavepacket or the universe in general is the Coulomb energy. Relating this extra energy to spin, since spin is really a frame-dependent quantity (at least in his mind – he makes it clear that he believed that the idea that it had to pass through two full rotations to return to its original state was a spurious notion), if angular momentum is to be conserved there must be a recoil spin inside the atom. It is the study of this recoil spin that leads to the derivation of the fine structure constant that is also equal to the multiplicity produced by the extra energy. As such conservation of momentum produces a recoil spin that produces the extra Coulomb energy term represented as interchange energy that was due to the ability of particles to exchange characteristics. In addition, this provides a direct relation between spin and the Coulomb energy, which is the basis of his reasoning for equating the Coulomb force with the exclusion principle.

# VIII

## *Exclusion*

Having discussed some of the conditions required for uniqueness including how this relates to the concept of interchange and the ability to represent composite particles in alternate ways, the robust development of these ideas will ultimately lead us down the path to the exclusion principle.

**Observations Involving Time**

As has been repeated numerous times, the process of measurement disturbs a system. However, the interference is not always intentional. It is possible that a system may spontaneously release information that is eventually detected by one or more of our five senses. In fact *anything* we naturally sense in our day-to-day lives is this type of 'measurement.' It does not eliminate the fact, however, that there is a selective bias in the measurement since the system under measurement is still interacting with its environment that happens to include ourselves. This seems to rule out purely objective theories (Eddington 1946).

      We can therefore not properly speak of 'undisturbed states' since there is no such thing. As such there should be a standard measure of the amount of interaction caused by observation in general, which is to say "the system is subjected to a conventional amount of probing" (Eddington 1946, p. 70). The only characteristics available for analysis from such a system are those revealed by this probing. Any sporadic information that pops into the picture cannot be included in the analysis since we're not then describing the normal state we set out to describe. In essence, we put limitations on measurements, i.e. we idealize experiments in order to filter out the noise of everyday experience. However, this means that the 'normal state' is not always the same since we are free to define it in any way we please. However, in doing this, new systems and particles will be defined. For example the standard hydrogen atom becomes the hydrocule when attempting to study its structure.

There is nothing, however, that is stopping us from introducing a standardised amount of interference just as we have introduced the standard background as the uranoid. Once again this is developed to help account for the transition from macroscopic to microscopic. Normally the generalised coordinates and momenta are enough to define a system where the number of these pairs that are produced is the multiplicity. But some level of standard interference really must be introduced to properly handle any measurement of such quantities. One of the problems here is that observables do not generally commute (e.g. coordinates and momenta). As such they are treated as a join-observable – they only make sense when observed in tandem. The joint-observable has a probability distribution that can describe either the coordinates or momenta. Since the uncertainty principle puts a lower bound on the joint knowledge of these two measurements, one way to define the level of interference (probing) here is to use this lower bound as the minimum fixed value for the observation. In this situation the system is described as being 'fully observed.'

What happens after an observation concludes and the system is no longer being disturbed (interfered with)? If the initial observation was made at time $t_0$, at some later time $t_0 + \tau$ the combined uncertainty is greater than it was at $t_0$. Essentially the uncertainty increases with $\tau$. This is really just the backwards interpretation of wavefunction collapse – a large amount of uncertainty exists prior to an observation. Once the observation is performed the uncertainty disappears since the wavefunction collapses to a single eigenvalue in a single eigenstate. Eddington qualitatively describes this in reverse – the wave packet is *expanding* as was described in the previous section. So, an observation reverses the process of expansion on a wave packet. In essence the observation furnishes the extra (negative) potential energy that inhibits and eventually reverses wave packet expansion. Eddington defines the time, $\tau$, as measured from the instant of cessation of the full observation, as a 'coefficient of under-observation' for the system (Eddington 1946). In essence any time an observation concludes the system is said to be under-specified. The only way to fully specify a system is to always observe it. The only way over-observation could occur is if measurement of the variables of some system continued after that system had completely vanished which is unlikely.

So the standard probing that corresponds to the zero-temperature uranoid that is the standard environment is a full observation. If the system is then observed from birth to death, so-to-speak, any interference is automatically *always* a part of that system. For instance, if one observes a particle for a tenth of its lifetime, nine-tenths of its lifetime exhibits different behaviour since it is not being observed during that time. However, if it is observed for its entire lifetime then any interference is always present and is automatically included in the full description of the particle. Practically speaking it is possible to do this with pair creation since the particle lifetimes are short. There are other instances of short-lived particles in nuclear situations that also could be fully observed.

Sometimes it is necessary to examine a particle in a different environment. In order to account for the change in environment from one uranoid to another during a single observation, the disturbance produced by the uranoid is supplied with the attributes of an object-field with the appropriate variables, usually potentials. A change in uranoid then simply requires a change in the values of the variables. Thus there is an environmental field carrier developed here by Eddington that is very similar to the standard carrier. Its purpose is to carry the information associated with the environment and it can change form depending on the situation. A similar field can be developed for the standard probing and is called the 'field of under-observation' (Eddington 1946). However Eddington does not mathematically develop this theory beyond a single case as he believed it had no practical application (with one exception).

The exception Eddington discusses is the abrupt cessation to some standard probing. The time $\tau$ is the only variable required to specify this situation and it is equal to the time interval between the moment the probing ceases and the moment being considered. If it is an observed time interval (e.g. if you stop observing something and then wish to consider what that something might be doing 5 minutes later, the 5 minutes is an observable time interval) then it is subject to uncertainty in its measurement simply due to the inability to develop wholly accurate measurement devices (and, by $\delta E \cdot \delta t \sim h$, it directly follows an uncertainty relation) and has an associated probability distribution. However, there is nothing that prevents $\tau$ from being stabilised either.

What does it mean, though, to have a probability distribution over time? Since in natural units times and lengths can be considered equivalent there is nothing that specifically prevents us from representing a given time by a probability distribution simply from this basic fact. But, in addition, there is a practical limit to the exactness of clocks. Normally a distribution function $f(x,y,z,t)$ gives a distribution over only *x*, *y*, and *z at* a given time *t*. If the distribution is to include $\tau$ then it should be represented as $f(x,y,z,\tau;t)$ however there is no homology (to use Eddington's own terminology) between $\tau$ and the spatial coordinates. In order to induce homology between these a physical origin for the time must be defined in a similar way as the physical origin for the spatial coordinates, i.e. "as the centroid of a large number of events uncorrelated in their time distribution" (Eddington 1946, p. 71). Eddington immediately discards this description as leading to mathematical terms with no corresponding physical explanation. As was developed in equation (6.24), the multiplicity provides a sign change for the time meaning that the relativistic homology occurs between space and *imaginary* time and thus a negative standard deviation is impossible.

Eddington resolves this apparent paradox by taking the scale and its accompanying phase dimension as the fourth dimension. Since the particles in the uranoid have uncorrelated phases (they must be non-interacting) one can simply write the physical origin as including Gaussian uncertainty in its phase coordinates as well as its spatial coordinates. The scale fluctuation as developed in chapter five (see equation (5.10) for example) is combined negatively. Therefore the scale and phase have a time-like relationship to the spatial measurements. By this line of reasoning Eddington argues that using time as the fourth coordinate in Dirac wavefunctions is incorrect. However, since the Dirac equation is Lorentz invariant he claims that invariance in phase coordinates is *analogous* to Lorentz invariance and the phase is a time analogue in any analogy between classical (time) and quantum (phase) mechanics (Eddington 1946).

**Structuralism v. Prediction**

Modern physics can generally be categorized as having two purposes – describing the structure of the universe as well as its constituents and making useful predictions regarding the objects within the universe. Obviously if a theory is to make a prediction then observation of the system under investigation has to cease before the moment of prediction (and subsequently restart) for the prediction to have any meaning (i.e. if you're holding a ball, predicting it will fall when you let it go requires actually letting it go). This relates directly to the cessation of observation that leads to $\tau$. On the other hand a theory can equally well describe the structure of various objects including the universe itself and such investigations include the determinations of fundamental constants, energy levels, etc. In these situations the system is either considered to be fully observable or $\tau$ is stabilised. Either way $\tau$ is not really involved in the process.

In predictive problems there is often an associated decay of some structure (including mathematical) and so such problems are sometimes described as problems of decay rather than prediction. Decay is a misleading term, though, as it is now most often associated with radioactive processes and the release of energy. This is not entirely what he describes despite his use of the term. For example, given an initial occupation for a set of eigenstates, one might wish to determine any changes to the system during a subsequent transition period given by $\tau$. The variables in this problem are the occupation factors, but there is no requirement that the system decay per se, though clearly Eddington reasons that the wavepacket will expand and not contract since it is at its full contraction (wavefunction collapse) while being observed.

The vast majority of *Fundamental Theory* is devoted to structuralism and prediction is only brought in near the end, well outside of the primary subjects considered in this monograph. As such $\tau$ is never involved in any of these considerations and all systems are considered to be fully observed. However, it is presented here as a philosophical point in order to emphasize once again the problem with a purely objective theory since considerations focusing on $\tau$ are as close as Eddington comes to something truly objective since *during* the time interval $\tau$ the observer is completely absent from the process. Even though the observer is present both before and after this interval, a ratio comparison between measurements made on either end of the interval still provides a reasonably good objective approximation for any changes that occurred during the interval.

In discussing structuralism, then, changes cannot be taken into account through predictions. In wave mechanics physical changes are represented by the changes in the occupation factors of steady states. Wavefunctions thus have to represent steady states with the occupation factor being the only variable that changes with time. If unsteady states are introduced physical changes are represented either by continuous motion or by discrete transitions, which are mutually exclusive. In order to only vary the occupation factor with time only two types of motion are allowed: steady circulation of probability within a given state and non-spatial transitions of probabilities between states where the states themselves can be steady or unsteady. Lorentz transformations introduce motion that does not fall under either of these categories. The time-like variable in the wavefunction is then the phase.

**Developing the Proper Mass**

Given that we now have a framework for working with wavefunctions in both structuralist as well as predictive situations we can bring the two together in the prediction of proper mass values. Equation (5.6) represents the physical distribution function while equation (5.7) represents the geometrical distribution function. A geometrical distribution function can be converted into a distribution of a correlated physical coordinate $\xi$, where $x = x_0 + \xi$, by:

$$h(\xi) = \int_{-\infty}^{\infty} g(x_0 + \xi) f(x_0) dx_0 . \tag{8.1}$$

Fourier integrals can be written for $h$, $g$, and $f$ as $H$, $G$, and $F$. Through separability a direct relation between $H$ and $G$ can be obtained:

$$H(q) = e^{-\frac{1}{2}\sigma^2 q^2} G(q) \tag{8.2}$$

where the factor $e^{-\frac{1}{2}\sigma^2 q^2} = 2\pi F(-q)$ is a result of the physical origin having a Gaussian distribution function as described in chapter five. In most cases momenta are of primary interest and distributions of physical momenta are converted into distributions of geometrical momenta (or vice versa). But, when dealing with coordinates, it helps to work with wavefunctions instead of probability distributions. As such two real distribution functions representing *x* and *p* are replaced by a single complex wavefunction that has two reciprocal forms (from Fourier transforms) that are mathematically derivable from one another. These Fourier transforms are analogous to similar transforms that can be derived from *H*, *G*, and *F* directly. In fact they are the square roots of the original reciprocal transforms and contain the term $p/\hbar$. As such, since the originals contained a *q* term in place of the $p/\hbar$ term, we find that $q = 2p/\hbar$ where the factor of 2 results from the squaring of the exponential that contains these terms. Equation (8.2) directly converts a distribution of geometrical momenta given by $G(2p/\hbar)$ into a distribution of physical momenta given by $H(2p/\hbar)$.

Given a wavepacket represented by a Gaussian distribution of momenta, with standard deviation *s*:

$$G(p) = \left(\pi\hbar^2/2s^2\right)^{\frac{1}{2}} e^{-2s^2 p^2/\hbar^2}, \tag{8.3}$$

equation (8.2) can be rewritten in general form as:

$$H(p) = C \cdot G(p) F(-p) \tag{8.4}$$

where *C* is a constant. This means that the probability of a physical momentum *p* is actually the combination of the probability of a geometrical momentum *p* of the object under consideration and an opposite momentum *–p* of the physical origin. The recoil of the physical origin has already been discussed. However, at the time we had not yet considered the physical origin as a two-particle system (say a hydrogen atom represented as an external and an internal particle) with two momenta represented by a single external momentum *P* and a single internal momentum $\varpi$ (where $\varpi$ is analogous to $\xi$ - both are relative measures). Since the two-particle system could alternately be resolved into its constituents (say the proton and the electron, for instance) each of these has their own momenta *p* and *p'*. The relationship between the distribution functions for the momenta of the two individual particles and the momenta of the external/internal particle representation is:

$$K(P)H(\varpi) = G(p)F(p'). \tag{8.5}$$

If *K(P)* is represented by a Dirac delta function such that for all states with a non-zero probability *P* = 0 (which results from the recoil momentum of the physical origin, i.e. conservation of total momentum or the probability of *p* without recoil is zero, hence the Dirac delta function representation) and it can be shown that $p = -p' = \varpi$, then (8.5) reduces to (8.4). The result is completely independent of the masses assigned to the particle and the physical origin since we're working with standard carriers. Ultimately by representing *K(P)* as a Dirac delta function the requirement of Gaussian distributions is relaxed and (8.4) can then be used at will to convert momenta between physical and geometrical systems. Changing *F(-p)* to *w(p)*, σ to *s*, and defining $\varpi = \hbar/2\sigma$, (8.4) can be rewritten as

$$H(p)dp = C \cdot G(p)w(p)dp \tag{8.6}$$

where

$$w(p) = (2\pi\varpi^2)^{-\frac{1}{2}} e^{-p^2/2\varpi^2} . \tag{8.7}$$

Equation (8.7) represents the 'weight function' *w* whereby a distribution of geometrical momenta is converted to a distribution of physical momenta by weighing the ranges *dp* with *w*. The idea of the weight function is to reduce the frequency of large momenta and Eddington gives the exact value for $\varpi$ as being roughly $200 m_e c^2$. He says that omitting *w* or, equivalently, setting σ = 0, introduces unwanted infinities (Eddington 1946). This is identical to Heisenberg's 'weight function' *ψψ\** but with slightly different notation.

So far I have only introduced the zero-temperature uranoid. What would happen, then, Eddington asks, if we considered an *infinite*-temperature uranoid where the particles are not at rest but rather have unlimited uniform probability distributions. The corresponding distribution of physical momenta is given by the weight function (in three dimensions), $wdp_1 dp_2 dp_3$. The mean values are given as:

$$\overline{p_1^2} = \overline{p_2^2} = \overline{p_3^2} = \varpi^2 = \hbar^2/4\sigma^2 . \tag{8.8}$$

If the momenta are suitably large, say relativistic, the correct energy is given by $E^2 = m^2 + p_1^2 + p_2^2 + p_3^2$ (employing natural units, of course). The mass term drops out if particles do not have a proper mass (e.g. a photon). By (8.8) the energy is then:

$$\overline{E^2} = 3\varpi^2 = 3\hbar^2/4\sigma^2 . \tag{8.9}$$

In fact regardless of whether or not the particles are photons, in an infinite temperature uranoid the particles would have a physically impossible infinite kinetic energy and thus velocity. As such their entire mass would be converted purely into energy. Eddington proves this result by applying his results for the standard deviation in macroscopic situations to the pressure-density equations for a steady uniform distribution of matter as given in the then-widely accepted text on relativity and thermodynamics by Richard C. Tolman (1881 – 1948, Tolman 1934).

In this way he shows that proper mass is a concealed form of energy (nothing new here) but says any fundamental theory should show *how* this conversion takes place. Lowering the temperature of the uranoid then should reverse the process and produce a proper mass *m* for particles (this is somewhat akin to what actually happened during the cooling of the universe, though the actual process is far more complicated). The proper mass here, however, is considered an invariant (and since it is the magnitude of four-momentum and four-momentum is conserved this is true) that is independent of the motion of the particle since motion is relative anyway. Thus it is the temperature of the *standard environment* that is to be considered here, not even the particle's *actual* environment. The trouble with Eddington's argument is that he is basing it partially on his belief that inertial and gravitational fields are different – i.e. on the falsity of equivalence – whereas equivalence has been experimentally proven many times to phenomenal accuracy. Regardless, he continues to derive various relations between quantities in a zero-temperature uranoid and an infinite-temperature uranoid. In these derivations he makes use of the fact that *K(P)* is a Dirac delta function that eliminates three degrees of freedom leaving the multiplicity at 3 (particles considered here are $V_3$ particles). As such "the probability distribution of momentum is uniquely determined by the probability distribution of coordinates and vice versa" (Eddington 1946, p. 79). The scale uncertainty can be accounted for directly in this process or replaced by curvature. Since wave mechanics is not well-suited to handle curvature the scale uncertainty ought to be accounted for directly. But, in doing so, *m* has a probability distribution that results from the uncertainty in the standard of mass used to measure it which is embodied in a comparison particle. If the scale uncertainty is included directly this adds a degree of freedom (extra variable or multiplicity) and the particle is a $V_4$. Equation (6.29) has already provided a conversion for the mass values of $V_3$ and $V_4$ particles. However, if rotational motion is to be considered the particles must really be $V_{10}$ with a mass $M = \tfrac{3}{10} m_3$.

In the derivations of the various physical constants in the zero-temperature and infinite-temperature uranoids he arrives at the following equation for mass

$$m = 2\varpi = \gamma/\sigma = 2\gamma\sqrt{N}/R_0 \tag{8.10}$$

where he has employed equation (5.13). He then derives the following for $\gamma$:

$$\gamma = \hbar\sqrt{M/\mu} = 136\hbar/\sqrt{10} \tag{8.11}$$

where $M$ and $\mu$ represent the masses of the external and internal particles in the hydrocule representation. The $V_{10}$ must then be the external representation of the hydrocule with the mass:

$$M = \frac{136}{10}\frac{3}{4}\frac{\hbar\sqrt{\frac{4}{5}N}}{R_0}. \tag{8.12}$$

But, since $M = \frac{3}{10}m_3$, the proper mass is given by:

$$m_0 = \frac{3}{4}\frac{\hbar\sqrt{\frac{4}{5}N}}{R_0}. \tag{8.13}$$

There is one step left in developing the proper mass. Since the move is being made from a theoretical to an observational system the factor $\beta$ must be included (see equation (7.5)). Equations (8.12) and (8.13) become:

$$m_0 = \frac{3}{4}\frac{\beta^{\frac{1}{6}}\hbar\sqrt{\frac{4}{5}N}}{cR_0}, \quad M = \frac{136}{10}\frac{3}{4}\frac{\beta^{\frac{1}{6}}\hbar\sqrt{\frac{4}{5}N}}{cR_0} \tag{8.14}$$

where $c$ has been reinserted in order to return any restrictions originally placed on the units. Eddington refers to (8.14) as the "central formula of unified theory" (Eddington 1946, p. 81). Equation (8.14) can be rearranged to solve for $R_0$, which then includes the term $\hbar/Mc$. Since $\hbar/e$ and $e/Mc$ are well known to great accuracy from experiment (the latter is Faraday's constant), $\hbar/Mc$ is also well-determined. One can then divide through by $\sqrt{N}$ to get a more accurate value for $\sigma$:

$$\sigma = \frac{R_0}{2\sqrt{N}} = \frac{136}{10}\frac{3}{4}\frac{\beta^{\frac{1}{6}}\hbar\sqrt{\frac{1}{5}}}{Mc} = 9.53657\times 10^{-14}\,cm. \tag{8.15}$$

The value obtained by applying (5.17) using the observed values used in (5.18) gives $9.5\times 10^{-14}\,cm$.

**Exclusion from Uncertainty**

Equation (8.14) can be derived in an entirely different way by utilizing the principle of exclusion the earliest version of which prevents more than one particle from occupying a given eigenstate in an atom. In the hydrocule model of the hydrogen atom, however, there is only one internal particle (the intracule) and thus this version of the exclusion principle is trivial (in fact it is trivial for the normal hydrogen atom as well since there is only one electron). Eddington thus states an alternative version of the exclusion principle as follows:

> If the 6-space obtained by taking $x, y, z, p_1, p_2, p_3$ as coordinates is divided into cells of volume $h^3$ ($h = 2\pi\hbar$), then in a steady state the maximum number of particles per cell is two electrons and two protons (Eddington 1946, p. 82).

A similar description can be given for extracules and intracules using relative coordinates and internal momenta for the intracules:

> In a steady state the maximum number of particles is two extracules per cell $h^3$ of $xp$-space and two intracules per cell $h^3$ of $\xi\varpi$-space (Eddington 1946, p. 82).

What exactly does this mean? Using the generalised coordinates and momenta as coordinates of a single space is simply using phase space in the classical sense (hence Eddington's introduction earlier of the phase dimension – c.f. Goldstein 1980). The maximum number of particles is then determined by the uncertainty principle! Let me reiterate this fascinating conclusion – Eddington has defined the exclusion principle *in terms of the uncertainty principle*! For example, in a single dimension the uncertainty principle is given by $\Delta p_x \cdot \Delta x \geq h/2\pi$ which can be rewritten as $2\pi \Delta p_x \cdot \Delta x \geq h$. Similar relations can be written for *y* and *z*. The factor of 2 (ignoring $\pi$ for a moment) means that two sets of generalised coordinates and momenta can be defined for a given direction. Thus two identical particles can occupy the given space within the bounds of uncertainty. In three dimensions this simply means that two identical particles can occupy the volume $h^3$ again within the bounds of uncertainty. Basically he lays claim to the fact that uncertainty prevents *perfect* exclusion since there's really no way of knowing for certain whether two particles share the same *exact* location (recall that to Eddington this was in configuration space and that quantum and classical descriptions were not as distinct as we usually treat them). Uncertainty then permits two identical particles to occupy the same volume in space, which does not *necessarily* mean that they will be in the exact same spot. This is analogous to the two particles in a box problem discussed previously (with the two particles being the proton and the electron). Initially both particles have an equal probability of being anywhere in the box. Regardless of whether or not they interact, their probability functions will always give them an equal probability of being anywhere in the box. If they are observed then the wavefunction collapses to a given location, but since the above argument applies right at the Planck scale any observation is inaccurate to within the given uncertainty thus allowing them to coexist, i.e. even after observation, the observation includes uncertainty meaning the wavefunction really didn't collapse 100%, it just shrank considerably. The probability for them to be anywhere in the volume (and thus not occupying the same spot) is minimal, but not zero. This is a profound result that brings up the thorny issue of completeness in quantum theory.

The exclusion principle as it is traditionally understood postulates quite simply a zero probability that two particles will share the same exact quantum numbers (which precludes them from existing at the exact same spot). This seems to imply an *exact* measurement in the sense that, even if their locations and momenta cannot be known exactly, it *is* known exactly that they are *not* in the same spot (or sharing the same quantum numbers). But the uncertainty principle implies that there are really no such things as exact measurements, per se. Ultimately it implies that we cannot have full knowledge of the state of a system, yet it can be argued that the exclusion principle *provides* full knowledge. There appears to be a contradiction here.

The contradiction can be rationalized away by considering that exclusion is still governed by probability even though it presents a null result. Thus, even though exclusion appears to provide an exact knowledge of a given state, it only gives a probability for that state that happens to be zero (and zero is no less real than 50%, for example). Eddington obviously had trouble with this, however, given that he alters the exclusion principle to explicitly account for uncertainty. But as a consequence there is nothing the prevents *any* particle, regardless of spin, from obeying this new version of exclusion. In Eddington's theory even photons would obey exclusion.

**Electrical and Mechanical Exclusion**

Since the intracule was defined as a $V_{137}$ particle it contains the extra degree of freedom given by the Coulomb energy and is thus electrical in nature while the extracule, being quite simply an electrically neutral hydrogen atom, is mechanical in nature (and by this I mean these particles are *observed* either mechanically or electrically). As such the second definition of the exclusion principle given by Eddington actually provides two separate versions – one for extracules and one for intracules – and in the process creates a mechanical exclusion principle (applying to extracules) and an electrical exclusion principle (applying to intracules). He uses this process, then, to replace gravitating (non-excluding) particles with excluding (non-gravitating) particles. As such he makes exclusion a wave-mechanical substitute for gravitation, or more correctly, he replaces the curvature of space with the mechanical exclusion principle noting that the electrical form corresponds to the version of the principle verified by experiment (he specifically mentions the super-dense matter in white dwarf stars). On the other hand, the concept is very narrow since exclusion assumes a steady state. Based on what we have developed so far, then, exclusion is thus also the same thing as the extraneous standard.

Mechanical exclusion as applied to extracules begins with the consideration of $n$ extracules per-unit volume in three-dimensional space. If their mass-constant (total mass) is $\mu_0$ then the total kinetic energy of an extracule is given by the classical form:

$$E = p^2/2\mu_0, \quad (p_0^2 = p_1^2 + p_2^2 + p_3^2). \tag{8.16}$$

In the zero-temperature uranoid this energy is obviously a minimum, though *not* necessarily zero due to the uncertainty principle in energy and time. The momenta $p_1$, $p_2$, and $p_3$ are distributed so that the total energy is at a minimum with a density of no more than two particles per cell (one is a comparison particle since all measurements are relative). Drawing on the description of spherical space given in chapter five, a volume of radius $r$ must be equal to half of the $n$ extracules per unit phase space volume, i.e. Eddington's exclusion rule has two extracules and two intracules per $h^3$ so the volumes in spherical space and phase space are related by:

$$\tfrac{4}{3}\pi \wp^3 = \tfrac{1}{2} n h^3 \tag{8.17}$$

with the radius of the spherical volume equivalent to the momentum, $\wp = p$ due to the application of Eddington's system of natural units. Substituting (8.17) into (8.16) then gives the energy $E$ as:

$$E = \frac{\wp^2}{2\mu_0} = \left(\frac{3n}{8\pi}\right)^{\tfrac{2}{3}} \frac{h^2}{2\mu_0} \tag{8.18}$$

that Eddington calls the 'top energy' corresponding to a total for that volume. The total number of extracules in the universe, being half the total *particles* in the universe, is given by:

$$\tfrac{1}{2} N = 2\pi^2 R_0^3 n \quad \text{or, solving for } n, \quad n = \tfrac{1}{2} N / 2\pi^2 R_0^3 \,. \tag{8.19}$$

Defining the term $\mu_1 = \mu_0 \left(\tfrac{3}{4} N\right)^{\tfrac{1}{3}}$ and using (8.19), (8.18) becomes:

$$E = \frac{3}{4} \frac{N}{2\mu_1} \left(\frac{h}{2\pi R_0}\right)^2 . \tag{8.20}$$

This can be further simplified to:

$$2\mu_1 E = \tfrac{3}{4} \varpi^2 \tag{8.21}$$

where by (8.8) the weight constant, $\varpi = \hbar/2\sigma = h\sqrt{N}/2\pi R_0$.

How is the above interpreted? Returning to a discussion of multiplicity, if $H^0$ is some characteristic (e.g. energy) dependent on the occupation factor of a large number of particles with a particle density, $s$, if one particle is removed, the amount of $H^0$ removed is $dH^0/ds$. If the assemblage is entirely in a single pseudo-discrete state, this amount is $dH^0/ds = -H^0/k \cdot s$ and is known as the $H$ of the 'top particle' (Eddington 1946, p. 28). The average $H^0$ per particle in a unit volume is $H^0/s$ and is called the $H$ of the mean (average) particle. The $H$ of the top particle, $\hbar$, in terms of the $H$ of the mean particle, $\overline{H}$, is $\hbar = -\overline{H}/k$. Since proper masses have the same ratio as proper densities this general form can be used for mass relations, $\mathfrak{M} = -\overline{m}/k$. If (8.20) defines the 'top energy' (energy of the top particle), (8.9) and (8.21) combine to give the mean (average) energy as $\overline{E} = \frac{3}{5}E$. Now, if the particles are in a zero-temperature uranoid, the top energy is just the rest mass of the top particle. The top energy is essentially the energy that the particle is boosted to due to the complete occupation of all the lower energy states, so it is associated with exclusion. So the rest mass is associated with exclusion since its value is given according to the occupation of all the lower energy states. Masses, of course, have gravitational fields and, in one interpretation of the uranoid scenario, the particles in a uranoid give an individual object particle its rest energy by determining the scale fluctuation that gives curvature to space. Both these interpretations of interactions must be equivalent and thus both exclusion and gravity (curvature) boost particle energy in the same way. Eddington, thus, interprets these as being equivalent thus concluding that exclusion acts as the wave-mechanical form of gravity (exclusion is derivable from gravity and vice versa).

The one major flaw that appears to jump out here is that, given Eddington's description, a naïve interpretation would assume that every pair of electrons, for instance, in the universe would have a different energy level. This doesn't make sense since we know that, as long as they are non-interacting, two electrons can share the same energy levels (quantum numbers) – e.g. it's easy to have two hydrogen atoms containing electrons in the same exact orbital as long as the electrons themselves don't interact. Ah, but Eddington has solved this problem earlier by making scale a variable on par with mass, energy, etc. His rationale for this comes out of the fact that gravity is the curvature of space-time itself and has energy, so a scale of measurement in space-time ought to have an associated energy since it is a 'chunk' so-to-speak of this curvature. Basically, coordinate locations contribute to the energy level just as mass, motion, and other characteristics do. So all the electrons could seemingly have the exact same energy levels, yet, since they are not in the same locations, they actually have *different* energy levels based solely on their coordinates. It doesn't mean they can't all have the same kinetic energy or even the same set of *traditional* quantum numbers. It simply means that they all have different locations. So just as gravity (curvature) provides an energy boost to particles, so does exclusion, and both can be interpreted in the same manner – as a change in location.

From the standpoint of the original formulation of the exclusion principle as applied to atoms, this makes sense from Eddington's point of view. In an atom, only two electrons are allowed in any given energy level – one with spin up and one with spin down. Once this level is filled, electrons start filling in the next level, and so on. Eddington viewed each change in energy level as being the same thing as a change in location. So after the lowest orbital was filled, filling the next orbital amounted to simply putting the new particles in a different location (which indeed is true). This idea of location as being the central point behind exclusion comes from Eddington's reliance on relativity to make sense of his viewpoint as an observer: moving locations was simply a relativistic transformation. When viewed in this light, since relativity implies that no viewpoint has special meaning and we can't really be absolutely certain of anything (at least anything objective), it is no surprise that he found a natural philosophical link between it and uncertainty and used this as his basis for the reasoning in *Fundamental Theory*.

**Dirac's Negative Energy Sea**

Relativity plays such a tremendous role in his interpretation of physics it leads to some unusual insights as I've shown. One that I have briefly pointed out was that negative energies were simply a result of a shift in the datum, especially when one considers the relation between energy and motion in kinetics and the fact that potential energies associated with interactions are really tied to exchange particles and can be interpreted in the same way. Theoretically one could always adjust the frame such that the datum moved, then, so that all previously negative energies were positive – or vice-versa. So Eddington considers the highest level of the top energy $E$ as the zero level meaning that all the other levels are below this. Thus the particles of a zero-temperature uranoid fill the series of negative energy states below the zero level just as Dirac described in 1929, interpreting the negative energy results of his own equation (Kragh 1999). Once again we see the fact that Dirac's equation, though not initially satisfactory to Eddington due to its non-tensor form, was continually on his mind throughout the development of *Fundamental Theory*.

Eddington further rationalizes that these particles filling the negative energy sea are really normal particles distributed in such a way that they do not interfere or interact with the 'positive-energy' particles governed by the normal laws of quantum mechanics. Since the particles are extracules there is a symmetry between positive and negative charge and the problem of infinite negative charge, as postulated by Dirac, does not arise since there are a finite number of particles in the universe. As such Eddington proposes that there ought to be negatrons (anti-protons) as well as positrons, and, of course, we know he was correct in this prediction. In fact, Eddington's theory, if extended to include any newly discovered particles, which would all obey his exclusion principle regardless of spin (since his version does not rely on spin in its formulation), *every* particle ought to have a corresponding anti-particle. Quantum field theory predicts exactly that and many of these have been found (n.b. neutral particles are their own anti-particle).

The formation of the *observable* particles in the universe is brought about through the excitation of top particles to a higher energy (i.e. above the zero level). The net addition to the uranoid is a particle with an energy $E$ above the zero level and a corresponding hole an energy $E$ below the zero level. The particle-hole combination is called a 'bi-particle' and carries the excitation energy. This relates directly to the discussion of mutual and self-energy conditions in the section on energy issues in chapter six. Expanding on this idea Eddington says that an introduction of a particle with a comparison hole (which is just a particle with a negative mass in the equations – again just a relativistic transformation) produces no net change in particle density since every new particle is balanced with a new hole. The bi-particle is a $V_{136}$ and the excitation energy it carries produces the normal particle (electron or proton) as a $V_{10}$ and the hole as a $V_1$. If one wants an observable *anti*particle a hole is 'excited' to a lower energy state where the net addition to the universe is a bi-particle (thus keeping the particle density the same) with a hole at a lower level and a comparison particle filling the original hole. Since the same equations are used with a simple sign reversal the masses of antiparticles must be the same as their corresponding normal matter partners. Top level particles in this formulation of exclusion can be used as comparison particles by changing their energies. Thus $E$ is identified with previous descriptions using $m_0$. Here we see that matter can be created, in essence, from nothing (in essence, this is Eddington's description of pair creation). I will discuss this further in chapter nine.

Eddington's comparison of exclusion and gravitation does not end there. He takes the zero level as being a threshold and defines studies of particles below this level, the Dirac sea, 'sub-threshold' theory and studies of particles above this level 'super-threshold' theory. Since $\bar{E} = \frac{3}{5}E$, there are slightly differing values for the various constants in sub-threshold theory in terms of the known (super-threshold) values, namely $G_1 = \frac{5}{3}G$ and $\hbar_1^2 = \frac{3}{5}\hbar^2$.

**Gravitation from Exclusion**

In comparing two uranoids that have the same microscopic constants but a different top quantum number, $\Im$, we find that both $E$ and $\mu_1$ are constant and $\mu_0 = \mu_1/\Im$. In addition (8.20) takes the form:

$$2\mu_1 E = \Im^3 (\hbar_1/R_0)^2. \tag{8.22}$$

A quantum particle, then, is defined as being an addition to a rigid environment. If it is a top particle that can be added or removed without changing the completely filled energy levels underneath then this definition is fulfilled in the exclusion representation. Replacing normal gravitating particles by excluding particles thus replaces the gravitational field of those particles with an 'exclusion field' that is automatically rigid (Eddington 1946). When $N$ is very small there actually are two forms of exclusion (not to be confused with the electrical and mechanical version) – lateral and vertical. Thusfar I have only introduced vertical exclusion where particles gain or lose energy. If, for example, the uranoid consists solely of two particles that mutually exclude each other from being in state $E = 0$, then either particle can be considered the top particle (this is built from the idea of interchange). This is not a rigid environment, however, since removal of either particle would allow the other to occupy the $E = 0$ state. If the energy of each particle is $E = \mu_0 + p^2/2\mu_0$ and since these energies in uranoids are actually independent of quantum number, (8.22) becomes:

$$E^2 = \Im^3(\hbar_1/R_0)^2 = \tfrac{3}{4}N(\hbar_1/R_0)^2 \tag{8.23}$$

which is valid in super-threshold theory since it contains the cosmological values $N$ and $R_0$. Since $\hbar_1^2 = \tfrac{3}{5}\hbar^2$, (8.23) becomes:

$$E^2 = m_0^2 = \frac{3}{4}\frac{3}{5}\frac{\hbar^2 N}{R_0^2}. \tag{8.24}$$

This is precisely the same as (8.13) which was derived from standard relativity equations given by Tolman. Eddington has thus derived the same exact formula for proper mass from both the exclusion principle (albeit his version) and relativity.

**The Planoid**

Several equations have thus been developed over the past few chapters that connect the cosmological numbers $N$ and $R_0$ with the microscopic constants of physics. A simplification can now be made in order to study objects and systems on a much smaller level (very small compared to $R_0$). Since curvature can be represented by the scale fluctuation, object-systems and their environments can be considered in flat space. The standard environment is then a uniform distribution of particles in flat space. In order to represent infinite plane wave functions the distributions can be considered to be infinite themselves (meaning they continue indefinitely) but the environment of the object-system can now be limited to a sphere of radius $R_1$ containing $N_1$ particles with these values chosen to give the correct value for $\sigma$. This form of standard environment is called the 'planoid' by Eddington (Eddington 1946). It is *not* a mathematical transformation but a separate distribution. When the scale and phase dimension are then used in place of curvature, as we've developed, the process of projecting spherical space onto flat space as developed in chapter six is now useful. Neither the uranoid nor the planoid are representative of the actual universe, but in treating object-systems locally the remote environment of the universe can be neglected just as we neglect the affects of the Andromeda galaxy when calculating gravitational interactions between the Sun and Earth. Local phenomena include the integrated effect of the universe in $g_{\mu\nu}$.

Distinguishing between the curvature of space-time and the shape of the environment, the planoid is, like the uranoid, spherical. It is still Euclidean meaning space-time is still flat, but we simply consider a spherical volume of it (in fact space-time on a large-scale *is* flat and not curved as was thought for most of the twentieth century and yet we deal with spherical objects on a regular basis). Boundary conditions are accounted for in the extraordinary fluctuation that replaces curvature. As such we can now consider a zero-temperature planoid of radius $R_1$ with $N_1$ particles. It is also assumed that the same units of mass and length are used in both the uranoid and the planoid so that the quantum-specified standards are the same in each meaning that $\sigma$ must be the same in both. From the standard deviation of a given coordinate (and coordinate of the centroid) in the planoid, the $\sigma$ must have the following relation (thus relating planoid and uranoid values):

$$\frac{R_1^2}{5N_1} = \sigma^2 = \frac{R_0^2}{4N}. \qquad (8.25)$$

Using the new constants for the planoid the proper mass is:

$$m_0 = \frac{3}{4}\frac{\hbar\sqrt{N_1}}{R_1}. \qquad (8.26)$$

Rather, the proper mass in the uranoid given by (8.13) can be rewritten as:

$$m_0 = \left(\frac{3}{4}\right)^{\frac{1}{2}}\frac{\hbar_1\sqrt{N}}{R_0}. \qquad (8.27)$$

Generally $N_1$ and $R_1$ are fixed by the ratio in (8.26), but for some applications it is necessary to fix each separately and this process creates the 'special planoid' that I will introduce presently.

**Interchange for Extracules**

Averaging over the volume of the planoid with $r$ being the distance from its centre and introducing the mass-constant $\mu = m_0/136$ for an intracule, (8.26) can be written as:

$$m_3 = \tfrac{136}{3}m_0 = \sum_s \frac{p_s^2}{2\mu} \quad (p_s = \tfrac{1}{2}\hbar/r_s), \qquad (8.28)$$

where $m_3$ is the mass of a $V_3$ extracule. Physically if the $V_3$ particle has half-quantum of angular momentum about the $s$th planoid extracule then $p_s$ in (8.28) is the corresponding linear momentum. If these particles are the extracules of the planoid then (8.28) is a single particle in an assemblage of similar particles allowing rigid-field theory to be used. If a $V_3$ particle makes a transition to a state of momentum $p_x$, $p_y$, and $p_z$ then its energy is given as $E_3 = m_3 + \left(p_x^2 + p_y^2 + p_z^2\right)/2\mu$. Substituting (8.28) into this gives:

$$E_3 = \left(p_1^2 + p_2^2 + ... + p_{\frac{1}{2}N_1}^2 + p_x^2 + p_y^2 + p_z^2\right)/2\mu. \tag{8.29}$$

This is interpreted as eliminating the initial energy in the form of rest mass thus making the entire energy due to the transition of a system with $\frac{1}{2}N_1 + 3$ degrees of freedom. All degrees of freedom except for the last three are covered by the standard planoid and so the final three are the only observables.

Here Eddington gets a bit lost in speculation. The initial or rest mass $m_3$ is really energy as demonstrated by (8.29) and he describes it as arising from the $V_3$ extracule having a half-quantum angular momentum about every other extracule in the assemblage of the planoid. But the particles are all at rest so the momentum must be extra-spatial and the half-quantum value represents interchange circulation (Eddington 1946). His complicated explanation for this phenomenon boils down to the idea that since only the three momenta $p_x$, $p_y$, and $p_z$ are observable the mass is $m_3$ and the particle is a $V_3$. Everything else is accounted for in the standard planoid, but the interchange circulation essentially is a fancy way of saying that the extracules can randomly interchange with each other. This is yet another way of accounting for curvature: the resultant energy produced by the constant interchange of the object-particle under consideration with the other particles in the planoid. In the exclusion interpretation, particles force each other to be individuals at differing energy levels while in this interpretation these same particles constantly interchange with each other producing the same total energy that the exclusion interpretation does. Eddington puts off a more detailed explanation of each until developing his $E$-number theory in the latter half of *Fundamental Theory*. At this point the topic begins to stray from the fundamental foundational issues of interest here and thus I will not discuss them. However, an analysis of Eddington's $E$-number theory is in the works.

**Modifications to the Planoid**

The special planoid mentioned above is a cross between the planoid and the uranoid that employs the convention $N_1 = N$ and $R_1 = R_0$ but retains the sub-threshold constant $\hbar_1$. Particles are no longer $V_3$ particles, rather they are $V_4$. From (8.25) we have:

$$\sigma_1^2 = R_1^2/5N_1 = R_0^2/5N = \tfrac{4}{5}\sigma^2 . \tag{8.30}$$

Combining (8.30) with (8.8) the new weight constant is:

$$\varpi_1 = \frac{\hbar_1}{2\sigma_1} = \left(\frac{3}{5}\bigg/\frac{4}{5}\right)^{\frac{1}{2}} \frac{\hbar}{2\sigma} = \left(\frac{3}{4}\right)^{\frac{1}{2}} \varpi . \tag{8.31}$$

In the case of the special planoid the standards for lengths and masses (energies) are no longer their usual values since they depend directly on $\sigma$ and $\varpi$ respectively. Equations (8.30) and (8.31) provide a direct transformation for each. The length transformation can be eliminated (changed to a 1:1 ratio) if (8.30) is applied to $R_1$ assuming the same number of particles is contained in the uranoid and planoid.

    The special planoid is used in the analysis of the exclusion principle, Dirac's negative energy states, and the derivation of the proper mass from exclusion theory, among other things. Eddington's primary argument in favour of the special planoid is that it introduces the factor $\tfrac{4}{3}$ rather than the $\tfrac{5}{3}$ in the standard uranoid. The usefulness of this is evident in the selection of top particles that give the same change of density as the stabilisation of scale that then transforms a $V_4$ particle into a $V_3$ particle. This, he claims, results from transforming from a flat planoid to a curved uranoid with a stabilisation of scale (thus reducing the multiplicity from 4 to 3). Unfortunately, at this point, it feels a bit like Eddington is simply playing with numbers, adjusting formulae in order to produce the desired result. His physical justification breaks down to pure mathematical manipulation. His motivation likely lies in his desire to maintain a multiplicity of 4 that would keep the scale free from stabilisation since his version of exclusion considers only the three spatial momenta as classifying state characteristics (thus a multiplicity of 3) and his solution for the mass in extracule interchange produces an $m_3$. The special planoid allows him to accept these results while at the same time

keeping his scale unfixed. It is a process all too familiar with Eddington – results arrived at from two different vantage points are tantalizingly close and thus a fudge factor (or, in many cases, an entire fudge theory) is introduced to connect the two. He clearly does this numerous times to convert between 136 and 137 since the former is a natural number that arises in many dimensional considerations while 137 is the approximation of the fine-structure constant. I would conclude from this that Eddington did not believe in coincidences nor did he believe in artificially difficult processes (although he seems strongly in favour of artificially simple ones).

**Some Remaining Energy Issues**

The electron-proton and extracule-intracule two-particle problems have both been analyzed so far. In both situations all the interactions considered (which in Eddington's case is only gravity and electromagnetism) were attractive. But how could the developed version of this theory be applied to particles of *like* charge? If the system consists solely of two like charges, they become the source of an extended electric field that presumably induces opposite charges somewhere else in the environment. In terms of generally neutral systems of the type thusfar considered, they technically form an incomplete half of a four-particle system. For proton-electron systems the Coulomb of interchange energy has been calculated but it would be difficult to calculate this for a proton-proton system (or electron-electron system – for simplicity we will simply consider a proton-proton system here). One can deduce the result, however, from the proton-electron results.

A charged particle is said to have no Coulomb energy in a neutral environment since Coulomb energy is really interchange energy and there would be nothing for the particle to interchange with. The mutual energy is then purely mechanical by definition. Obviously equal distributions of positively and negatively charged particles produce neutral matter so for all values of $r$ the proton-proton electrical energy should be equal and opposite to the proton-electron energy. This is easily verified using the well-known classical equation for Coulomb (electrical) potential energy. Since the proton-electron energy is proportional to $-e^2/r$ the proton-proton system will be proportional to $e^2/r$.

When placed in a neutral environment a charge of $-2e$ is induced in the environment. If the system is treated as isolated a term must be introduced into the Hamiltonian in order to represent the 'ignored' induced charges and Eddington equates this term with the Debye-Hückel energy that frequently appears in astrophysical (plasma), high-energy (Yukawa potential), and solid-state (Thomas-Fermi potential) situations (see footnote Liboff 1998, p. 804). In treating an object-system in the standard uranoid it is simply the non-Coulombian energy.

As I have shown Eddington made the assumption that charge was a frame-dependant quantity and so the classical concept of charge is referred to as relativity charge. These charges induce opposite charges in the environment. However, quantum charges, which are not considered frame-dependent quantities, are simply superimposed on a rigid environment. This is the analysis used in the present formulation.

Consider, then, a single proton as an object-system that is superimposed on an undisturbed uranoid containing $\frac{1}{2}N$ protons and $\frac{1}{2}N$ electrons. If $V$ is the volume of the uranoid and each particle has a probability $dV/V$ of being in some volume element $dV$ the mutual energy of the proton and the uranoid is:

$$\frac{1}{V} \int E dV = -\Omega \tag{8.32}$$

and:

$$\frac{1}{V} \int E dV = \Omega + B/V \tag{8.33}$$

for a proton-proton where $B$ is a constant. The total mutual energy for the proton object-particle and the uranoid is then $\frac{1}{2} N B/V = -\Omega = -\frac{3}{2}\left(\frac{4}{3}\right)^{\frac{1}{2}} e^2/R_0$ where the last part of the equality results from the consideration of the interchange of extracules and the 4/3 conversion factor for moving from a planoid to normal measure. Since the volume is $V = \frac{4}{3}\pi R_0^2$ the constant $B$ is:

$$B = -\left(\tfrac{4}{3}\right)^{\frac{1}{2}} 4\pi e^2 \, R_0^2/N = -\left(\tfrac{4}{3}\right)^{\frac{1}{2}} 16\pi e^2 \sigma^2 . \tag{8.34}$$

Equation (8.33) can be rewriting in terms of a Dirac delta function multiplying $B$.

Consider then that this proton exists in a large assemblage of indistinguishable protons and one of these other protons is referenced as the origin (and remember that this is Eddington's definition of indistinguishability). Due to their indistinguishable nature there is a probability that any of them could be that origin. Various coordinate transformations along with (8.34) provide the equation for the energy distribution in a uranoid if the origin is randomly placed in one of the indistinguishable protons:

$$(B\delta(r') - \Omega')\frac{dV}{V}$$

where $dV/V$ is the probability distribution of the carrier, $B\delta(r') - \Omega'$ is the amount of energy carried, $B = -2\Omega V/N$, and $\Omega' = (1 - 2/N)\Omega$. The carriers are quantum protons meaning that they are superimposed on the neutral background environment and have a uniform distribution. Ultimately, then, the energy of a quantum proton in the field of another proton that is taken as an indistinguishable origin in an assemblage of protons in a uranoid contains the energy $B\delta(r')$ as well as the normal Coulomb energy. One final coordinate transformation is required to restore the original frame and, given the singular point $r' = 0$ that corresponds to a Gaussian probability distribution in a relative coordinate frame we have:

$$\delta(r') = (4\pi\sigma^2)^{-\frac{3}{2}} e^{-r^2/4\sigma^2}. \tag{8.35}$$

Combining (8.35) and (8.34) the final result for the non-Coulomb portion of the energy is:

$$B\delta(r') = -\left(\frac{16}{3\pi}\right)^{\frac{1}{2}} \frac{e^2}{\sigma} e^{-r^2/4\sigma^2}. \tag{8.36}$$

Equation (8.33) can be rewritten as simply:

$$E = e^2/r + B\delta(r') = e^2/r - \left(\frac{16}{3\pi}\right)^{\frac{1}{2}} \frac{e^2}{\sigma} e^{-r^2/4\sigma^2}. \tag{8.37}$$

Integrating (8.37) with respect to $r$ yields the modified form of Coulomb's law that I argued was necessary from a statistical argument for the identification of exclusion with Coulomb repulsion.

The explanation for (8.37) is that when a system of two like charges is treated as a superposition on an undisturbed environment the energy has to be adjusted to compensate for not including the induced charges. So the additional term takes the place of the induced charges to some degree. The Dirac delta function appears to account for the fact that this term is zero when there is no separation. This is because the two protons are then physically touching and can now be considered to be a single +2*e* charge that then induces a –2*e* charge in the usual way as in a proton-electron system.

The result for (8.36) is modified in some situations by adding a multiplicity factor. Basically the question becomes, where in this process are quantum protons substituted for classical protons? This point is where the non-Coulomb energy term is added. Eddington reasons that this point is the point at which the mass of the hydrocule is $m_0$ meaning that the adjustment can be made directly in the special planoid where this is the mass of top particles. Comparison particles in this situation come in two types: those that include the extra energy term and those that don't. Remember that this additional energy term simply allows the normal four-particle system of two proton-electron pairs to be simplified to a two-particle system of a single proton-proton pair. The adjustment of the mass energy is determined by manipulating (6.28) something that will be studied further in the next chapter. Ultimately, however, Eddington finds the standard deviation from the Rydberg constant and combines this to find the multiplicity for the non-Coulomb energy term, $k = 1.9208 \times 10^{-13} cm$. This is on the order of the range nuclear forces which is precisely what should be found by any good quantum field theory (see the end of chapter five).

We are now ready to apply Eddington's theory to several other situations where we will find many curious numerical results that led to its labelling as 'cosmonumerology.' In reality, no mystical notions were ever put forward in the theory. It simply was of a form that derived numerous physical constants. Eddington's philosophical reasoning for this is discussed in chapter three but it still leaves as unanswered the question of whether or not the theory was subconsciously (not likely consciously) designed to be this way.

# IX

## *Numerical Considerations and Applications*

The striking numerical coincidences (or planned results, depending on interpretation) in *Fundamental Theory* begin early with Eddington's calculation of the range of nuclear forces and the recession of the galaxies. The results continue to be peppered throughout the work as the theory becomes more developed and opens new avenues for consideration. Some results are definitely striking and some definitely appear artificial. Others clearly are pure coincidences.

**Nuclear Forces and Galactic Recession**

The earliest numerical results produced in *Fundamental Theory* appear very early in his consideration of the range of nuclear forces and the recession of the galaxies. I showed in chapter five how, simply through an application of probabilistic methods, Eddington arrived at the following equations in terms of $R_0$ and $N$:

$$R_0/N = Gm_h/\pi c^2 = 3.95 \times 10^{-53} cm \quad \text{and} \quad R_0/\sqrt{N} = k = 1.921 \times 10^{-13} cm. \quad (9.1)$$

The latter, as I've shown, is in clear agreement with values calculated by Yukawa and others. This is potentially a striking result as it links the range of nuclear forces with cosmological parameters. Eddington's arrival at it arises simply through the basic foundations of the theory: the uncertainty in the physical origin, fluctuations in the scale, and the application to spherical space. In addition, as a bedrock of his thinking, Eddington adopts a truly relativistic viewpoint where coordinates and other characteristics are quantities that are solely relative to other physical quantities and not to random, observer-based quantities. Essentially he establishes early on the concept of comparison particles.

The equations in (9.1) provide a system for uniquely determining $N$ and $R_0$ and thus, $R_0$ can be applied to the speed of recession of the galaxies (as derived by George Lemaître (1894 – 1966)), $V_0 = c/R_0\sqrt{3}$. The result of applying (9.1) then gives $V_0 =$ 572.4 $km/s/MPc$. At the time Eddington was reasonably close to the experimentally determined value of $V_0 = 560$ $km/s/MPc$ given by Hubble and Humason (Eddington 1946). This parameter, now known as the dimensionless Hubble parameter, is actually a function of the Hubble constant, $H_0$:

$$h = H_0/100 \, km/s \cdot MPc. \qquad (9.2)$$

This is, of course, not to be confused with Planck's constant. The notation used here is standard in current texts. Its value now ranges from 50 to 100 with a best estimate being around 72 (see any modern cosmology text such as Peacock 1999). The first part of (9.1) is also (5.18). It appears from this that Eddington's results would have to depend on the values for the parameters in (5.18) and since $G$, $m_h$, and $c$ are not appreciably different than they were in the 1940s, the new value for $N$ would need to produce the proper value for $R_0$. Logically (9.1) would have to be a pair of internally consistent simultaneous equations – i.e. changing $N$ would *have* to change $R_0$ since if their definitions are identical in both parts of (9.1) simple mathematics requires the relationship to hold. Just for the sake of argument let's verify this with current values.

Current values for $N$ are most frequently given as being on the order of $10^{80} - 10^{85}$, though there are estimates that range as high as $10^{120} - 10^{130}$. Using these values with the second part of (9.1), values for $V_0$ would range from a high of 282 to a low on the order of $10^{-18}$, which certainly encompasses the currently accepted values for $h$. Using the 'best estimate' value of 72 for $h$ (or $V_0$ in Eddington's notation) and knowing that the range of nuclear forces, $k$, has not changed, we find $N$ to be on the order of $10^{81}$, which is well within the more conservatively accepted range of $10^{80} - 10^{85}$. Using these values for $R_0$ and $N$ in (5.18) produces a value of $4.863 \times 10^{-54}$ $cm$ – still tantalizingly close to the known value of $3.95 \times 10^{-53}$ $cm$. So, even when adjusted to include currently accepted values of $h$ the relationship still holds – (9.1) appears to be more than a simple coincidence. The radius of the universe and the number of protons and electrons (and neutrons) in the universe appear to combine to give a relation between macroscopic cosmological quantities and microscopic nuclear quantities. On the other hand, since the universe is expanding and thus $R_0$ is changing, this implies $N$ is continually growing which implies continuous matter creation, a hallmark of steady state theories. However, he solves the matter creation problem simply by changing the datum thus shuffling the Dirac sea a bit as I've described – the conservation laws still hold.

**The Proton-Electron Mass Ratio**

One of the most studied results in *Fundamental Theory* was his derivation of the mass ratio of the proton to the electron. This particular discussion arises out of the discussion of external and internal particles. Working from equation (7.2) the total particle energy for the atom is $E = m_0 + p^2/2\mu$ where $\mu = m_0/k = m_0/136$. In the external/internal treatment the rest energy of the internal particle is *not* $\mu$ (the mass of the internal particle) but, rather, zero which is a highly useful result. The energy, $E$, then can be decomposed into its constituent pieces labelled:

$$E_e = m_0, \quad E_i = p^2/2\mu \tag{9.3}$$

where the subscript $e$ is for the external particle and the subscript $i$ is for the internal particle. Rather than a single energy tensor now applying to the atom as a whole, the external and internal particles now have their own energy tensors that reduce their multiplicities from 136 to 10 (the standard carrier $V_{136}$ now is broken into two $V_{10}$'s which are now carriers of vectors rather than tensors – the vector quantity here is the momentum). The change in multiplicity does not affect the transition energy thus leaving $E_i$ unchanged. The energy of the external particle is now $E_e = M_1$ where $M_1 = \frac{136}{10} m_0$. For the internal particle $\mu = \frac{1}{136} m_0$. If this given definition is accepted as a potential solution for the hydrogen atom the ratio of the mass of the external particle to that of the internal particle is:

$$\eta_1 = \frac{M_1}{\mu} = \frac{136^2}{10} = 1849.6. \tag{9.4}$$

If the masses of the proton and electron are, respectively, $m_p$ and $m_e$, (7.2) gives $m_p + m_e = M_1$ and $m_p m_e = M_1 \mu$. The proton and electron masses then become the roots of the quadratic equation $m^2 - mM_1 + M_1\mu = 0$ where $m$ is the generic variable that gives both $m_p$ and $m_e$. However, given $M_1 = \frac{136}{10} m_0$ and $\mu = \frac{1}{136} m_0$, this becomes:

$$m^2 - m\left(\frac{136}{10}m_0\right) + \left(\frac{136}{10}m_0\right)\left(\frac{1}{136}m_0\right) = 10m^2 - 136mm_0 + m_0^2 = 0. \qquad (9.5)$$

Solving this equation gives the following slightly different result:

$$\eta_2 = \frac{m_p}{m_e} = 1847.6. \qquad (9.6)$$

All Eddington is really doing here is showing that the ratio provided by the external/internal particle method (9.4) is roughly the same as that provided by taking the proton and electron (9.6), which simply means the external/internal particle representation looks physically about the same. There is nothing terribly unusual about anything Eddington has done here. Similar results can be derived from the treatments given in any standard physics text. Part of the reason for this is that the "reduced mass $\mu$ differs from the electron mass by very little in the hydrogen atom" (Gasiorowicz 1996, p. 280). Since the mass of the hydrogen atom as a whole differs very little from the mass of a proton (since the electron is so small in comparison) it stands to reason the ratio of the mass of the atom (external particle) to the reduced mass (internal particle) should be roughly the same as the mass ratio of the proton and the electron. It is curious, then, that historians and physicists have jumped on this particular result as being cosmonumerological. The reasoning he employed leading to (9.5), particularly in regard to the various particle multiplicities, is unusual, but since it is purely mathematical it is not physically disprovable, per se. The result is unremarkable. As such all Eddington has done is devise a new way of *mathematically* describing the hydrogen atom that experimentally yields the correct results. Certainly by Ockham's Razor one could eliminate Eddington's theory when compared to others simply by its complicated nature. But there have been other theories far more complicated, bold, and completely incorrect, that did not receive as much bad press (though this is a purely anecdotal observation).

However, Eddington does take this a step further, as is his trademark. He interprets (9.4) as being the more experimentally direct value where (9.6) is the listed value. Masses given by (9.5) are then referred to as the 'standard masses' of the proton and the electron. When splitting his theory into electrical and mechanical components via the transition from $V_{136}$ to $V_{137}$ particles he finds a different ratio. The first strictly Eddingtonian step taken here is in his definition of the Rydberg constant:

$$\mathbb{R} = \frac{1}{2}\left(\frac{1}{137}\right)^2 \frac{\mu c}{2\pi\hbar}. \tag{9.7}$$

A 1935 edition of Pauling and Wilson's *Introduction to Quantum Mechanics* contains the following expression:

$$R = \frac{2\pi^2 \mu e^4}{h^3 c} \tag{9.8}$$

(Pauling and Wilson 1935, p. 41). When comparing this to (9.7) there are two items to notice here. First, he substitutes the exact value of 1/137 in for the fine-structure constant though it is known from experiment to be closer to 1/137.037. Second, in performing the algebra to transform (9.8) to (9.7) (maintaining, for the sake of argument, Eddington's value for the fine-structure constant) one ends up with an extra $e^2$ in the numerator:

$$R = \frac{1}{2}\left(\frac{1}{137}\right)^2 \frac{\mu c e^2}{2\pi\hbar}. \tag{9.9}$$

Eddington specifically excluded this term, saying that the derivation in *Fundamental Theory* leads directly to (9.7) "there being no occasion to introduce $e^2$ except as an abbreviation for $\hbar c/137$" (Eddington 1946, p. 58). With that single sentence, appearing in parentheses in the text, he manages to gloss over a not-so-trivial point.

In an early draft of *Fundamental Theory* Eddington does little to clear up this vague statement, but does describe why it is necessary:

> By accepting [9.7] as the formula for the experimentally determined Rydberg constant, we tie the resulting value of $m_p/m_e$ to a system that is certainly not B [see discussion of systems A and B in chapter seven, beginning p. 160]. The transformation from B to the observational system corresponds to the recognition in classical theory of an e.m. aether as part of the environment of every object-system (Eddington as quoted in Slater 1957, p. 139).

Focusing first on the reasoning behind his definition of the Rydberg constant, we find that he seeks to tie the ratio $m_p/m_e$ to an experimentally determinable quantity. I am about to demonstrate that *this* version of that ratio distinguishes the standard masses as given in (9.6) from what are known as the 'current masses.' The intent is admirable but given that he begins the derivation with (9.7) he has left a gaping hole in the proof – how exactly can $e^2$ be eliminated. There is nothing else in any early draft that further elucidates this move.

Continuing, then, under the assumption that (9.7) is correct, the empirical Rydberg constant, $\mathfrak{R}$, is measured according to his assumptions in system $A$ which is the standard particle uranoid with bound intracules. The formula, however, is obtained from the theory of free intracules that corresponds to system $B$, a hydrocule uranoid with free intracules. How does Eddington, then, define an observational system exactly?

> We can now give a formal definition of the 'observational system' … masses of neutral particles are molarly [macroscopically or relativistically] controlled, and quantum masses or energies optically controlled. *We therefore define the observational system to be that in which these two conditions are satisfied simultaneously* (Eddington as quoted in Slater 1957, p. 139).

The empirical Rydberg constant, then, must account for the conversion factors used in transforming between systems $A$ and $B$. It is given as:

$$\mathfrak{R} = \frac{1}{2}\left(\frac{1}{137}\right)^2 \frac{\mu_A c}{2\pi\hbar} \qquad (9.10)$$

where $\mu = \mu_B = \beta^{\frac{1}{6}} \mu_A$. I have already given several instances where $\beta$ is used in converting between systems $A$ and $B$. Recall that its value is 137/136 and it represents the addition of an extra degree of freedom in the intracule to account for the Coulomb energy contained within it. It is not purely an invention of Eddington's (c.f. Bond 1934) but is employed far more liberally by him than anyone else. In (9.10) $\mu_A$ is the 'current mass' of the intracule (Eddington 1946). He gives this as the *observed* reduced mass value. Unlike intracules, extracules transform their masses as measured densities: $M_{1B} = \beta M_{1A}$. Equation (9.1) then becomes:

$$\eta_1' = \left(\frac{M_1}{\mu}\right)_A = \beta^{-\frac{5}{6}}\left(\frac{M_1}{\mu}\right)_B = \frac{136^2}{10}\beta^{-\frac{5}{6}} = 1838.34 . \tag{9.11}$$

The quadratic (9.5) becomes:

$$10m^2 - 136mm_0 + \beta^{\frac{5}{6}}m_0^2 = 0 \tag{9.12}$$

Solving this yields:

$$\eta_2' = \frac{m_p}{m_e} = 1836.34 . \tag{9.13}$$

Masses determined in this way are referred to as 'current masses' for the proton and electron and are distinguished from the standard masses derived above. Again, the difference between the values is a result of which system one is investigating – the observational system or the nearly objective theoretical system. Once again he is taking into account the problem of interference from the observer and the difference here is that (9.11) and (9.13) account for the transition between systems $A$ and $B$ (which amounts to considering the observational system to be subjective and the theoretical system nearly objective) while (9.4) and (9.6) do not.

Eddington actually proceeds to correct (9.11) which is the 'corrected' observed value. Following on the preceding quote from Slater he determines that the masses of extracules are molarly controlled while the masses of intracules are spectroscopically (optically) controlled. The Faraday constant for hydrogen $\mathfrak{J} = e/m_h c$ can be found by measuring the charge that results from the electrolysis of a known mass of water. Given that $\hbar = 137e^2/c = 137\mathfrak{J}^2 m_h^2 c$. Substituting this into (9.10) gives:

$$\mu_A = 4\pi \cdot 137^3 \mathfrak{R}\mathfrak{J}^2 m_h^2 . \tag{9.14}$$

A correction must be made to this since the definition of the Faraday constant assumes that $e$ is molarly defined. Eddington finds that the value in quantum *theory* is slightly different and, as such, when converting from a quantum system $B$ to a molar system $B'$ lengths and times are multiplied by $\beta^{-\frac{1}{12}}$. Or one could instead convert from a quantum system $A$ to a molar system $B'$ using $\beta^{-\frac{1}{4}}$. Note that the origin of all the exponents for $\beta$ is simple dimensional analysis and the corrections it provides are very small. Hence even without this correction factor Eddington's theory would still be surprisingly close to observed values.

Similar conversion factors are found for $e$, $\hbar$, and $\Im$. Using accepted observational data from 1942 for these constants, Eddington makes the proper corrections to arrive at an observed value of $1838.56 \pm 0.51$ for the extracule/intracule mass ratio. Converting this to a proton/electron mass ratio gives $1836.56 \pm 0.51$ which is considerably closer to the value given in (9.13) further supporting Eddington's use of the extracule/intracule model in place of the proton/electron model. Using the ratios of the Rydberg constants for deuterium and hydrogen and the ratio of the atomic weights, both of which can be determined experimentally from mass-spectroscopic methods, a ratio of $1836.14 \pm 0.22$ is obtained. Numerous other alternative methods for determining the various ratios of these physical constants are explored by Eddington, all with the same theme: begin with a set of measurements and move through corrections to see how close the data comes to (9.13).

**A Brief Note on the Aether**

The quote on page 197 justifying the Rydberg constant opens yet another Pandora's Box, however: he clearly advocates for an aether. He actually prefaces the above quote with the following:

> Although separation of mechanical and electrical energies is important … for analytical treatment, they are not separated in anything that the observer handles – or supposes he handles. The elementary particles, protons and electrons, carry both electrical and mechanical characteristics; the aether is a carrier of electrical and gravitational waves (Eddington as quoted in Slater 1957, p. 139).

It is almost unfathomable that so strong a supporter of relativity would also support an idea whose debunking actually *led* to the development of relativity. This quote is not an anomaly, either. In a 1932 letter to Larmor (see chapter six, p. 112) he introduces an 'aether displacement' to represent a link between an electron and its environment. I have previously argued that the uranoid was his manifestation of the aether (Durham 2003b). However, further investigations have led me to change my view on this. The uranoid is a standardized environment and can be made up of not just space-time itself but also other particles in a large assemblage, as I've already shown. Eddington's use of the aether most likely stems from the fact that he recognized space-time as quantisable (essentially foreshadowing the graviton). Since light travels on geodesics in space-time and gravitational waves are propagated through the 'fabric' of space-time, it is well-known that space-time has some sort of structure to it. Eddington recognized this structural aspect and applied the term 'aether' to it. His definition of the aether, then, was not necessarily the same as the classical definition. However, as I have previously stated (see p. 112), he did not feel this aether was necessarily a field. He obviously struggled with exactly *what* it was, physically, though it can be easily described as the 'fabric' of space-time. But just what *is* the fabric of space-time? I will leave that for quantum gravity to answer.[40] Suffice it to say, Eddington's use of the term 'aether' was not in keeping with its classical use, but, nonetheless, demonstrated a belief in a structural vacuum.[41]

**The Fine-Structure Constant**

Unlike most other physical constants given in *Fundamental Theory* the fine-structure constant is not derived from any first principles, rather, it is assumed to *be* a first principle which seems to violate, again, the deductive, non-arbitrary spirit of the theory. It could even be said to be a stabilised quantity given as 1/137. In fact, Eddington would say it is fixed by the theory – its value is not derivable, per se, but the given value is the only one that works within the theory. This seems to be a rather odd conclusion and one cannot help but speculate that he was driven to force this to be true by the oddly coincidental closeness of 136 and 137. In deference to Eddington, I have already mentioned that the ratio 137/136 was not his own discovery (see Bond 1934 – $\beta$ is called the Bond factor). Nonetheless, as I will show, finding a need for a multiplicity of 137 in many situations results in his definition of the fine-structure constant as 1/137.

---

[40] Steven French has pointed me to a PhD thesis currently being written by Dean Rickles at the University of Leeds as being an excellent source for a discussion of this topic.

[41] The aether concept is still debated and it has been clear for several decades that (from quantum field theory) the vacuum has an energy of its own and particles can spontaneously pop into existence (and subsequently annihilate one another) in vacuum. The aether is, in fact, still discussed in the literature (see Wilczek 1999).

Eddington's own explanation for the value of the fine-structure constant derives from the addition of the interchange energy that I outlined in chapter seven. The internal wavefunctions of the hydrogen atom imply half-integer spin in any plane of rotation. But viewing this relativistically in the sense that there are no preferred frames of reference, the implied (or rather *applied*) frame of reference introduces chirality into the problem – either particle could have right-handed or left-handed spin. The atom does not naturally possess any such preference since a relativistic transformation can easily flip right-to-left and vice versa. The only requirement is that the two particles remain in the same orientation relative to each other. This is just another way to explain Eddington's view that spin is a frame-dependant quantity.

Spin, then, is formally added in order to remain consistent within our own analytical system. In essence we're constrained by our own subjective perspective that forces chirality into our observations. Since in the hydrogen atom, regardless of how it is viewed (extracule/intracule or proton/electron) the particles each contribute half-integer spin. If these spins are in the same direction the entire atom contains integer spin. If they are in the opposite direction the entire atom contains zero spin, though Eddington does not mention the zero spin situation. Since there is an interchange circulation – a circulation due to the constant interchanging of indistinguishable particles in the planoid – the interchange energy can be interpreted as being provided by the spin. Interchange is really a quantum number, then, and since Eddington equates it with the Coulomb energy, then the Coulomb energy must also be the same quantum number. The interchange circulation arises for the same reason – our limited viewpoint – and that is the real reason it can be related to spin. So interchange angular momentum is a full quantum, $\hbar$ (or zero) rather than a half-integer quantity. If this system is analyzed in rigid coordinates, the rigid coordinate time is $k^{-1}$ times the Galilean time. In working with interchange energy, which is just Coulomb energy, the multiplicity is taken to be 137. The interchange angular momentum provides a linear momentum to the intracule of $\hbar/r$ that appears in an extra-spatial dimension that is normal to $r$. If this extra-spatial dimension is actually the time dimension in normal (3 + 1) space-time then the linear momentum reduces to the Galilean coordinate representation $\hbar/kr = \hbar/137r$. Given that the classical electron radius is given by $r_0 = e^2/mc^2$, we can write $E = mc^2 = e^2/r_0$. Dividing through by $c$ gives $mc = e^2/cr_0$. Using Eddington's system of natural units this has the same units as $\hbar/kr = \hbar/137r$. Setting them equal to each other and rearranging with the knowledge that $e^2/\hbar c = \alpha$ is the fine-structure constant, we find:

$$\frac{1}{137} = \frac{e^2}{\hbar c} = \alpha. \tag{9.15}$$

Eddington's theory thus requires the fine-structure constant be the inverse of 137 due to the correction factor introduced in moving from rigid to Galilean coordinates in an intracule. The inverse of the fine-structure constant then appears as the multiplicity of an intracule ($V_{137}$) rather than a coincidental $k + 1$ relation to a standard carrier ($V_{136}$). In this way he also directly relates it to electrical situations since it represents the mechanical degrees of freedom plus an extra degree of freedom from the Coulomb (interchange) energy contained within the intracule representation. In modern interpretations, the fine-structure constant is the coupling constant for electromagnetism meaning it measures the strength of the electromagnetic interaction. So Eddington is at least consistent in that he essentially finds the same thing – it is produced because the energy of the bound electron and proton in the intracule that produces an extra degree of freedom leading directly to the fine structure constant happens to be the Coulomb energy.

**Non-Coulombian Energy**

I ended the previous chapter with a discussion of the non-Coulomb energy term that appears when like charges interact. A deeper analysis leads to some additional numerical results. One method that I introduced for adding protons and electrons to the universe involved taking the hydrocule mass as a comparison particle. As such protons and electrons are then individually added to the environment together with their comparison holes. Comparison particles will then consist of an equal number of two different types: those that have the additional non-Coulomb energy and those that don't. Eddington associates those with the additional energy with protons and those without it with electrons. When a proton is introduced its comparison hole eliminates a comparison particle that has the additional energy term whereas the electron's comparison hole eliminates a comparison particle *without* the extra energy term. The comparison hole is simply the regular hole created by exciting the proton to an energy level above the datum. The additional energy term actually adds a corrective element to the energy of a proton, $m_p$ assuming its rest energy is $m_0$: $(m_p/m_0)B\delta(r')$. Again the non-Coulombian energy is given as an adjustment to the initial energy that simplifies a four-particle system to a two-particle system. It appears on the same footing as a particle's rest mass, which is an adjustment that simplifies a $\frac{1}{2}N$-particle system to a one particle system. Basically it accounts for the larger mass of the proton (or the difference between the masses of the proton and the electron). Another way of looking at this is to say that since there is technically only one mathematical form for comparison particles regardless of whether or not the original particle is an electron or a proton. Since these comparison particles, if they are internal, must contain only a single variate (the scale uncertainty) then the difference in the mass of the proton has to be accounted for somewhere. As such the excess mass supplied by the proton is converted to this non-Coulomb energy term in the comparison particle. But then why does this non-Coulomb energy have anything to do with charge at all? There is no clear explanation for this in the texts, in any of the early drafts, or in *Relativity Theory of Protons and Electrons* (1936).

Nonetheless (8.36) can have the factor $m_p/m_0$ added to it to give:

$$\frac{m_p}{m_0}B\delta(r') = -\left(\frac{16}{3\pi}\right)^{\frac{1}{2}}\frac{m_p}{m_0}\frac{e^2}{\sigma}e^{-r^2/4\sigma^2}. \tag{9.16}$$

This is then the corrected non-Coulomb energy of two protons. Since we know that $k = 2\sigma = R_0/\sqrt{N}$, if we define:

$$A = \left(\frac{16}{3\pi}\right)^{\frac{1}{2}} \frac{m_p}{m_0} \frac{e^2}{\sigma}$$

(9.16) can be rewritten as:

$$E_{nc} = -Ae^{-r^2/k^2} \tag{9.17}$$

where the subscript $nc$ indicates that this is the non-Coulomb energy term. Using the Rydberg constant and previously described methods to determine the values for $k$ and $\sigma$ the value for $A$ is found to be $52.01 m_e c^2$ and $A$ is called the nuclear-energy constant. Both $k$, interpreted here as the range of nuclear forces, and $A$, which is understood to be an optically (spectroscopically) controlled energy, are found experimentally in proton-proton scattering experiments. The observed value given by Eddington is $52.26 m_e c^2$. Notice that this term is negative indicating that it is an attractive energy. When combined with the normal Coulomb repulsion term one finds regions where the total energy and resultant force are attractive and repulsive. Naturally one could determine a ratio of the values of the Coulomb to non-Coulomb energy and Eddington gives its value as 15.20.

    Despite his claim that the extra energy term is only associated with proton comparison particles due to their larger mass, he applies the preceding argument to the scattering of electrons since this extra energy term also arises when there are like charges. Substituting the mass of the electron into (9.16) in place of the mass of the proton gives the ratio of Coulomb to non-Coulomb energy as very nearly unity with the same said for the related forces. As such the non-Coulomb energy is negligible for the electron. It is perhaps here that he rationalizes using the non-Coulomb energy term in both a mass situation and a charge situation. It actually *does* arise in any situation involving like charges, but because it can be equated with a rest mass it is negligible in the case of the comparatively tiny electron.

    In a nutshell, Eddington is attempting to adjust the standard Coulomb interaction to account for anomalies experienced in scattering experiments. Of course, what he is presaging here without knowing it is the strong interaction. Why strong and not weak? He assumed that the standard Coulomb repulsion might possibly be overcome at high enough energies and some other interaction or energy would have to take over and produce the observed scattering. Electron-electron scattering was more predictable since

electrons are fundamental particles, but because protons, unbeknownst to anyone in the early 40s, are made up of quarks and gluons, proton-proton scattering experiments contain anomalies not found in electron-electron scattering. So, in essence, he was trying to explain away effects we now usually attribute to the strong interaction (though the weak interaction also plays a role).

As I will discuss below, particles can be created and annihilated in the vacuum in Eddington's theory just as in quantum field theory. This is as to be expected since Dirac's equation essentially predicted antimatter as early as 1928, though the realization that the antielectron (positron) was the correct solution to these problems did not come until 1931 when Dirac formally introduced it as a new kind of particle (Kragh 1999). Eddington simply reinterprets Dirac's negative energy sea. The direct link between this and an increasing $N$ is not explicitly discussed, however.

**Newton's Gravitational Constant**

The preceding discussion of proton-proton scattering provides a way to determine $R_0/\sqrt{N}$. By measuring $k$ one obtains $R_0/\sqrt{N}$ directly. Measuring $A$ then provides $\sigma$ which gives $R_0/2\sqrt{N}$ directly. It also possible to determine $R_0/N$ and $R_0$ from observation. Later in *Fundamental Theory* Eddington gives a derivation of the value for $N$ that gives $N = \frac{3}{2} \cdot 136 \cdot 2^{256}$. The principle formulae for these various quantities are then:

$$\frac{R_0}{N} = \frac{Gm_h}{\pi c^2} \tag{9.18}$$

$$\frac{R_0}{\sqrt{N}} = \frac{136}{10}\left(\frac{9}{20}\right)^{\frac{1}{2}} \frac{\beta^{\frac{1}{6}} h}{2\pi c m_h} \tag{9.19}$$

$$N = \tfrac{3}{2} \cdot 136 \cdot 2^{256}. \tag{9.20}$$

Combining (9.18) and (9.19) with the fine-structure constant and Faraday's constant gives:

$$\frac{G}{\mathfrak{I}'^2 c^2} = \frac{136 \cdot 137}{10}\left(\frac{9}{20}\right)^{\frac{1}{2}} \frac{\pi \beta^{\frac{1}{4}}}{\sqrt{N}}. \tag{9.21}$$

This then can be solved for G, which gives a value of $6.6665 \times 10^{-8}$ which he claims is accurate to one part in five-thousand. Conversely, using an experimentally determined value for G will provide an alternative derivation for N that does not require the far more complex derivation that is a result of Eddington's E-number theory that is not discussed here. The physical interpretation of this is that the number of particles in the universe is determined by a ratio of locally measurable constants. Since he assumes that his later value of N given by (9.20) is correct he then assumes that his value of G must be the correct one.

The force constant, then, which is the ratio of the electrical to gravitational forces (interactions) between a proton and an electron is given by:

$$F = \frac{e^2}{Gm_p m_e} = \frac{2}{3\pi\beta^2}\sqrt{5N} \qquad (9.22)$$

where $m_p m_e = \beta^{\frac{5}{6}} m_0^2/10$ and was substituted into (9.21). This, of course, is known to be an enormous number and is often calculated in introductory physics courses as an example of just how weak gravity is.

One further result Eddington immediately jumps to that could have been derived much earlier (see chapter five) is the relationship between limiting speed of recession of the galaxies and the range of nuclear forces. The recession velocity, which was introduced before, is $V_0 = c/R_0\sqrt{3}$. Since $R_0 = k\sqrt{N}$ we can write $kV_0 = c/\sqrt{3N}$ "so that the recession-constant can be derived from the range-constant of nuclear forces, or vice-versa, with no other observational data except the velocity of light" (Eddington 1946, p. 105). One more time the microscopic and macroscopic are linked but the point can now be better emphasized that macroscopic (molar) interactions are governed by gravity while microscopic (quantal) interactions are governed by electricity. In order to build the universe up from the very small to the very large there must be some transition region where the two interactions can be linked. This transition region is given by Eddington as being the scale-free theory that makes up the majority of the statistical portion of *Fundamental Theory*.

**Degenerate Matter**

The final application of Eddington's statistical theory that has yet to be discussed is the application of his work with exclusion to degenerate matter such as that contained within white-dwarf stars. The densities of such stars are tremendous. At such densities the electrons and protons (or nuclei) are in a constant state of collision. Separating carriers of electrical energy, $V_{137}$, from those of mechanical energy, $V_{136}$, is vastly different than anything as yet encountered, which has been mostly isolated two-particle systems separated into extracules and intracules.

      Both mechanical and electrical systems include waves as a fundamental physical representation. In mechanical systems, waves include such forms as sound waves, waves on a string, etc. In electrical systems, waves include changes of electric polarisation (with the resulting currents) and various magnetic effects. For the most part the electrical and mechanical aspects are independent. However, if the amplitudes of the various waves are not infinitesimal, cross-terms arise that create an energy equilibrium between the two forms. Consider, for instance, a system at non-zero temperature. Since the temperature is non-zero there must be a field of radiation that can be determined by Planck's law. This field induces electrical waves in the material. But, there is a slow transfer of energy from the field to the material that can induce mechanical waves in the system (by adding kinetic energy).

      Now comes the Eddingtonian part. If the system is dropped to zero-temperature the waves do not vanish since uncertainty predicts that there isn't necessarily perfect uniformity at zero-temperature, i.e. uncertainty really says that *exactly* zero-temperature uranoids (no motion) are impossible. As such there are residual fluctuations resulting purely from the uncertainty principle. This idea is not so far fetched, actually. As Zee tells us straight off, "In quantum mechanics the uncertainty principle tells us that the energy can fluctuate wildly over a small interval of time" (Zee 2003, p. 3) and since relativity tells us energy and mass can be converted into each other, the wildly fluctuating energy can turn into mass. These fluctuations exist in the vacuum and particles pop into and out of existence from the vacuum itself. Ignoring, for a moment, the cosmic microwave background radiation (which, at 2.9 K, is pretty close to zero anyway), the

vacuum is basically a zero-temperature system. As I've described before, Eddington explained the possibility of the production of particles (and even antiparticles) as excitations in the Dirac sea, which is simply a group of particles below a subjective datum. Where do these particles *receive* the energy necessary to excite them into existence? The answer is, quite simply, the random zero-temperature fluctuations produced by uncertainty. Eddington's study of degenerate matter is actually an attempt to determine the energy and pressure of the residual waves or fluctuations. This is the same reason given by quantum field theory in the union of special relativity and quantum mechanics.

Microscopically the polarisation provides a distribution function of electrical coordinates while macroscopically displacements produced as a result of sound waves provide a distribution function of mechanical coordinates. The electrical coordinates are provided by doublets and thus the state of the material is described by probability distributions of electrical coordinates of unidentified doublets in a large assemblage and the mechanical coordinates of an unidentified neutral particle. In order for this description to work the system must be in statistical equilibrium so that the fluctuations are of a completely random nature.

Now calling on the methods developed for transforming two-particle systems where a proton and electron, for instance, are replaced by an extracule and an intracule, we can assume that the material actually consists of a superposition of positively and negatively charged matter and thus, rather than consider the mechanical and electrical waves separately, we can bind them together and rather consider two different wave types: displacement of positively charged matter and displacement of negatively charged matter. Electrical waves can be described by these two wave types in opposite phase while mechanical wave can be described by these two wave types in the same phase. Microscopically the state of a material is described by a probability distribution of the displacement of an unidentified positively charged particle and a probability distribution of the displacement of a negatively charged particle. Now as the mechanical and electrical waves are largely independent, these two new wave types actually have intense interaction since they represent charged particles. In terms of the wave descriptions, whenever an electron and proton collide, energy is transferred from one to the other meaning energy can be transferred from one wave type to the other. The entire energy can be said to have passed from one wave type (system) to the other when each particle in one type has had at least one collision and thus the relaxation time is roughly the time between collisions. But in degenerate matter, the protons and electrons are in a constant state of collision and so neither wave type can be said to exist separately even for a brief moment, though at low (really mid) densities there comes a transition point where they can be treated separately using formulae similar to the two-particle transformation.

The electrical coordinates are given by $\xi_\alpha$ and the mechanical coordinates are given by $X_\alpha$ and, since we're using the *second* set of wave types (displacement of positive and negative charges) each particle has both electrical and mechanical coordinates. These coordinates have the conjugate momenta $\varpi_\alpha$ and $P_\alpha$. Wavefunctions (in the quantum sense) are then obtained in electrical and mechanical space and thus the exclusion principle is then applied: there are no more than two particles per cell $h^3$ of $\xi\varpi$-space and not more than two particles per cell $h^3$ of $XP$-space. The latter substitutes for gravitation and inertia and is concealed in the rest masses of the extracules. As such the former is the only form of exclusion that is necessarily dealt with. The doublets mentioned above turn out to be intracules and we can reanalyze the matter as being extracules and intracules – again a tie to the two-particle transformation. Also, the only matter considered here is hydrogen since a far more complex transformation is required for other elements since we're no longer working in the two-particle paradigm.

The average energy per intracule is $\overline{E} = \tfrac{3}{5} E$ where $E$ is given by (8.18). This introduces another definition of scale uncertainty as a particle density for the intracules. In astronomical situations the pressure is often also required but can be calculated independently of the energy. The average contribution of each intracule to the normalisation volume $V_0$ is $\overline{\Delta T_{11}} = \overline{p_1^2}/V_0\mu$. The total number of particles in the volume is then $\sigma V_0$ and thus the total pressure is $P = \sigma \overline{p_1^2}/\mu$. The spherical form of the momentum distribution is given as $\overline{p_1^2} = \tfrac{1}{5}\wp^2$ (see the section on spherical space in chapter five). By (8.17) the total pressure is then:

$$P = K\sigma^{\frac{5}{3}} \qquad \left( K = \frac{1}{5}\left(\frac{3}{8\pi}\right)^{\frac{2}{3}} \frac{h^2}{\mu} \right). \tag{9.23}$$

This is what is known as the degeneracy pressure of white-dwarf matter and is interpreted as the pressure at zero-temperature. Modern notation for *K* (Carroll and Ostlie 1996, p. 588) is:

$$K = \frac{(3\pi^2)^{\frac{2}{3}}}{5} \frac{\hbar^2}{m_e} \tag{9.24}$$

which is entirely consistent with (9.23) assuming the electron mass and reduced mass are nearly identical and noting that Eddington uses $h$ rather than $\hbar$. Eddington has thus derived the correct formula for total degeneracy pressure from the basic tenets of *Fundamental Theory*, another remarkable or perhaps remarkably coincidental result, though it should be noted that the basic form of (9.23) was first given by Eddington's old mentor Fowler in 1926 (Fowler 1926). But it is not the result that is remarkable, since it is the standard equation used for nearly eighty years. Rather, it is the fact that the sometimes exceedingly complex arguments in *Fundamental Theory* were able to produce it. This still does not indicate that *Fundamental Theory* has any legitimacy as a theory as, once again, it could be eliminated by Ockham's Razor, but certainly lends additional credit to the methods since they have turned out to be, at the very least, consistent.

The wavefunctions of steady states are standing waves where the $p_\alpha^2$ reduce to eigenvalues which puts this analysis squarely in scale-free physics. A scale transformation between the two uranoidal systems, $B \to A$, must be applied for large values of $p_\alpha^2$, but in doing so this allows

$$E = m + p^2/2m \tag{9.25}$$

to be used as the Hamiltonian energy for standing waves (where $p^2$ is not small). This then also serves as the energy provided by the zero-temperature fluctuations that cause pair creation in the vacuum to occur. This is *not* a field theoretic approach in that it is single-dimensional whereas modern quantum field theory assumes, obviously, a field theoretic approach that includes a canonical momentum density (Zee 2003). It is thus a gross simplification of the actual physical process and ultimately not correct in its application. However, (9.25) can be used to represent a particle in a potential where the potential is given by the first term, *m*, using natural units and the mass-energy relation. Free particles are represented as having standing waveforms and the interaction that results from the presence of the potential will not necessarily change this. So Eddington is partially correct, but his application is far too simplified to correctly account for true physical processes.

**Motivation**

That concludes a detailed study of Eddington's statistical theory contained in the first part of *Fundamental Theory* (Chapters I-VI which are a merely an expansion of *Relativity Theory of Protons and Electrons* published a decade earlier). I will summarize and analyze it in the context of modern quantum field theory in the next chapter but I wish to conclude with a remark on the apparent motivation for Eddington's entire quest. I say 'apparent' since it is semi-speculative, there being no direct evidence to support the idea in any of Eddington's writings.

Eddington's career was spent largely as an astronomer. This begs the question: where was he introduced to quantum mechanics? Reading *Fundamental Theory* it presents itself as an early quantum field theory and tends to get bogged down in formalism much of the time. Why would an astronomer take on a purely physics project seemingly so far removed from his field of study? The answers to these questions, particularly the latter, are likely obvious to us in hindsight, but were not necessarily so obvious to Eddington. Primarily the motivation appears to lie in the discovery of super-dense astronomical objects. Since he was influenced directly by Fowler who studied degenerate matter in white dwarfs as early as 1926, and having been one of the early pioneers of the theory of stellar structure, in extreme astronomical conditions he found quantum mechanics and relativity colliding head-on: quantum mechanically degenerate objects that had tremendous gravitational fields. In order to further study stellar structure he found it necessary to find a way to bring the two seemingly incompatible theories together in a single unified theory. Ultimately, then, it was his interest in stellar structure that gave him the motivation for developing *Fundamental Theory* in the first place in order to explain internal stellar processes particularly in stellar oddities such as white-dwarfs.

Despite acknowledging the potentiality of an unlimited number of types of particles created simply by plugging in different values to the standard carrier, he appears oddly stuck in the two-particle paradigm despite the discovery of numerous particles by his death in 1944. This is a result of the fact that his theory only admitted two types of particles – electrical and mechanical. Any normal particles (electrons, protons, neutrons, positrons, etc.) could theoretically be included in either of these two types depending on the situation (neutral particles likely would only appear as mechanical particles). So he was not necessarily stuck in the electron-proton paradigm, which would have been absurd, but rather in an electrical – mechanical paradigm that was *not* shattered for certain until after his death, with the introduction of the weak and strong interactions. Nonetheless, most likely without ever knowing it, he developed an early quantum field theory that is largely outmoded and incorrect in many places, but is remarkably predictive of what was to come. This is the subject of the next chapter.

# X
## *Clarity of Perception*

Having reviewed Eddington's statistical theory in great detail it is now prudent to both summarize it and to ask how correct, relevant, and foresighted it was. This must be described in the context of both his motivation and methods. We return, then, to 1916 with Eddington poring over large data sets at the Greenwich Observatory. This was in line with the observatory's chief mission of improving observations in general, both through improvements to equipment as well as improvements in methods of analysis. We have already seen in his obituary of Schwarzschild that this work led him to consider the general structure of the universe from a stellar point-of-view. In addition the methods of analysis used in this case were statistical. Statistical methods include methods of probability and, in fact, statistical *mechanics* (better known as thermodynamics) was the forefather of the quantum revolution. So the stage was partially set when, early on, he applied statistical methods to general astronomical applications. The stage was completely set once general relativity had made its appearance in Britain when a copy of Einstein's paper was smuggled to Eddington via de Sitter (French 1979)[42]. This circumstance led to Eddington becoming the foremost expert on relativity in the English-speaking scientific community. The collision of these two historical events in Eddington's life – his assigned research involving statistical methods as Greenwich and his being on the receiving end of a smuggled paper by Einstein – led to the broad formation of Eddington's scientific outlook in part presented in *Fundamental Theory*.

**Tracing the Roots**

Precisely when the two collided in Eddington's own mind is debatable. It did not occur immediately upon his receipt of Einstein's paper. Clearly it could *not* have occurred prior to his receiving this paper from de Sitter since he had no knowledge of general relativity at this point. The most obvious confluence of the two methods was the eclipse expedition of 1919. In attempting to experimentally prove a (theoretical) prediction of general relativity he was required to employ statistical methods of data reduction. In addition there had to be the recognition that the results were only accurate to a point. This limit in accuracy begs the question of whether it is a limit imposed by the experimental and/or data reduction methods, or whether it is inherent in nature.

---

[42] Germany and Britain were, of course, at war at the time thus requiring the use of the "smuggler" (de Sitter).

This did not turn Eddington directly onto the path leading to *Fundamental Theory*. He was still a career astronomer at the time and upon his return to Cambridge, was immersed almost wholly in astronomy. But it was also around this time that astronomy began to move away from the realm of solely positional and dynamical methods to include studies of the actual *structure* of astronomical objects. Eddington followed the trend, turning his interests to stellar structure. The motivation, again, is explicitly stated in his comment in Schwarzschild's obituary which I will repeat here for purposes of clarity: the "task of determining accurate data for a large number of stars inevitably leads the mind to consider the great problems of the structure of the stellar universe" (Eddington 1916b). It was in 1919 that Fowler returned to Cambridge from the war. Three years later, in collaboration with Darwin, Milne, and others, he began work on his seminal studies of statistical mechanics and thermodynamics that included applying these methods to problems of stellar structure. Eddington had already been Plumian Professor for nearly a decade and, since Fowler worked at the Cavendish Laboratory (under his father-in-law Lord Rutherford), they were in close proximity along Madingley Road. In 1926 (yet another brilliant discovery in that amazing year) Fowler published his most seminal paper linking gaseous degenerate states (a short time later discovered to obey Fermi-Dirac statistics) to white dwarf stars that were not yet well understood (Fowler 1926). Somewhat simultaneously Eddington and Chandrasekhar were publicly arguing over just how far stellar collapse could go (this is the discovery of the famous Chandrasekhar limit for white dwarfs).[43]

The problem that presented itself as a result of the observation of the very strange companion star to Sirius A that is now known as Sirius B. Sirius B's existence was predicted in 1844 by Friedrich Wilhelm Bessel (1784 – 1846) and first observed by Alvan Clark (1804 – 1887) in 1862 (Carroll and Ostlie, 1996). But it wasn't until 1915 when Walter Adams (1876 – 1956) working at Mt. Wilson Observatory discovered that Sirius B's temperature was huge (modern values are around 27,000 K). But using the Stefan-Boltzmann law this predicts a size smaller than the Earth! So it had a tremendous luminosity but a very small surface area. No hitherto known laws of physics could seemingly explain how something so small could produce so much energy until Fowler proposed that white dwarfs were, in fact, in a degenerate state and thus obeyed quantum statistics which were just then being developed. Specifically, in that *same year* (1926) Dirac proposed his groundbreaking idea that wavefunction symmetry was related to the statistics developed just a year before by Bose, Einstein, and Fermi. Einstein had proposed that quantum gases and molecular gases were completely analogous to one another after having pondered Bose's derivation of Planck's radiation law that was based purely on the statistics of photons. Since Eddington was in the throes of arguing about white dwarfs, Fowler's paper could not have escaped his attention. In addition Dirac, Chandrasekhar, Darwin, and Milne were all at Cambridge around this time along with Fowler and Eddington and contact between them could scarcely have been avoided.[44] Thus, by the end of 1926 quantum mechanics had entered the realm of astronomy.

---

[43] In the philosophy of science literature some recent discussions of Eddington's observations have been made, cf. Deborah G. Mayo's *Error and the Growth of Experimental Knowledge*, University of Chicago Press, 1996.

      Being stellar objects, white dwarfs were also ripe for relativistic examinations and their gravitational aspects had been well-studied, particularly the orbital mechanics of Sirius A and B. In addition the new debate over the Chandrasekhar limit brought the relation between mass and pressure to the fore. A link was thus developed between mass and quantum statistics in degenerate matter and, since mass is at the heart of general relativity, a natural link between general relativity and quantum mechanics was found in white dwarf stars. This is where Eddington's statistical background in astronomy, that had led him to study stellar structure and introduced him to quantum statistics, collided head-on with his extensive knowledge of general relativity: there *had* to be some link between the two, some theory that independently recovered *both* in non-degenerate situations. One major key to this interpretation, however, is that Eddington equated the microscopic quantum (or 'quantal' as he says) world with electricity and the macroscopic

---

[44] This truly marked the pinnacle of the golden years of Cambridge physics and astrophysics.

('molar') world with gravity. What he's really doing is following up on his 1921 generalization of Weyl's unifying theory of gravity and electromagnetism employing extra-dimensional methods similar to those employed by Kaluza and Klein in their earlier attempts at such a unification. Ultimately, Eddington really saw quantum mechanics as an extension of electromagnetic theory. This, then, is the earliest seedling of *Fundamental Theory*: the unification of electromagnetism and gravity, finally appearing to bear some fruit in the study of white dwarf stars.

**Unique Extensions of Perception**

In employing Kaluza-Klein methods Eddington added a fifth dimension in a rather unique way that philosophically calls into question our ability to objectively measure it: he assumes that one can take normal 4-*D* space and rotate it. In order to rotate, it must have a dimension within which to move. This is the added dimension. In doing this, however, the added dimension is then subject to uncertainty. In addition this assumes that there would be some way to objectively observe this rotation in the extra dimension. The philosophical nature of this assumption is not addressed, but endowing regular 4-*D* space with the ability to rotate gives Eddington the relativistic method by which different particle types can be interchanged. Also, since the fifth dimension, like other dimensions, must have some coordinate structure and, since relativity implies all observables can be reduced to coordinates, relativity ought to be a generalization of quantum mechanics. Uncertainty plays no adverse role in this since it is completely causal and only limits knowledge of the present.

So, since a set of observables is simply a coordinate system, 4-*D* space is a 'hyperplane' in 5-*D* Euclidean space (see chapter five). This is the origin, then, of the 4/5 ratio that appears in Equation 8.25 developed to transition between a planoid and a uranoid. There is also a direct link here to his interpretation of multiplicity. Multiplicity is the number of dimensions in the space under consideration, a definition I've already shown is consistent with current interpretations. Since a coordinate system can represent a manifold and a manifold is a continuously parameterisable set where the number of independent parameters is the number of dimensions, the multiplicity describes the number of dimensions in the manifold. In phase space this number is either 136 or 137, but in regular space this can either be 4 in normal 4-*D* space or 5 in the 5-*D* space that includes the rotation dimension. So the transition between the planoid and the uranoid involves a change in multiplicity and the addition or subtraction of the rotation of the entire frame. In this way the 4/5 ratio is macroscopically much like the 136/137 ratio (or its inverse) in microscopic theory.

Eddington also makes the very valid point that when we speak of relative quantities we must ask the question, relative to what? Eddington introduced comparison particles in order to answer this question. Generally this means that every measurement requires a minimum of two observables (two endpoints, two like particles, etc.). This is why there must be a fifth dimension: the rotation of a 4-*D* space must be a rotation relative to something and that something is the extra dimension. Again, the generalization of this requires every measurable quantity to be a relative measurement between two things (in length measurements this means there must be two locations or endpoints). This is a very valid observation since our specification of any measurement quantity and origin location is purely arbitrary. For instance, in length measurements, Eddington felt it made more sense to place the origin at another particle since measuring relative distances is often more useful than measuring distances relative to some arbitrary origin. In a direct measurement, the units themselves may be arbitrary, but the physical distance is not. Measurements are all then just relationships between objects. In mathematics, relations such as this are treated as ratios, hence Eddington's prodigious use of ratios in *Fundamental Theory*. To further simplify things Eddington uses relativity to reduce all units to length measurements and reinterprets physical scenarios so that any uncertainty appears in measurements involving lengths (as opposed to densities, etc. – see Equation 5.8 and the discussion preceding it for an example). The idea of using a ratio as a unitless measure of length then suggests that anything that can be reproduced from some quantum specification can serve as a standard since quantum specifications are ratios to the fundamental unit of length $h/mc$ that appears in various fundamental quantum mechanical equations.

It is clear from this discussion that relativity played an important role in his interpretation of physical phenomena beyond the standard view I discussed in the previous section. Eddington took any phenomena that involved motion or coordinates to be interpretable in a relativistic sense, including spin and charge. I will discuss charge, which may not seem to be obvious in this sense, a bit later. Focusing on spin, however, Eddington inserts a co-rotating frame in the electron in order to Lorentz boost that frame to rest. Goudsmit and Uhlenbeck had already found in 1925 that if the electron had some finite size, the rotational velocity at the surface would exceed the speed of light several times over. Presumably this spelled the end of co-rotating frames for point particles. But there is nothing that requires there to be a *surface* to attach the frame to. Even if the particle is a dimensionless point there is nothing that prevents one from inserting a co-rotating frame into the picture. So if spin is simply rotation then Eddington's interpretation is rather unremarkable. However, spin-1/2 particles have the unusual property that they must be rotated through two complete rotations before returning to their original state. This is counterintuitive in a classical sense since anything rotating through *one* complete rotation of $2\pi$ is returned to its original state. For the non-classical quantum spin case this would mean that given two reference frames $A(x,y,z)$ and $A´(x´,y´,z´)$, $A \neq \Re A´$ where $\Re$ is the usual rotation matrix. There obviously would need to be some mechanism by which this could be reconciled, i.e. perhaps a transformation of the rotation matrix that would turn $A$ into $-A$ upon a single complete rotation of $2\pi$. The solution cannot be a *new* rotation matrix, it simply must be a relativistically transformed version of the original rotation matrix, otherwise the operation would not make sense. Essentially, at $2\pi$ the rotation matrix must become *minus* the identity matrix while at $4\pi$ it returns to the identity matrix. In short for any rotation through $(4n-2)\pi$ where $n$ is any integer (*not* the number of rotations) the sign must flip. It turns out that finding a scheme by which this happens under a simple transformation is not trivial. The problem gets even more difficult when one considers that this only works for particles with a half-integer multiple of spin. No solution is readily obvious but, on the other hand, there is no proof to the contrary. So Eddington's assumption that spin could be treated this way (i.e. Lorentz boosted to rest) *may* not be incorrect, but it has yet to be proven to be *correct*. Once again, this calls on the nature of permutation invariance (PI) for a thorough understanding and is something I discussed at length in chapter six. The main thing to remind yourself of is the fact that Eddington blurred the distinction between classical and quantum phenomena thus making this a slightly more complex situation.

The relativistic nature of charge is an even thornier issue. Unlike spin, charge seemingly has no relation whatsoever to coordinates! However, as I tell my introductory physics students, signs in physics inevitably indicate nothing more than direction. In the most basic interpretation, charge is simply an indication of the direction of the electric field lines. On a deeper level I mentioned earlier that charge is a measure of how the photon propagates. The current density term in the path integral can be written to account for positive and negative values and, since the current density can be related to a spatial Lagrangian, the positive and negative values can be related directly to spatial directions. But there is nothing here that explains *why* this is true – i.e. why can we write the current density this way? The deepest level of truth lies in electroweak unification and the ability of the weak interaction to exchange charge (really the weak interaction exchanges quarks with quarks and leptons with leptons). But the full relativistic application in Eddington's sense, where a simple transformation can change charge is an seems plausible as I discussed in chapter seven but remains unproven.

But the essence of both of these situations (spin and charge as relativistic quantities) is a haunting possibility. Though Eddington's own reasoning was naïve, rather than a cursory dismissal, a deeper exploration of these ideas only brings to light more questions rather than a resounding denial of their truth.

**Perceiving Measure**

Again, relating Eddington's concepts directly to the Standard Model, the perception of everything as relative was combined with the realization that objects cannot be considered apart from their environments – the universe is nothing but particles and their interactions, and studying particles without any interactions (n.b. there *are* self-interaction schemes) would be rather boring anyway. Thus in determining a quantum-specified standard one must recognize, as Eddington did, that there are two major limitations, both of which I discussed in chapter five. The first is that the standard is not fully reproducible in strong fields since the standard would be based on some particle (quanta) that could be altered by a strong field (even neutrinos could theoretically be altered by an extremely strong weak-field). Clearly Eddington assumes any such standard would be based on the photon since it sets an upward bound for velocity. But, the photon is the carrier particle of the electromagnetic interaction itself, which creates a problem when one encounters neutral particles such as the neutrino as well as in photon-photon interactions. Since electromagnetism is actually united with the weak interaction in electroweak theory there ought to be some relation here to W and Z bosons. But why should the standard be necessarily based on electroweak theory? Gravitons and gluons also move at the speed of light (and, though unlikely in present theories, so might neutrinos which are not even carrier particles!). Eddington's mathematical solution falls again on the shoulders of the standard carrier that is a single rigid form that can be altered depending on the particle desired. Out of this he constructs the standard deviation model that that provides a standard based on aggregate number alone and no other property. The standard deviation model has the advantage that it is independent of scale and thus can be equally well applied to both very large and very small scales. Though other values and quantities can be altered through relativistic transformations and actual physical interactions, pure number is somewhat unalterable (I say 'somewhat' since, for example, Hawking radiation predicts that not all spontaneous pair creations end in pair annihilations).

The other problem with the quantum-specified standard is that it needs to be suitably short. Eddington's point is an excellent one. Certainly we have compensated for the meter by subdividing it so that we use lengths such as Angstroms, for example, on very small scales (where Angstroms are $10^{-10}$ $m$). But Eddington envisioned a base standard that was ultimately unitless and very tiny[45]. Logically, for quantum-specified standards one might start with the Planck length, the theoretical lower limit to size (or, rather, *knowledge* of size). But this could become quite cumbersome with normal, everyday values and is not as forceful as Eddington's argument in favour of a standard impervious to potential field interactions.

---

[45] One might question how something could be considered 'tiny' if it is unitless. Eddington's point was simply that it would take an extraordinarily large number of these things to make anything macroscopic.

The idea of using a system of pure numbers as a measurement standard (and as the basis of the derivations of physical constants normally found only by experiment) smacks to some physicists and historians as cosmonumerology. On the other hand, Eddington had a point. Pure number is one of the only unalterable aspects of the universe – two is two regardless of where you are in the universe (even if you can't see 'two' – double stars, for example). Certainly there are the post-Eddingtonians who continue this line of reasoning, most notably Clive Kilmister and Ted Bastin who, along with John Amson, Frederick Parker-Rhodes, and Pierre Noyes founded the Alternative Natural Philosophy Association (ANPA) in the 1970s. They have been rather successful in many of their predictions and they do readily admit Eddington was incorrect in some areas of his work. But their basic philosophical premise is the same and, unfortunately, they tend to be marginalized in some circles. The perception of pure number as a root foundation in measuring (and ordering) the universe makes logical sense from its unalterable perspective and is the primary point of *Fundamental Theory*. The entire research program is built around finding ways in which various structures relate to this measurement standard. Since Eddington uses $\sigma$ as his standard measure, the entire treatise is a justification of that choice, something I mentioned in chapter five. So if *Fundamental Theory* could be boiled down to a single theme it would be the use of a standard for measurement that is either unitless or based on a single unit, i.e. one standard unit like length for every quantity in the universe which is what Eddington initially did before eliminating it entirely through ratios. The latter point he justified by the fact that all measurements are really comparing two things (endpoints, objects, etc.) so it didn't really matter what the single measurement unit was as long as everything could initially be expressed in it (that way seemingly unrelated characteristics could be directly compared). It could later be eliminated entirely in the ratio of comparison (it's a bit analogous to the idea that one can only measure a *difference* in potential energy – and thus total energy – rather than measure it at a singular point).

As I mentioned earlier Eddington viewed the universe as being fragmented by the various units of measure and felt that a single unit of measure for everything was more elegant. By making everything relativistically transformable any unit could be translated into a length that would then be completely eliminated by a ratio. In his 1951 book

*Quantum Theory* Bohm supplied three Hamiltonians for any observation (observer, object, interaction). I discussed this in chapter five but would add to this, in regard to the present discussion, the observation that this endows the interaction with particle-like characteristics. Certainly since Yukawa proposed his meson hypothesis in the late '30's it is not an unusual action taken by Bohm, but further solidified the growing vision that interactions really were particles themselves (or were mediated by them). This being the case it solidifies Eddington's argument that a standard based on only *one* of these exchange particles was fraught with problems. Bohm's Hamiltonian for the actual interaction between observer and observed could be equally well applied to *any* exchange particle since visual observation is not the only type of observation. In fact, since Hamiltonians can be applied to *any* particle the Hamiltonian is a bit like Eddington's standard carrier – a mathematical form that holds information about a particle. Eddington argued the observational point as well by asking why one should take visual observations as having a greater value than other observations (gravitational, etc.). What's the difference? Since any gravitational interaction in interstellar space is too small an effect for us to notice we do not regard this as hard evidence of anything. Since strong and weak interactions have very limited ranges they generally do not appeal to our senses either. That leaves electromagnetism as the only interaction that appeals to our senses, at least astronomically (certainly in gravitational interactions with the Earth there is a case to be made that we have sensory reactions, but this is an isolated case, at least at present while human space travel remains limited). But why, for instance, should electromagnetic data *not* collected directly by our senses (e.g. from CCD cameras, voltmeters, etc.) be more acceptable than gravitational data? Perhaps it is due to the fact that humanity's primary quantifier, number, is best recognized visually which means everything gets reduced to sensory data in the end.

That being the case we are limited to the use of the photon as a basis for a quantum-mechanical measurement standard. But Eddington's point about the photon's reaction to strong fields is a valid one – the standard is technically not fully reproducible in this case. Hence the case can be made for unitless ratios, particularly when combining this argument with the argument that all measurements are really comparisons between two objects. In the case of length it is a comparison between two endpoints, i.e. all coordinates are relative and all measurements *ought* to run between two objects without reference to an arbitrary origin. Another way of looking at this is to realize that measurements only make sense when made *between* two objects (at least most of the time). For instance, it is usually pointless to measure the distance from some star to some piece of empty space thousands of light-years away.

There are two major problems with this, however, that harken, again, back to basic physics. As I warn my students, reference frames should not be placed on or in any object that one wishes to study directly since that frame runs the rather high risk of being non-inertial. Since the endpoints of a length measure could be argued to be part of that measure, an origin placed in either endpoint *could* be interpreted as including the endpoints thus making the frame non-inertial. This is even more obvious when measuring a distance between two objects if that distance is in any way affected by one of the two objects (this can be ignored, then, in most cases, but two gravitationally strong sources near each other will affect the nature of the length measure between them). In addition, upon occasion one actually *does* care about points in empty space. The simplest example of this is in centre-of-mass problems where the centre-of-mass may be a point in empty space. A more specific and applicable example of this is the Genesis spacecraft that recently returned to Earth with samples of the solar wind: it orbited the Sun at the Sun-Earth Lagrangian point which is an empty point in space (if anything were actually there it would ultimately be pure coincidence or intentional as in the case of the spacecraft itself).

**Relative Measure and Flatness**

As I've just demonstrated Eddington's perception of the universe was not terribly heterodoxical: a natural background energy field exists (the essence of the universe) and natural fluctuations in it caused by the combination of uncertainty and special relativity produce the particles that make up the mass found in the universe. Mass produces curvature meaning that curvature is simply a result of the natural background fluctuations. So a universe devoid of fluctuations (mass) would be flat. In a modern interpretation one could say that since mass is the measure of how gravitons propagate then so is curvature. This is very nearly the truth, though his interpretation does not account for the evolution of the universe from the Big Bang on, but the Big Bang was a relatively recent and not wholly accepted concept at the time Eddington was working. In fact it did not see wide-spread acceptance until the mid-1960s. Lemaître had proposed the first model resembling a big bang in 1931 after Hubble's discovery a year before that the universe was expanding. Carl Friedrich von Weizsäcker (1912 – present) suggested a more refined model in 1938 that began from nuclear principles. Gamow and other nuclear physicists built upon this model but, though the Big Bang model gained momentum through the early 1940s, the war distracted most physicists from thinking about such matters. By the time the war was over Eddington was gone (Kragh 1999).

So considering the Big Bang was so new at the time, Eddington still came remarkably close to the truth in describing the fundamental nature of the universe as a 'blank sheet' of energy where random background fluctuations give rise to matter that produces *local* curvature while on a large-scale the universe remains flat (again, see de Bernardis, et. al. 2000). In tying this to the standard of measure, the transformation of relative coordinates was really the only difference between curved and flat space (see Equation 6.1). So even curvature *itself* was a frame dependent quantity! This makes sense in Eddington's context since mass gives rise to curvature and mass is frame dependent (i.e. if you Lorentz boost an object to such a degree that its mass is fully converted to energy – which is, of course, impossible – you would eliminate curvature altogether). As more and more fluctuations give rise to more and more particles the relevant effects of quantization are reduced thus explaining the lack of quantum effects on a large scale. In a slightly different argument he suggests that $g_{\mu\nu}$ as it appears in large-scale structures differs from that which is used in the wave equation. Since there is such a wide deviation in the two instances of $g_{\mu\nu}$ this implies frequent transitions between eigenstates making wave analysis useless much like reducing the persistence time of dynamic integrals reduces quantization effects. But, since curvature is introduced via fluctuations rather than purely through matter (essentially he reinterprets matter as fluctuations) $g_{\mu\nu}$ this suggests a third form of relativity intermediate to special and general where the curvature is uniform. Again, he has some very valid points about transitioning between macroscopic and microscopic situations and his statistical approach helps him in this regard.  Recognizing that an energy tensor is both tensorially and dimensionally a product of two momentum vectors he designs his system so that particle densities, which are statistical quantities, are defined as momentum vectors. As such particle densities are conjugate to spatial dimensions making them subject to uncertainty.  Again, this makes sense when one considers that either the volume is held constant while the particles move in and out of it thus making the density uncertain, or the density is held constant causing the volume to fluctuate.  Essentially, density is conjugate to volume.  Eddington accomplishes this with his system of natural units.  In terms of *SI* units this does not appear to work unless one considers total number rather than density and then considers the volume in spherical coordinates as pseudo-one-dimensional represented by an angle. In condensed matter physics it is well known that number and *phase* angle are conjugate to one another.  More generally Heisenberg showed that momentum *density* was conjugate to the phase *field* (Zee 2003).  Phase angles can be used in complex spaces to create circles.  A phase *field* then should be able to create a sphere in complex space. Eddington also introduces a phase *coordinate* as an extra dimension whose conjugate is the scale uncertainty that is manifested as an extraneous momentum in an extra dimension.  When these are reduced to eigenvalues the scale uncertainty becomes a momentum and the phase coordinate becomes a coordinate.

   Now since all measurement standards are based on units of length in *Fundamental Theory* they must be transformed from their original units into units of length.  This transformation has an associated uncertainty associated with it as described in chapter six.  The dimensionality constant *y* associated with this transformation is directly associated with the physical dimensions of the space, which in the particle density case is either 4 or 5 depending on the method of investigation.  This seems obvious but these numbers also provide multiplicity values that give the dimensions in phase space and as I just showed this is conjugate to the scale uncertainty that is manifested as a momentum. The relation to wave mechanics is then brought to bear by describing discrete wavefunctions as particle densities that rapidly decrease outwards forcing the integral over space to converge, further solidifying the momentum-particle density relation.

**An Exercise in Natural Unit Manipulation**

Upon stepping back to look objectively at this, however, one still is left with the nagging question, *how in the world could a particle density be physically related in any way to a momentum vector?* Perhaps surprisingly (perhaps not) this is not as strange as it sounds. In the case of photons one can simply integrate the Poynting vector to show that momentum can be defined in terms of the photon *number* (see Peacock 1999, p. 181 or van Fraassen 1991, pp. 438-442). This becomes a density when the analysis refers to a specific volume in space. Let's examine this case in greater detail but in a more heuristic way. For electromagnetic waves momentum density, **P**, is simply the momentum per unit volume of an electromagnetic wave. Finding the momentum for photons is fairly simple. It can be derived from the well-known relativistic formula $E^2 = p^2c^2 + m^2c^4$. Solving this for the momentum, $p$, we find $p = E/c$. This holds true for individual photons so the same must be true for electromagnetic radiation as a whole and thus the momentum density of electromagnetic radiation is its energy *density* divided by the speed of light. Let us say a volume of space has a photon density, $n$, and contains a single amalgamated electromagnetic field related to these photons (so the volume is filled with electromagnetic radiation in the form of photons). Though massless, photons are particles in keeping with the Standard Model and thus a photon density can be interpreted as a particle density.

The total number of photons in the volume is $N = nV$. Each photon has some amount of energy, $u$, that it contributes to the total energy of the volume, $E$. We define this to be $E = \langle u \rangle N = \langle u \rangle nV$ where <$u$> is the average individual particle energy for the volume. Define the energy density for the volume as $U = E/V = \langle u \rangle n$. The magnitude of the momentum density can then be defined in terms of the energy density as:

$$P = \frac{p}{V} = \frac{1}{V} \cdot \frac{E}{c} = \frac{U}{c}. \tag{10.1}$$

This defines the momentum density for a field of electromagnetic radiation. Since the final result requires a relation between momentum *vectors* and photon densities, though, we need to modify (10.1) in order to find the momentum in a vector format. Since the electric field of an electromagnetic wave oscillates rapidly it is useful to calculate average values. Thus, for the oscillating total electric field in this volume we can write (using brackets to indicate averages and assuming, momentarily, that it oscillates in one direction):

$$\langle \mathbf{p} \rangle = \langle \mathbf{P} \rangle V = \frac{\langle U \rangle V}{c} \hat{\mathbf{k}} = \frac{\langle u \rangle n V}{c} \hat{\mathbf{k}} = Cn\hat{\mathbf{k}} \qquad (10.2)$$

where $C$ is a constant with *SI* units of $kg \cdot m^4/s$ which is simply units of momentum multiplied by units of volume. This also assumes that the average photon energy <u> and the volume $V$ do not change. So, at this point I have shown that a momentum vector is linearly proportional to particle (photon) density. The goal is to show that the proportionality constant, $C$, is both unitless and unit-valued (1) valued since this would show that $\mathbf{p} = n\hat{\mathbf{k}}$ or, in scalar form, $p = n$.

Employing the usual system of natural units (not Eddington's) such that $c = 1$ and $\hbar = 1$ we can show that $kg \propto s$ as follows:

a. From $c = 1$ we find that $1s = 3 \times 10^8 \, m$.
b. From $\hbar = 1$ we find that $1s = 1.054 \times 10^{-34} \, kg \cdot m^2$.
c. Dividing a. into b. we find that $1 = 3.513 \times 10^{-43} \, kg \cdot m$ which implies that $m^{-1} = 3.513 \times 10^{-43} \, kg$.
d. The combination of a. and c. implies that $1 kg = 8.540 \times 10^{50} \, s$.

Expressing (10.2) in scalar form for the moment we write $p = Cn$. Obviously the units on both sides must cancel, but do they cancel in natural units such that $C = 1$? The *SI* units for the relation $p = Cn$ are $kg \cdot m/s = (kg \cdot m^4/s) \cdot (1/m^3)$. These units cannot be shuffled around without changing the basic nature of the quantities involved (i.e. without changing $n$ to $N$ or something similar). Using the values derived in the steps above for natural units we can, however, change everything to units of $s^i$. Given that $C = pn^{-1}$, which solely in terms of the natural unit relations derived above is 
$(1.054 \times 10^{17} \, s^4) = (2.847 \times 10^{42} \, s) \cdot (3.704 \times 10^{-26} \, s^3)$, we find that $C$ appears to have units

of $s^4$ which does not meet our criterion for unitlessness. However, crunching the numbers *does* show that its *value* is indeed 1! Thus $C = 1s^4$. Is the requirement that this be unitless necessary? What does the simple fact that $C$'s value is 1 tell us? It tells us that when represented in natural units the numerical value for particle (photon) density and the magnitude of the momentum are *always identical*! Now, technically I still haven't shown the relationship between $p$ and $n$ in vector form. Since $n$ is not a vector quantity we assume that it must be multiplied by some vector quantity in order to extend the relationship with $p$ beyond simple magnitude discussions. We might first try to define a unit vector (not an operator) that carries the correct units of $s^4$. This is essentially a redefinition of $C$ as a unit vector. In any case, in vector notation we might write $\mathbf{p} = n(\mathbf{1})$ where $(\mathbf{1})$ is a unit vector with units of $s^4$. In this context the unit vector serves two purposes: to change the units of $n$ and convert it to a vector representation. *But*, recovering the relationship $p = n$ out of this turns out to be impossible without turning $(\mathbf{1})$ into a function of some sort. It appears that *an operator is required* in this situation. Since $n$ is the magnitude of $\mathbf{p}$, which is simply $\mathbf{p} \cdot \mathbf{p}$ or $p$, we would have to show that the fourth-order derivative of $\mathbf{p}$ is equal to $\mathbf{p}$'s magnitude:

$$\nabla \cdot \left( \frac{d^3}{d(x^\mu)^3} \mathbf{p} \right) = \frac{d^3}{d(x^\mu)^3} (\nabla \cdot \mathbf{p}) = p \qquad (10.3)$$

where the divergence is required to eliminate the vector nature. In order to tidy (10.3) up a bit let us define

$$\Xi^s = \nabla \cdot \frac{d^{s-1}}{d(x^\mu)^{s-1}} = \frac{d^{s-1}}{d(x^\mu)^{s-1}} \nabla \cdot \qquad (10.4)$$

and choose $s = 4$. (10.3) can then be rewritten in the more compact form:

$$\Xi^4 \mathbf{p} = p \qquad (10.5)$$

The simplest solution to this is $\mathbf{p} = e^{-kt}\hat{\mathbf{k}}$. Since $\mathbf{p}$ automatically has natural units of seconds and $k$ has units of inverse seconds, the third derivative of the divergence of $\mathbf{p}$ (which we have defined as only being in a single direction) leads to $e^{-kt}/k^4$ which has units of inverse cubic seconds. Setting $k = 1$ preserves the original form and yet brings the units in line with the units of particle density. The initial divergence converts the vector $\mathbf{p}$ to a scalar and the regular derivatives simply add derivative permutations to the function. Should the divergence or the ordinary derivatives come first? That depends on *when* the momentum needs to (or *can*) be converted to a scalar. In the case considered here it doesn't matter since the end result is the same. That ultimately means the constant in (10.2) is really an operator of the form (10.4) and thus the combination of (10.2) and (10.5) gives:

$$\Xi^4 \mathbf{p} = n. \qquad (10.6)$$

Recall, however, that the present derivation was for *photons* and we arbitrarily set $k = 1$ in order to preserve form. In addition we defined the constant $C$, which represents the average individual photon energy divided by $c$, to be an operator, although this is hardly unusual in quantum mechanics since momentum itself is an operator. In fact this implies that the momentum operator is really just an integer multiple (to some extent, disregarding units at the moment) of some other operator. Whether this relationship, (10.6), would hold for other particles, in particular massive ones, is questionable, but the point is that Eddington was not necessarily *wrong* in his assumption that particle density and momentum are equivalent since it at least holds in one specific (and very important) case. Incidentally, when one considers relative measure and compares ratios of $p$ to ratios of $n$ one need not worry about units and the stated *ratios* will always be equal (e.g. a ratio $p_2/p_1$ for two fields is always equal to the ratio $n_2/n_1$ for the particles producing those fields).

So, in conclusion, intriguingly, Eddington may have been correct. He obviously recognized that there was some natural relationship between the numerical values of momentum and particle density in natural units. Once again he clearly perceives an underlying relationship that is not readily obvious and that could have profound implications including the fact that the particle density in some sense could be conjugate to spatial dimensions. This is made even more intriguing by the known fact that momentum density is conjugate to phase field and pure number is conjugate to phase angle; if a relationship exists between phase angles and phase fields one ought to exist for momentum densities and pure number thus suggesting one exists for momentum and number density. Finally, since curvature is introduced by the fluctuations observed as particles then the greater the particle density, the greater the curvature, meaning the momentum of the field is correlated to curvature (assuming the particles have mass – this obviously doesn't work for photons).

**From Natural Units to No Units**

Eddington established both a scale-free and a scale-fixed theory as I have described. Eddington's system of natural units brought everything down to a single unit (length). In order to establish the equivalence of scale-free (mechanical) and scale-fixed (electrical) systems, ratios can be used thus eliminating units entirely and holding to the philosophical idea that the most meaningful measurements are comparisons (ratios). This also has the effect of reducing the importance of quantisation which only appears in scale-fixed theory. In essence Eddington was simplifying everything by putting it into his 'native language' of relativity (scale-free theory).

Now, since units are no longer an issue and since relative measurements are made through tensors or vectors, the field has the same variables as a distribution of particles which is essentially what I argued above in equations (10.2) – (10.5). Eddington therefore argued that since true fields were equivalent to the average characteristics of an ensemble of particles *field theory was reducible to statistical studies of datasets for large numbers of particles*.

This is a profound conclusion and harkens, once again, back to Eddington's statement (which I will repeat for a third time) that the "task of determining accurate data for a large number of stars inevitably leads the mind to consider the great problems of the structure of the stellar universe" (Eddington 1916b). The stars were the first objects to be considered particles in Eddington's statistical studies of the universe, and as an understanding of the nature of galaxies matured in the wake of Hubble's discovery of the nature of M31, galaxies became the particles. In Eddington's cosmology fields acquired their own characteristics since these characteristics simply arose from the statistics of the distribution. Changes in the fields came about through changes to the occupation states of the individual particles in the distribution where the occupation states appeared as generalized coordinates or momenta. Eddington later converted these discrete occupation factors to a continuous occupation factor that is a function of the coordinates. This satisfies Eddington's criteria that the particles be unidentified members of an assemblage.

Generalized characteristics of the particle assemblage are normally additive. Since Eddington works within a rigid-field framework, he is assuming the overall characteristics of the field do not change, meaning the average characteristics of the particle assemblage do not change which is reminiscent of basic conservation laws. Essentially the rigid-field condition is his conservation law.

**Multiplicity and the Assemblage**

The rigid-field can be used as an inexhaustible reservoir, then, since its average characteristics remain the same. Since all measurements must be relative measures between two values, the rigid-field acts as a convenient reference point in measurement. For instance, equation (6.22) gives the probability that some particle in a quantum state with energy $E$ is in contact with the reservoir or, in this case, rigid-field. Another interpretation of (6.22) was given in chapter six stating that it is the probability that a system will be in a microstate (quantum state) with an energy $E$ relative to the macrostate's (rigid-field's) energy. The total number of possible microstates we know is the multiplicity. As such, the multiplicity tells us the number of possible quantum states of energy $E$ that a particle could be in while (6.22) gives the probability for each state. So, for instance, say the multiplicity is 137; then there are 137 possible microstates. Each of these 137 quantum states is not necessarily equally possible, however, and thus (6.22) is employed to determine just how likely each of these 137 states is. This is less than the actual number of dimensions in phase space since some aspects of the system are presumably stabilised (i.e. assumed values from prior data). In fact, as given in (6.11), the multiplicity $k$ is the dimension index (number of dimensions) of an individual particle's energy, i.e. the number of possible states it can have (corresponding to the number of independent components of the energy tensor), while there is a separate dimension index, $l$, for the total energy that is often set to 1 as in (6.12) and (6.13). Since the total energy is set to 1 positive values for $k$ imply negative values of $-(k+1)$ for the rigid-field's possible energy states in order to balance positive and negative (recalling

Eddington's assertion that a given datum for positive and negative measurements is completely arbitrary). The reason the total energy is set to 1 is because the universe is assumed to only have a single macrostate and thus a single possible overall state that fluctuates internally.

Using the rigid-field as an inexhaustible reservoir as developed in general statistical mechanics situations (see (6.17)) allows energy to explain away Dirac's sea of negative energy states. These states are really just the states that make up the rigid-field. Rigid fields are the primary domain of quantum mechanics and the coordinates describing states in rigid fields are slightly different than those in non-rigid domains as given in (6.24). Eddington uses this transformation to explain the covariant nature of momentum in wave mechanics. Even the word 'covariant' suggests rigidity since it implies that quantities change together. Out of this argument he then derives the origin of $i$ in quantum theory. A closer examination of his argument, however, shows that the true nature of his reasoning is embedded in relational considerations. As he separates out the particle and field energies he inserts a datum to explain an extra minus sign (present in the covariant momentum expression in quantum theory). The extra minus sign is contained in $i$ and thus the relational consideration explains its existence. Now relational considerations are merely relative considerations based on one's choice of reference origin. Such a choice is governed by special relativity meaning *the presence of i in quantum theory is really a relativistic phenomenon.*

**Profound Results**

Certainly the meeting of relativity and statistical methods in Eddington's mind, brought to an apex with the developments related to white dwarfs in the 1920s, played the most critical role in developing the foundation of *Fundamental Theory*. However, the adherence to fairly strict interpretations of relativity and statistics present unique conclusions. To reiterate two of the most important points I have just made, Eddington showed that field theory is reducible to the statistics of a large assemblage of particles and all measurable quantities are frame-dependent and can be altered simply by a change in reference. Starting with his 'blank sheet' hypothesis, this leads to a remarkably modern conclusion.

From a structuralist standpoint, Eddington is quite clear on his interpretation of the nature of particles: they are simply conceptual objects that carry information in the form of variables. This is the root of the concept of a standard carrier where there is a single mathematical form that manifests itself differently depending upon which variables are filled. Since Eddington also held that all measurable quantities (mass, charge, spin, etc.) are frame-dependent, a transformation can easily shift a standard carrier from one particle into another. From a physical standpoint, the universe consists of this 'blank sheet' of energy that is initially undisturbed. The uncertainty principle combined with special relativity, as is well known, can disturb this vacuum energy and create particles and anti-particles – particles literally pop into and out of existence in the vacuum. Since all particles are different manifestations of the standard carrier and a Lorentz boost can theoretically transform one into another by changing the values in the standard carrier's matrix of characteristics, *all particles in the universe are ultimately indistinguishable from one another*. This is the current accepted view of many field theorists (Zee 2003) though for different reasons including the fact that the exchange of massive bosons is really responsible for transferring characteristics from one particle to another. Philosophers of physics, as we have seen, have recently shied away from this blanket statement demonstrating through permutation invariance that particles can actually distinguishable in cases where they previously were thought not (cf. French and Redhead 1988). But, the idea was discussed as early as the late 1920's with Dirac's work. Eddington simply took a holistic view by assuming that there was a single, continuous, uniform energy field that, due to the uncertainty principle, is endowed with various particle characteristics mathematically represented by altering the standard carrier. Cassirer, as I discussed previously, has suggested that particles are simply intersection points in certain relations which is a similar point of view. Another way to look at this is to define particles as being the intersection points of interacting fields, though I do not personally find that description elegant. If a unified theory is ever found, theoretically all four fundamental interactions and their fundamental fields would be manifestly identical. Thus I find Eddington's original description to be simpler – all particles are fluctuations in a single field and the fluctuations arise from uncertainty (the fact that zero fluctuations has a near-zero probability of occurrence). Taking this a step further, the fields associated with the four fundamental interactions are merely different manifestations of a single unified field. In the interest of pure simplicity the four manifestations of this single field are associated with the type of fluctuation (particle) that is created by uncertainty (as the weak field in neutrinos, for example, or as all four in a proton). So, the unified field manifests itself in different ways depending on how it is disturbed by the uncertainty principle. Since Eddington was ultimately a statistician at heart he transformed field properties into the properties of a large assemblage of particles. In deference to his relativistic sensibilities, however, he also adopted the view that everything in the universe is simply energy. All of existence is reduced to random fluctuations in the uniform background or energy field. Eddington's conception here is far from heterodoxical.

The line of reasoning pertaining to the exchange of characteristics between particles leads us tantalizingly close to the weak interaction again. Since I have shown in chapter seven that the frame-dependence of charge only works with a global change in reference frame that would always flip pairs of charges (thus conserving overall charge) this might be used to explain why exclusion works in pairs in *Fundamental Theory* – flip one particle or the other, but relativity says you can never really tell which one flipped. Eddington's explanation is rooted in comparison particles, but by his own reasoning, this can be explained by recalling that exclusion's job is really to make indistinguishable particles distinguishable. In *Fundamental Theory* it is the job of interchange to mediate the exchange of information such that exclusion works. In fact it allows properties to be transferred from one particle to another, essentially substituting a quantity in one standard carrier for a quantity in another standard carrier. The weak interaction by contrast exchanges quarks with quarks and leptons with leptons thus allowing particles to change identity.

Another remarkable conclusion Eddington reached via this line of reasoning was his version of the exclusion principle which, as I described in chapter eight, is *defined in terms of the uncertainty principle*. The root of this lies in the above discussion of the statistics of large data sets and the indistinguishability of all particles. Exclusion's job is to make indistinguishable particles distinguishable. Since Eddington works in a phase space of coordinates and momenta, common even in classical physics, he opens the door to uncertainty since it is defined in terms of commuting physical properties and coordinates and momenta are the most common commuting pair in quantum problems. Thus, since one could easily choose a cell of phase space having a volume $h^3$ Eddington simply rearranges the uncertainty principle to find the maximum number of particles in this volume. Since uncertainty ultimately prevents complete knowledge of particle properties Eddington is implying that it prevents *perfect* exclusion (meaning two particles at the same exact spot in space or sharing the same exact energy levels) since it prevents complete knowledge of a particle's exact location. So rather than at a point, particles are limited in a certain volume. As I discussed in chapter eight this also implies that wavefunction collapse is never 100%. This problem has been discussed in depth by a number of physicists and philosophers and is often referred to as the GRW interpretation of wavefunction collapse after the authors of one of the original and most seminal papers in this area (Ghirardi, Rimini, and Weber 1986). Yet it could be argued that exclusion *provides* full knowledge (or nearly full knowledge) which would appear to be a contradiction. Of course we know full well that neither exclusion nor anything else, for that matter, provides full knowledge of the state of a system. That was precisely Heisenberg's point – our limitation is in our knowledge of the present, not our ability to predict the future. Nonetheless, Eddington's result is remarkable in that it strongly suggests a relation between uncertainty and exclusion that has been hitherto unexplored.

**Trouble in Paradise**

However, the strict adherence to the relativistic worldview appears to have one major exception in Eddington's formulation. One of relativity's greatest triumphs is the equivalence of gravitational and inertial situations (e.g. one cannot distinguish between an inertial frame and a frame in free-fall in a gravitational field). This triumph of reasoning is actually what led to the development of general relativity from special relativity and was really built on the equivalence of inertial and gravitationally measured masses that was developed by Galileo! Eddington, however, clearly distinguishes between the two situations in a way that appears to violate this principle.

The problem is that Eddington attributes the potential energy to the object (or appears to, based on the wording in *Fundamental Theory*). Potential energy is really associated with an interaction. If an interaction is not present then there is no potential energy. So, for instance, consider two isolated particles each with some amount of kinetic energy. In order for these particles to dump some of their kinetic energy they have to interact with something, say each other. During the interaction some kinetic energy is taken from each *by the interaction* (which in the Standard Model is just another particle). The interaction then gains potential energy. The potential energy does not exist without the interaction. The interaction could then impart some kinetic energy back to the particles, often in the form of an exchange of some sort (e.g. particle *A* loses 10 J of energy, particle *B* gains 10 J of energy).

The fact that the interaction is represented by an exchange particle can also help explain the equivalence principle. One could imagine that, in the case of the two particles just described, the interaction's potential energy is really, on a fundamental level, the kinetic energy of the exchange particle (since most exchange particles are massless, this is a bit more complicated than I'm describing, but is an illustrative analogy). So, one could imagine that a given particle with some kinetic energy interacts with some other particle by exchanging a massless boson. The amount of kinetic energy it loses goes into the boson which has to gain kinetic energy in order to move. Once this exchange particle comes in contact with the second particle it hands over this kinetic energy to the new particle. In fact, this is the basis of Feynman diagrams and is very easily understood via the photoelectric effect. In the photoelectric effect a surface of some material is interacting electromagnetically with something else (the Sun, a lamp, a laser, etc.). The photons, which are the exchange particles for the electromagnetic interaction, have a certain amount of energy which can be interpreted as kinetic (though, note that all *massless* particles including three of the four types of fundamental exchange bosons move at the speed of light, so this is not understood in the classical sense but rather through relativistic means). The amount of energy associated with the photon's motion has to be equal to the minimum energy required to dislodge an electron – there's a fundamental equivalence there not to mention a *clear* indication of a transfer of momentum.

So returning once again to the equivalence principle itself, a gravitational potential simply adds an extra particle to the mix (the graviton) ferrying kinetic energy from one object to another. Since all massive particles can be brought to rest through a Lorentz boost, one could attempt to bring the graviton to rest. Doing so would stop any gravitational information from being transferred between the objects and seemingly eliminate gravity. Thus it appears gravity is associated with the motion of gravitons. So a reference frame attached to a graviton, for instance, which can be thought of as freely falling in some sense, can be brought to rest by a Lorentz boost. Thus it is impossible to tell the difference between an inertial frame and one that is freely falling in a gravitational field. The presence of the gravitational potential has no affect on the internal constitution of a system in this sense and simply accelerates it in reference to some other frame of reference.

There is another problem in *Fundamental Theory* that is perhaps not quite as serious when understood in its historical context. In order to stop wave packet dispersal and in order to fully explain energy relativistically he introduces negative energy states that can seemingly give rise to matter spontaneously. Historically there are two points to be made here. The first is that, as I have already pointed out, the Big Bang scenario was not fully accepted until after Eddington's death and thus the 1930s (post-Hubble) marked the first shot-in-the-dark, so to speak, for modern cosmology. Prior to Hubble's discovery the universe was generally thought to be static. Once it was found to be expanding the logical conclusion was that, if time were run backward, the universe would begin with a singularity. That leads to the second historical point: singularities were viewed as headaches to be eliminated in the 1930s and 1940s. Singularities are nothing more than infinities present in an equation and even today infinities in theories are usually considered a pox.[46] The fact that the universe had historically been viewed as static and the fact that singularities were not viewed in a positive light at the time both contributed to the acceptance of continuous matter creation as a possibility for quite some time. Recall that the steady-state theory of Hoyle, Bondi, and Gold did not appear until well after Eddington's death. Thus, *at the time* this aspect of Eddington's theory was not very heterodoxical and in hindsight was no more or less so than theories developed by Milne and Dirac, neither of whom share the same historical treatment as Eddington.

---

[46] As a modern example the presence of infinities in the original theory of weak interactions was what led Glashow, Weinberg, and Salam to develop electroweak unification as a way to eliminate those infinities.

**Overlooked Subtleties**

In truth Eddington's 'steady state' theory, as one might call it, is far more subtle than that. From a structuralist point of view changes cannot necessarily be taken into account via predictions such as those offered by probabilistic methods. In wave mechanics changes are represented by changes to the occupation factors of steady states meaning wavefunctions must represent steady states where the occupation factor is the *only* variable that changes with time. Eddington took this to be a principle upon which a full theory could be developed. As such he utilized relativity and uncertainty in union to change occupation factors, either by a change in the reference datum for energy or in a random fluctuation designed to create or annihilate particles.

With the occupation factor interpretation of energy states as a foundation, as all lower energy states become occupied, each particle has what Eddington called a 'top energy' (see chapter eight) that happens to be the energy a particle is boosted to due to the occupation of all the lower energy states. So clearly there is a link between the occupation factor and the exclusion principle. Since exclusion limits the number of particles in a volume, if the volume is completely filled as dictated by exclusion, then the occupation of that energy level is complete. Eddington also associated the rest mass of a particle with exclusion since its value results from the occupation of lower energy states. So, in simplified terms, the way to build the universe is to start with random fluctuations in the background field. Uncertainty not only *creates* these fluctuations but also limits how many can exist in a given volume forcing any new fluctuations to be spatially separate. Fluctuations in the scale of measurement give space curvature and endow individual particles with their rest energies. Since both descriptions must be equivalent Eddington concluded that exclusion was a wave-mechanical form of gravity. In order to validate this link Eddington derived the same formula for proper mass from both the exclusion principle and relativity. If this is true and if exclusion is the weak interaction in disguise then the W and Z bosons would *have* to have mass (and indeed they do!). Of course we now know this is more subtle than described here. The origin of mass lies in spontaneous symmetry breaking and the Anderson-Higgs mechanism but Eddington's idea is fairly similar (though without the reliance on symmetry principles). Mass can be interpreted as the measure of how a *graviton* propagates just as charge is the measure of how a *photon* propagates. Nonetheless, uncertainty is at the deepest level of the heart of spontaneous symmetry breaking and so in a way Eddington was correct in assuming it played some role in the origin of mass (and thus curvature).

Returning momentarily to Eddington's problem with the equivalence principle, there is a somewhat new interpretation of the origin of mass that was put forth in 2000 by Albrecht Giese that says the inertial behaviour of the mass of extended objects can be traced to the time of transmission for exchange particles in an interaction and his reasoning has valid points to it (Giese 2000). Unfortunately he does not explain the origin of mass for point particles. But his notion is intriguing and, on the surface, also appears to question the equivalence principle. Tests to show the equivalence between gravitational and inertial frames have overwhelmingly proven the two are equivalent to a phenomenal degree of accuracy. But Eddington has a curious addition in his theory that may be the root of this seeming difference.

In *Fundamental Theory* coordinate locations are endowed with energy just as mass, motion, and other characteristics are – different locations indicate different energy levels. Based on Eddington's interpretation of the exclusion principle and occupation factors given above this makes complete sense. Consider a particle with a constant velocity moving in space. In the above interpretation, if the coordinate location is constantly changing the energy associated with it (the coordinate location) must be changing. Since energy must be conserved, the mass must be changing (since the velocity isn't) which makes no sense! But, bring in a gravitational potential and another source (or sink) of energy has been added such that rather than the energy of the coordinate locations being exchanged with the *mass* energy, it can be exchanged with the *gravitational* energy (potential). As such, gravity would be interpreted as a mass-stabilization energy term. Since all massive particles have (albeit miniscule) gravitational potential wells associated with them, there is a gravitational self-energy term that stabilizes the mass – *so mass stabilizes itself.* Unfortunately Eddington seems to forget, despite his liberal use of it elsewhere, that relativity tells us that coordinate systems are entirely relative so the coordinate location energy must really be an energy associated with the fact that a certain particle, for instance, is not in the same location (regardless of coordinate system) as another particle. In addition this does not adequately explain how massless particles remain massless.

Nonetheless, both gravity and exclusion provide an energy 'boost' to particles by changing their location that is manifested in wave mechanics as a changed in occupation. Either way it simply amounts to a change in location which is ultimately a relativistic phenomenon and might even suggest that gravity and electromagnetism were different manifestations of the same phenomenon to Eddington. In this way *Fundamental Theory* is truly a unified theory. Regardless, there is such a thing, then, as 'exclusion energy' in *Fundamental Theory*. Continuing with my line of reasoning regarding the relation between the weak interaction and exclusion this again makes sense – the 'exclusion energy' is simply the energy of the weak interaction.

Another remarkable subtly in Eddington's theory is the nature of the particles filling the negative energy sea. Since the fact that their energy is negative is simply due to a choice of datum, they have mass as any other particle does. Also, since exclusion, particularly when combined with the fact that most of the negative energy states are filled, forces extended objects into being and ostensibly, due to repulsive interactions, forces them apart, it could be interpreted as the mechanism driving expansion in Eddington's theory. The negative energy sea, then, has a very real total mass and energy and this could be interpreted in this context as being the mysterious dark energy of the universe especially considering Eddington explicitly includes the cosmological constant in his theory. Of course, with today's knowledge we know this is not the case (even though we do not know precisely what dark energy and dark matter are). But, the picture I am painting here of *Fundamental Theory* is that, despite being filled with numerous problems, it is internally self-consistent *and* in keeping with experimental evidence. In fact it is remarkably prescient of subsequent discoveries many of which I have already pointed out. Yet another example related to the present discussion is that it predicts that every particle *must* have a corresponding anti-particle, something we know quantum field theory predicts (and something that is troublesome for its lack of experimental confirmation – where's all the anti-matter?).

**Subtly Weak Suggestions**

As a large portion of the past few chapters has indicated Eddington found that exclusion and interchange were complementary concepts.  In his exclusion interpretation of the universe that I have just analyzed particles force each other to be individuals at varying energy levels.  Using interchange as an interpretive model instead, these same particles are constantly interchanging with one another producing the same total energy as the exclusion interpretation.  We already know that the weak interaction is responsible for the interchange of particle characteristics for quarks and leptons so, in a way, this once again validates my assertion that exclusion and the weak interaction are intimately related.  His discussion of the non-Coulomb repulsion terms adds yet more fuel to this argument as this repulsion amounts to an additional force.  Since the weak and strong interactions were not only not well understood but thought to be one and the same at that time, Eddington is essentially advocating for their existence or the existence of something similar.  Basically he acknowledges that electromagnetism and gravity cannot be the only two interactions in the universe.  There must be at least one more.

      The one major flaw in the suggestion that the weak interaction and the exclusion principle are the same phenomenon is that mesons participate in weak interactions but do *not* obey exclusion.  On the other hand, if one realizes that the weak interaction is not really manifesting itself between mesons but rather between the quarks and anti-quarks *within* the interacting mesons themselves, the problem is partially solved.  Why then would baryons, which are also compound particles composed of quarks, obey exclusion while mesons do not?  Spin is the obvious (and universally accepted) answer, but I believe the truth is more subtle.  The question really should be (if one is assuming that exclusion and the weak interaction are so closely related), if the weak interaction is there and its occurrence suggests exclusion, where is the exclusion phenomenon?  What makes mesons different from baryons?  Well, first, baryons have three quark-like particles while

mesons have two. Second, while all three particles in a baryon are quarks, in a meson one is an anti-quark. So in terms of matter (as opposed to anti-matter), there is three times as much in a baryon. Perhaps the meson's internal weak interactions cancel each other out such that there is no residual weak field outside the meson. Adding a third quark-like particle would break the 1:1 symmetry forcing the residual weak interaction energy to search for a 'new partner' as-it-were. As such it is reasonable to assume that the new pentaquark discovered last year would *obey* the exclusion principle since its weak interactions are unbalanced.[47]

**Numerological Results and Extra Energy Terms**

As chapter nine demonstrated *Fundamental Theory* was also remarkable in the number of physical constants derivable from it. There is still a great debate over the derivability of physical constants, especially as certain new theories and experimental evidence suggest these constants are of greater importance and more mysterious than first thought. Eddington's method went further by exploring the relations *between* physical constants and it established relationships between constants seemingly at opposite ends of the size spectrum. Some of Eddington's conclusions appear to be more than pure coincidence, however. (9.1), for instance, is still remarkably close to accepted values despite major changes in value for the dimensionless Hubble parameter since the 1940s.

      Perhaps the most studied numerical result of *Fundamental Theory*, the mass ratio of the proton to the electron, has been offered up by critics as an example of his 'cosmonumerology' but, as I demonstrated in chapter nine, he simply utilizes a mathematical trick or two to arrive at the result. In addition, at least one of these mathematical devices, the derivation based on external and internal masses (atomic and reduced masses) makes sense when performed for hydrogen since in this case the reduced hydrogen mass differs very little from the electron mass and the atomic hydrogen mass differs very little from the proton mass. Analyzed in this light it is not an unexpected

---

[47] Subsequent to the writing of this treatise but before it was submitted I completed a phenomenological proof that exclusion is *not* equivalent to any of the four fundamental interactions (forces). I have inserted it as an appendix but have left my previous writing intact since it raises important philosophical issues and demonstrates the difficulty this problem presents.

result. It is entirely possible that the poor reception of it was due to the obscure language of *Fundamental Theory*.

Another very studied result of *Fundamental Theory* is the value for the fine-structure constant. Eddington held that its inverse was exactly equal to 137 due to the correction factor required when moving from rigid to Galilean coordinates. Essentially it is the inverse of the multiplicity of an intracule. Ignoring the specific value for a moment, how does this compare to modern interpretations of the nature of the fine-structure constant? The fine-structure constant is nothing more than the coupling constant for electromagnetic interactions and appears at vertices in Feynman diagrams for such processes. It is essentially a measure of the strength of the electromagnetic interaction and can be expanded in a series expansion. In a way Eddington was close to this description since he viewed the fine-structure constant as being related to the degrees of freedom of the intracule and the intracule's motion is constrained by its internal electromagnetic interaction (the internal binding energy of the intracule).

The non-Coulombian energy term that arises in these situations is essentially an accounting of the extra rest mass given by the proton over the electron. Eddington recognized that if all particles in the universe are ultimately indistinguishable something must account for the difference between the rest masses of the proton and electron. This was a very clever insight and, though incorrect, indicates that Eddington was able to see the inherent problem in the idea of indistinguishability. Of course, we now can attribute the excess mass of the proton to the fact that it is actually a composite particle made of three quarks. Quantum field theory can now explain the meaning and origins of such characteristics thus explaining indistinguishability. Again, less important than Eddington's method was the mere fact that he recognized the underlying problem. In addition, his solution, though incorrect, is self-consistent since it explains why the non-Coulombian energy term is only associated with the proton. His work actually presages the discovery of internal structure in the proton. His methods outlined in equations (9.16) and (9.17) are actually based on scattering experiments. Anomalies had arisen in such experiments, some of which have since been shown to be a result of the three-quark structure of the proton, and Eddington was simply attempting to account for them. For instance he develops this extra energy term using a Dirac delta function such that it appears when the separation distance is zero (and he suggests that this implies the existence of a singularity, the only real mention of singularities in *Fundamental Theory*). Since the internal structure of the proton is primarily governed by the strong interaction he was also suggesting the existence of another internal interaction. In fact, in a way, the non-Coulombian energy in Eddington's theory is directly related to the strong interaction since it is this that binds the quarks together providing the extra mass. The determination of the various mass terms is then the primary goal of Eddington's statistical theory and, indeed, he calls (8.14) the "central formula of unified theory" (Eddington 1946, p. 81). In any case, the problem of the excess mass combined with the anomalies in the scattering experiments clearly indicated the existence of some unknown process and Eddington's attempt at an explanation is certainly as admirable as any at that time.

**Philosophical Implications**

The major philosophical theme behind *Fundamental Theory* is the search for a truly objective theory of physics. Eddington struggles mightily with this throughout, vacillating between truly attempting to develop one and realizing it is most likely impossible. The most obvious manifestation of this objectivity requirement is a measurement standard since, at the moment, all measurement standards are inherently subjective. Eddington concluded that for a standard to be truly objective it must be derivable from the theory itself and not imposed by an observer. Subsequent objective physical theories have appealed to hidden variables as one solution to this problem, a solution that immediately calls into questions the completeness of quantum mechanics as a description of nature. These questions were nothing new to Eddington since they go back to the work of von Neumann and others in the 1930s (though it should be noted Eddington would have had to have read von Neumann in German since his works were not translated into English until the 1940s). But Eddington does not appeal to a hidden variable scheme in locating a solution to this problem but rather attempts to develop an objective theory that is consistent with accepted quantum mechanics principles such as uncertainty and complementarity. For this very reason *Fundamental Theory* should be viewed with less derision than hidden variable theories which continue to be disproved.

In fact Eddington relied on the fundamental philosophical tenets of the founders of quantum theory employing such basic principles as the interference of observer and observed, something not as clearly defined in all hidden variable theories.

Eddington did, of course, introduce novel ideas, but they were nonetheless based on fundamental principles. In the case of the search for objectivity it is well understood that uncertainty plays a role in our inability to develop a truly objective theory. Eddington, however, took this a step further and placed an uncertainty in the fabric of space-time that was a *result* of the inability to separate the reference frame from the object under investigation. So, in a way, he applied this fundamental principle in reverse: rather than uncertainty being the root cause of subjectivity, it is our subjective viewpoint that actually leads to uncertainty in the fabric of space-time. In addition it is our subjectivity that introduces chirality into physics (rather than spontaneous symmetry breaking as it is now understood). He then took the unusual step of trying to determine the evolution of a system *after* any observation had ended. Uncertainty, in this case, increases the further from the end of the observation one gets. Since an observation can be modelled as a collapse of the wavefunction, this is essentially the reverse: the wavefunction spreads out and the number of possible eigenvalues increases. Again, Eddington is not introducing anything terribly novel but rather is simply working backwards from a point of knowledge to see how one might arrive at that piece of knowledge. In doing this he introduces a time coordinate that measures the time difference between the end of the observation and the moment in question. Statistically this must be added to the distribution function. The full description of this process gets bogged down in pedantic labelling and transformations between ordinary and 'imaginary' time since a distribution function is taken at a *point* in time but would include in it a *change* in time. Eddington may have unnecessarily complicated this process since one can resolve this by a simple comparison of distribution functions for the end of observation and the moment in question. Theoretically one could then create a third distribution function for the comparison of the first two and one would then have the answer.

Eddington never delves into greater depth with this problem clearly emphasizing the inherent difficulty in objective theories. The vast majority of *Fundamental Theory* is devoted to studies of actual observations where the observer clearly interferes and the wavefunction is collapsed. Any hope for a truly objective theory in this instance is in vain and perhaps this was Eddington's way of acknowledging the limitations of normal theories. Nonetheless, he is striving for a working and applicable theory rather than one that is untestable or speculative and thus he sticks to building a theory around experimental results, though one should notice on studying his methods more closely that he attempts to greatly simplify things (as many physicists do). But we know that experimental results are far from simple in their verification. For instance, he assumes in equations such as (8.19) that all protons and electrons are bound. This is likely just his way of representing the fact that the majority of the mass in the universe is made up of hydrogen, but it is a major simplification for experimental purposes particularly when one considers that if $N$ is really a count of hydrogen only, how could it have any particular usefulness in general equations used for free particles, larger atoms, and the like?

In addition to building a theory around experimental evidence he also builds a theory around the notion that everything can be reduced to mathematics. He clearly takes the stance that the nature of mathematics is such that it is inherent and in some way prior to the physical universe. In a way this is his route to objectivity. Since mathematics are inherent in the fabric of nature and, in fact, are independent of nature itself, a purely mathematical theory should be a purely objective theory. So in his view, a perfect theory would be purely mathematical and would yield numerical results consistent with experiment. Part of this desire for a purely objective theory based solely on mathematics is a result of his dissatisfaction with the rather arbitrary nature of our choice of just what type of observational evidence is acceptable as solid proof. Visual evidence is nearly always accepted as solid proof yet visual evidence is simply electromagnetic in nature (notice the strong influence of observational astronomy in Eddington's thinking here). He asks why gravitational evidence is not equally acceptable as solid proof. As I mentioned earlier the difference here is that in most cases we do not have direct sensory experience of gravitational evidence. Regardless, the choice of a certain type of evidentiary proof as the most acceptable is somewhat arbitrary. A purely mathematical theory would blur the distinction, providing numerical results for both types. On the other hand, in his attempt to link microscopic and macroscopic phenomena he indicates that electromagnetic evidence is the realm of the microscopic while gravitational evidence is the realm of the macroscopic and *both* are transmitted to us via light since most gravitational solutions are observed via light since we don't physically experience them. His most forceful example of this is the companion to Sirius that was inferred by gravitational effects but not 'proven' until it was discovered optically. Since Eddington had proven in 1919 that light reacted to gravitational fields he did not necessarily recognize the purely electromagnetic nature of light, though it is also fair to say photons were not recognized as the exchange particles for electromagnetism only, on par with gravitons, W and Z bosons, and gluons until after Eddington's death. It was obviously known that they were related to electromagnetism but the eclipse expeditions implied that there might be more to them than that.

**Eddington in Context**

In critiquing *Fundamental Theory* too many historians oddly seem to have forgotten some of the basic interpretations of quantum mechanics including those developed at the onset of the theory's development. One important conclusion that was first reached by Heisenberg was that quantum theory was not limited to statistical conclusions – it could make precise and accurate predictions of physical phenomena. This then relates to the second important conclusion that the inability to make certain valid predictions arose from the inability to know everything about the *present*. Famously this is the question of whether or not quantum theory is complete. But it also implies that uncertainty is *causal*. Since the universe is then completely causal Eddington tries to avoid the incomplete knowledge problem by developing a system that provides all fundamental constants in a derived manner. Eddington's methods were certainly a bit abstract and he is often overly pedantic, but his philosophical reasoning was sound and he did not deny any major foundational principle of physics (with the possible exception of the equivalence principle). It is rather curious, then, that he should receive the treatment he does in historical literature. Further, my analysis shows that much of his work was remarkably similar to modern quantum field theory and, though outdated and incorrect, leads to new questions such as the nature of exclusion and the relationship between particle density and field momentum. It is my hope that this monograph will serve as a partial vindication of Eddington's cosmonumerological reputation and a stimulation of further research.

## *Appendix A*

### *The Zoo Puzzle*

The Zoo Puzzle first appeared as 'The Looking-glass Zoo' in Hubert Phillips' 1937 book *Question Time: An Omnibus of Problems for a Brainy Day* published by J.M. Dent & Sons (London).  Phillips often wrote under the pseudonym Caliban and Eddington himself submitted this puzzle to Phillips.  Notice the reference to Lewis Carroll (as Charles Dodgson).  There are numerous solutions.  I present (verbatim) the solution that appears in the terrific collection on rec-puzzles.org.  This is an excellent introduction to Eddington's work in combinatorics and relates to some of his work in group theory.

**The Zoo Puzzle as posed by Eddington in Phillips (1937)**

*I took some nephews and nieces to the Zoo, and we halted at a cage marked*

*Tovus Slithius, male and female.*
*Beregovus Mimsius, male and female.*
*Rathus Momus, male and female.*
*Jabberwockius Vulgaris, male and female.*

*The eight animals were asleep in a row, and the children began to guess which was which. "That one at the end is Mr Tove." "No, no!  It's Mrs Jabberwock," and so on.  I suggested that they should each write down the names in order from left to right, and offered a prize to the one who got most names right.*

*As the four species were easily distinguished, no mistake would arise in pairing the animals; naturally a child who identified one animal as Mr Tove identified the other animal of the same species as Mrs Tove.*

*The keeper, who consented to judge the lists, scrutinised them carefully.  "Here's a queer thing.  I take two of the lists, say, John's and Mary's.  The animal which John supposes to*

*be the animal which Mary supposes to be Mr Tove is the animal which Mary supposes to be the animal which John supposes to be Mrs Tove. It is just the same for every pair of lists, and for all four species.*

*"Curiouser and curiouser! Each boy supposes Mr Tove to be the animal which he supposes to be Mr Tove; but each girl supposes Mr Tove to be the animal which she supposes to be Mrs Tove. And similarly for the other animals. I mean, for instance, that the animal Mary calls Mr Tove is really Mrs Rathe, but the animal she calls Mrs Rathe is really Mrs Tove."*

*"It seems a little involved," I said, "but I suppose it is a remarkable coincidence." "Very remarkable," replied Mr Dodgson (whom I had supposed to be the keeper) "and it could not have happened if you had brought any more children."*

*How many nephews and nieces were there? Was the winner a boy or a girl? And how many names did the winner get right?*

**The Answer (rec-puzzles.org)**

*Given that there is at least one boy and one girl (John and Mary are mentioned) then the answer is that there were 3 nephews and 2 nieces, the winner was a boy who got 4 right.*

**Detailed Solution (rec-puzzles.org)**

*Number the animals 1 through 8, such that the females are even and the males are odd, with members of the same species consecutive; i.e. 1 is Mr. Tove, 2 Mrs. Tove, etc.*

*Then each childs (sic) guesses can be represented by a permutation. I use the standard notation of a permutation as a set of orbits. For example: (1 3 5)(6 8) means 1 -> 3, 3 -> 5, 5 -> 1, 6 -> 8, 8 -> 6 and 2,4,7 are unchanged.*
    1.    *Let P be any childs (sic) guesses. Then P(mate(i)) = mate(P(i)).*

2. *If Q is another childs (sic) guesses, then [P,Q] = T, where [P,Q] is the commutator (sic) of P and Q (P composed with Q composed with P inverse composed with Q inverse) and T is the special permutation (1 2) (3 4) (5 6) (7 8) that just swaps each animal with its spouse.*
3. *If P represents a boy, then P*P = I (I use * for composition, and I for the identity permutation: (1)(2)(3)(4)(5)(6)(7)(8)*
4. *If P represents a girl, then P*P = T.*

*(1) and (4) together mean that all girl's guesses must be of the form: (A B C D) (E F G H) where A and C are mates, as are B & D, E & F G & H.*

*So without loss of generality let Mary = (1 3 2 4) (5 7 6 8) Without to (sic) much effort we see that the only possibilities for other girls "compatible" with Mary (I use compatible to mean the relation expressed in (2)) are:*

```
g1:   (1 5 2 6) (3 8 4 7)
g2:   (1 6 2 5) (3 7 4 8)
g3:   (1 7 2 8) (3 5 4 6)
g4:   (1 8 2 7) (3 6 4 5)
```

*Note that g1 is incompatible with g2 and g3 is incompatible with g4. Thus no 4 of Mary and g1-4 are mutually compatible. Thus there are at most three girls: Mary, g1 and g3 (without loss of generality) By (1) and (3), each boy must be represented as a product of transpostions (sic) and/or singletons: e.g. (1 3) (2 4) (5) (6) (7) (8) or (1) (2) (3 4) (5 8) (6 7).*

*Let J represent John's guesses and consider J(1). If J(1) = 1, then J(2) = 2 (by [1]) using [2] and Mary J(3) = 4, J(4) = 3, and g1 & J => J(5) = 6, J(6) = 5, & g3 & J => J(8) = 7 J(7) = 8 i.e. J = (1)(2)(3 4)(5 6)(7 8). But the [J,Mary] <> T. In fact, we can see that J must have no fixed points, J(i) <> i for all i, since there is nothing special about i = 1.*

*If J(1) = 2, then we get from Mary that J(3) = 3. contradiction.*

If J(1) = 3, then J(2) = 4, J(3) = 1, J(4) = 2 (from Mary) => J(5) = 7, J(6) = 8, J(7) = 5, J(8) = 6 => J = (1 3)(2 4)(5 7)(6 8)  (from g1)

But then J is incompatible with g3.

A similar analysis shows that J(1) cannot be 4,5,6,7 or 8; i.e. no J can be compatible with all three girls. So without loss of generality, throw away g3.

```
We have Mary = (1 3 2 4) (5 7 6 8)
        g1   = (1 5 2 6) (3 8 4 7)
```

The following are the only possible boy guesses which are compatible with both of these:

```
  B1: (1)(2)(3 4)(5 6)(7)(8)
  B2: (1 2)(3)(4)(5)(6)(7 8)
  B3: (1 3)(2 4)(5 7)(6 8)
  B4: (1 4)(2 3)(5 8)(6 7)
  B5: (1 5)(2 6)(3 8)(4 7)
  B6: (1 6)(2 5)(3 7)(4 8)
```

Note that B1 & B2 are incombatible (sic), as are B3 & B4, B5 & B6, so at most three of them are mutually compatible. In fact, Mary, g1, B1, B3 and B5 are all mutually compatible (as are all the other possibilities you can get by choosing either B1 or B2, B3 or B4, B5 or B6. So if there are 2 girls there can be 3 boys, but no more, and we have already eliminated the case of 3 girls and 1 boy.

The only other possibility to consider is whether there can be 4 or more boys and 1 girl. Suppose there are Mary and 4 boys. Each boy must map 1 to a different digit or they would not be mutually compatible. For example if b1 and b2 both map 1 to 3, then they both map 3 to 1 (since a boy's map consists of transpositions), so both b1*b2 and b2*b1 *map 1 to 1. Furthermore, b1 and b2 cannot map 1 onto spouses. For example, if b1(1) = a and b is the spouse of a, then b1(2) = b. If b2(1) = b, then b2(2) = a. Then b1*b2(1) = b1(b) = 2 and b2*b1(1) = b2(a) = 2 (again using the fact that boys are all transpositions(sic)). Thus the four boys must be:*

```
B1: (1)(2)...    or (1 2)....
B2: (1 3)...    or (1 4) ...
B3: (1 5) ...    or (1 6) ...
B4: (1 7) ...    or (1 8) ...
```

*Consider B4. The only permutation of the form (1 7)... which is compatible with Mary ( (1 3 2 4) (5 7 6 8) ) is:*

```
    (1 7)(2 8)(3 5)(4 6)
```

*The only (1 8)... possibility is:*

```
    (1 8)(2 7)(3 6)(4 5)
```

*Suppose B4 = (1 7)(2 8)(3 5)(4 6)*

*If B3 starts (1 5), it must be (1 5)(2 6)(3 8)(4 7) to be compatible with B4. This is compatible with Mary also.*

*Assuming this and B2 starts with (1 3) we get B2 = (1 3)(2 4)(5 8)(6 7) in order to be compatible with B4. But then B2*B3 and B3*B2 moth map 1 to 8. I.e. no B2 is mutually compatible with B3 & B4.*

*Similarly if B2 starts with (1 4) it must be (1 4)(2 3)(5 7)(6 8) to work with B4, but this doesn't work with B3.*

*Likewise B3 starting with (1 6) leads to no possible B2 and the identical reasoning eliminates B4 = (1 8)...*

*So no B4 is possible! i.e. at most 3 boys are mutually compatiblw (sic) with Mary, so 2 girls & 3 boys is optimal. Thus:*

```
Mary = (1 3 2 4) (5 7 6 8)
Sue  = (1 5 2 6) (3 8 4 7)
John = (1)(2)(3 4)(5 6)(7)(8)
Bob  = (1 3)(2 4)(5 7)(6 8)
Jim  = (1 5)(2 6)(3 8)(4 7)
```

*is one optimal solution, with the winner being John (4 right: 1 2 7 & 8).*

# *Appendix B*

## *The Burying Ground at St. Giles*

**A Most Unusual Cemetery**

Certainly Britain has its famed cemeteries, burial grounds, and catacombs with Westminster Abbey topping the list. But there exists a curious little cemetery in Cambridge that, for its small size, has what seems to be a disproportionate number of "Cambridge's greatest talents" (Goldie 2000). To quote from Mark Goldie's *A Cambridge Necropolis*,

> There you will find the graves of the physicist who split the atom, the biochemist who discovered vitamins, the astronomer who discovered Neptune, the anthropologist who explored the roots of religion, the architect who designed more of the university's buildings than did any other, a son and granddaughter of Charles Darwin who made their own marks as a scientific instrument-maker and a poet, and two of the most important philosophers of the twentieth century.
>
> Here lie two Nobel Prize-winners, seven members of the Order of Merit, eight masters of colleges, fifteen knights of the realm, and thirty-nine people who appear in the *Dictionary of National Biography* (Goldie 2000, p. 3).

Remarkably these graves lie on only one-and-a-half acres currently known as Ascension Cemetery and formerly known as the St. Giles Burial Ground. There are nearly 2500 people buried there in 1500 plots and they represent numerous religions (as well as none).

I had the good fortune to visit there during the workshop *Arthur Eddington: Interdisciplinary Perspectives* that took place in March of 2004. Eddington himself is buried at Ascension as are several people who play peripheral roles in this monograph. I have already included photographs of several headstones but wish to include photographs of a few others representing people who had some connection to Eddington, astronomy, physics, philosophy, etc. I also include the full list given by Goldie (2000) but refer readers to that work for maps, directions, and biographical sketches.

**The Photographs**

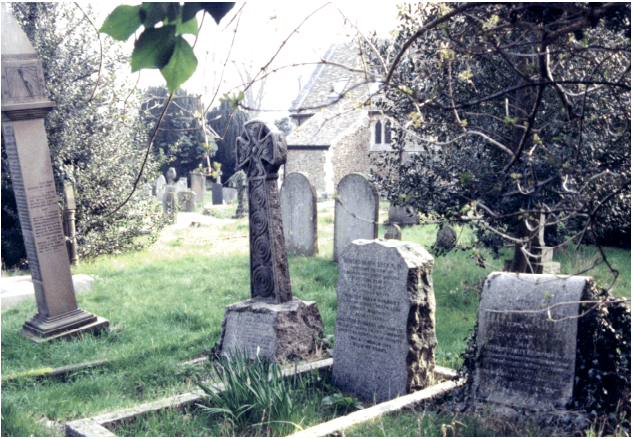 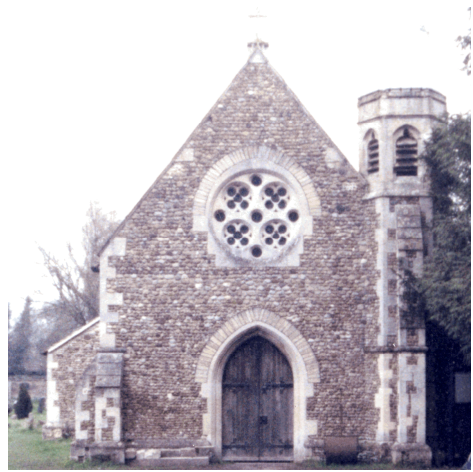

**Figure B1.** Ascension Cemetery with St. Giles Chapel in the background and the Eddington family plot at bottom Right.

**Figure B2.** St. Giles Chapel, now privately owned.

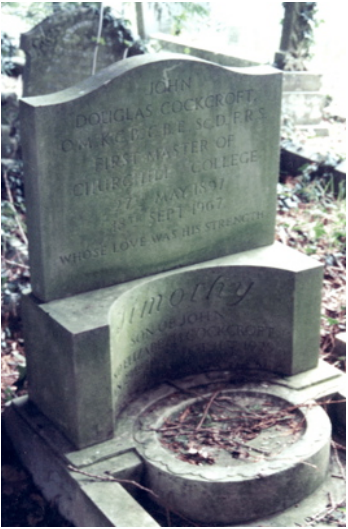 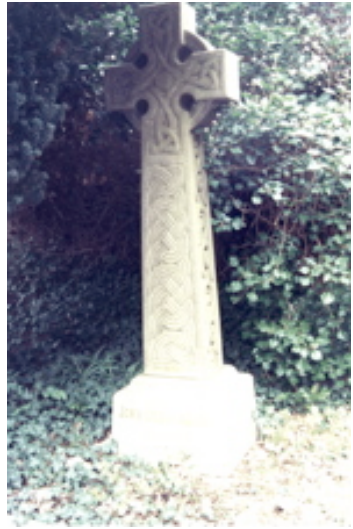 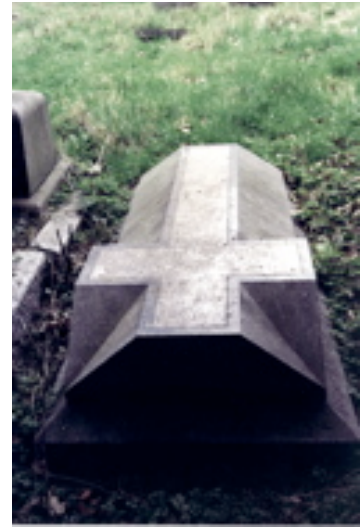

**Figure B3.** A trio of Cambridge luminaries, from left to right: Nobel laureate physicist Sir John Cockcroft (1897 – 1967), former Cambridge Observatory director and discoverer of Neptune John Couch Adams (1819 – 1892), and mathematician and historian W.W. (Walter William) Rouse Ball (1850 – 1925).

**The List**

This is the list of luminaries given by Goldie (2000). It is always difficult to define greatness so perhaps it is better to say that the following people are recognizably public figures in their fields or in the Cambridge University community. Nonetheless, it attests to the phenomenal influence of Cambridge in British Society.

Adams, John Couch (1819 – 1892), see Figure B3 above.
Anderson, Sir Hugh (1865 – 1928), physiologist, Master of Caius.
Appleton, Rev. Richard (1849 – 1909), Master of Selwyn.
Ball, Sir Robert, FRS (1840 – 1913), see Figure 3 (in Chapter II).
Benson, Arthur (1866 – 1925), Master of Magdelene.
Bethune-Baker, James, FBA (1861 – 1951), theologian.
Brink, Charles, FBA (1906 – 1994), classicist.
Brogan, Sir Denis, FBA (1900 – 1974), historian, political scientist.
Brooke, Zachary, FBA (1883 – 1946), historian.
Bushnell, Geoffrey, FBA (1903 – 1978), archaeologist.
Clark, Sir William, KCB (1876 – 1952), civil servant.
Cockcroft, Sir John, OM, FRS (1897 – 1967), see Figure B3 above.
Cornford, Frances (1886 – 1960), poet.
Darwin, Sir Francis, FRS (1848 – 1925), botanist, biographer.
Darwin, Sir Horace, FRS (1851 – 1928), scientific instrument maker.
Darwin, Ida (1854 – 1946), mental health pioneer.
Eddington, Sir Arthur, OM, FRS (1882 – 1944), the subject of this book!
Frazer, Sir James, OM, FBA, FRS (1854 – 1941), anthropologist.
Gwatkin, Henry (1844 – 1916), historian, theologian, conchologist.
Hopkins, Sir Frederick, OM, FRS (1861 – 1947), biochemist, Nobel laureate.
Hopkinson, Bertram, FRS (1874 – 1918), engineer.
Hutchinson, Arthur, FRS (1866 – 1937), mineralogist.
Jackson, Henry, OM (1839 – 1921), classicist.
Jebb, Sir Richard, OM, FBA, MP (1841 – 1905), classicist.
Kenny, Courtney Stanhope, FBA, MP (1847 – 1930), legal scholar.
Lamb, Sir Horace, FRS (1849 – 1934), see Figure 1 (in Chapter II).
Lubbock, Hugh Roger (1951 – 1981), cell biologist.
MacAlister, Sir Donald (1854 – 1934), Vice-Chancellor of Glasgow.
McCarthy, Sir Desmond (1877 – 1952), literary and drama critic.
McLean, Norman, FBA (1865 – 1947), orientalist, Master of Christ's.
Marshall, Alfred, FBA (1842 – 1924), economist.
Mayor, John (1825 – 1910), antiquarian.
Moore, G.E., OM, FBA (1873 – 1958), philosopher.
Newall, Hugh, FRS (1857 – 1944), astrophysicist.
Newton, Alfred, FRS (1829 – 1907), ornithologist.
Ramsey, Frank (1903 – 1930), philosopher and mathematician.
Roberts, David (1911 – 1982), architect.
Rouse Ball, Walter William (1858 – 1925), see Figure B3 above.
Sandys, Sir John (1844 – 1922), classicist and orator.

Scott, Charlotte (1853 – 1921), pioneer female student.
Selwyn, Rt. Rev. John (1844 – 1898), Bishop, Master of Selwyn.
Skeat, Walter (1835 – 1912), philologist, Anglo-Saxonist.
Spufford, Bridget (1967 – 1989), namesake of Bridget's hostel.
Stern, Peter, FBA (1920 – 1991), Germanist.
Taylor, Henry, FRS (1842 – 1927), mathematician, Braille expert.
Verrall, Arthur (1851 – 1912), classicist and literary scholar.
Wisdom, John (1904 – 1993), philosopher (how great a name is *that*?).
Wittgenstein, Ludwig (1889 – 1951), see Figure 6 (in Chapter IV).
Wood, Charles (1866 – 1926), composer.
Wright, William (1831 – 1914), Shakespearean and biblical scholar.

# *Appendix C*

## *A Dialogue Concerning the Nature of Exclusion and its Relation to Force*

I present the following argument that the exclusion principle is *not* equivalent to any of the four known fundamental interactions via a dialogue. The dialogue in scientific writing has been a lost art for many years. I present it in this form since that is the form it took in my head as I was working out the details. I find I often play 'devil's advocate' to myself in my own 'internal' dialogues concerning my research. I give the role of the scientist to Peter Higgs of the University of Edinburgh who first formulated the idea of the Higgs mechanism as the source of mass in the Standard Model of particle physics (Higgs 1963). My reasons for this choice should be apparent by the end. The 'advocate,' as it were, remains … the 'advocate.'

*HIGGS*: You will agree, will you not that the four fundamental interactions are gravity, electromagnetism, the weak interaction, and the strong interaction?

*ADVOCATE*: Indeed, that is clearly understood.

*HIGGS:* Would you also agree that exclusion would be repulsive if it were an interaction?

*ADVOCATE*: Yes, that does appear to be its nature – exclusion, repulsion – minor semantics, really.

*HIGGS*: Well then, since gravity and the strong interaction are attractive we can immediately rule them out (thus blowing a hole in Eddington's *Fundamental Theory*).

*ADVOCATE*: Yes, well Eddington wasn't *really* a numerologist anyway – none of this prognostication like that fellow Nostradamus. I like prognosticators.

*HIGGS*: No numbers necessary here, at least until the end. Anyway, that leaves us with the weak and electromagnetic interactions.

*ADVOCATE*: Which are really one-and-the-same according to those GWS fellows.

*HIGGS*: You mean Sheldon, Steven, and Abdus? Yes, well at normal energies those two interactions at least look different.

*ADVOCATE*: Indeed, but I think at least one of them can be eliminated.

*HIGGS*: Yes, well I was getting to that. I suppose you were thinking of our rather elusive friend the neutrino.

*ADVOCATE*: Yes, he is a bit of a bother; about the only thing that regularly visits our toasty little haven 'beneath' and heaven knows we like our privacy (or at least its occupants do).

*HIGGS*: Precisely. That is because neutrinos only participate in the weak interaction! But, being leptons, they *do* obey the exclusion principle!

*ADVOCATE*: Putting the final nail into Eddington's coffin I suppose.

*HIGGS*: I'm ignoring your puns, by the way.

*ADVOCATE*: Yes, most people do, but it's not the puns that get them…

*HIGGS*: Yes, well, as I was saying at the most fundamental level, all quarks *and* leptons interact weakly *and* obey the exclusion principle. It seems that the weak interaction is the only universally common attribute of particles with half-integer spin.

*ADVOCATE*: Lest you move too hastily in your assumptions, recall that both exclusion and the weak interaction have antisymmetric properties.

*HIGGS*: Oh, yes, nearly forgot that. Indeed the weak interaction Lagrangian is chiral while excluding wavefunctions are antisymmetric. But, if we build composite particles out of quarks and antiquarks we find one family of composite quark particles – mesons – participate in weak interactions but do not exclude! That should do it, eh?

*ADVOCATE*: If you say so. Personally I find it logically unsatisfying since someone is bound to argue that the nature of mesons is a bit different since they contain one quark and one antiquark whereas other composite particles generally contain three such fundamental particles (or five in the case of the newly discovered pentaquarks). In addition one might argue that composite particles really don't count since we're interested in the most fundamental nature of exclusion (i.e. mesons may not exclude as a whole but their constituents surely do and it is these constituents that ultimately interact weakly with each other).

*HIGGS*: Ah, good point.

*ADVOCATE*: I have them occasionally.

*HIGGS*: So perhaps the nature of the bosonic exchange is different inside a meson. Well, if that is the case then let us look at it another way. Exclusion is clearly related to spin which can be related to charge via magnetic moment, but neutrinos exclude yet have no charge (as do neutrons but based on our current line of reasoning their constituents are what we care about).

*ADVOCATE*:  Tut-tut.  Think outside the box, man.  What is charge anyway but a measure of how a photon propagates?  It is reasonable to assume that spin simply tells photons what to do *if they are present* but their presence is, by no means, required.  Don't get bogged down in classical descriptions.  Recall that the Wigner (spin conservation) rule says that for any transition, be it radiative or radiationless, transitions between terms that have the same multiplicity are 'spin-allowed' while those between terms that have different multiplicity are 'spin-forbidden.'  Exclusion depends solely on the spin quantum number.

*HIGGS*:  Hmmm … I guess we'll have to think some more about the weak interaction.  Approaching this, then, from the standpoint of the Standard Model, the weak interaction exchanges *W* and *Z* bosons.

*ADVOCATE*:  So exclusion would have to be equivalent in energy to these bosons.

*HIGGS*:  But the *W* boson exchanges electrical charge.  Since exclusion does not affect charge the only boson left is the *Z* that transfers nothing but momentum (all its quantum numbers are zero since it represents a neutral current interaction and is its own antiparticle).  This should immediately close the door on this line of reasoning since spin is a quantum number.

*ADVOCATE*:  A truly sharp mind would ask the following question: what if we suppose that the *Z carries* quantum numbers (not its own, since it has none) from one *identical* particle to another?  Essentially two identical particles interacting weakly would constantly exchange *Z* bosons meaning they would be constantly exchanging the same information.  This would mean, for instance, that an electron pair with opposite spin could coexist in the same orbital because they cannot exchange a *Z* boson since they do not have identical quantum numbers.  *Z* bosons *only* propagate between particles with *identical* quantum numbers.  Momentum is conserved in the transfer since each of the two interacting particles emits (and then receives) a *Z* at the same time as the other.  This appears to be a realistic model for exclusion as a neutral current interaction.

*HIGGS*:  Interesting, but it must be subjected to a test.

*ADVOCATE*: You scientists and your tests. Can't anything be accepted on mere faith anymore? I'm clearly in the wrong business these days.

*HIGGS*: Imagine that an electron was attempting to reach the lowest orbital in a hydrogen atom and that it had quantum numbers that were identical to the electron already occupying that orbital. If this invading electron had enough energy to overcome electromagnetic repulsion it still could not get any closer than the second lowest orbital due to quantisation and exclusion (recall it has the same spin as the electron that is already there). As such the *Z* boson, if it represented exclusion, would have to transmit an equivalent amount of energy in the given distance between the two orbitals. But in very simple atoms the distance between orbitals can be on the order of $10^{-10}$ *m* while the *Z* can only have a range of up to $10^{-17}$ *m* which is the maximum range of the weak interaction. To put it another way, if exclusion is represented by the *Z* boson then atoms would be considerably smaller than they are. For exclusion to be represented by any boson at all that boson would have to be less massive than the *Z* as given by (5.21).

*ADVOCATE*: What in the Devil's name is (5.21)? Is that some Biblical reference again? Please say otherwise. I never win any arguments with Him.

*HIGGS*: Never mind.

*ADVOCATE*: Well, look here my dear fellow, your argument rules out your very own mechanism as well since the Higgs particle is theoretically more massive than the *Z*. One should always strive to maximize the applicability of one's own theories.

*HIGGS*: Ha! You really *don't* understand scientists do you? Vanity is often cured by a good puzzle. So, there, I've proved it! Exclusion is *not* represented by any of the known fundamental interactions or forces, if you will!

*ADVOCATE*: But, as Margenau has argued, it clearly displays quasi-force behaviour in some theoretical situations. Since charge is a measure of how photons propagate and mass is a measure of how gravitons propagate, one would expect spin to be a measure of how some third type of boson propagates. But we have ruled out gluons, Higgs particles, and even *W* and *Z* bosons despite the very attractive similarities between exclusion and the weak interaction (antisymmetric nature, rules for identical particles, etc.).

*HIGGS*:  It simply means that the nature of the exclusion principle may be on par with the nature of the conservation laws that are immutable rules of design in the universe with no satisfactory explanation outside of symmetry.

*ADVOCATE*:  Yes, well, regardless, it is clearly a more complex phenomenon than one might expect.

*HIGGS*:  Again you underestimate us scientists.  That's a reason to celebrate – it's another puzzle to solve!

French, S.R.  2004.  "Patterns of Interweaving: Structuralism for the Quantum Age," in K. Price, ed., pre-print proceedings of the Cambridge University workshop *Arthur Stanley Eddington: Interdisciplinary Perspectives*.

Fry, A.R.  1926.  *A Quaker Adventure*.  London: Nisbet.

Gamow, G.  1930.  *PRSLA* **126**: 632.

Gasiorowicz, S.  1996.  *Quantum Physics*.  New York: John Wiley & Sons.

Ghirardi, G.C., A. Rimini, and T. Weber.  1986.  *Physical Review D* **34**: 470.

Giese, A.  2000.  First presented at the spring conference of the German Physical Society and reported at http://www.ag-physics.de/rmass.

Goldie, M.  2000.  *A Cambridge Necrology*.  Unpublished, copyrighted guide distributed at the Cambridge workshop *Arthur Eddington: Interdisciplinary Perspectives*.

Goldstein, S.  1980.  *Classical Mechanics*, 2$^{nd}$ *Ed*.  Reading: Addison-Wesley.

Heisenberg, W.  1927.  *Zeitschrift für Physik* **43**: 172.

Heisenberg, W.  1960.  "Erinnerungen an die Zeit der Entwick-lung der Quantenmechanik," in M. Fierz and V.F. Weisskopf, eds., *Theoretical Physics in the Twentieth Century*.  New York: Interscience.

Higgs, P. W. 1964.  *Physical Review Letters* **12**: 132.

Jammer, M.  1966.  *The Conceptual Development of Quantum Mechanics*.  New York: McGraw-Hill.

Kaluza, T.  1921.  *Sitzungsberichte der Preussische Akademie der Wissenschaften* **1921**: 966.

Kapteyn, J.C.  1915.  Letter to G.E. Hale (23 September).  Hale Papers.

Kilmister, C.W. and B.O.J. Tupper.  1962.  *Eddington's Statistical Theory*.  Oxford: Clarendon.

Kilmister, C.W.  1994.  *The Fundamental Theory: Eddington's search for meaning*.  Cambridge: Cambridge University Press.

Klein, O.  1926.  *Nature* **118**: 516.

Kragh, H.  1999.  *Quantum Generations: A History of Physics in the Twentieth Century*.  Princeton: Princeton University Press.

Liboff, R.L.  1998.  *Introductory Quantum Mechanics*.  Reading: Addison Wesley.

Linklater, A.  2002.  *Measuring America: How an Untamed Wilderness Shaped the*
286